\documentclass[
	paper = a4,
	twoside,
	fontsize = 11pt,
	numbers = noendperiod,
	headings = small,
	headsepline,
	open = right,
	cleardoublepage = empty,
	chapterprefix=false,
	draft = false,
	final = true
]{scrreprt}

\usepackage[british]{babel}
\ifdefined\directlua
	\usepackage{fontspec}
\else
	\pdfoutput=1
	\usepackage[T1]{fontenc}
	\usepackage[utf8]{inputenc}
	\usepackage[nomath]{lmodern}
	\usepackage{textcomp}
\fi

\usepackage[a4paper, width=460pt, height=660pt, footskip=40pt, centering]{geometry}

\usepackage[table, hyperref]{xcolor}
\usepackage{graphicx}

\usepackage{scrhack}
\RequirePackage{xspace}

\usepackage[centertags, sumlimits, nointlimits, namelimits]{amsmath}
\usepackage[fixamsmath, disallowspaces]{mathtools}
\usepackage{siunitx}
\usepackage{amsfonts, amssymb, dsfont, slashed}
\usepackage[makeroom]{cancel}

\usepackage{array, booktabs, multirow}

\usepackage[xspace]{ellipsis}
\usepackage{url}
\usepackage{setspace}
	\renewcommand{\arraystretch}{1.213}
\usepackage[bottom, stable, multiple]{footmisc}
\usepackage[autostyle=true, english=british]{csquotes}
\usepackage[title, titletoc]{appendix}

\usepackage[
	style=numeric-comp,
	natbib=true,
	sorting=none,
	backend=biber,
	bibwarn=true,
	texencoding=auto,
	bibencoding=auto,
	isbn=false,
	doi=false,
	hyperref=true,
	url=false,
	maxnames=8,
	minnames=1,
	giveninits=true,
	eprint=true,
	arxiv=abs
]{biblatex}
	\bibliography{references}

\usepackage{floatrow}
\usepackage{caption}
\usepackage[labelfont=rm]{subcaption}

\usepackage[headsepline, automark, pagestyleset=KOMA-Script, markcase=ignoreuppercase]{scrlayer-scrpage}
	\automark[chapter]{chapter}
	\automark*[section]{}

\usepackage[babel=true, draft=false, final=true, verbose=true]{microtype}
	\AtBeginEnvironment{verbatim}{\microtypesetup{activate=false}}

\usepackage[hyperfootnotes=false, bookmarks=true, pdfpagelabels=true]{hyperref}
	\hypersetup{
		linktocpage = true,
		colorlinks = true,
		linkcolor = DarkBlue,
		citecolor = DarkBlue,
		urlcolor = DarkBlue,
		pdfstartview = Fit,
		bookmarksopen = false,
		breaklinks = true,
		pdfpagelayout = TwoColumnRight,
		pdfcreator = \LaTeX{},
		pdfauthor = {Benjamin~Wallisch},
		pdftitle = {Cosmological Probes of Light Relics -- PhD Thesis of Benjamin Wallisch},
		pdfsubject = {Cosmological Probes of Light Relics},
		pdfkeywords = {Benjamin~Wallisch,~PhD~Thesis,~Cosmological~Probes~of~Light~Relics}
	}
	\ifdefined\directlua
		\hypersetup{
			pdfproducer = LuaLaTeX,
		}
	\else
		\hypersetup{
			pdfproducer = pdfLaTeX,
		}
	\fi
\usepackage[depth=2]{bookmark}

\usepackage{calc, ifthen, xparse, xpatch}

\DeclareDocumentCommand\d{}{\operatorname{d}\!}
\newcommand{\ii}{\mathrm{i}}
\newcommand{\e}{\mathrm{e}}
\newcommand{\const}{\mathrm{const}}
\def\H{\mathcal{H}}
\def\k{{\boldsymbol{k}}}
\def\n{{\boldsymbol{n}}}
\def\p{{\boldsymbol{p}}}
\def\q{{\boldsymbol{q}}}
\def\vecv{{\boldsymbol{v}}}
\def\x{{\boldsymbol{x}}}
\def\Nf{N_\mathrm{eff}}
\def\Nn{N_\mathrm{fluid}}
\def\c{c_s}
\def\rec{\mathrm{rec}}

\def\As{A_\mathrm{s}}
\def\ns{n_\mathrm{s}}
\def\Neff{{N_\mathrm{eff}}}
\def\NeffLSS{\tilde{N}_\mathrm{eff}}
\def\thetasLSS{\tilde{\theta}_s}
\def\aeq{a_\mathrm{eq}}
\def\Tr{T_R}
\def\Tf{T_F}
\def\Tff{T_{\tilde F}}
\def\Tfl{T_\mathrm{fluid}}
\def\Trec{T_\rec}

\def\Teq{T_\mathrm{eq}}
\def\Td{T_D}
\def\mt{m_t}
\def\Mp{M_\mathrm{pl}}
\def\L{\mathcal{L}}
\def\M{\mathcal{M}}
\def\O{\mathcal{O}}

\def\Pnw{P^\mathrm{nw}}
\def\Pw{P^\mathrm{w}}
\def\Pobs{P_g}
\def\Oobs{O_g}

\def\fsky{{f_\mathrm{sky}}}
\def\lmin{\ell_\mathrm{min}}
\def\lmax{\ell_\mathrm{max}}
\def\kmin{k_\mathrm{min}}
\def\kmax{k_\mathrm{max}}
\def\zmax{z_\mathrm{max}}

\def\beq{\begin{equation}}
\def\eeq{\end{equation}}

\DeclareSIUnit{\parsec}{pc}
\DeclareSIUnit{\Mpc}{\mega\parsec}
\DeclareSIUnit{\Gpc}{\giga\parsec}
\DeclareSIUnit{\h}{\mathit{h}}
\DeclareSIUnit{\hPerMpc}{\h\per\Mpc}
\DeclareSIUnit{\MpcPerh}{\per\h\Mpc}
\DeclareSIUnit{\muKelvin}{\mu\kelvin}

\definecolor{DarkBlue}{rgb}{0.0,0.0,0.4}
\definecolor{Blue}{rgb}{0.25, 0.41, 0.88}
\definecolor{Red}{rgb}{0.92, 0., 0.}
\definecolor{DarkOrange}{rgb}{1.0, 0.549, 0.}
\definecolor{Purple}{rgb}{0.5, 0., 0.5}
\definecolor{darkgreen}{rgb}{0.075,0.302,0.047}
\definecolor{pyBlue}{RGB}{31, 119, 180}
\definecolor{pyRed}{RGB}{214, 39, 40}
\definecolor{pyGreen}{RGB}{44, 160, 44}
\definecolor{pyGray}{rgb}{0.5, 0.5, 0.5}
\definecolor{pyLightBlue}{RGB}{30, 144, 255}
\definecolor{pyDarkBlue}{RGB}{0, 0, 128}
\definecolor{pyBlue2}{RGB}{0, 111, 237}
\definecolor{pyRed2}{RGB}{224, 52, 36}
\definecolor{gray80}{gray}{0.8}
\definecolor{gray60}{gray}{0.6}

\newenvironment{myappendix}{
								\addtocontents{toc}{\protect\setcounter{tocdepth}{1}}
								\begin{appendices}
									\bookmarksetupnext{level=-1}
									\pdfbookmark[chapter]{Appendices}{appendices}
							}{
								\end{appendices}
								\addtocontents{toc}{\protect\setcounter{tocdepth}{0}}
								\bookmarksetupnext{level=-1}
							}

\setkomafont{chapter}{\normalfont\LARGE\scshape}
\addtokomafont{chapterprefix}{\usekomafont{chapter}}
\addtokomafont{chapterentry}{\normalfont\bfseries}
\addtokomafont{section}{\normalfont\bfseries}
\addtokomafont{subsection}{\normalfont\bfseries}
\addtokomafont{subsubsection}{\normalfont\bfseries}
\addtokomafont{paragraph}{\normalfont\bfseries}
\addtokomafont{subparagraph}{\normalfont\bfseries}
\addtokomafont{pagehead}{\itshape}

\renewcommand*{\chapterformat}{%
	\mbox{\chapappifchapterprefix{\nobreakspace}\thechapter
		\autodot\IfUsePrefixLine{}{}}}

\DeclareRobustCommand{\SkipTocEntry}[4]{}
\allowdisplaybreaks[0]
\numberwithin{equation}{chapter}
\numberwithin{figure}{chapter}
\numberwithin{table}{chapter}

\makeatletter
\renewcommand{\@pnumwidth}{3em} 
\renewcommand{\@tocrmarg}{4em}
\makeatother

\AtEveryBibitem{
  \clearfield{booktitle}
  \clearfield{issue}
  \clearfield{month}
  \clearfield{number}
  \clearfield{pagetotal}
}

\DeclareFieldFormat{pages}{\mkfirstpage[]{#1}}
\renewbibmacro{note+pages}{%
  \printfield{pages}%
  \setunit{\addspace}%
  \printfield{note}%
  \newunit
}

\DeclareFieldFormat[article,thesis,inproceedings,incollection]{title}{\mkbibemph{#1}}
\DeclareFieldFormat[article]{volume}{\mkbibbold{#1}\space}
\DeclareFieldFormat[inproceedings]{date}{\iffieldundef{journaltitle}{}{#1}}

\DeclareFieldFormat{doilink}{%
	\iffieldundef{doi}{%
		\iffieldundef{url}{%
			#1%
		}{%
			\href{\thefield{url}}{#1}%
		}
	}{%
		\href{http://dx.doi.org/\thefield{doi}}{#1}%
	}%
}
\xpatchbibdriver{article}
  {\usebibmacro{journal+issuetitle}%
   \newunit
   \usebibmacro{byeditor+others}%
   \newunit
   \usebibmacro{note+pages}}
  {\printtext[doilink]{%
     \usebibmacro{journal+issuetitle}%
     \newunit
     \usebibmacro{byeditor+others}%
     \newunit\nopunct
     \usebibmacro{note+pages}}}
  {}{}

\DeclareSourcemap{
 \maps[datatype=bibtex,overwrite=true]{
  \map{
    \step[fieldsource=Collaboration, final=true]
    \step[fieldset=usera, origfieldval, final=true]
  }
 }
}
\renewbibmacro*{author}{\iffieldundef{usera}{\printnames{author}}{\printnames{author} (\printfield{usera} Collaboration)}}

\xpatchbibmacro{name:andothers}{%
	\bibstring{andothers}%
	}{%
	\bibstring[\emph]{andothers}%
	}{}{}

\renewbibmacro{in:}{}

\makeatletter
\DeclareFieldFormat{eprint:arxiv}{%
	\ifhyperref{%
		\ifthenelse{\equal{#1}{1610.02743}}{%
			\href{http://arxiv.org/\abx@arxivpath/#1}{%
				\nolinkurl{arXiv:161}\textbf{\nolinkurl{0.027}}\nolinkurl{43}%
				\iffieldundef{eprintclass}{%
				}{%
				\addspace\mkbibbrackets{\thefield{eprintclass}}%
			}%
		}%
			}{%
			\href{http://arxiv.org/\abx@arxivpath/#1}{%
				\nolinkurl{arXiv:#1}%
				\iffieldundef{eprintclass}{%
					}{%
					\addspace\mkbibbrackets{\thefield{eprintclass}}%
					}%
				}%
			}%
		}{
		\ifthenelse{\equal{#1}{1610.02743}}{%
				\nolinkurl{arXiv:161}\textbf{\nolinkurl{0.027}}\nolinkurl{43}%
				\iffieldundef{eprintclass}{}{\addspace\mkbibbrackets{\thefield{eprintclass}}}%
			}{%
				\nolinkurl{arXiv:#1}%
				\iffieldundef{eprintclass}{}{\addspace\mkbibbrackets{\thefield{eprintclass}}}%
			}%
		}%
	}
\makeatother

\vfuzz \hfuzz
\begin{document}
		\pagestyle{empty}
	\pagenumbering{roman}
	\pdfbookmark[0]{Frontmatter}{frontmatter}
\newgeometry{top=20mm, bottom=20mm, left=20mm, right=20mm}
\begin{titlepage}
	\begin{figure}[h!]
		\raggedright
		\includegraphics[scale=1.3]{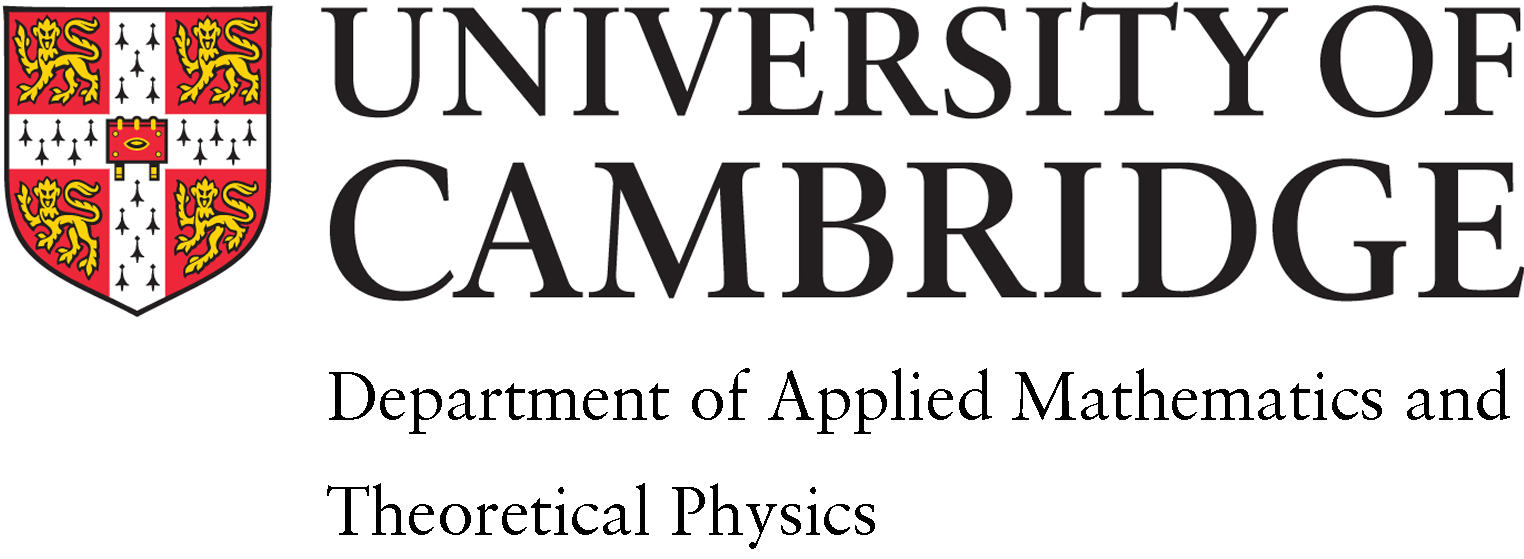}
	\end{figure}
	
	\vfill\vfill
	
	\begin{center}
		{\fontsize{24.88}{24}\selectfont {\textsc{Cosmological Probes\\[14pt]of Light Relics}}}
		
		\vskip 60pt
		
		{\fontsize{20.74}{24}\selectfont \textsc{Benjamin Wallisch}}
		
		\vfill\vfill\vfill\vfill\vfill
				
		{\Large This dissertation is submitted for the degree of\\[4pt]
			\textit{Doctor of Philosophy}\\[15pt]
			June 2018}
	\end{center}

	\vskip-12pt
	\begin{figure}[h!]
		\raggedleft
		\includegraphics[scale=0.95]{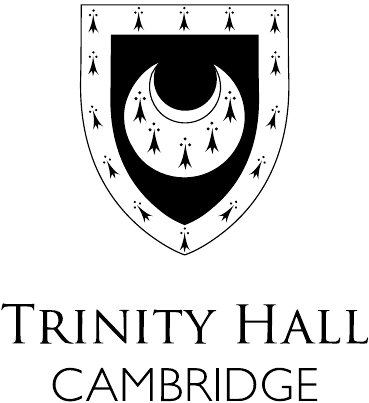}
	\end{figure}
\end{titlepage}
\restoregeometry
	\cleardoublepage

	\pagestyle{plain}
	\pdfbookmark[1]{Abstract}{abstract}
\thispagestyle{empty}
	
\begin{center}
	~\\\bigskip\bigskip\bigskip\bigskip\medskip
	{\LARGE \textsc{Cosmological Probes of Light Relics}}\\\bigskip\medskip
	{\Large \textsc{Benjamin Wallisch}}\\\bigskip\bigskip\bigskip
\end{center}

\noindent
One of the primary targets of current and especially future cosmological observations are light thermal relics of the hot big bang. Within the Standard Model of particle physics, an important thermal relic are cosmic neutrinos, while many interesting extensions of the Standard Model predict new light particles which are even more weakly coupled to ordinary matter and therefore hard to detect in terrestrial experiments. On the other hand, these elusive particles may be produced efficiently in the early universe and their gravitational influence could be detectable in cosmological observables. In this thesis, we describe how measurements of the cosmic microwave background~(CMB) and the large-scale structure~(LSS) of the universe can shed new light on the properties of neutrinos and on the possible existence of other light relics.\medskip

These cosmological observations are remarkably sensitive to the amount of radiation in the early universe, partly because free-streaming species such as neutrinos imprint a small phase shift in the baryon acoustic oscillations~(BAO) which we study in detail in the CMB and LSS power spectra. Building on this analytic understanding, we provide further evidence for the cosmic neutrino background by independently confirming its free-streaming nature in different, currently available datasets. In particular, we propose and establish a new analysis of the BAO~spectrum beyond its use as a standard ruler, resulting in the first measurement of this imprint of neutrinos in the clustering of galaxies.

Future cosmological surveys, such as the next generation of CMB experiments~(\mbox{CMB-S4}), have the potential to measure the energy density of relativistic species at the sub-percent level and will therefore be capable of probing physics beyond the Standard Model. We demonstrate how this improvement in sensitivity can indeed be achieved and present an observational target which would allow the detection of any extra light particle that has ever been in thermal equilibrium. Interestingly, even the absence of a detection would result in new insights by providing constraints on the couplings to the Standard Model. As an example, we show that existing bounds on additional scalar particles, such as axions, may be surpassed by orders of magnitude.
	\cleardoublepage
	
	\pagestyle{empty}
	\pdfbookmark[1]{Dedication}{dedication}
\vspace*{\stretch{2}}
\begin{flushright}
	\Large \textit{For my parents}
\end{flushright}
\vspace{\stretch{3}}
	\cleardoublepage
	
	\pagestyle{plain}
	\pdfbookmark[1]{Declaration}{declaration}
\chapter*{Declaration}
\label{chap:Declaration}
This dissertation is the result of my own work and includes nothing which is the outcome of work done in collaboration except as declared below and specified in the text. Following the tendency of modern research in theoretical physics, most of the material discussed in this dissertation is the result of collaborative research. In particular, Chapters~\ref{chap:cmb-axions} and~\ref{chap:bao-forecast}, and Appendices~\ref{app:cmb-axions_appendices}, \ref{app:bao-forecast_appendices} and~\ref{app:broadband+phaseShiftExtraction} are based on work done in collaboration with Daniel Baumann and Daniel Green, published in~\mbox{\cite{Baumann:2016wac, Baumann:2017gkg}}, Chapter~\ref{chap:cmb-phases} and Appendix~\ref{app:cmb-phases_appendices} are the result of the research done in collaboration with Daniel Baumann, Daniel Green and Joel Meyers, published in~\cite{Baumann:2015rya}, and Chapter~\ref{chap:bao-neutrinos} is mostly based on the work done in collaboration with Daniel Baumann, Florian Beutler, Raphael Flauger, Daniel Green, An\v{z}e Slosar, Mariana Vargas-Maga\~{n}a and Christophe Y\`{e}che, published in~\cite{Baumann:2018qnt}. I have made major contributions to the above, both in terms of results and writing. All figures presented in this dissertation were produced or adapted by myself.\bigskip\medskip

\noindent
I hereby declare that my dissertation on
\begin{center}
	\textit{Cosmological Probes of Light Relics}
\end{center}
is not substantially the same as any that I have submitted or is being concurrently submitted for a degree, diploma or other qualification at the University of Cambridge, any other University or similar institution. I further state that no substantial part of my dissertation has already been submitted or is being concurrently submitted for any such degree, diploma or other qualification at the University of Cambridge, any other University or similar institution.

\bigskip\bigskip
\begin{flushright}
	Benjamin Wallisch\\
	June 2018
\end{flushright}
	\cleardoublepage

	\pagestyle{scrheadings}
	\pdfbookmark[1]{Acknowledgements}{acknowledgements}
\chapter*{Acknowledgements}
\label{chap:Acknowledgements}
\markboth{Acknowledgements}{Acknowledgements}
First and foremost, I wish to express my sincere gratitude to my supervisor Daniel Baumann for his phenomenal guidance, tireless effort and unwavering support. His enthusiasm and passion for physics continue to amaze me and are truly contagious. He is a remarkable mentor with endless patience and dedication. I cannot thank him enough for all his contributions at many different levels to this work and beyond.\medskip

Second, I am most grateful to Daniel Green who was my other main collaborator and has essentially become a second supervisor in a number of ways. I could always count on his help and support, and the fact that our many discussions were enlightening and insightful. His imagination, excitement and approach to physics are an inspiration. It has been a true privilege and great pleasure to work with and learn from both Daniels throughout my~PhD.\bigskip

Thanks are also due to my other collaborators, Florian Beutler, James Fergusson, Raphael Flauger, Helge Gruetjen, Joel Meyers, Paul Shellard, An\v{z}e Slosar, Mariana Vargas-Maga\~{n}a and Christophe Y\`{e}che, for their valuable input, amazing ideas and enriching discussions. Working~with this great group of cosmologists has helped me to understand many different aspects of our field.\medskip

Furthermore, I thank my departmental colleagues in both Cambridge and Amsterdam, especially Valentin Assassi, Matteo Biagetti, Horng Sheng Chia, Garrett Goon, Hayden Lee, Guilherme Pimentel and John Stout, for numerous conversations and providing such a friendly and stimulating environment. Special thanks go to my Cambridge officemates Chandrima Ganguly and Will Cook.\bigskip

I am thankful to my friends and fellow students in Heidelberg, Cambridge, Amsterdam and elsewhere for spending countless fun and enriching moments together, for their understanding when I have neglected them, and for sharing numerous adventures.\medskip

Finally, I want to thank my brother. More than anyone else, I however owe my deepest gratitude to my parents for their continuous and unconditional support in so many ways. I would not be where I am now without them.\newpage

I acknowledge support by a Cambridge European Scholarship of the Cambridge Trust, by the Department of Applied Mathematics and Theoretical Physics, by a Research Studentship Award of the Cambridge Philosophical Society, from a Starting Grant of the European Research Council (ERC~STG Grant~279617), by an STFC Studentship, by Trinity Hall and by a Visiting PhD Fellowship of the Delta-ITP consortium, a program of the Netherlands Organisation for Scientific Research~(NWO) that is funded by the Dutch Ministry of Education, Culture and Science~(OCW). I am also grateful to the CERN theory group, the Institute for Theoretical Physics at the University of Heidelberg and, in particular, the Institute of Physics at the University of Amsterdam for their hospitality.\medskip

This work uses observations obtained by the Planck satellite (\href{http://www.esa.int/Planck}{http:/\!/www.esa.int/Planck}), an ESA science mission with instruments and contributions directly funded by ESA Member States, NASA and Canada. This research is also partly based on observations obtained by the Sloan Digital Sky Survey~III (\mbox{SDSS-III}, \href{http://www.sdss3.org/}{http:/\!/www.sdss3.org/}). Funding for SDSS-III has been provided by the Alfred P.\ Sloan Foundation, the Participating Institutions, the National Science Foundation and the U.S.~Department of Energy Office of Science. Parts of this work were undertaken on the COSMOS Shared Memory System at DAMTP (University of Cambridge), operated on behalf of the STFC DiRAC HPC Facility. This equipment is funded by BIS National E-Infrastructure Capital Grant ST/J005673/1 and STFC Grants ST/H008586/1, ST/K00333X/1. Some analyses also used resources of the HPC cluster At\'ocatl at IA-UNAM, Mexico, and of the National Energy Research Scientific Computing Center, which is supported by the Office of Science of the U.S.\ Department of Energy under Contract No.~DE-AC02-05CH11231.\medskip

The results presented in this thesis made use of \texttt{CAMB}~\cite{Lewis:1999bs}, \texttt{CLASS}~\cite{Blas:2011rf}, \texttt{CosmoMC}/\texttt{GetDist}~\cite{Lewis:2002ah}, \texttt{FORM}~\cite{Kuipers:2012rf}, \texttt{IPython}~\cite{Perez:2007ipy}, \texttt{MontePython}~\cite{Audren:2012wb}, and the Python packages \texttt{Astropy}~\cite{Robitaille:2013mpa}, \texttt{emcee}~\cite{ForemanMackey:2012ig}, \texttt{Matplotlib}~\cite{Hunter:2007mat}, \texttt{nbodykit}~\cite{Hand:2017pqn} and \texttt{NumPy}/\texttt{SciPy}~\cite{Walt:2011num}.
	\cleardoublepage
	
	\pdfbookmark[1]{\contentsname}{toc}
	\microtypesetup{protrusion=false}
\tableofcontents
	\microtypesetup{protrusion=true}
	\cleardoublepage

	\pagestyle{scrheadings}
	\pagenumbering{arabic}
	\chapter{Introduction}
\label{chap:introduction}
Cosmology is a sensitive probe of particle physics, both within the Standard Model and beyond~it. In fact, cosmological observations have now become precise enough to start complementing laboratory and collider experiments. For example, by measuring the radiation density of the universe, future observations may provide further insights into the properties of neutrinos. Moreover, if these measurements reach sub-percent level, they have the potential to discover particles that are more weakly coupled than neutrinos, which are predicted in many interesting models of physics beyond the Standard Model. In this thesis, we are searching for these elusive particles by identifying and extracting their robust signatures in cosmological observables.\medskip

Experiments at particle accelerators have established the Standard Model~(SM) of particle physics as the description of the elementary building blocks of matter and their non-gravitational interactions. At the same time, observations of the anisotropies in the cosmic microwave background~(CMB), the fossil radiation from the beginning of the universe, have led to the standard model of cosmology, which captures the entire evolution of the universe from the hot big bang until today. Despite the great successes of the standard models of both cosmology and particle physics, many questions remain unanswered. Strikingly, solutions to some of the cosmological puzzles may influence those in particle physics and vice versa. For instance, some ingredients of the cosmological model, such as dark matter and inflation, ask for new microscopic descriptions. Concurrently, many extensions of the Standard Model give rise to new particles that can be efficiently produced at the high temperatures in the early universe and may therefore be detectable in cosmological observables.\medskip

About \num{373000}~years after the big bang, photons decoupled from the rest of the primordial plasma and the cosmic microwave background was released carrying a treasure-trove of information. Most of our knowledge about the early universe comes from observations of these relic photons. As a matter of fact, cosmologists generally study the history and contents of the universe by detecting relics from the past or extracting their imprints. To probe the time before the epoch of recombination, we therefore rely either on theoretical extrapolations or the existence of further relics to get a snapshot of our cosmos. This has been very successfully employed by measuring the relic abundances of the light elements, which were synthesized about three minutes after the beginning of the universe during big bang nucleosynthesis~(BBN).

Roughly one second after the big bang, a thermal background of relic neutrinos was released when the rate of neutrino interactions dropped below the expansion rate of the universe and neutrinos were no longer in thermal equilibrium with the rest of the Standard Model. Measuring this cosmic neutrino background~(C$\nu$B) would establish a window back to this time, when the universe was at nearly nuclear densities. Since these neutrinos were a dominant component of the energy density in the early universe, they played an important role in the evolution of cosmological perturbations. Extracting the imprints of neutrinos in observations may therefore provide new insights into the least understood sector of the Standard Model. In fact, as we will show in this thesis, one of the most remarkable results of the Planck satellite is the detection of cosmic neutrinos and a confirmation of their free-streaming nature. At present, we can therefore use cosmological measurements to explore the Standard Model and investigate the history of the universe back to a time when it was one second old.\medskip

Probing even earlier times requires detecting new particles that are more weakly coupled than neutrinos. Since neutrinos are the most feebly interacting SM~particles, these new species necessarily lie beyond the Standard Model~(BSM). There is indeed a lot of circumstantial evidence from both theoretical considerations and experimental measurements that the Standard Model is incomplete. In addition to new massive particles, an interesting consequence of many proposals for BSM~physics are extra light species~\cite{Essig:2013lka}, such as axions~\cite{Peccei:1977hh, Weinberg:1977ma, Wilczek:1977pj}, axion-like particles~(ALPs)~\cite{Arvanitaki:2009fg}, dark photons~\cite{Holdom:1985ag, Galison:1983pa} and light sterile neutrinos~\cite{Abazajian:2012ys}. The search for these particles is one of the main objectives of particle physics, but detecting them could be difficult in terrestrial experiments because their couplings might be too small or their masses too large. Interestingly, the temperature in the early universe was high enough to make the production of weakly-coupled and/or massive particles efficient. Their gravitational influence could then be detected if the energy density carried by these particles was significant. This will be the case for light relics which were in thermal equilibrium with the Standard Model at early times and subsequently decoupled from the SM degrees of freedom. This sensitivity to extremely weakly interacting particles is a unique advantage of cosmological probes of BSM physics.\medskip

Another advantage of cosmological observations is that they can provide broad constraints on phenomenological descriptions, whereas particle physics searches can be blind to unknown or incompletely specified forms of new physics. This means that terrestrial experiments may give strong constraints on specific scenarios, while cosmological measurements are less sensitive to the details of the models and can compress large classes of BSM~physics into broad categories. This approach has led to important discoveries in the past: by comparing observations against simple phenomenological parametrizations, the existence of dark matter~($\Omega_m$) and dark energy~($\Omega_\Lambda$) was established, the baryon asymmetry~($\eta$) was identified, and evidence for cosmological inflation~($\ns$) was presented. We will take a similar path by theoretically describing light thermal relics within an effective field theory~(EFT) framework~\cite{Brust:2013xpv} and experimentally searching for their contribution to the radiation density in the early universe~($\Neff$).

Rather remarkably, future cosmological observations, in particular the next generation of CMB~experiments, such as the CMB~\mbox{Stage-4}~(CMB-S4) mission~\cite{Abazajian:2016yjj}, have the potential to determine the value of~$\Neff$ at the sub-percent level. We find it intriguing that this level of sensitivity exactly corresponds to the thermal abundance of light relics in minimal scenarios. An important aspect of this thesis is the realization that the improvement of current constraints by one order of magnitude would allow us to either detect any particles which have ever been in thermal equilibrium with the Standard Model, or put strong constraints on their SM~couplings. This sensitivity would therefore not only provide precision tests of the Standard Model, but also a new window into the very early universe and BSM~physics.\medskip

Since reaching this threshold may have far-reaching implications, we contribute analytical and numerical insights into the main cosmological observables to facilitate the optimization of upcoming experiments. The cosmic microwave background anisotropies, which are the remnants of sound waves in the primordial plasma, are particularly interesting in this respect. The presence of neutrinos and other light relics is imprinted in these fluctuations in two notable ways: Their contribution to the background energy density leads to a characteristic damping of the CMB~power spectra on small scales~\cite{Hou:2011ec} and their free-streaming nature causes a coherent shift in the locations of the acoustic peaks~\cite{Bashinsky:2003tk}. This subtle shift in the temporal phase of the primordial sound waves has recently been extracted from Planck data~\cite{Follin:2015hya}. We will present an alternative detection of this effect and establish a precise link to the underlying particle properties, which will in turn allow us to probe SM~extensions in a complementary way. Furthermore, it will become clear that future CMB~observations will be extremely sensitive to both the damping and the phase shift of the anisotropy spectrum mainly through measurements of the small-scale anisotropies and polarization.

The same physics that is imprinted in the~CMB also contributed to the initial conditions for the clustering of matter and may therefore be observable in the large-scale structure~(LSS) of the universe as well. The sound waves in the primordial plasma manifest themselves as baryon acoustic oscillations~(BAO), which we can observe in the distribution of galaxies. This implies that we should be able to extract the same neutrino-induced phase shift in the BAO~spectrum. As a matter of fact, we will present the first such measurement based on a newly developed BAO~analysis. Moreover, near-future LSS~surveys are projected to map more than ten times as many objects in a much larger cosmic volume than currently operating telescopes. With such remarkable improvements in sensitivity on the horizon, it is timely to re-assess how the wealth of incoming CMB and LSS~data could sharpen our understanding of the early universe and, particularly, how they will inform our view of extensions of the standard models of particle physics and cosmology.

\clearpage
\subsection*{Outline of the thesis}
The rest of this thesis is organized as follows. We first provide a review of the important aspects of both cosmology and particle physics which underlie the research presented in this work. This also serves as background material for the later chapters. In Chapter~\ref{chap:background_cosmology}, we focus on the cosmological standard model. We discuss the thermal history of the universe and introduce the two main observational windows employed in this thesis: the anisotropies of the cosmic microwave background and the large-scale structure of the universe with its distinct BAO~signal. In Chapter~\ref{chap:background_species}, we concentrate on the Standard Model of particle physics and some of its well-motivated extensions. We collect a few current hints for BSM~physics and present an effective field theory of light species as an efficient tool to study the additional particles that arise beyond the Standard Model. Furthermore, we unveil the main cosmological parameter studied in this thesis, the effective number of relativistic species~$\Neff$, and analyse the primary signatures of light relics in cosmological observables.

The remaining chapters consist of the main research results. In Chapter~\ref{chap:cmb-axions}, we derive new constraints on light thermal relics from precise measurements of the radiation density in the early universe and explicitly demonstrate the sensitivity of future cosmological observations to the SM~couplings of light scalar particles, such as axions. The constraints achievable from cosmology have the potential to surpass existing bounds from laboratory experiments and astrophysical searches by orders of magnitude. In Chapter~\ref{chap:cmb-phases}, we examine the phase shift in the acoustic peaks of the~CMB as a robust probe of both free-streaming Standard Model neutrinos and new physics. We find that the physical origin of this signature is limited to either free-streaming relativistic particles or isocurvature fluctuations. In addition, we provide observational constraints from Planck data which establish the free-streaming nature of cosmic neutrinos. In Chapter~\ref{chap:bao-forecast}, we explore to what degree these CMB~observations can be enhanced by upcoming large-scale structure surveys. We carefully isolate the information encoded in the shape of the galaxy power spectrum and in the spectrum of baryon acoustic oscillations. In particular, we propose a new analysis of the BAO~signal and show that the neutrino-induced phase shift can be detected at high significance in future experiments. In Chapter~\ref{chap:bao-neutrinos}, we implement this analysis and report on the first measurement of this coherent shift in the peak locations of the BAO~spectrum at more than \SI{95}{\percent}~confidence in galaxy clustering data collected by the Baryon Oscillation Spectroscopic Survey~(BOSS). Besides being a new measurement of the cosmic neutrino background and its free-streaming nature, it is also the first application of the BAO~signal to early universe physics. In Chapter~\ref{chap:conclusions}, we conclude with a brief summary of our results and an outlook.

A series of appendices contains technical details underlying the results presented in the main part of this thesis. In Appendix~\ref{app:cmb-axions_appendices}, we provide further details on the calculations underlying the findings of Chapter~\ref{chap:cmb-axions}. In Appendix~\ref{app:cmb-phases_appendices}, we comment on extensions of the analytic treatment of the phase shift in Chapter~\ref{chap:cmb-phases}. In Appendix~\ref{app:bao-forecast_appendices}, we collect details of the CMB and LSS~forecasts conducted in Chapter~\ref{chap:bao-forecast}. Appendix~\ref{app:broadband+phaseShiftExtraction} finally includes our methods for extracting the broadband power spectrum of matter fluctuations and the neutrino-induced phase shift as employed in Chapters~\ref{chap:bao-forecast}~and~\ref{chap:bao-neutrinos}.

\subsection*{Notation and conventions}
For ease of reference, we provide a collection of the employed notation and conventions. Table~\ref{tab:abbreviations}%
\begin{table}[b!]
	\centering
	\begin{tabular}{l l} 
			\toprule
		Acronym					& Expression										\\
			\midrule[0.065em]
		BAO						& Baryon acoustic oscillations						\\
		BBN						& Big bang nucleosynthesis							\\
		BOSS					& Baryon Oscillation Spectroscopic Survey			\\
		BSM						& Beyond the Standard Model	(of particle physics)	\\
		CDM						& Cold dark matter									\\
		CMB						& Cosmic microwave background						\\
		CMB-S4					& CMB Stage-4 (a future CMB~experiment)				\\
		C$\nu$B					& Cosmic neutrino background						\\
		CVL						& Cosmic variance limit								\\
		DES						& Dark Energy Survey								\\
		DESI					& Dark Energy Spectroscopic Instrument				\\
		EFT						& Effective field theory							\\
		EWSB					& Electroweak symmetry breaking						\\
		$\Lambda\mathrm{CDM}$	& Standard cosmological model						\\
		LSS						& Large-scale structure	(of the universe)			\\
		LSST					& Large Synoptic Survey Telescope					\\
		pNGB					& Pseudo-Nambu-Goldstone boson						\\
		QCD						& Quantum chromodynamics							\\
		SM						& Standard Model (of particle physics)				\\
			\bottomrule 
	\end{tabular}
	\caption{Common acronyms used throughout this thesis.}
	\label{tab:abbreviations}
\end{table}
contains a list of abbreviations that will commonly be used. Throughout this thesis, we work in natural units in which the speed of light~$c$, the reduced Planck constant~$\hbar$ and the Boltzmann constant~$k_B$ are set to unity, $c = \hbar = k_B = 1$, and the reduced Planck mass is given by $\Mp^2 = (8\pi G)^{-1}$, with Newton's constant $G$. Our metric signature is~$(-+++)$ and we use Greek letters for four-dimensional spacetime indices, $\mu,\nu,\ldots=0,1,2,3$. Spatial vectors are written in boldface,~$\x$, or in index notation, $x^i$, with Latin letters, $i,j,\ldots=1,2,3$. Repeated indices are summed over except in cases
where the same index appears unpaired on the other side of the equation as well. The magnitude of a vector is defined as $x \equiv |\x|$ and unit vectors are hatted,~$\hat{\x} \equiv \x/x$. Our Fourier convention is
\beq
\tilde{f}(\k) = \int\! \d^3x\, f(\x)\, \e^{-\ii \k \cdot \x}\, ,	\qquad f(\x) = \int\! \frac{\d^3k}{(2\pi)^3}\, \tilde{f}(\k)\, \e^{\ii \k \cdot \x}\, ,
\eeq
where we commonly drop the tilde, $\tilde{f}(\k) \to f(\k)$, for ease of notation. Overdots and primes denote derivatives with respect to conformal time~$\tau$ and physical time~$t$, respectively. We use~$\tau_0$ for the present time, $\tau_\mathrm{rec}$ for the time of recombination and photon decoupling, $\tau_\mathrm{eq}$ for matter-radiation equality, and $\tau_\mathrm{in}$ for the time at which we set the initial conditions. We will use a subscript~`$\alpha$' to denote quantities evaluated at the time $\tau_\alpha$, such as the radiation energy density today, $\rho_{r,0} = \rho_r(\tau=\tau_0)$. As for the radiation content, individual components of the universe (like photons, matter, neutrinos, etc.)~will be denoted by a subscript $b=\gamma, m, \nu, \cdots$. The conformal and physical Hubble parameters are $\H \equiv \dot a/a$ and $H \equiv a'\!/a$, respectively, with the scale factor~$a$ normalized to unity today, $a_0 \equiv 1$. The dimensionfull and dimensionless power spectra~$P_{\hskip-0.5ptf}(k)$ and~$\mathcal{P}_{\hskip-0.5ptf}(k)$ of a Fourier mode~$f(\k)$ are defined by
\beq
\left\langle f(\k)\, f^*\!(\k') \right\rangle = (2\pi)^3 \, P_{\hskip-0.5pt f}(k)\, \delta_D^{(3)}(\k - \k') = \frac{(2\pi)^3}{k^3} \, \mathcal{P}_{\hskip-0.5pt f}(k)\, \delta_D^{(3)}(\k - \k')\, ,
\eeq
where $\delta_D^{(3)}$ is the three-dimensional Dirac delta function. Finally, statistical error bars are quoted as one Gaussian standard deviations ($1\sigma$, corresponding to about \SI{68}{\percent}~c.l.), unless stated otherwise.
	\chapter{Review of Modern Cosmology}
\label{chap:background_cosmology}
Cosmology is the quantitative study of the structure and evolution of the universe. In the last few decades, it has emerged as a data-driven field of study which has revolutionized our understanding of the cosmos. The analysis of observations of type~Ia supernovae~\cite{Riess:1998cb, Perlmutter:1998np}, measurements of the temperature anisotropies in the cosmic microwave background (especially by the satellite missions COBE~\cite{Smoot:1992td, Bennett:1996ce}, WMAP~\cite{Spergel:2003cb, Hinshaw:2012aka} and Planck~\cite{Ade:2013zuv, Ade:2015xua}) and maps of the large-scale structure~\cite{Colless:2001gk, Stoughton:2002ae, Alam:2016hwk} have contributed decisive insights. Together with important theoretical advances, this has led to the standard model of cosmology, which describes the roughly 13.8~billion years of cosmic expansion in terms of just six parameters.\medskip

In this chapter, we discuss both the theory and the observations underlying modern cosmology. In Section~\ref{sec:homogeneousCosmology}, we consider spatially homogeneous and isotropic spacetimes, collect the basic equations governing the universe on the largest scales, and introduce the $\Lambda\mathrm{CDM}$~model. In Section~\ref{sec:thermalHistory}, we study the thermal history of the universe, including the cosmic neutrino and microwave backgrounds. In Section~\ref{sec:inhomogeneousCosmology}, we move beyond homogeneity and consider small fluctuations around the smooth universe. Apart from the mechanism underlying the growth and presence of structure, we will also encounter the sound waves in the primordial plasma which are of tremendous observational significance. In Section~\ref{sec:cosmologicalObservables}, we finally review the cosmic microwave background anisotropies, the large-scale structure of the universe and the baryon acoustic oscillations. These are the main cosmological observables employed in the rest of this thesis.

\section{Homogeneous Cosmology}
\label{sec:homogeneousCosmology}
Cosmological observations indicate that the universe is both spatially homogeneous and isotropic on large scales and/or at early times. This is the basis of modern cosmology. In the following, we study the implications of these profound findings for the shape and content of the universe~(\textsection\ref{sec:geometryAndDynamics}), and introduce the $\Lambda\mathrm{CDM}$~model which has emerged as the standard cosmological model~(\textsection\ref{sec:StandardModelCosmology}).

\subsection{Geometry and Dynamics of the Universe}
\label{sec:geometryAndDynamics}
A spatially homogeneous and isotropic spacetime can be described by the Friedmann-Lema\^{i}tre-Robertson-Walker~(FLRW) metric
\beq
\d s^2 = \bar{g}_{\mu\nu} \d x^\mu \d x^\nu = -\d t^2 + a^2(t) \gamma_{ij} \d x^i \d x^j\, ,	\label{eq:FLRW}
\eeq
where $\gamma_{ij}$ is the induced metric on the spatial hypersurfaces. The FLRW~metric~\eqref{eq:FLRW} depends on one time-dependent function, the scale factor~$a(t)$. To see this, it is instructive to decompose~$\bar{g}_{\mu\nu}$ into a scalar,~$\bar{g}_{00}$, a three-vector,~$\bar{g}_{i0}$, and a three-tensor,~$\bar{g}_{ij}$. Homogeneity requires that the mean value of any scalar can only be a function of time. Absorbing this function into a redefinition of the time coordinate, we get $\bar{g}_{00} = -1$. Isotropy on the other hand implies that the mean value of any three-vector has to vanish, i.e.\ $\bar{g}_{i0} \equiv 0$. Finally, the mean value of the three-tensor~$\bar{g}_{ij}$ has to be proportional to the three-metric~$\gamma_{ij}$ based on isotropy. In turn, homogeneity restricts the three-curvature to be the same everywhere and the proportionality coefficient to be only a function of time. We therefore have $\bar{g}_{ij} = a^2(t)\gamma_{ij}$, with $\gamma_{ij}$ being restricted by a constant three-curvature~$R^{(3)}$. Since there are three unique three-metrics that lead to $R^{(3)} = \const$, the spatial geometry of the universe can only be positively curved~($R^{(3)}>0$), flat~($R^{(3)}=0$) or negatively curved~($R^{(3)}<0$).

Since we mostly rely on observations of photons to study the universe, we have to consider the propagation of light. It is therefore useful to work with a metric that is conformally invariant to Minkowski space. This can be achieved by introducing conformal time $\d\tau \equiv \d t/a$. As a consequence, we can express the FLRW~metric as
\beq
\d s^2 = a^2(\tau) \left( -\d\tau^2 + \d\x^2 \right) , \label{eq:FLRWmetric}
\eeq
where we defined $\d\x^2 \equiv \gamma_{ij} \d x^i \d x^j$. Observations of the~CMB have constrained the three-curvature to be close to zero, which implies that the flat FLRW~metric describes the global geometry of the universe to very good approximation. For the rest of this thesis, we will therefore neglect spatial curvature, $R^{(3)}=0$, in which case $\gamma_{ij} = \delta_{ij}$ in Cartesian coordinates.\medskip

Of the four fundamental forces of Nature, only the gravitational force is relevant on cosmological scales because the other three are shielded by opposite charges or confined to subatomic scales. According to General Relativity, the dynamics of the universe is therefore governed by the Einstein field equations~\cite{Einstein:1916vd},
\beq
G_{\mu\nu} = 8 \pi G\, T_{\mu\nu}\, ,	\label{eq:EinsteinEquations}
\eeq
where~$G$ is Newton's constant, and the Einstein tensor~$G_{\mu\nu}$ depends on the metric~$g_{\mu\nu}$ and its first two derivatives. The energy-momentum tensor~$T_{\mu\nu}$ on the right-hand side captures the entire content of the universe. The fact that energy-momentum is conserved, $\nabla_\mu T^{\mu\nu} = 0$, is automatically ensured by the Bianchi identities.\medskip

The form of the energy-momentum tensor in a homogeneous and isotropic universe can be derived by following a similar argument as for the metric. It is again convenient to decompose~$T_{\mu\nu}$ into a scalar,~$T_{00}$, a three-vector,~$T_{i0}$, and a three-tensor,~$T_{ij}$. As above, homogeneity implies that the scalar can only depend on time, i.e.\ $\bar{T}_{00} = \bar{\rho}(t)$, with an arbitrary function~$\bar{\rho}(t)$. Isotropy again requires $\bar{T}_{i0} \equiv 0$ since there would otherwise be a non-zero energy flux. By isotropy we also have $T_{ij} \propto g_{ij}$, with homogeneity restricting the proportionality coefficient to be a function of time alone. The energy-momentum tensor of a homogeneous and isotropic background can therefore only take the following form:
\beq
\bar{T}_{00} = \bar{\rho}(t)\, , \qquad \bar{T}_{i0} = \bar{T}_{0j} = 0\, ,	\qquad \bar{T}_{ij} = \bar{P}(t)\, g_{ij}\, .	\label{eq:perfectFluid}
\eeq
This is the energy-momentum tensor of a perfect fluid as seen in the reference frame of a comoving observer. The content of a spatially homogeneous and isotropic universe can therefore be characterised by the energy density $\bar{\rho} = \bar{\rho}(t)$ and the pressure $\bar{P}=\bar{P}(t)$ in the rest frame of the cosmic fluid. 

We obtain the evolution equations of the cosmic fluid by plugging the conformal FLRW~metric~\eqref{eq:FLRWmetric} and the energy-momentum tensor~\eqref{eq:perfectFluid} into the Einstein equations~\eqref{eq:EinsteinEquations}. Given the imposed symmetries, they simplify dramatically to the first and second Friedmann equations~\cite{Friedmann:1922kd},
\beq
3\H^2 = 8 \pi G a^2\, \bar{\rho}\,, \qquad 2\dot{\H} + \H^2 = -8 \pi G a^2\, \bar{P}\, ,	\label{eq:FriedmannEquations}
\eeq
where we introduced the (conformal) Hubble rate $\H = \dot{a}/a$.\footnote{For convenience, we invert the otherwise standard notation and use overdots (primes) to denote derivatives with respect to conformal (physical) time~$\tau$~($t$), i.e.\ $\dot{a} \equiv \d a/\!\d\tau$ and $a' \equiv \d a/\!\d t$. The physical Hubble rate is therefore given by $H = a'\!/a$.} By combining these equations, or by directly employing energy-momentum conservation, we obtain the continuity equation,
\beq
\dot{\bar{\rho}} + 3 \H\, (\bar{\rho} + \bar{P}) = 0\, ,
\eeq
from which we can read off the evolution of the cosmic fluid with a given equation of state~$ w \equiv \bar{P}/\bar{\rho}$. In this thesis, we will mostly be concerned with two types of cosmological content: radiation with $w_r=1/3$, i.e.\ $\bar{\rho}_r \propto a^{-4}$, and (pressureless) matter with $w_m \approx 0$, i.e.\ $\bar{\rho}_m \propto a^{-3}$. It is easy to see that radiation dominates at early times in a universe with these two components, but matter takes over as the main constituent after matter-radiation equality at $a_\mathrm{eq} \equiv \bar{\rho}_{r,0}/\bar{\rho}_{m,0}$, with $\rho_{b,0} = \rho_{b}(\tau=\tau_0)$. The third component, which is important in our universe and dominates at late times, is dark energy. Many observational clues point towards dark energy being described by an equation of state of $w_\Lambda=-1$ with $\rho_\Lambda = \const$ and, hence, the cosmological constant~$\Lambda$ (see e.g.~\cite{Weinberg:2012es}).

\subsection{Standard Model of Cosmology}
\label{sec:StandardModelCosmology}
It is rather remarkable that all current cosmological data (e.g.~\cite{Hinshaw:2012aka, Ade:2015xua, Alam:2016hwk, Abbott:2017wau}) is fit by a simple six-parameter model---the $\Lambda\mathrm{CDM}$~model. As the name suggests, at late times, the universe is dominated by the cosmological constant~$\Lambda$ and cold dark matter~(CDM). At the same time, these are the components that we know least about. The radiation energy density is very small today, but photons and neutrinos still vastly dominate the entropy of the universe due to their large number densities of about~\SI{411}{\per\cubic\centi\meter} and~\SI{112}{\per\cubic\centi\meter}, respectively  (cf.~\textsection\ref{sec:cnub} and~\textsection\ref{sec:cmb}). Finally, cosmologists refer to the visible matter comprised of the known Standard Model particles as baryonic matter, which, for most of the universe's history, consists mainly of electrons and protons.\medskip

The $\Lambda\mathrm{CDM}$~model includes two parameters characterising the initial conditions, namely the amplitude~$\As$ and the tilt~$\ns$ of the almost scale-invariant spectrum of primordial curvature perturbations,
\beq
\mathcal{P}_{\hskip-0.5pt\zeta}(k) = \As \left( \frac{k}{k_0} \right)^{\!\ns-1}\, ,	\label{eq:primordialSpectrum}
\eeq
where the arbitrary pivot scale is commonly set to $k_0 = \SI{0.05}{\per\Mpc}$. Such a spectrum is notably predicted by generic inflationary models~\cite{Baumann:2009ds}. 

\begin{figure}[b!]
	\centering
	\includegraphics{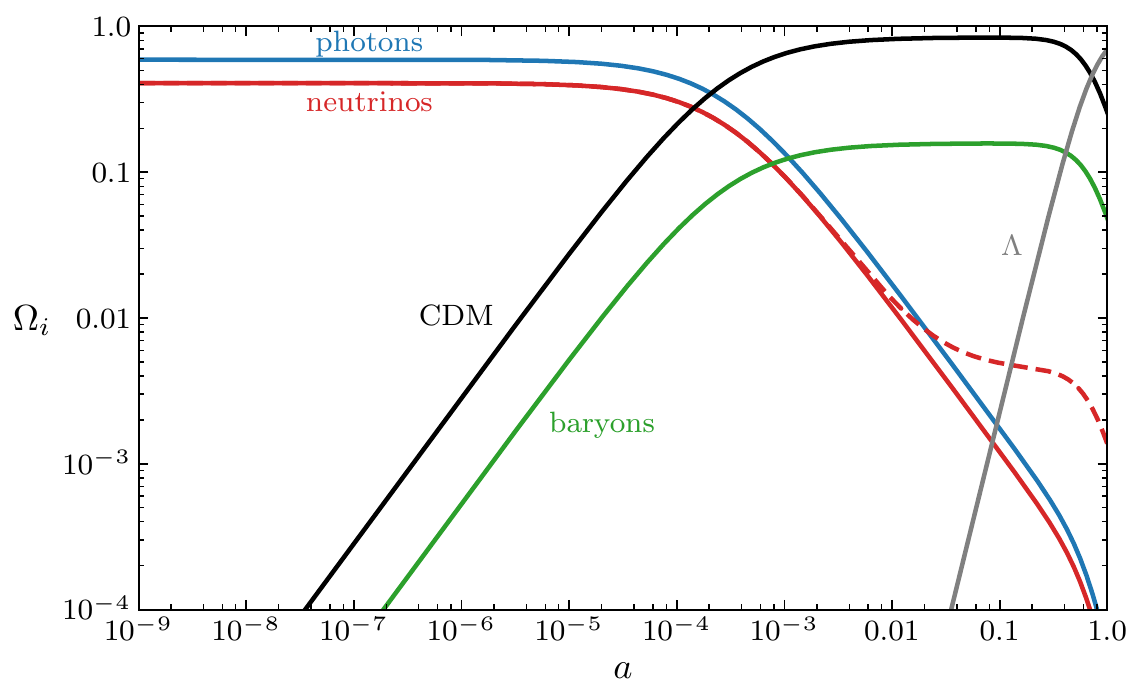}
	\caption{Evolution of the fractional energy densities~$\Omega_i$ for photons, neutrinos, cold dark matter, baryons and dark energy described by a cosmological constant within the standard $\Lambda\mathrm{CDM}$~model. The dashed \textcolor{pyRed}{red}~line assumes massive neutrinos with $\sum_i m_{\nu_i} \approx \SI{58}{\milli\electronvolt}$, whereas the solid \textcolor{pyRed}{red}~line takes these particles to be massless.}
	\label{fig:energyDensity}
\end{figure}
The remaining four parameters are associated with the geometry and composition of the universe. The matter content of the universe is described by the physical baryon and cold dark matter densities, $\omega_b \equiv \Omega_b h^2$ and $\omega_c \equiv \Omega_c h^2$, where $\Omega_a \equiv 8\pi G/(3 H_0^2)\,\rho_a$, with reduced Hubble constant $h \equiv H_0/\!\left(\SI{100}{\kilo\meter\per\second\per\Mpc}\right)$. Sometimes the dark matter density~$\Omega_c$ is traded for the total matter density $\Omega_m = \Omega_b + \Omega_c + \Omega_\nu$, where $\Omega_\nu \approx \sum_i m_{\nu_i}/\SI{94.1}{\electronvolt}$ is the (small) contribution of massive neutrinos to the matter density.\footnote{The minimal sum of masses from neutrino oscillation experiments is about \SI{58}{\milli\electronvolt}~\cite{Patrignani:2016xqp} and in particular cosmological measurements are closing in on this value with a current upper bound of $\sum_i m_{\nu_i} < \SI{0.23}{\electronvolt}$ (\SI{95}{\percent}~c.l.)~\cite{Ade:2015xua}. As the masses are so small that neutrinos have been (ultra-)relativistic for a large part of cosmological history, especially around the time of photon decoupling ($\Trec \approx \SI{0.26}{\electronvolt}$),  their effect on the aspects of interest in this thesis is small. For simplicity, we will therefore treat neutrinos as massless particles throughout, except in the BOSS cosmology of Chapter~\ref{chap:bao-neutrinos}.} Instead of the Hubble constant~$H_0$, we often use the angular size of the sound horizon at photon decoupling (see below), $\theta_s \equiv r_s(z_\rec)/D_A(z_\rec)$, where~$r_s$ is the physical sound horizon and~$D_A$ is the angular diameter distance, both evaluated at the redshift of decoupling,~$z_\rec$. The parameter~$\theta_s$ receives a contribution from the dark energy density $\Omega_\Lambda \equiv \Lambda/(3H_0^2)$. We note that $\Lambda\mathrm{CDM}$ assumes the universe to be exactly flat, $R^{(3)} \equiv 0$, i.e.\ $\Omega_r + \Omega_m + \Omega_\Lambda \equiv 1$, where the radiation density is comprised of photons at a temperature of $T_0 = \SI{2.7255}{\kelvin} = \SI{0.235}{\milli\electronvolt}$~\cite{Fixsen:2009ug} as well as three relativistic neutrino species. We refer to Section~\ref{sec:neutrinosDarkRadiation} for further details on the radiation density and neutrinos in particular, but note that neutrinos carry a sizeable fraction of the energy density in the early universe. As illustrated in Fig.~\ref{fig:energyDensity}, the history of the universe can therefore be divided into three large epochs: the radiation-dominated, matter-dominated and dark energy-dominated eras. The standard six-parameter model is completed by the optical depth~$\tau$ due to reionization in the late universe.\footnote{Although we use the same symbol to denote the optical depth and conformal time, its meaning will always be clear from context.} In Table~\ref{tab:cosmologicalParameters}, we list our fiducial values of the $\Lambda\mathrm{CDM}$~parameters, based on the Planck best-fit cosmology~\cite{Ade:2015xua}. Unless otherwise stated, these values will be used throughout this thesis.
\begin{table}
	\sisetup{group-digits=false}
	\begin{tabular}{c S[table-format=1.5] l} 
			\toprule
		Parameter 				& {Fiducial Value}	& Description															\\
			\midrule[0.065em]
		$\omega_b$ 				& 0.02230 			& Physical baryon density $\omega_b \equiv \Omega_b h^2$				\\
		$\omega_c$				& 0.1188			& Physical dark matter density $\omega_c \equiv \Omega_c h^2$			\\
		$100\,\theta_s$ 		& 1.04112 			& $100\,\times\,$angular size of the sound horizon at decoupling 		\\ 
		$\tau$ 					& 0.066 			& Optical depth due to reionization										\\
		$\ln(\num{e10}\As)$		& 3.064 			& Log of scalar amplitude (at pivot scale $k_0 = \SI{0.05}{\per\Mpc}$)	\\
		$\ns$					& 0.9667 			& Scalar spectral index (at pivot scale $k_0 = \SI{0.05}{\per\Mpc}$)	\\
		\midrule[0.065em]
		$\Neff$					& 3.046 			& Effective number of (free-streaming) relativistic species				\\
		$Y_p$					& 0.2478			& Primordial helium fraction 											\\ 
			\bottomrule 
	\end{tabular}
	\caption{Parameters of our reference cosmological model and their fiducial values based on~\cite{Ade:2015xua}.}
	\label{tab:cosmologicalParameters}
\end{table}

\section{Thermal History}
\label{sec:thermalHistory}
We now turn our attention to the precise evolution of the different species in the universe. This is mainly a story of (local) thermal equilibrium, production and decoupling of particles. Before we provide any details, we give a brief history of the main events in the thermal history of the universe~(\textsection\ref{sec:briefHistory}). We then discuss its thermodynamic description at early times and the notion of particles freezing out~(\textsection\ref{sec:equilibrium}). We conclude the section by reviewing the formation of the cosmic neutrino and microwave backgrounds~(\textsection\ref{sec:cnub} and~\textsection\ref{sec:cmb}). They are of particular importance in this thesis.

\subsection{Brief History of the Hot Big Bang}
\label{sec:briefHistory}
We take the beginning of the hot big bang to be the end of inflation when the particles of the Standard Model (and possibly its extensions) were produced during reheating. The associated energy scale might be as large as \SI{e16}{\giga\electronvolt} and all (known) particles were massless. When the temperature of the universe dropped to about $\SI{100}{\giga\electronvolt}$, the electroweak symmetry of the~SM became spontaneously broken. As a consequence, the SM~particles acquired their mass through the Higgs mechanism, but most particles were still relativistic. During the quark-gluon transition, quarks and gluons became confined in composite hadronic states. This event occurred around the temperature of the non-perturbative QCD~scale, $T \sim \SI{150}{\mega\electronvolt}$, and is usually denoted as the QCD~phase transition although it might be a cross-over. When the universe was about one second old, corresponding to a temperature of $T \sim \SI{1}{\mega\electronvolt}$, neutrinos decoupled from the rest of the primordial plasma and the cosmic neutrino background was released because the weak interactions were no longer efficient enough to maintain local thermal equilibrium~(cf.~\textsection\ref{sec:cnub}).

Around the same time, the interactions between neutrons and protons became inefficient leading to a relic abundance of neutrons. After about three minutes, the light elements, in particular hydrogen and helium, were synthesized from these neutrons and protons during big bang nucleosynthesis. By numerically solving coupled Boltzmann equations, the primordial helium fraction is predicted to be $Y_p = 4n_\mathrm{He}/n_b \sim 0.25$, with the precise value depending on the baryon density and the amount of radiation in the universe. This estimate agrees well with both the observations of primordial abundances (see e.g.~\cite{Aver:2015iza}) and the value inferred from CMB~measurements (see e.g.~\cite{Ade:2015xua}). In fact, BBN~has become one of the main tools to constrain the evolution of the universe above the \si{\mega\electronvolt}~scale. Moreover, the theoretical predictions for the primordial abundances as a function of~$\omega_b$ and~$\omega_r$ can be used to infer the value of~$Y_p$ that is consistent within the $\Lambda\mathrm{CDM}$~model. We refer to this procedure as imposing consistency with~BBN.

The last major event in the thermal history of our universe was the formation of the first hydrogen atoms, which is referred to as recombination (cf.~\textsection\ref{sec:cmb}). Since the number density of free electrons dropped sharply as a consequence, photons decoupled from matter at $\Trec \approx \SI{0.26}{\electronvolt}$ about \num{373000}~years after the beginning of the universe. They have been free-streaming ever since and we observe these relics of the big bang today as the cosmic microwave background.

\subsection{In and Out of Equilibrium}
\label{sec:equilibrium}
Throughout the thermal evolution of the universe, the interaction rate of particles, $\Gamma \sim n\, \sigma$, with number density~$n$ and thermally-averaged cross section~$\sigma$, competes with the expansion rate~$H$. As long as $\Gamma \gg H$, thermal equilibrium can be maintained locally because there are many particle interactions per Hubble time and the evolution is quasi-stationary. On the other hand, if the universe at some point expands faster than these particles can interact with each other, $\Gamma \ll H$, they are no longer in equilibrium and evolve separately. In the following, we will discuss both regimes, including the transition period when $\Gamma \sim H$.

\subsubsection{Thermal equilibrium}
At early times, all SM~particles were in local thermal equilibrium, i.e.\ the interactions between them were efficient and kept these particles locally in close thermal contact. In phase space, every particle species~$a$ can be described by its distribution function~$f_a(t,\x,\p)$, the number of particles per unit phase space volume, with momentum~$\p$. Homogeneity and isotropy dictate that the distribution function can neither depend on the position~$\x$ nor the direction of the momentum~$\hat{\p}$, which implies that $f_a(t,\x,\p) \to \bar{f}_a(t,p)$. The thermal Bose-Einstein and Fermi-Dirac distribution functions are given by
\beq
\bar{f}_a(p) = \left[\e^{(E_a(p)-\mu_a)/T_a} \mp 1 \right]^{\!-1}\, ,	\label{eq:thermalDistributions}
\eeq
for bosons~($-$) and fermions~($+$), respectively, where~$E_a$ is the relativistic energy which includes the mass~$m_a$. The chemical potentials~$\mu_a$ are likely small for all SM~species and, in particular, vanish for photons (since the number of photons is not conserved). For electrons, for instance, we can estimate $\mu_e/T \sim \num{e-9}$ because the universe is electrically neutral, i.e.\ the proton number density is equal to the difference in the number densities of electrons and positrons, $n_p = n_e - \bar{n}_e$, the baryon-to-photon ratio is $\eta = n_b/n_\gamma \approx n_p/n_\gamma \sim \num{e-9}$ and $\mu_e/T \sim (n_e - \bar{n}_e)/n_\gamma$. For simplicity, we therefore set the chemical potentials to zero, $\mu_a \equiv 0$, from now on. All species which are in thermal equilibrium with one another of course share the same temperature $T_a = T$.\medskip

Often, we are only interested in the momentum-integrated quantities. Integrating the distribution function yields the number density
\beq
n_a = g_a \int\! \frac{\d^3 p}{(2\pi)^3}\, f_a(\x,\p)\, ,
\eeq
where $g_a$ is the number of internal degrees of freedom of species $a$. Similarly, we obtain the energy density and pressure from the distribution function via the weighted integrals
\beq
\rho_a = g_a \int\! \frac{\d^3 p}{(2\pi)^3}\, E_a(p) f_a(\x,\p)\, , \qquad
P_a = g_a \int\! \frac{\d^3 p}{(2\pi)^3}\,  \frac{p^2}{3E_a(p)} f_a(\x,\p)\, .
\eeq
In the non-relativistic limit, $m_a \gg T_a$, it is easy to see that $\rho_a \approx m_a n_a$ and $P_a \ll \rho_a$, i.e.\ a non-relativistic gas of particles behaves like pressureless matter. If they become non-relativistic while being in thermal contact with other species, their number and energy densities get exponentially suppressed, $f_a \to \e^{-m_a/T}$. Physically speaking, this arises because particles and anti-particles annihilate while the reverse process is kinematically forbidden below the mass threshold. As a consequence, the primordial plasma is dominated by relativistic species. If particles freeze out, on the other hand, they retain their equilibrium distribution function and do not get further depleted which results in a finite relic abundance. This explains why dark matter may be cold and comprised of thermal relics. 

In the relativistic limit, $m_a \ll T_a$, we recover the equation of state of radiation, $P_a = \frac{1}{3} \rho_a$. Taking the thermal distributions~\eqref{eq:thermalDistributions} and defining the temperature of the universe as the photon temperature, $T \equiv T_\gamma$, the total radiation energy density can be written as
\beq
\bar{\rho}_r = \sum_a \bar{\rho}_a = \frac{\pi^2}{30} g_*(T)\, T^4\, ,	\label{eq:relativisticEnergyDensity}
\eeq
where we summed over all relativistic species and introduced the effective number of relativistic degrees of freedom
\beq
g_*(T) = \sum_{a=b} g_{*,a}(T) + \frac{7}{8} \sum_{a=f} g_{*,a}(T) = \sum_a g_a \left(\frac{T_a}{T}\right)^{\!4} \left(1 - \frac{1}{8} \delta_{af}\right) .	\label{eq:gStar}
\eeq
Here, the Kronecker delta $\delta_{af}$ vanishes for bosons, $a=b$, and equals unity for fermionic species, $a=f$, to account for their relative Fermi-Dirac suppression factor of~$7/8$. We reiterate that all particles which are in thermal equilibrium with photons have the same temperature $T_a = T$, but the temperature of decoupled species may be different, $T_a \neq T$. In the Standard Model, this is only relevant for neutrinos after electron-positron annihilation (cf.~\textsection\ref{sec:cnub}). However, this difference will play a prominent role in Section~\ref{sec:neutrinosDarkRadiation} when we discuss additional light relics which might appear in BSM~models. We also note that the effective number of relativistic degrees of freedom $g_*$ is approximately constant away from mass thresholds, $T \sim m_a$, but decreases when a species becomes non-relativistic and its contribution to the energy density becomes negligible.

\begin{figure}
	\centering
	\includegraphics{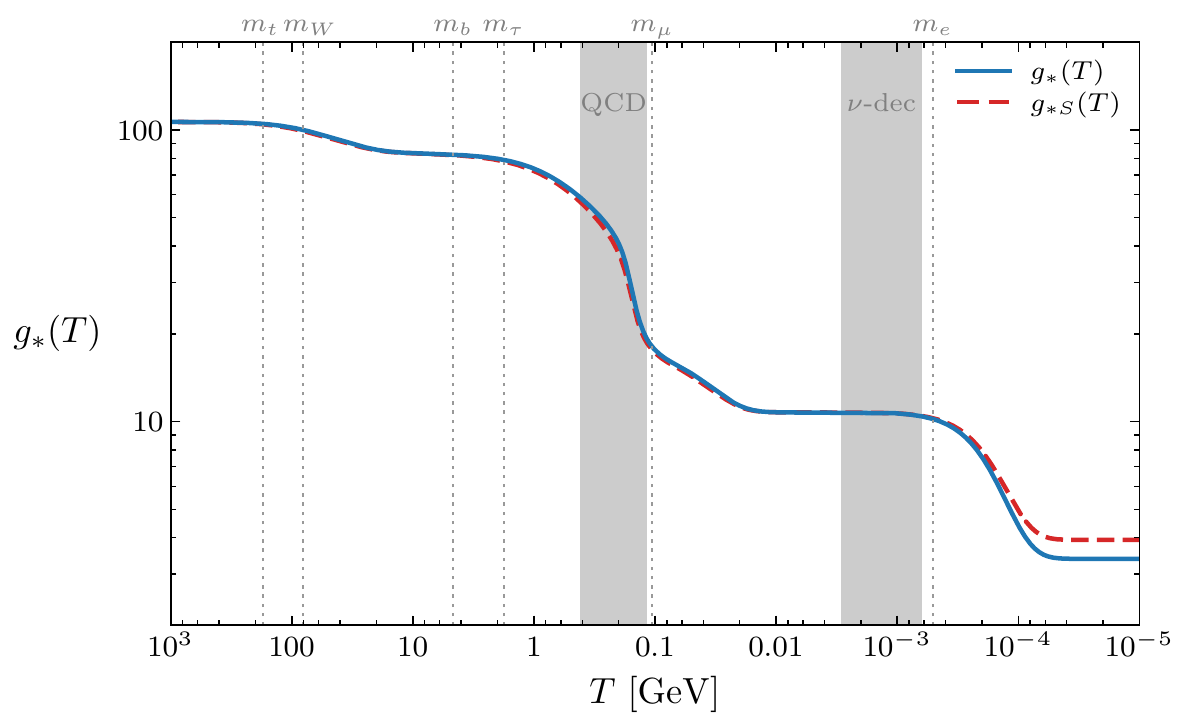}\vspace{-3pt}
	\caption{Evolution of the effective number of relativistic degrees of freedom and those in entropy, $g_*(T)$ and $g_{*S}(T)$, assuming the SM~particle content. We used the state-of-the-art lattice QCD~calculations of~\cite{Borsanyi:2016ksw} for $\SI{20}{\mega\electronvolt} \lesssim T \lesssim \SI{150}{\giga\electronvolt}$ and numerically evaluated the left-hand sides of~\eqref{eq:relativisticEnergyDensity} and~\eqref{eq:entropyDensity} otherwise. The small differences between~$g_*(T)$ and~$g_{*S}(T)$ for $T \gtrsim \SI{1}{\mega\electronvolt}$ arise from non-perturbative QCD~effects which are not captured in~\eqref{eq:gStarS}. The gray bands indicate the QCD~phase transition and neutrino decoupling, and the dotted lines denote some of the mass scales at which SM~particles and anti-particles annihilate.\vspace{-2pt}}
	\label{fig:gStarEvolutionSM}
\end{figure}

The number of internal degrees of freedom $g_a$ of a particle species depends on some of its properties (cf.\ Table~\ref{tab:degreesOfFreedom}).%
\begin{table}[b!]
	\newlength{\wordlength}
	\setlength{\wordlength}{1.5cm}
	\newcolumntype{x}{>{\centering\arraybackslash\hspace{0pt}}p{\wordlength}}
	\begin{tabular}{c x x x x x x} 
			\toprule
		Spin 	& \multicolumn{2}{c}{$0$}	& \multicolumn{2}{c}{$1/2$}	& \multicolumn{2}{c}{$1$}	\\
			\midrule[0.065em]
		Type	& Real		& Complex		& Weyl		& Dirac			& Massless	& Massive		\\
		$g_a$	& 1			& 2				& 2			& 4				& 2			& 3				\\
			\bottomrule 
	\end{tabular}
	\caption{Number of internal degrees of freedom,~$g_a$, for scalar particles, spin-$1/2$~fermions and vector bosons.}
	\label{tab:degreesOfFreedom}
\end{table}
For example, real scalar particles carry one degree of freedom, $g_s=1$, while Weyl fermions have two spin states, $g_f = 2$. Since massless vector bosons are transversely polarized, they provide $g_v = 2$ degrees of freedom, while their massive counterparts have an additional longitudinal polarization, resulting in $g_v = 3$. Accounting for all SM~particles, there are effectively $g_{*,\mathrm{SM}} = 106.75$ relativistic degrees of freedom at high temperatures. After electron-positron annihilation around $T \sim \SI{0.5}{\mega\electronvolt}$, the only relativistic SM~particles are photons and neutrinos with $g_\gamma = 2$ and $g_\nu = 3\cdot2 = 6$. The entire evolution of~$g_*(T)$ within the Standard Model is displayed in Fig.~\ref{fig:gStarEvolutionSM}.%
Since the quarks and gluons are confined into hadrons during the QCD~phase transition, and pions are the only relativistic composite particles afterwards, the value of $g_*(T)$ is reduced by about one order of magnitude around $T \sim \SI{150}{\mega\electronvolt}$. This will be of further importance in subsequent chapters and dramatically impact the detectability of light BSM~relics which might have decoupled at very early times.\medskip

When describing the thermal history of the universe, it is convenient to track conserved thermodynamic quantities. According to the second law of thermodynamics, the entropy of a system can only be constant or increase. For relativistic species, the entropy density is given by
\beq
s = \sum_a s_a = \sum_a \frac{\bar{\rho}_a + \bar{P}_a}{T_a} = \frac{2\pi^2}{45} g_{*S}(T)\, T^3 \, ,	\label{eq:entropyDensity}
\eeq
where we defined the effective number of relativistic degrees of freedom in entropy
\beq
g_{*S}(T) = \sum_a g_a \left(\frac{T_a}{T}\right)^{\!3} \left(1 - \frac{1}{8} \delta_{af}\right) ,	\label{eq:gStarS}
\eeq
in analogy to the definition of~$g_*(T)$ in~\eqref{eq:gStar}. In equilibrium, the entropy in a comoving volume is, in fact, conserved which implies $s a^3 = \const$. Moreover, the entropy is approximately constant~even out of equilibrium because any non-equilibrium entropy production, e.g.\ from decaying particles,~is usually small compared to the large entropy in photons. We can therefore treat the expansion of the universe as basically adiabatic. When particles and anti-particles annihilate, the released entropy is then redistributed among all species in thermal equilibrium. This implies that the temperature of the thermal bath redshifts slightly less than without the annihilation events,~$T \propto g_{*S}^{-1/3} a^{-1}$.

\subsubsection{Beyond equilibrium}
If the SM~particles had remained in thermal equilibrium throughout the history of the universe, $\Gamma \gg H$, any past events would be irrelevant. In this case, our universe would almost entirely consist of photons and would not be an interesting place. Deviations from equilibrium, $\Gamma \lesssim H$, are therefore crucial. The full evolution of a species~$a$ is determined by the Boltzmann equation, which is schematically given by
\beq
\frac{\d f_a}{\d t} = C\!\left[f_a,\{f_b\}\right] .
\eeq
This provides the time evolution of the distribution function~$f_a(t,\x,\p)$ of each particle species~$a$ as a function of its interactions with all other particles in the system. These interactions are captured by the collision term~$C$ on the right-hand side. Solving the time evolution of an entire system may therefore involve a set of Boltzmann equations which can become computationally involved. If an exact treatment is not necessary, it is therefore advantageous to follow an approximate scheme.

At early times, the interactions are frequent enough that they keep the SM~particles in thermal equilibrium. At some point, however, the interaction rate of a species may become of equal size to the Hubble rate,
\beq
\Gamma(T) \sim H(T)\, ,
\eeq
and these particles freeze out, i.e.\ they loose their thermal contact with other species and decouple. This is, of course, not an instantaneous phenomenon at a specific freeze-out temperature~$\Tf$ defined by $\Gamma(\Tf) = H(\Tf)$. Having said that, freeze-out usually happens faster than a few Hubble times and the instantaneous decoupling limit often provides rather accurate estimates for the relic abundances at temperatures $T \ll \Tf$ when $\Gamma \ll H$. We will therefore usually work within this approximation and avoid solving a set of Boltzmann equations.

Given the instantaneous decoupling limit, we can get qualitative insights into the competition between the interaction and the expansion rate in the early universe. By the first Friedmann equation~\eqref{eq:FriedmannEquations}, the Hubble rate in the radiation-dominated epoch is given by
\beq
H(T) = \sqrt{\frac{\bar{\rho}_r(T)}{3\Mp^2} } = \sqrt{\frac{\pi^2}{90} g_*(T)}\, \frac{T^2}{\Mp}\, ,
\eeq
where we inserted~\eqref{eq:relativisticEnergyDensity} and defined the reduced Planck mass $\Mp \equiv (8\pi G)^{-1/2} \approx \SI{2.4e18}{\giga\electronvolt}$. Away from mass thresholds, in particular for $T\gtrsim\SI{100}{\giga\electronvolt}$, the Hubble rate therefore has a quadratic temperature dependence, $H \propto T^2$. This means that any interaction with rate $\Gamma \propto T^m$ and $m>2$ has the chance to be in thermal equilibrium at high enough temperatures and freeze out at some later point. In the SM, the particles generally follow $\Gamma \propto n \propto T^3$ above the electroweak symmetry breaking~(EWSB) scale which implies that all particles (except gravitons which only have Planck-suppressed couplings) are in thermal equilibrium in the early universe. On the other hand, particles whose interactions were governed by rates $\Gamma \propto T^m$ with $m<2$ would not be in thermal equilibrium at early times, but might have the possibility to come into thermal equilibrium later. This phenomenon is often referred to as freeze-in.

\subsection{Cosmic Neutrino Background}
\label{sec:cnub}
We are now in the position to discuss the cosmic neutrino background which is one of the main subjects of this thesis. Neutrinos interact with the rest of the Standard Model only through the weak force. Because neutrinos are, in fact, the most weakly interacting SM~particles, they are the first to decouple from the primordial plasma. There are in particular two processes, which keep these particles in thermal equilibrium at high temperatures: pair conversion of neutrinos into leptons (especially electrons and positrons), $\nu+\bar{\nu} \leftrightarrow l+\bar{l}$, and neutrino scattering off of leptons, $\nu+l \leftrightarrow \nu+l$. For energies far below the mass of the weak gauge bosons, $T \ll \SI{80}{\giga\electronvolt}$, the corresponding cross section is $\sigma \sim G_F^2 T^2$, with Fermi's constant $G_F \approx \SI{1.2e-5}{\per\giga\electronvolt\squared}$. The weak interaction rate for neutrinos is therefore given by $\Gamma_\nu \sim G_F^2 T^2 \bar{n}_e \propto G_F^2 T^5$, where~$\bar{n}_e$ is the electron number density in equilibrium. Neutrinos thus freeze out around a temperature of
\beq
T_{F,\nu} \sim \left( \frac{\sqrt{g_*}}{G_F^2 \Mp} \right)^{\!-1/3} \sim \SI{1}{\mega\electronvolt}\, ,
\eeq
when photons, electrons/positrons and neutrinos were the only relativistic particles left in the primordial plasma. After neutrinos decouple, they maintain their relativistic Fermi-Dirac distribution function $f_\nu(p) \approx [\exp(p/T_\nu)+1]^{-1}$. Since the momentum~$p$ redshifts according to $p \propto a^{-1}$, the neutrino temperature has to have the same scaling and evolve as $T_\nu \propto a^{-1}$.

Since we have measured the temperature of photons~$T$ very well from the black-body spectrum of the~CMB (see below), we want to relate the neutrino temperature~$T_\nu$ to this quantity. Until electrons and positrons annihilate shortly after neutrinos decouple, the temperatures are of course the same, $T_\nu = T$. As noted above, however, the annihilation process releases the entropy of electrons and positrons. As a consequence, over some short period, the photon bath cools more slowly than $T_\nu \propto a^{-1}$, which implies $T > T_\nu$ henceforth. In the limit of perfect and instantaneous neutrino decoupling,\footnote{We will come back to this assumption and correct for it in Section~\ref{sec:neutrinosDarkRadiation}.} electrons/positrons are the only relativistic particles remaining in thermal equilibrium with photons before the annihilation. The effective numbers of the relativistic degrees of freedom in entropy, which are in thermal equilibrium at the temperatures~$T_>$ and~$T_<$ before and after the annihilation, are therefore given by 
\beq
g_{*S}(T_>) = g_\gamma + \frac{7}{8} g_e = 2 + \frac{7}{8} \cdot 4 = \frac{11}{2}\, , \qquad g_{*S}(T_<) = g_\gamma = 2\, ,
\eeq
respectively. Since the entropy~\eqref{eq:entropyDensity} of the thermal bath is conserved in a comoving volume and $T_\nu \propto a^{-1}$, we can easily relate the photon and neutrino temperatures:
\beq
g_{*S}\, \frac{T^3}{T_\nu^3} = \const	\quad \Rightarrow \quad	\frac{T_\nu}{T_\gamma} = \left(\frac{g_{*S}(T_<)}{g_{*S}(T_>)}\right)^{\!1/3} = \left(\frac{4}{11}\right)^{\!1/3} \,.	\label{eq:neutrinoTemperature}
\eeq
The measured photon temperature, $T_0 = \SI{2.7255}{\kelvin}$, implies that the C$\nu$B has a thermal spectrum with a temperature of $T_{\nu,0} \approx \SI{1.95}{\kelvin}$ today. In consequence, the number density of cosmic neutrinos is very large, $\bar{n}_\nu \approx \SI{112}{\per\centi\meter\cubed}$, and, in fact, exceeds the flux from astrophysical neutrino sources, such as our Sun. Nevertheless, the direct detection of the~C$\nu$B is extremely challenging because the neutrino distribution peaks at the very small energy of $T_{\nu,0} \approx \SI{0.17}{\milli\electronvolt}$. Having said that, as we will also show in this thesis, there is more and more indirect evidence for its existence.

\subsection{Cosmic Microwave Background}
\label{sec:cmb}
A very important event in the thermal history of the universe is the formation of the first atoms and the decoupling of photons. The observation of these photons, which make up the cosmic microwave background, has ultimately led to a number of major breakthroughs in modern cosmology and is still one of the main sources of information about our universe.\medskip

From the formation of the C$\nu$B, we fast forward about \num{250000}~years in cosmic history. In the meantime, the light elements, in particular helium, were synthesized during big bang nucleosynthesis and the universe entered into the matter-dominated epoch. The primordial plasma at that time consisted of (many) photons, free electrons and ionized nuclei (mostly protons). Compton scattering, $e^- + \gamma \leftrightarrow e^- + \gamma$, tightly coupled photons and electrons resulting in a small mean free path for photons. In turn, electrons strongly interacted with protons via Coulomb scattering, $e^- + p^+ \leftrightarrow e^- + p^+$. Finally, electromagnetic reactions such as those forming/ionizing neutral hydrogen, $e^- + p^+ \leftrightarrow H + \gamma$, kept the baryons and photons in equilibrium. However, once the universe cooled to a temperature of about \SI{0.4}{\electronvolt}, the ionization of neutral hydrogen became less and less efficient. This is usually referred to as recombination although it is the first time that electrons and protons combined without being ionized again immediately.

By this time, photon-electron scattering is governed by Thomson scattering, which is the low-energy limit of Compton scattering and has an interaction rate of $\Gamma_\gamma \sim n_e \sigma_T$, with constant Thomson cross section~$\sigma_T$. However, the number density of free electrons~$n_e$ got reduced dramatically by the increasing amount of neutral hydrogen and the Thomson scattering rate~$\Gamma_\gamma$ dropped. When $\Gamma_\gamma \lesssim H$, the mean free path of photons became longer than the horizon size and the photons decoupled from matter. This happened at a temperature of about $\Trec \approx \SI{0.26}{\electronvolt}$ corresponding to a redshift of $z \approx 1090$. These photons have since freely streamed through the universe effectively unimpeded and are what we observe as the CMB today.\medskip

Recombination is not instantaneous, but requires a finite time, i.e.\ some photons encountered their last scattering event earlier than others while the plasma was still hotter.\footnote{Recombination and (photon) decoupling are often used synonymously, in particular when referring to the time of decoupling, but actually are different processes. We will keep these two notions distinct in those cases where it is important and it will be apparent in all other cases.} This results in the so-called last-scattering surface to have a finite width. However, although these photons last-scattered at slightly different temperatures, we observe the CMB with an almost perfect black-body spectrum at a single temperature of $T_0 = \SI{2.7255}{\kelvin}$ today. We do not measure a spectrum comprised of a set of black-body spectra because the photons which decoupled earlier redshifted according to $T \propto a^{-1}$, which is exactly the decrease in temperature that the photons which last-scattered later experienced as well. It was this uniform background that Penzias and Wilson discovered in~1965~\cite{Penzias:1965wn} and whose black-body spectrum was measured exquisitely by the FIRAS~instrument on the COBE~satellite in the early 1990s~\cite{Fixsen:1996nj}. In the main chapters of this thesis, we will get a glimpse of the wealth of information about our universe that is transmitted by these CMB~photons.

\section{Inhomogeneous Cosmology}
\label{sec:inhomogeneousCosmology}
So far, we have discussed the perfectly homogeneous and isotropic universe. This is an extremely good approximation on large scales, but breaks down at smaller distances. On the cosmological scales of interest to us, however, the spacetime is well described by perturbation theory around FLRW. For our applications, it will, in fact, be sufficient to work at linear order in perturbations as the departure from spatial homogeneity and isotropy is small. For instance, the CMB temperature varies across the sky at the level of one part in \num{e4} reflecting small spatial variations in the density of the primordial plasma (see below). Of course, this approach breaks down when high-density regions, such as galaxies, form through the gravitational instability of these small fluctuations. In the following, we will introduce first-order cosmological perturbation theory, and show how the large-scale structure of the universe formed and evolved~(\textsection\ref{sec:structureFormation}). Studying the perturbations in the primordial photon-baryon fluid will reveal the presence of sound waves which have since been imprinted in several cosmological observables~(\textsection\ref{sec:cosmicSoundWaves}).

\subsection{Structure Formation}
\label{sec:structureFormation}
Metric and matter fluctuations are coupled by the Einstein equations and, therefore, have to be treated simultaneously. We write the perturbations of the FLRW~metric~\eqref{eq:FLRWmetric} and of the energy-momentum tensor of a perfect fluid~\eqref{eq:perfectFluid} as
\beq
g_{\mu\nu}(\tau,\x) = \bar{g}_{\mu\nu}(\tau) + \delta g_{\mu\nu}(\tau,\x)\, , \qquad T_{\mu\nu}(\tau,\x) = \bar{T}_{\mu\nu}(\tau) + \delta T_{\mu\nu}(\tau,\x)\, .
\eeq
Due to the coordinate independence of general relativity, these perturbations are not uniquely defined. For example, the metric perturbations can be non-zero, $\delta g_{\mu\nu} \neq 0$, even though the spacetime is described by a perfect FLRW~universe, just in a different set of coordinates, $g_{\mu\nu}(\tau,\x) = \bar{g}_{\mu\nu}(\tau',\x')$. We therefore choose a particular coordinate system (or `fix the gauge') when defining the metric perturbations. It is useful to decompose the perturbations in purely scalar, vector and tensor components which, at linear order, evolve separately under the Einstein equations. In this thesis, we will focus on the scalar degrees of freedom and, therefore, neglect vector and tensor perturbations from now on. We refer to the seminal papers~\cite{Bardeen:1980kt, Kodama:1985bj} (see also~\cite{Mukhanov:1990me, Ma:1995ey, Bashinsky:2003tk, Mukhanov:2005sc, Malik:2008im}, for instance) for further details on these points and general treatments of cosmological perturbation theory. We choose to work in (conformal) Newtonian gauge where the scalar part of the metric is given by
\beq
\d s^2 = a^2(\tau) \left[-(1+2\Phi) \d \tau^2 + (1-2 \Psi) \delta_{ij} \d x^i \d x^j \right] .
\eeq
The name of this gauge stems from the fact that the perturbations $\Phi$ and $\Psi$ are related to the (Newtonian) gravitational potential with~$\Phi$ controlling the motion of non-relativistic particles and~$\Psi$~being determined by the Poisson equation on small scales.

Every energy-momentum tensor has four scalar degrees of freedom which are related to the density $\rho$, the pressure $P$, the bulk velocity $v_i$ and the anisotropic stress $\Sigma_{ij}$. Since the contributions of different species are simply added, we define the perturbed energy-momentum tensor separately for each species $a$ as
\beq
T^0{}_{0,a} = - (\bar \rho_a+\delta\rho_a)\, , \quad T^0{}_{i,a} = (\bar \rho_a+ \bar P_a) v_{i, a}\, , \quad T^i{}_{j,a} = (\bar P_a + \delta P_a)\delta^i_j + (\bar \rho_a + \bar P_a) \Sigma^i{}_{j,a} \, .	\label{eq:stressEnergyTensor}
\eeq
The scalar part of the velocity can be written as $v_{i,a} = -\nabla_i u_a$, where $u_a$ is the velocity potential. Similarly, the anisotropic stress tensor~$\Sigma_{ij,a}$ can be expressed as $\Sigma_{ij,a} =\frac{3}{2} (\nabla_i\nabla_j - \frac{1}{3} \delta_{ij} \nabla^2) \sigma_a$, where~$\sigma_a$ is the scalar potential of the anisotropic stress and the factor of~$\frac{3}{2}$ was introduced for later convenience. Instead of the density perturbation $\delta\rho_a$ we often employ the dimensionless overdensity
\beq
\delta_a \equiv \frac{\delta\rho_a}{\bar \rho_a}\, .
\eeq
The previously introduced equation of state $w_a \equiv \bar P_a/\bar\rho_a$ and the speed of sound $c_a^2 \equiv \delta P_a/\delta\rho_a$ relate the (adiabatic) pressure~$P_a$ to the density~$\rho_a$, which effectively removes the pressure as a free variable for adiabatic fluctuations.\medskip

The evolution of the remaining three matter and two metric perturbations can be derived using the conservation of the energy-momentum tensor and the (linearised) Einstein equations. Energy-momentum conservation for each decoupled species, i.e.\ those without energy and momentum transfer, implies the continuity and Euler equations,
\begin{align}
\dot{\delta}_a	&\,=\, (1+w_{a})\left(\nabla^2 u_a + 3\dot{\Psi}\right) - 3\H\left( \frac{\delta P_a}{\delta\rho_a} - w_{a}\right) \delta_a														\\[4pt]
\dot{u}_a 		&\,=\, - \left[\H(1-3w_{a})  + \frac{\dot{w}_{a}}{1+w_{a}}\right]u_a + \frac{1}{1+w_{a}}\hskip0.5pt\frac{\delta P_a}{\delta\rho_a}\, \delta_a + \nabla^2 \sigma_a + \Phi\, .
\end{align}
These two equations can be combined into a second-order differential equation for the density contrast with a source term comprised of the other three perturbations. For adiabatic fluctuations, this evolution equation simplifies to
\beq
\ddot{\delta}_a + \chi_a \dot{\delta}_a - c_a^2 \nabla^2 \delta_a = (1+w_a)\left(\nabla^4 \sigma_a + \nabla^2\Phi + 3 \ddot{\Psi} + 3 \chi_a \dot{\Psi}\right) ,	\label{eq:combinedConservationEquations}
\eeq
where $\chi_a \equiv \H(1-3 c_a^2)$ is the Hubble drag rate. The two metric potentials $\Phi$ and $\Psi$ are determined by the following first-order Einstein equations:
\begin{align}
\nabla^2 \Psi - 3\H(\dot\Psi + \H \Phi) 													&= 4 \pi G a^2\, \delta \rho \, ,	\label{eq:Poisson}	\\[4pt]
\ddot\Psi + \H(2\dot\Psi + \dot\Phi) + (2\dot\H+\H^2)\Phi + \frac{1}{3}\nabla^2(\Phi-\Psi)	&= 4 \pi G a^2\, \delta P \, ,		\label{eq:evolution}
\end{align}
where $\delta\rho \equiv \sum_a \delta \rho_a$ and $\delta P \equiv \sum_a \delta P_a$ are the total density and pressure perturbations, respectively. Equation~\eqref{eq:Poisson} is known as the relativistic Poisson equation with the density perturbation sourcing the metric potentials. Finally, the spatial trace-free part of the Einstein equations results in the constraint equation
\beq
\Phi - \Psi = - 12\pi G a^2\, (\bar \rho+\bar P) \sigma\, ,	\label{eq:EinsteinC}
\eeq
where $(\bar \rho + \bar P)\hskip1pt \sigma \equiv \sum_a (\bar \rho_a + \bar P_a)\hskip1pt \sigma_a$. This implies that the metric potentials are equal for vanishing~$\sigma_a$.  If a finite anisotropic stress potential~$\sigma_a$ is present, its evolution equation can be obtained from the corresponding (linearised) Boltzmann equation (cf.~\textsection\ref{sec:examples}). In the standard cosmological model, free-streaming neutrinos notably induce a small anisotropic stress, but are essentially the only such source. This closes the system of equations and we can solve the entire evolution at first order in cosmological perturbation theory. We commonly decompose each variable into Fourier modes denoted by the same symbol, e.g.
\beq
\delta_\k(\tau) = \int \!\d^3x\, \e^{-\ii \k \cdot \x}\, \delta(\tau,\x)\, ,
\eeq
and often suppress the mode index, $\delta(\tau) = \delta_\k(\tau)$, for convenience. This decomposition is particularly helpful as each Fourier mode evolves separately under the linear evolution equations.\medskip

When applying these evolution equations to our universe, we attempt to solve an initial value problem. It is convenient to set the initial conditions at sufficiently early times when all scales of interest in current observations were outside the Hubble radius, $k \ll \H$. As current observations strongly suggest adiabatic initial conditions, we will usually assume these in this thesis.\footnote{There may also exist isocurvature perturbations for which the density fluctuations of one species do not necessarily correspond to density fluctuations in other species. These are disfavoured by current observations, in particular those of the CMB anisotropies.} Adiabatic fluctuations are characterised by the fact that the initial overdensities of all species are related according to $(1+w_b)\,\delta_{a,\mathrm{in}} = (1+w_a)\,\delta_{b,\mathrm{in}}$, i.e.\ for example $\delta_{r,\mathrm{in}} = 4\,\delta_{m,\mathrm{in}}/3$. They can equivalently be described as perturbations induced by a common local shift in time~$\delta\tau(\x)$ of all background quantities, $\delta \rho_a(\tau,\x) = \bar{\rho}_a(\tau+ \delta\tau(\x)) - \bar{\rho}_a(\tau)$. From this point of view, it might not be too surprising that generic single-field slow-roll models of inflation provide adiabatic initial conditions for the hot big bang given by 
\beq
\delta_{a,\k}(\tau_\mathrm{in}) = -3(1+w_a)\zeta_\k\, ,
\eeq
where $\tau_\mathrm{in}$ is the initial time and $\zeta$ is the primordial curvature perturbation. The latter is conserved on super-Hubble scales and predicted to follow the almost scale-invariant power spectrum~$\mathcal{P}_{\hskip-0.5pt\zeta}(k)$ defined in~\eqref{eq:primordialSpectrum}. Importantly, the equation of state and the speed of sound in an adiabatically perturbed fluid are approximately equal, $c_a^2 \approx w_a$.\medskip

The growth of these primordial density fluctuations is determined by a competition between gravity and pressure. While gravity attracts matter into overdense regions in the universe, pressure pushes matter out of these regions. This means that gravity leads to a growth of the initial inhomogeneities, whereas pressure will inhibit this growth. Specializing to matter perturbations with $c_m^2 \approx w_m \approx 0$, the evolution equation~\eqref{eq:combinedConservationEquations} implies in the subhorizon limit, $k \gg \H$, after time-averaging the gravitational potentials over a Hubble time\hskip1pt\footnote{The radiation perturbations oscillate on small scales (cf.~\textsection\ref{sec:cosmicSoundWaves}). After time-averaging over a Hubble time, these perturbations can however be neglected and the potentials are only sourced by the matter fluctuations~\cite{Weinberg:2002kg}. We can therefore neglect the time derivatives of the potentials on subhorizon scales, $k^2\Phi \gg \ddot{\Psi},\H\dot{\Psi}$.} that the density perturbations only grow logarithmically during the radiation-dominated era (due to the large photon pressure), $\delta_m \propto \ln a$, but linearly while the universe is dominated by matter, $\delta_m \propto a$.\footnote{Once dark energy takes over as the main component of the universe, the clustering of matter stops and the growth of structures is halted by the accelerated expansion of the universe.} The sub-horizon gravitational potential $\Phi_\k$, on the other hand, oscillates with a decaying amplitude~$\propto a^{-2}$ during radiation domination, but approaches a constant after matter-radiation equality. In contrast, the super-horizon modes of the gravitational potential do not evolve in either epoch. The combination of these power laws gives rise to the characteristic shape of the matter power spectrum (cf.~\textsection\ref{sec:lss}). Before we provide further details on the growth of structure, we will discuss photon perturbations because they also leave a small, but distinct imprint which is a key observable.

\subsection{Cosmic Sound Waves}
\label{sec:cosmicSoundWaves}
In the following, we study perturbations in the photon-baryon fluid of the early universe. It will turn out that the initial fluctuations excited sound waves in the primordial plasma which are observed today as the so-called baryon acoustic oscillations in both the anisotropies of the cosmic microwave background (see~\textsection\ref{sec:cmbAnisotropies}) and the clustering of galaxies (see~\textsection\ref{sec:lss}). We will give an approximate description of the main features of these observables and refer to the excellent reviews~\cite{Hu:1995em, Hu:2001bc, Challinor:2009tp} for a more detailed treatment.\medskip

Prior to recombination, photons, electrons and protons were tightly coupled through Thomson and Coulomb scattering in the photon-baryon fluid. The evolution equation for the density perturbations in this fluid can be obtained from~\eqref{eq:combinedConservationEquations} and is given by
\beq
\ddot{\delta}_\gamma + \frac{\H R}{1+R} \dot{\delta}_\gamma - c_s^2 \nabla^2 \delta_\gamma = \frac{4}{3}\nabla^2 \Phi + 4 \ddot{\Psi} + \frac{4 \H R}{1+R} \dot{\Psi} \, ,	\label{eq:masterEquation}
\eeq
where we introduced the momentum density ratio of baryons to photons $R \equiv 3\bar{\rho}_b/(4\bar{\rho}_\gamma)$. We also defined the sound speed in the fluid $c_s^2 \equiv 1/[3 (1+R)]$ which is smaller than the standard value for a relativistic fluid, $c_s^2 = 1/3$, because the presence of baryons adds inertia to the fluid. The forced harmonic oscillator equation~\eqref{eq:masterEquation} essentially governs the entire BAO phenomenology. The metric potentials on the right-hand side evolve as determined by the matter in the universe (including neutrinos and dark matter). They source the fluctuations in the photon-baryon fluid on the left-hand side which are in turn supported by photon pressure and damped by Hubble friction. Note that the anisotropic stress of photons vanishes as it can only develop effectively after decoupling when the photons begin to stream freely.\medskip

To extract the general phenomenology of solutions to the master equation~\eqref{eq:masterEquation}, we make a few simplifying assumptions.\footnote{Accurately computing the evolution of all perturbations in the universe requires solving many coupled equations as the interactions between the various species have to be captured by a set of Boltzmann equations. This can only be done numerically which is achieved in the current state-of-the-art Boltzmann solvers \texttt{CAMB}~\cite{Lewis:1999bs} and~\texttt{CLASS}~\cite{Blas:2011rf}. Nevertheless, it is instructive to obtain approximate analytic solutions and get analytic insights in order to deepen our understanding of the underlying physics (cf.\ e.g.\ Chapter~\ref{chap:cmb-phases}).} However, we will revisit these considerations in~\textsection\ref{sec:cmbAnisotropies} and especially in Chapter~\ref{chap:cmb-phases}. First, we neglect the small anisotropic stress due to neutrinos, i.e.\ we set $\Phi = \Psi$ according to~\eqref{eq:EinsteinC}. Moreover, we also ignore the time dependence of~$\Phi$ for now.\footnote{This is a good approximation in the matter-dominated era. During radiation domination, the time evolution of the gravitational potential around sound-horizon crossing leads to the radiation-driving effect which we will discuss below.} Defining $\Theta \equiv \frac{1}{4}\delta_\gamma + \Phi$, equation~\eqref{eq:masterEquation} can then be written as
\beq
\ddot{\Theta} + \frac{\H R}{1+R} \dot{\Theta} - c_s^2 \nabla^2 (\Theta+ R\Phi) = 0\, .
\eeq
Since the baryon-photon ratio evolves as $R \propto a$, the damping term is proportional to $\H R = \dot{R}$. Similar to the time evolution of the gravitational potentials, the baryon-photon ratio $R$ also changes on much larger time scales than $\Theta$. Treating~$R$ as approximately constant, the evolution equation is therefore simply given by the differential equation of a harmonic oscillator with mode-dependent frequency~$c_s k$,
\beq
\ddot{\vartheta}_\k + c_s^2 k^2 \vartheta_\k = 0\, ,
\eeq
where we introduced $\vartheta \equiv \Theta + R \Phi = \frac{1}{4}\delta_\gamma + (1+R)\Phi$. The solutions to this equation are of course sound waves:\footnote{Note that we could of course rewrite this solution in terms of an amplitude~$A_\k$ and a non-zero phase~$\phi_\k$, i.e. $\cos(c_s k \tau) \to \cos(c_s k \tau + \phi_\k)$, with $B_\k = 0$ implying $\phi_\k = 0$.}
\beq
\vartheta_\k(\tau) = A_\k \cos(c_s k \tau) + B_\k \sin(c_s k \tau)\, .
\eeq
Imposing adiabatic initial conditions on superhorizon scales sets $B_\k \equiv 0$ and $A_\k = 3\zeta_\k$. The quantity~$\Theta$ consequently evolves according to
\beq
\Theta(\k,\tau) = 3 \zeta_\k\, \cos(c_s k \tau) - R\Phi_\k\, .	\label{eq:soundWaves}
\eeq
The presence of baryons, $R \neq 0$, therefore not only changes the sound speed~$c_s$ of the photon-baryon fluid, but also moves the equilibrium point of the oscillations from~$0$ to~$-R\Psi$. This effect is often referred to as `baryon loading' because baryons change the balance of pressure and gravity. Since photons decouple during recombination, we should evaluate these solutions at $\tau = \tau_\rec$.\footnote{As decoupling happens during the epoch of recombination, we use~$\tau_\rec$ instead of~$\tau_F$ or~$\tau_\mathrm{dec}$ to specify the time of photon decoupling. This will allow easier discrimination of other freeze-out events in later parts of this thesis.} At this time, modes with wavenumbers $k_n = n \pi/r_s$ had their extrema. This implies that the sound horizon at decoupling, $r_s \equiv c_s \tau_\rec \approx \tau_\rec /\sqrt{3}$,\footnote{To be precise, the sound horizon is given by $r_s(\tau) = \int_0^\tau \!\d\tau\, c_s(\tau)$, which also captures the small time dependence of~$c_s$ that we however neglect in our analytic treatment.} is imprinted as a fundamental scale in the photon fluctuations that we can still observe today.

Up to now, we have considered the background to be essentially fixed. However, both the baryon-photon ratio and the gravitational potentials actually evolve. Including the evolution of the background densities in~$R$ leads to the photon fluctuations being damped over time. While the gravitational potentials remain constant in the matter era, their oscillating amplitude decays proportional to $a^{-2}$ inside the horizon during the radiation-dominated epoch. Because the decay is due to photon pressure and happens when the photon-baryon fluid is in its most compressed state, the fluid bounces back without a counterbalancing effect from the gravitational potential. Since those fluctuation modes that entered the horizon during matter domination do not experience this radiation-driving effect, the amplitude of the photon perturbations in the subsequent rarefaction stage is enhanced in comparison.

Finally, we have to take the finite mean free path of photons into account. In the master equation~\eqref{eq:masterEquation}, we assumed that photons and baryons are so tightly coupled that we can treat them as a single fluid. In reality, however, the mean free path of photons, which is given by the inverse Thomson scattering rate of photons with electrons, $\lambda_\mathrm{mfp} = 1/(a \sigma_T n_e)$ (in comoving coordinates), is small but finite even before decoupling. This diffusion process results in an additional damping of fluctuation modes. (Note that this damping effect is completely separate from the damping related to $\dot{R} \neq 0$ which we have just discussed.) Intuitively, this is caused by photons washing out inhomogeneities in the primordial plasma that are smaller than their mean free path. Small scales (large $k$) are therefore exponentially more damped than large scales according to $\exp[-(k/k_d)^2]$, where $k_d$ is the wavenumber associated with the mean squared diffusion distance at decoupling. A careful treatment, which includes corrections from the polarization of these photons, gives~\cite{Zaldarriaga:1995gi}
\beq
k_d^{-2} \equiv \int^{a_\rec}_0 \frac{\d a}{a^3 \sigma_T n_e H} \frac{R^2 + \tfrac{16}{15}(1+R)}{6(1+R)^2} \, ,	\label{eq:damping}
\eeq
where $a_\rec$ is the scale factor at decoupling. Since the diffusion scale depends on an integral over the Hubble parameter $H$, we see that the damping of the fluctuations in the photon-baryon fluid is sensitive to the expansion history of the early universe.\medskip

\begin{figure}
	\centering
	\definecolor{baryon}{RGB}{69, 95, 181}
	\definecolor{photon}{RGB}{205, 30, 37}
	\definecolor{neutrino}{RGB}{203, 89, 30}
	\includegraphics{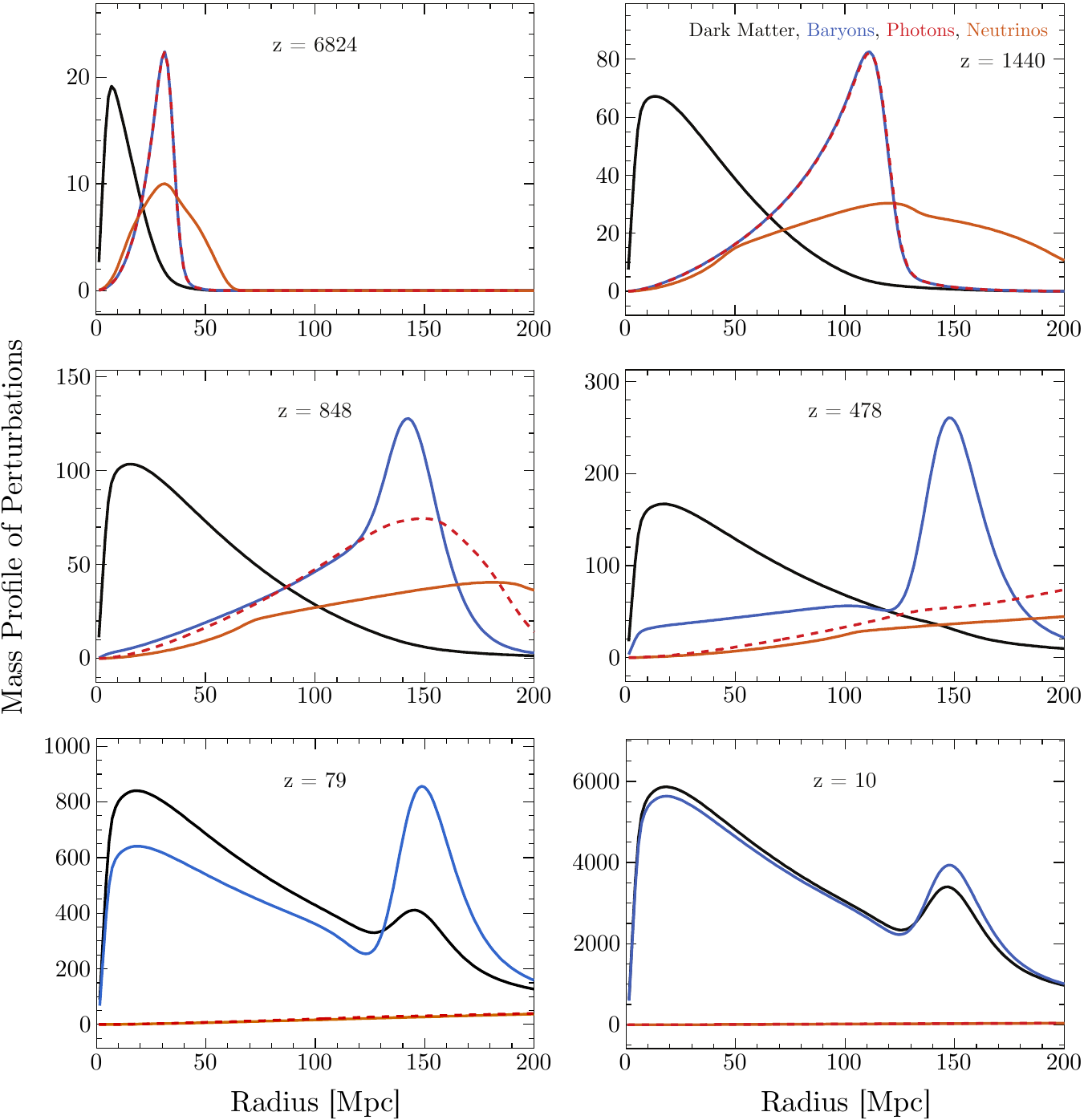}
	\caption{Evolution of the radial mass profile (density times radius squared) of initially point-like dark matter, \textcolor{baryon}{baryon}, \textcolor{photon}{photon} and \textcolor{neutrino}{neutrino} overdensities located at the origin as a function of the comoving radius (adapted from~\cite{Eisenstein:2006nj}). All perturbations are fractional for that species and the photon and neutrino fluctuations were divided by~$4/3$ to put them on the same scale. The units of the mass profile are arbitrary, but are correctly scaled between the panels for synchronous gauge~\cite{Weinberg:2008zzc}. We observe photons and baryons travelling outwards due to radiation pressure. After the decoupling of photons, they stream freely at the speed of light, whereas the baryon perturbation is left behind in a shell. Dark matter remains concentrated at the origin, but partly falls into the gravitational potential created by the photons and baryons. The baryon and dark matter fluctuations grow as $\delta_m \propto a$ due to gravitational instability in the matter-dominated epoch, and finally trace each other. Today, we can observe the location of the matter shell at about \SI{150}{\Mpc} in the distribution of galaxies. Since neutrinos have been free-streaming close to the speed of light since they decoupled around $z \sim \num{e10}$, they may travel ahead of the sound horizon.}
	\label{fig:overdensities}
\end{figure}

So far, we have focussed on photon fluctuations, but baryons inherit the same variations since they are tightly coupled to photons before the CMB is released. It is instructive to discuss the evolution of the baryon perturbations in real space instead of Fourier space and consider a single initial overdensity (following~\cite{Bashinsky:2002vx, Eisenstein:2006nj}) because, for adiabatic fluctuations, the primordial density field is a superposition of such point-like overdensities. As illustrated in Fig.~\ref{fig:overdensities},%
the overdensities of photons and baryons spread out as spherical shells, while the dark matter perturbation does not move much and is left behind at the centre. After photon decoupling around $z\sim1100$, the sound speed drops dramatically and the pressure wave slows down, producing a shell of gas at about~\SI{150}{\Mpc} from the point of the initial overdensity. Subsequently, baryons fall into the dark matter potential well. At the same time, the baryonic shell also attracts the dark matter which therefore develops the same density profile with a peak at the same radius.

Going back to Fourier space, the presence of the baryonic shell corresponds to oscillations whose frequency is determined by the distance of propagation of the primordial sound waves, i.e.\ the sound horizon~$r_s$.\footnote{The size of the sound horizon imprinted in the baryon perturbations is slightly larger than the size observed in the~CMB anisotropies. The latter is set at the time when most photons had decoupled from the baryons. At this point, the baryons however still feel the drag of photons and essentially remain coupled to the photons because there are about \num{e9}~times more photons than baryons. The end of the so-called drag epoch, $\tau_\mathrm{drag} > \tau_\rec$, marks the time when baryons finally loose this contact. The two sizes of the sound horizon inferred from the latest CMB~measurements are $r_s(z_\rec \approx 1090) \approx \SI{145}{\Mpc}$ and $r_s(z_\mathrm{drag} \approx 1060) \approx \SI{148}{\Mpc}$~\cite{Ade:2015xua}, i.e.\ they are relatively close, but different at a significance of more than~$8\sigma$ at the current level of precision.} At late times, galaxies formed preferentially in the regions of enhanced dark matter density. For the most part, these are located where the initial overdensities were, but there is a small (about \SI{1}{\percent}) enhancement in the regions roughly $r_s \sim \SI{150}{\Mpc}$ away from these positions. Consequently, there should be a small excess of galaxies \SI{150}{\Mpc}~away from other galaxies. This is how the acoustic scale is imprinted in the two-point correlation function of galaxies (see below) as the so-called BAO~peak. As a result, we can observe the remnants of cosmic sound waves today in both the CMB~anisotropies and in the large-scale structure of the universe. We will discuss these cosmological observables in the next section.

Finally, let us briefly comment on the mathematical form of the BAO~signal. From our discussion so far, one might expect the same pure cosine shape for the matter oscillations as for the photon perturbations in~\eqref{eq:soundWaves}. However, we actually observe a pure sine solution, $\delta_m \sim \sin(k r_s)$. Heuristically, this can be understood as follows. Assuming decoupling to be instantaneous, we match the baryon perturbation and its time derivative at the last-scattering surface onto the growing and decaying mode of the matter fluctuations. Since $\dot{\delta}_b \sim c_s k\, \sin(k r_s)$, with $r_s = c_s \tau$, the time derivative dominates over $\delta_b \sim \cos(k r_s)$ for large wavenumbers~$k$. In this way, the pure sine solution gets imprinted in the matter perturbations, with the initial conditions being fixed at the time of recombination (see e.g.~\cite{Slepian:2015zra} for an analytic treatment).

\section{Cosmological Observables}
\label{sec:cosmologicalObservables}
So far, we have explained how initially small fluctuations grow and evolve throughout the history of the universe. In particular, we have seen that cosmic sound waves are excited. In the following, we will relate the photon and matter perturbations to the quantities that we actually observe. This includes temperature anisotropies as well as polarization in the CMB sky and the locations of galaxies in the universe. These are the observables that we will use in the rest of the thesis to gain new insights into the early universe and particle physics. Although this is only a subset of the possible and currently employed observables, they dominate the cosmological information that has been inferred to date. We refer to the review literature~\cite{Bartelmann:1999yn, Carlstrom:2002na, Furlanetto:2006jb, Weinberg:2012es, Kilbinger:2014cea} for a comprehensive discussion of additional observables such as weak gravitational lensing, galaxy clusters, the \mbox{Lyman-$\alpha$}~forest and \SI{21}{\centi\metre}~tomography.\medskip

Understanding the physics behind these observables is important in the quest to uncover the laws of Nature. At the same time, we also have to know how the data can be best characterised and compared to theoretical expectations in a quantitative way. The most important statistic when studying both the cosmic microwave background and the large-scale structure is the two-point correlation function in real space or, equivalently, the power spectrum in Fourier space. These quantities contain all of the statistical information if the perturbations are drawn from a Gaussian distribution function. Since the initial conditions predicted by inflation are very nearly Gaussian, the primordial perturbations are almost entirely described by the primordial correlation function $\left\langle \zeta(\x)\, \zeta(\x') \right\rangle \equiv \xi_\zeta(\x,\x') = \xi_\zeta(|\x -\x'|)$, where we employed statistical homogeneity and isotropy in the second equality. The primordial power spectrum~$\mathcal{P}_{\hskip-0.5pt\zeta}(k)$ is then defined as the Fourier transform of~$\xi_\zeta(x)$ and given by\vspace{-1pt}
\beq
\left\langle \zeta_{\k\vphantom{\k'}}^{\vphantom{*}}\, \zeta^*_{\k'} \right\rangle = \frac{(2\pi)^3}{k^3} \, \mathcal{P}_{\hskip-0.5pt\zeta}(k)\, \delta_D^{(3)}(\k - \k')\, ,
\eeq
where $\delta_D^{(3)}$ is the three-dimensional Dirac delta function. As long as density perturbations can be described in linear theory, their modes evolve independently and the power spectrum still captures most of the information. Having said that, the Einstein equations are inherently non-linear and the gravitational evolution will always couple different modes which introduces non-Gaussianities and, consequently, non-vanishing (connected) higher-point functions. Nevertheless, linear perturbation theory is reliable for many scales of interest, in particular for those modes measured in the~CMB. As a consequence, the statistical analysis of~CMB and LSS~data heavily relies on the power spectrum at the moment.

\subsection{CMB Anisotropies}
\label{sec:cmbAnisotropies}
Since the cosmic microwave background anisotropies were first observed by the DMR~instrument on the COBE~satellite in~1992~\cite{Smoot:1992td}, they have proven to be a treasure-trove of information and have played a pivotal role in establishing the standard cosmological model introduced in~\textsection\ref{sec:StandardModelCosmology}. In the following, we will establish the relation between the photon fluctuations on the last-scattering surface and the measured CMB power spectra, and illustrate how general properties of our universe can be deduced from its characteristic shape. We will also discuss the generation of CMB polarization and mention gravitational lensing. More detailed derivations and an overview of current developments can be found in~\cite{Hu:1995em, Hu:1997hv, Hu:2001bc, Dodelson:2003cos, Staggs:2018gvf}.

\subsubsection{Temperature anisotropies}
\begin{figure}
	\centering
	\includegraphics{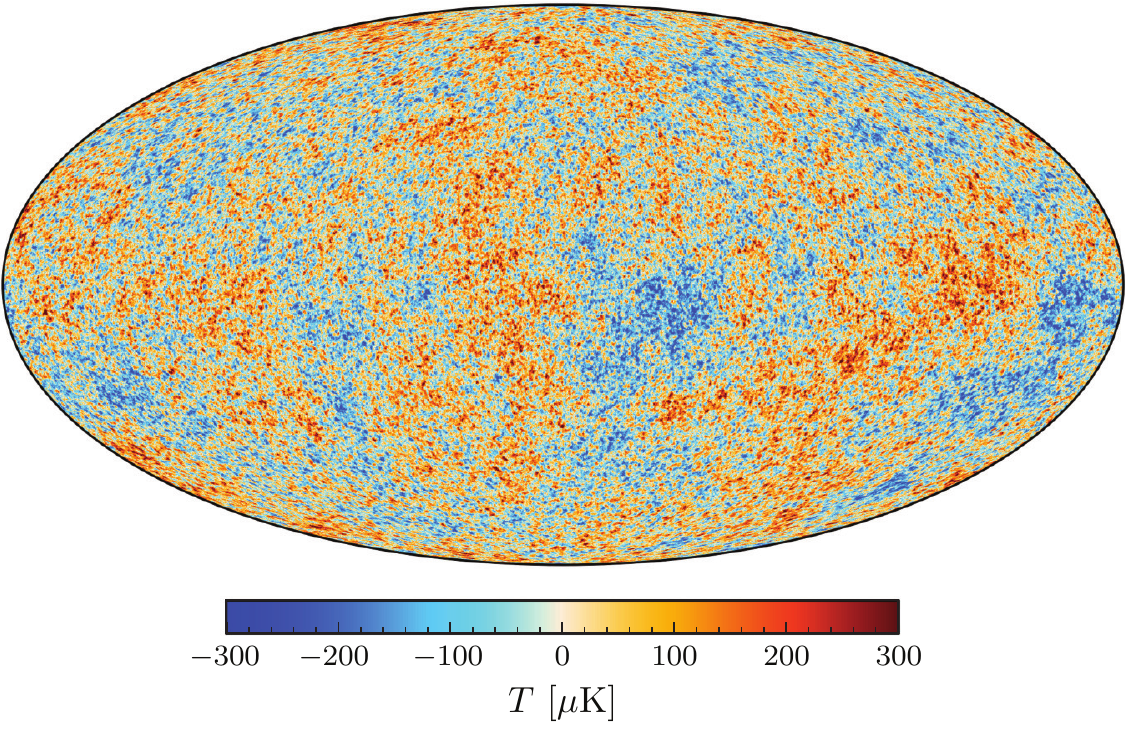}
	\caption{CMB intensity map at \SI{5}{arcmin} resolution based on Planck observations using the SMICA~component separation algorithm (adapted from~\cite{Adam:2015rua}). A small strip of the Galactic plane was masked and subsequently filled in by a constrained realization with the same statistical properties as the rest of the sky. The characteristic spot size is about \SI{1}{\degree}.}
	\label{fig:CMBmap}
\end{figure}

Generally speaking, CMB~experiments map the sky at microwave lengths\hskip1pt\footnote{In this context, CMB~experiments are surveys which map the CMB~anisotropies in the sky. These measurements are usually not taken at a single, but at several frequencies. This is to reliably subtract galactic and astrophysical foregrounds, which are other sources of microwave emission and polarization originating, in particular, from galactic dust. We will generally assume that these foregrounds have been accounted for (or their effects can easily be marginalized over) so that we have direct access to the primordial signal.} and measure the intensity and polarization of the incident photons. As discussed in~\textsection\ref{sec:cmb}, these photons have a mean temperature of $\bar{T} = \SI{2.7255}{\kelvin}$. We are now interested in the deviations $\delta\tilde{T}(\hat{\n}) = T(\hat{\n}) - \bar{T}$, where~$\hat{\n}$ indicates the line-of-sight direction in the sky. The dominant contribution to~$\delta \tilde{T}$ comes from the motion of our solar system with respect to the CMB~rest frame. The induced Doppler effect gives rise to an overall dipole anisotropy with $\delta \tilde{T} \approx \SI{3.4}{\milli\kelvin} \sim \num{e-3}\,\bar{T}$~\cite{Adam:2015vua} which we generally subtract to get the primordial anisotropies~$\delta T(\hat{\n})$. The resulting CMB~temperature map shows fluctuations of the order of $\delta T/\bar{T} \sim \num{e-4}$ as illustrated in Fig.~\ref{fig:CMBmap}.%
To relate the perturbations in the photon density at the last-scattering surface to the observed temperature inhomogeneities, we follow the free-streaming evolution from decoupling to today and project the acoustic oscillations onto the observer's celestial sphere. Both effects lead to additional modulations of the primordial density field that we have to take into account.\medskip

To simplify the discussion, we assume that recombination happened instantaneously. This will capture most of the phenomenology, but the finite width of the last-scattering surface of course has to be taken into account when comparing to data. Integrating the Boltzmann equation of photons along their corresponding line-of-sight from decoupling,~$\tau_\rec$, to today,~$\tau_0$, we find
\beq
\frac{\delta T}{\bar{T}}(\hat{\n}) = \left(\frac{1}{4}\delta_\gamma + \Phi+ \hat{\n} \cdot \vecv_e\right)_{\!\rec} + \int_{\tau_\rec}^{\tau_0}\!\d\tau \left(\dot{\Phi}+\dot{\Psi}\right) ,
\eeq
where we dropped the term $\Phi(\tau_0)$ since it only affects the monopole perturbation. The first term,~$\frac{1}{4}\delta_\gamma$, captures the intrinsic temperature variations~$\theta$ from perturbing the thermal distribution~\eqref{eq:thermalDistributions} of photons, $f_\gamma = \bar{f}_\gamma + \delta\hskip-0.5pt f_\gamma = \left( \exp\left\{ p/\!\left[\bar{T} (1+\theta)\right] \right\} - 1\right)^{\!-1}$, whereas the presence of~$\Phi$ accounts for the gravitational redshifting that occurs when photons climb out of a potential well at decoupling. The combination $\Theta = \frac{1}{4}\delta_\gamma + \Phi$ is called the Sachs-Wolfe~(SW) term and can be thought of as the effective temperature fluctuation of the primordial CMB. Note that an overdense region at decoupling, which has $\Phi<0$, leads to a cold spot in the large-scale CMB sky because photons climbing out of this potential loose more energy than they had at the bottom of the well. Analogously, hot spots are observed at the locations of underdense regions. Since the electrons in the photon-baryon fluid are not at rest when the photons scatter off of them, the third term,~$\hat{\n} \cdot \vecv_e$, is induced and describes the associated Doppler effect. The integral over the evolution of the gravitational potentials is denoted the integrated Sachs-Wolfe~(ISW) term and has both an early and a late component because $\Phi,\Psi \neq \const$ in the presence of either radiation or dark energy. Overall, the dominating component of~$\delta T$ is the Sachs-Wolfe term~$\Theta$, in particular on scales below the sound horizon.

\begin{figure}
	\centering
	\includegraphics{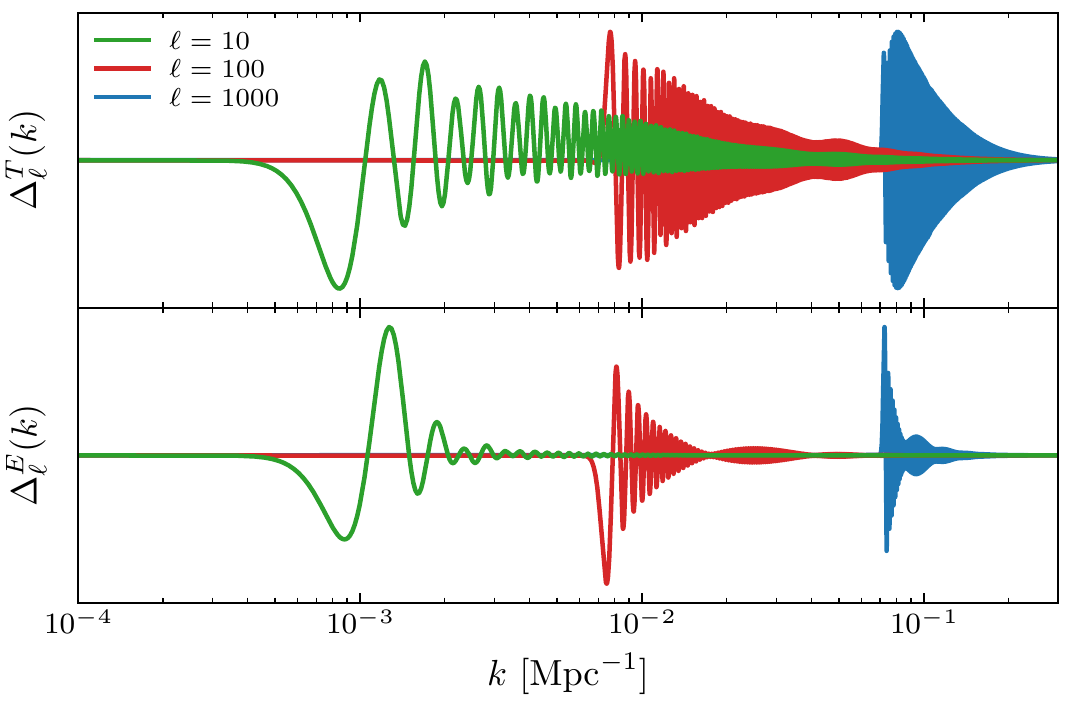}
	\caption{Transfer functions $\Delta_\ell^X(k)$ in temperature and polarization, $X = T,E$, for $\ell = 10,\ 100\text{ and }1000$ (normalized to the same maximum amplitude). Both functions peak around $\ell \sim k \chi_\rec$ and decay towards larger wavenumbers. This decay is however much more pronounced in polarization and, in general, a smaller number of wavenumbers~$k$ contribute to the same multipole~$\ell$. We also note that the acoustic peaks in the two transfer functions are out of phase. Taken together, this explains why the polarization transfer is both cleaner than and complementary to the temperature transfer.}
	\label{fig:transferFunctions}
\end{figure}

To project the primordial sound waves onto the two-dimensional sky, it is useful to work in Fourier space and extract the multipole moments of the temperature anisotropies~$\delta T$, which are defined by
\beq
\delta T_\ell = \frac{1}{(-\ii)^{\ell}} \int_{-1}^{1}\!\frac{\d\mu}{2} P_\ell(\mu)\, \delta T(\mu)\, ,
\eeq
with Legendre polynomials $P_\ell(\mu)$. We then find that the projection results in the SW and ISW terms being reweighted by Bessel functions~$j_\ell(k \chi_\rec)$, which arise via a Rayleigh plane-wave expansion. Similarly, the Doppler term is multiplied by the first derivative of these Bessel functions. Here, we introduced the comoving scale to the last-scattering surface~$\chi_\rec$, which equals $\tau_0 - \tau_\rec$ in a flat universe. This means that each multipole moment~$\ell$ in principle gets contributions from many different momentum modes~$k$. Since the Bessel functions~$j_\ell(x)$ are highly peaked near $x\approx\ell$ for large~$\ell$, this effect is less pronounced on small scales than on large scales. In multipole space, the acoustic peaks are therefore located at $\ell_n \sim k_n \chi_\rec = n\pi\,\chi_\rec/r_s$. Finally, it is convenient to introduce the transfer function $\Delta_\ell^T(k) \equiv \delta T_\ell(\tau_0,k)/\zeta_\k$ which captures the entire linear evolution of the initial perturbations and includes these projection effects. The upper panel of Fig.~\ref{fig:transferFunctions}%
illustrates this function for three representative multipoles.\medskip

\begin{figure}
	\centering
	\includegraphics{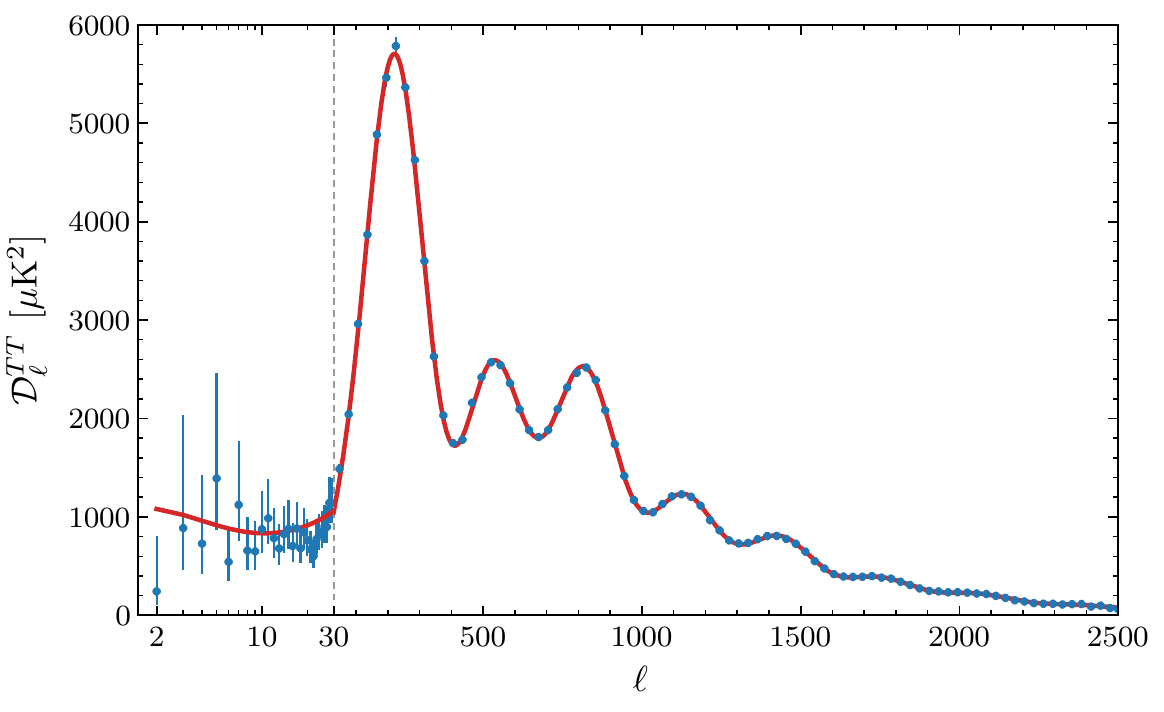}\vspace{-5pt}
	\caption{Planck 2015 temperature power spectrum $\mathcal{D}_\ell^{TT} \equiv \ell(\ell+1)/(2\pi)\, C_\ell^{TT}$ (based on data from~\cite{Aghanim:2015xee}). The error bars of the binned data at high multipoles are smaller than the data points. The red line shows the best-fit theoretical spectrum of the six-parameter $\Lambda\mathrm{CDM}$~model inferred from the Planck~TT likelihood.\vspace{-2pt}}
	\label{fig:PlanckTT}
\end{figure}

Finally, we turn to the two-point correlation function $\left\langle \delta T(\hat{\n})\, \delta T(\hat{\n}') \right\rangle$ which is the quantity that we ultimately extract from CMB temperature maps. Assuming the initial conditions are statistically isotropic, we can expand this two-point function as
\beq
\left\langle \delta T(\hat{\n})\, \delta T(\hat{\n}') \right\rangle =  \sum_{\ell} \frac{2\ell+1}{4\pi}\, C_\ell^{TT}\, P_\ell(\hat{\n} \cdot \hat{\n}')\, ,	\label{eq:cmbPowerSpectrum}
\eeq
where the Legendre polynomials~$P_\ell(\hat{\n} \cdot \hat{\n}')$ only depend on the relative orientation of~$\hat{\n}$ and~$\hat{\n}'$. The expansion coefficients $C_\ell^{TT}$ in~\eqref{eq:cmbPowerSpectrum} are the famous angular (temperature) power spectrum and given by\hskip1pt\footnote{We could have equivalently obtained the angular power spectrum by decomposing the temperature fluctuations~$\delta T$ into spherical harmonics and computing the correlation function of the expansion coefficients. This is how measurements of $\delta T$ are commonly processed to obtain the power spectrum $C_\ell^{TT}$.}
\beq
C_\ell^{TT} = \frac{4\pi}{(2\ell+1)^2} \int\!\d\hskip1pt\ln k\, \left(\Delta_\ell^T(k)\right)^2 \mathcal{P}_{\hskip-0.5pt\zeta}(k)\, .
\eeq
For convenience, we usually show the rescaled spectrum~$\mathcal{D}_\ell^{TT} \equiv \ell(\ell+1)/(2\pi)\, C_\ell^{TT}$ which is displayed together with the latest measurement of the Planck satellite in Fig.~\ref{fig:PlanckTT}.%
The shape of~$\mathcal{D}_\ell^{TT}$ is very characteristic: The SW~effect dominates on large scales (low multipoles), the acoustic oscillations are observed on intermediate scales (scales smaller than the projected sound horizon which corresponds to an angular scale of about~\SI{1}{\degree} or~$\ell \sim 200$) and the smallest scales (large multipoles) are exponentially damped. Because the initial power spectrum is almost scale-invariant, all features in the CMB power spectrum arise from the evolution of the cosmic sound waves being captured at the moment of last-scattering, i.e.\ the observed oscillations are a snapshot of these waves caught at different phases in their evolution, and subsequently projected onto the sky. The first acoustic peak was discovered by the Toco experiment~\cite{Miller:1999qz}. The next peaks were tentatively detected by several experiments, but measured decisively by the WMAP~satellite~\cite{Hinshaw:2003ex}. Nowadays, Planck has measured the temperature power spectrum to the cosmic variance limit~(CVL)\hskip1pt\footnote{Cosmic variance refers to the statistical uncertainty inherent in cosmological measurements since we are only able to measure one realization of the true model underlying the universe. In a cosmic variance-limited measurement, the statistical error is dominated by this uncertainty, which is given by $\Delta C_\ell = \sqrt{2/(2\ell+1)}\,C_\ell$ for a CMB~auto-spectrum.} for multipoles up to $\ell \approx 1600$~\cite{Aghanim:2015xee}. This measurement is complemented by the results of many ground-based experiments which mainly target large multipoles as a consequence of their better angular resolution.\medskip

Armed with these insights, we can infer the general dependence of the shape of the power spectrum on the cosmological parameters. The overall amplitude of the spectrum depends not only on the primordial amplitude~$\As$, but on the combination~$\As\, \e^{-2\tau}$. The reason for this is that photons scattered off of electrons again after the neutral hydrogen was reionized by the large amounts of ultraviolet radiation that were emitted after the first stars and galaxies had formed. As a result, the optical depth increased and scales smaller than the horizon at that time are suppressed by a factor of $\e^{-\tau}$. This degeneracy between~$\As$ and~$\tau$ can be broken to some extent by CMB~lensing measurements (see below). From the measured value $\As \approx \num{2.1e-9}$ we can infer the amplitude of the primordial perturbations (at $k_0=\SI{0.05}{\per\Mpc}$): $\sqrt{\mathcal{P}_{\hskip-0.5pt\zeta}(k_0)} \approx \num{5e-5}$. Moreover, the spectrum is measured to be slightly red-tilted with a spectral index of $\ns = \num{0.968 \pm 0.004}$, i.e.\ there is a bit more power on large scales than on small scales. This departure from scale invariance, $n_s-1 < 0$, which is a natural consequence of many inflationary models, has now been measured at a significance of more than~$7\sigma$~\cite{Ade:2015xua}. The spectral tilt and the optical depth are somewhat degenerate in the temperature power spectrum. This degeneracy can however be lifted by including information from polarization (see below) since reionization leads to a distinct bump at low multipoles in the polarization spectrum.

The positions of the peaks in the CMB spectrum, $\ell_n$, are particularly sensitive to the distance to last-scattering. In fact, the angular size of the sound horizon at decoupling,~$\theta_s$, is a direct measure of the first peak location at an angular scale of about~\SI{1}{\degree}, which is the characteristic size of the spots in the CMB~map of Fig.~\ref{fig:CMBmap}. Since curvature is exactly zero within~$\Lambda\mathrm{CDM}$,\footnote{The location of the first peak is very sensitive to the curvature of the universe via the distance to last-scattering. The measurement of this peak famously led to the conclusion that our universe has a geometry that is very close to flat~\cite{deBernardis:2000sbo}, which laid the groundwork for the $\Lambda\mathrm{CDM}$~model. Today, the famous \mbox{$\Omega_m$-$\Omega_\Lambda$}~plot shows that the confidence regions inferred from~CMB, BAO and supernovae data, which all have different degeneracy lines, intersect in a single small region that is consistent with a flat universe, $\Omega_M+\Omega_\Lambda+\Omega_r = 1$.} the peak positions become a precise measure of the expansion history and, therefore, of the Hubble parameter~$H_0$ and the physical matter density~$\omega_m$. The overall peak heights relative to the large-scale plateau are however a much more sensitive probe of~$\omega_m$ since the amplitude of the cosmic sound waves depends on the time of matter-radiation equality through the radiation-driving effect. As a consequence, the small-scale modes, which entered the horizon in the radiation era, are enhanced in comparison to the modes which started evolving only later during matter domination. The relative peak heights, on the other hand, are directly related to the baryon density~$\omega_b$ as the odd peaks are larger than the even peaks due to baryon loading, cf.~\eqref{eq:soundWaves}. Moreover, we can already anticipate from~\eqref{eq:damping} that the damping tail of the spectrum is particularly sensitive to the early expansion history and, consequently, the radiation density (see Section~\ref{sec:signaturesLightRelics} for a detailed discussion, including the related degeneracies). Finally, the late~ISW~effect on large scales is sensitive to dark energy, as we previously mentioned. This list of dependencies as derived from our relatively simple analytic treatment of the CMB~phenomenology only indicates the potential of uncovering cosmological information encoded in the temperature anisotropies. In Section~\ref{sec:analytics}, we will see how much more information we can deduce when adopting a slightly more rigorous (but still analytic) treatment of cosmic sound waves.

\subsubsection{Polarization}
\begin{figure}
	\centering
	\includegraphics{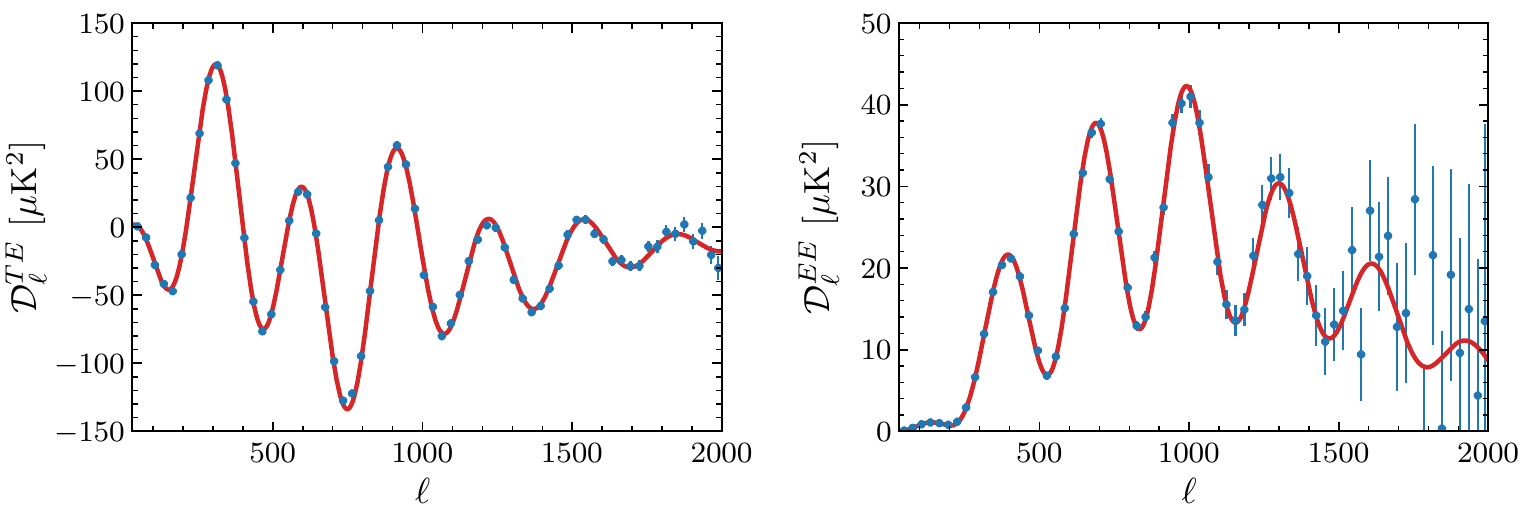}
	\caption{Planck 2015 high-$\ell$~TE and EE~power spectra,~$\mathcal{D}_\ell^{TE}$ and~$\mathcal{D}_\ell^{EE}$ (based on data from~\cite{Aghanim:2015xee}). The red line shows the best-fit~$\Lambda\mathrm{CDM}$ theoretical spectrum inferred from the Planck~TT likelihood. The smallness of the residuals with respect to this model indicates a very good fit providing a non-trivial confirmation of the standard cosmological model and CMB~phenomenology.}
	\label{fig:PlanckTEEE}
\end{figure}

We do not only expect that the temperature varies across the CMB sky, but also that this ancient radiation is linearly polarized at the level of a few~$\si{\muKelvin}$ due to Thomson scattering at the time of decoupling (and reionization). As we will explain below, the scattering of photons with electrons is the only possibility to generate CMB~polarization. The polarization signal therefore tracks free electrons and is a particularly clean probe of the physics at the last-scattering surface (and reionization). DASI~detected CMB~polarization about ten years after~COBE announced their discovery of the temperature anisotropies~\cite{Kovac:2002fg}. Precise polarization measurements provide a non-trivial consistency check for the standard cosmological model because the temperature anisotropies and the polarization signal are directly related. As illustrated in Fig.~\ref{fig:PlanckTEEE},%
the Planck~2015 temperature-polarization cross- and polarization auto-spectra show this impressively. In addition, these spectra help to break degeneracies between cosmological parameters and provide complementary information. In the following, we give a lightning review of the main aspects of CMB~polarization and refer to the seminal papers~\cite{Bond:1984fp, Polnarev:1985pai, Seljak:1996ti, Zaldarriaga:1996xe, Kamionkowski:1996ks} and the pedagogical review~\cite{Hu:1997hv} for further details.\medskip

The generation of CMB~polarization is best described in the rest frame of a free electron in the primordial plasma. If the incident radiation field is isotropic, the Thomson-scattered radiation remains unpolarized since orthogonal polarization directions cancel out. The same statement holds if the incoming photons have a dipolar anisotropy. However, a net linear polarization arises if the radiation field around the electron has a non-zero quadrupole moment. Put differently, a non-vanishing quadrupole of the temperature anisotropy generates the linear polarization of the~CMB. However, prior to decoupling, Thomson scattering keeps the CMB~radiation very nearly isotropic in the rest frame of the electrons. A local temperature quadrupole can therefore only develop from a gradient in the velocity field once the photons have acquired an appreciable mean free path just before they decouple. CMB~polarization is therefore only generated in the very last scattering events and results from the velocities of the electrons on scales smaller than the photon mean free path. Since both the temperature inhomogeneities and the velocity field of the photons originate from primordial density fluctuations and are out of phase, we expect the polarization peaks of the~CMB to be both correlated and out of phase with the temperature peaks. This is exactly what we see in the data (see the left panel of Fig.~\ref{fig:PlanckTEEE}). Moreover, cosmological parameters can be independently constrained from the temperature and polarization spectra because the CMB~temperature primarily traces the density perturbations, whereas the polarization is effectively induced by the velocity fluctuations. In addition, the polarization spectrum should have less power than the temperature spectrum because the quadrupole moment is suppressed compared to the monopole and dipole moments. This prediction has also been confirmed quantitatively as displayed in the right panel of Fig.~\ref{fig:PlanckTEEE}.

In general, a linearly-polarized radiation field can mathematically be described by three variables: the temperature~$T$, and the Stokes parameters~$Q$ and~$U$. Whereas the temperature can be conveniently decomposed in terms of scalar spherical harmonics, the convenient complex combinations~$Q \pm \ii\hskip0.5ptU$ of the Stokes parameters are spin-2 quantities and have to be expanded in the more complicated tensor spherical harmonics. It is however possible to construct two scalar quantities which are invariant under coordinate transformations and commonly referred to as~$E$ and~$B$.\footnote{In analogy with the properties of the electric and magnetic fields in electrodynamics, E-mode and B-mode polarization is curl- and divergence-free, respectively.} Importantly, scalar/density perturbations only create E-modes and no B-modes, while tensor perturbations (i.e.\ gravitational waves) induce both E- and B-modes. In this thesis, we therefore neglect B-modes and focus on E-modes, which we will generally refer to as polarization,\footnote{A fraction of the E-modes is converted to B-modes in the late universe through gravitational lensing (see below). In contrast to the primary B-modes, these induced B-modes have been detected and can in principle be used to revert the effects of lensing on the temperature and E-mode spectra in a process referred to as delensing.} i.e.\ we consider the following temperature and polarization auto- and cross-spectra: TT, TE and EE (as displayed in Figs.~\ref{fig:PlanckTT} and~\ref{fig:PlanckTEEE}).

Although the generation mechanism of polarization is somewhat more involved and the power spectrum is suppressed compared to the temperature anisotropies, the transfer function~$\Delta^E_\ell(k)$ is simpler since there are no SW or ISW~effects, for instance. Moreover, the mapping between wavenumbers~$k$ and multipoles~$\ell$ is much sharper for polarization as illustrated in Fig.~\ref{fig:transferFunctions}. The underlying reasons are a slightly different projection onto the celestial sphere and the fact that the polarization signal is only generated effectively in the very last scattering events. The acoustic peaks in the polarization spectrum are therefore a more direct snapshot of the primordial sound waves and the peaks themselves are slightly sharper than in the temperature spectrum.

\subsubsection{Lensing}
There are a number of secondary effects that impact the observed~CMB on small scales. For the purpose of this thesis, only one of them will be of some relevance: the weak gravitational lensing of the~CMB (see~\cite{Lewis:2006fu} for a comprehensive review). On the long way from the last-scattering surface to our detectors, the CMB~photons pass through the increasingly inhomogeneous matter distribution which acts as gravitational lenses. In effect, the CMB~photons are deflected by gradients in the gravitational potential along the line-of-sight. This not only generates B-mode polarization on small scales, but also affects the temperature and E-mode power spectra. Since the photon paths are slightly perturbed, the location in the sky where we observe the photons is slightly offset from the location where they actually decoupled. The primordial hot and cold spots are therefore distorted and the acoustic peaks in the lensed power spectra are slightly smeared since power is transferred between multipoles. Because we will be interested in precisely measuring the acoustic peaks, this effect is a nuisance (although it is possible to `delens', i.e.\ revert the effects of lensing, to some extent~\cite{Hu:2001kj, Knox:2002pe, Hirata:2003ka, Green:2016cjr, Sehgal:2016eag, Carron:2017vfg}). At the same time, the lensing power spectrum has now been extracted at high significance from Planck data~\cite{Ade:2015zua} and CMB~lensing allows to infer the integrated matter distribution between us and the last-scattering surface.

\subsection{Large-Scale Structure}
\label{sec:lss}
\begin{figure}
	\centering
	\includegraphics{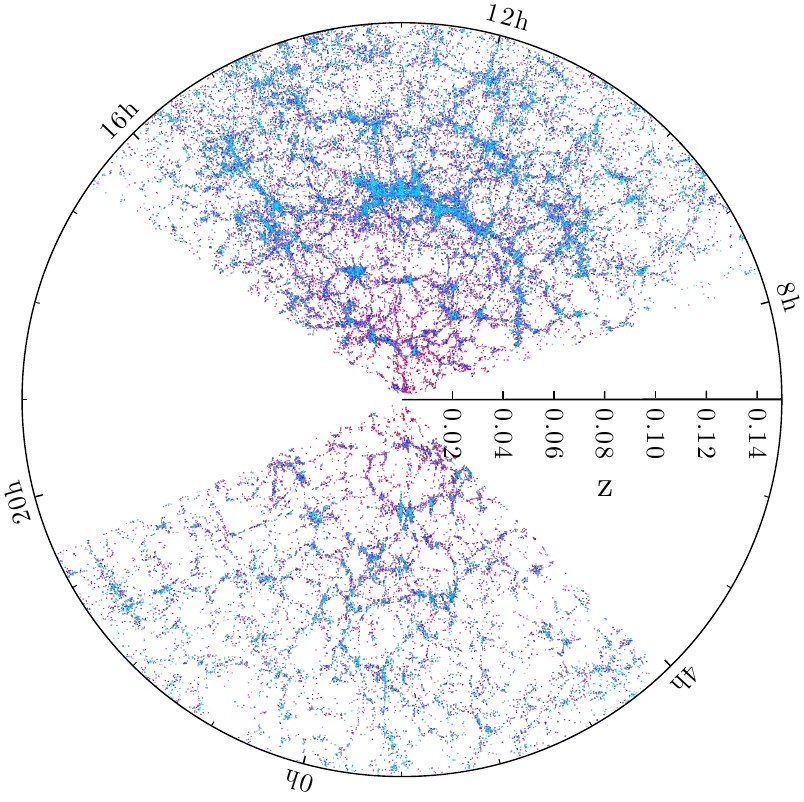}
	\caption{Map of galaxies from the 14th data release of the Sloan Digital Sky Survey~(SDSS; adapted from~\cite{sdssMap}).}
	\label{fig:SDSSmap}
\end{figure}

\begin{figure}
	\centering
	\includegraphics{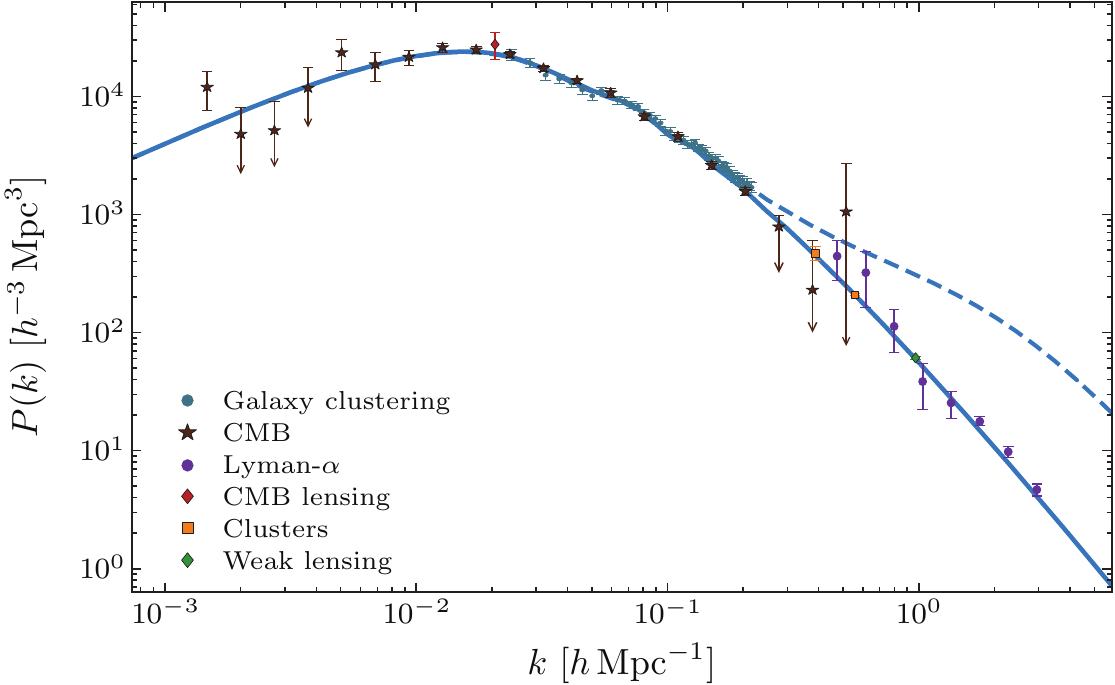}
	\caption{Linear matter power spectrum reconstructed from CMB~temperature, CMB~lensing, galaxy, cluster and \mbox{Lyman-$\alpha$}~forest measurements (adapted from~\cite{Hlozek:2011pc} with data from~\cite{McDonald:2004eu, Vikhlinin:2008ym, Reid:2009xm, Sehgal:2010ca, Das:2011ak, Tinker:2011pv, Hlozek:2011pc}). The solid and dashed lines display the linear and non-linear power spectra of the best-fit $\Lambda\mathrm{CDM}$~model inferred from the shown CMB~temperature data. The observed agreement highlights the consistency of the measurements conducted by an array of different cosmological probes over a large range of scales.}
	\label{fig:PkMeasurements}
\end{figure}

Having discussed the imprints of photon perturbations in the cosmic microwave background, we now return to the matter fluctuations, which grew under the influence of gravity. Eventually, galaxies formed which we can now observe in cosmological surveys (see Fig.~\ref{fig:SDSSmap}).%
The principle LSS~observable, inferred from both theory and data, is the two-point correlation function, $\xi(r) \equiv \xi_m(r)$, or the power spectrum of matter perturbations, $P(k) \equiv P_m(k)$, which is defined by
\beq
\left\langle \delta_m(\k)\, \delta_m(\k') \right\rangle = (2\pi)^3\, P(k)\, \delta_D^{(3)}(\k-\k')\, .
\eeq
The characteristic shape of the spectrum is easily derived by combining the Poisson equation on subhorizon scales, $\delta_m \propto k^2 \Phi$, which implies $P(k) \propto k^4 P_\Phi(k)$, and the evolution of the gravitational potential~$\Phi$ as discussed at the end of~\textsection\ref{sec:structureFormation}. Modes which entered the horizon after matter-radiation equality, $k < k_\mathrm{eq}$, remain constant as $\Phi = \const$, i.e.\ $P_\Phi(k)$ is scale-invariant and \mbox{$P(k) \propto k$.\footnotemark}\footnotetext{For simplicity, we set the primordial spectral tilt to unity, $\ns = 1$, in this discussion, i.e.\ assume a perfectly scale-invariant primordial power spectrum.} On the other hand, the sub-horizon potential decays as $\Phi \propto a^{-2} \propto \tau^{-2}$ during radiation domination. Its power spectrum is therefore suppressed according to $P_\Phi(k) \propto k^{-3} (k_\mathrm{eq}/k)^4$ because a mode~$k$ crosses the horizon at $\tau = 1/k$. This implies that the matter power spectrum peaks around $k \sim k_\mathrm{eq}$ and scales as $P(k) \propto k^{-3}\,\log^2(k/k_\mathrm{eq})$ for $k>k_\mathrm{eq}$, where we included the logarithmic growth of matter perturbations in the radiation era. In addition, the cosmic sound waves are imprinted on top of this power law as we discussed in~\textsection\ref{sec:cosmicSoundWaves}. Figure~\ref{fig:PkMeasurements} illustrates that we observe exactly this behaviour on linear scales. Finally, the power increases in time proportional to~$a^2$ since the matter perturbations grow linearly during matter domination. In order to capture the entire time evolution including the dark-energy era, we usually introduce the linear growth function~$D_1(z)$ such that $P(k,z) = \left[D_1^2(0)/D_1^2(z)\right] P(k,z=0)$.

As the density contrast grows, first linear perturbation theory and eventually all perturbative treatments break down with small scales being affected earlier than large scales. Understanding the non-linear evolution is one of the main challenges when trying to connect LSS~observables to fundamental physics. A description of non-linear effects is however important because they can mimic or distort primordial signals, but complicated by the fact that these effects are hard to characterise from first principles. Fortunately, scales corresponding to $k\lesssim\SI{0.1}{\hPerMpc}$ (at $z=0$ and larger for earlier times) can be treated well in linear perturbation theory throughout cosmic history. A lot of effort is currently being put into pushing the scale up to which we trust perturbative computations into the mildly non-linear regime (see e.g.~\cite{Bernardeau:2001qr, Crocce:2005xy, Baumann:2010tm, Carrasco:2012cv, Bartelmann:2014gma, Blas:2015qsi}). For smaller scales, we have to resort to numerical simulations. In general, however, the power on smaller scales is enhanced, in particular for $k\gtrsim\SI{0.1}{\hPerMpc}$. On the bright side, we are able to model and account for these non-linearities to some extent. Nevertheless, they clearly impose a limitation on how well we can use the small-scale information to infer properties of the primordial plasma at the present time.\medskip

Up to now, we have assumed that we can directly measure the matter density field in cosmological surveys. However, this is usually not the case because we generally observe tracers of the matter density which may be highly non-linear objects. Galaxy surveys for example measure the three-dimensional spatial distribution of galaxies which we subsequently have to relate to the underlying distribution of matter. This relation is described, in a statistical sense, by the galaxy bias $b=b(z)$. In the limit of linear bias, we have $\delta_{g,\k} = b\, \delta_{m,\k}$, where the galaxy distribution is usually captured by its number density field, $\delta_g(\x) = (n(\x) - \bar{n})/\bar{n}$, with the mean density of galaxies $\bar{n} = \langle n(\x) \rangle$. The linear-bias approximation may be sufficient on large scales, but the bias also picks up a scale dependence,~$b(k,z)$, on smaller scales. While there have been a number of advances in the recent past, this dependence is challenging to predict (cf.~\cite{Desjacques:2016bnm} for a comprehensive review). Although the wavenumbers at which this becomes important get smaller at higher redshifts, the observed objects tend to be more strongly biased since they are intrinsically brighter in order to be detected.

Further sources of uncertainty in the mapping between theory and observations arise, for instance, from redshift-space distortions due to the peculiar velocity of galaxies (the relative velocity with respect to the Hubble flow), and because we cannot measure the positions~$\x$ of objects in the universe, but only their redshifts and angular positions on the sky. All these points (and more) have to be accounted for when trying to link~LSS observables, such as the power spectrum of galaxies, to the physics in the early universe. The large number of potentially available modes is very encouraging in principle, but it seems as if Nature makes us work hard to harness this information.

\subsection{Baryon Acoustic Oscillations}
\label{sec:bao}
\begin{figure}
	\centering
	\includegraphics{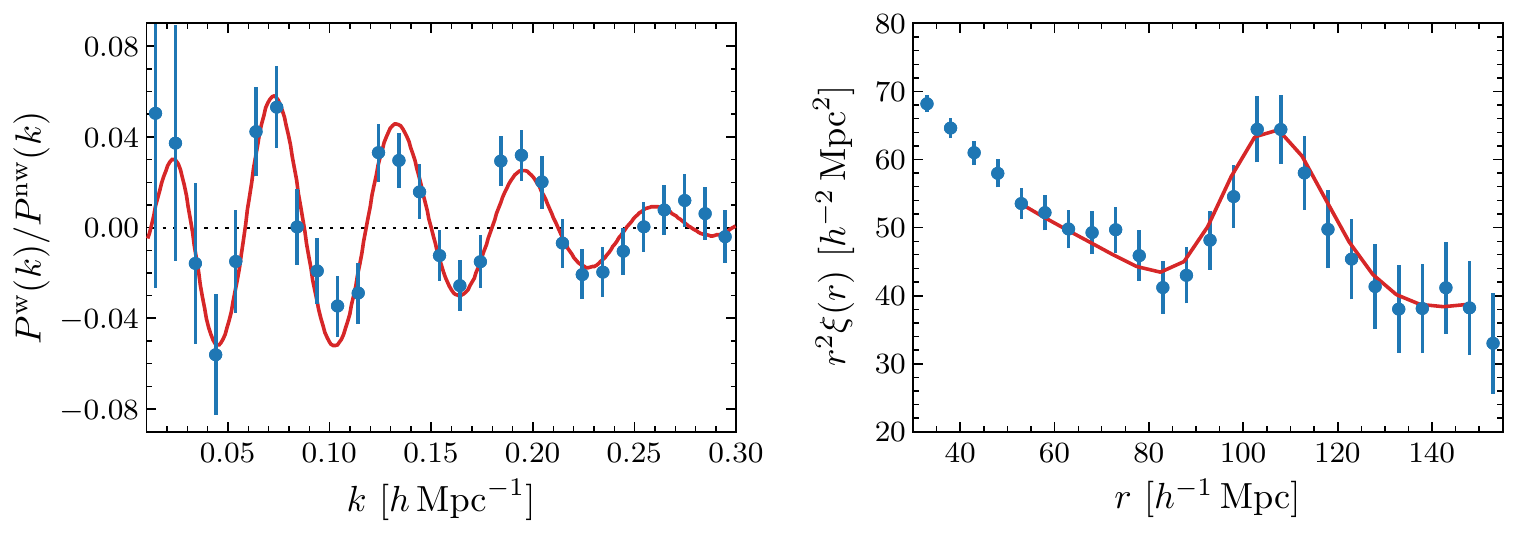}
	\caption{BAO~spectrum~(\textit{left}) and BAO~peak~(\textit{right}) measurements in the redshift bin $0.5 < z_3 < 0.75$ of the last data release of the Baryon Oscillation Spectroscopic Survey together with the respective best-fit model in the employed fitting range (adapted from~\cite{Beutler:2016ixs, Ross:2016gvb}).}
	\label{fig:baoMeasurements}
\end{figure}

As we have just discussed, several theoretical challenges in the galaxy power spectrum are related to its overall shape and amplitude. This is in particular the case for the issues of non-linear evolution and biasing. 
Subtracting this smooth (`no-wiggle') part,~$\Pnw(k)$, from the full spectrum, we are left with the oscillatory (`wiggle') part, $\Pw(k) \equiv P(k) - \Pnw(k)$. This contains the signal of the cosmic sound waves in the primordial plasma which is why we refer to the ratio of the oscillatory to the smooth spectrum,
\beq
O(k) = \frac{\Pw(k)}{\Pnw(k)} = \frac{P(k) - \Pnw(k)}{\Pnw(k)}\, ,
\eeq
as the BAO~spectrum.\footnote{Although we observe the BAO~signal as acoustic peaks both in~CMB and in LSS~measurements, we will usually refer to the latter when we mention BAO~observations. Strictly speaking, the spectrum of baryon acoustic oscillations is~$\Pw(k)$. For convenience, we will however also refer to~$O(k)$ as the BAO~spectrum.} In 2005, this BAO signal was first detected in both the two-point correlation function of the Sloan Digital Sky Survey~(SDSS)~\cite{Eisenstein:2005su} and the power spectrum measured by the 2dF Galaxy Redshift Survey~\cite{Cole:2005sx}. The currently highest signal-to-noise measurements are provided by the Baryon Oscillation Spectroscopic Survey and displayed in Fig.~\ref{fig:baoMeasurements}%
for one of their redshift bins.\medskip

As in the case of the matter power spectrum, gravity still non-linearly processed the BAO~signal from its primordial form~\cite{Eisenstein:2006nj, Crocce:2007dt, Sugiyama:2013gza}. However, we have been able to understand and get certain aspects of this observable under better theoretical control. We can schematically express the BAO spectrum as
\beq
O(k) = A(k) \sin\! \left[\hskip0.5pt \omega(k)\, k + \phi(k) \right] ,
\eeq 
with the amplitude~$A$, frequency~$\omega$ and phase~$\phi$ (see our discussion at the end of~\textsection\ref{sec:cosmicSoundWaves}). The amplitude depends on the wavenumber~$k$ because of the exponential damping both due to photon diffusion in the photon-baryon fluid and due to non-linear gravitational evolution (see e.g.~\cite{Eisenstein:2006nj, Seo:2007ns, Baldauf:2015xfa, Blas:2016sfa}). This is why the BAO~amplitude is currently not employed in the standard inference of cosmological parameters.

The frequency~$\omega$ of the BAO spectrum is also affected by non-linearities. These effects can however approximately be removed by BAO reconstruction, which is a well-tested technique to better estimate the initial (linear) density perturbations (see e.g.~\cite{Eisenstein:2006nk, Padmanabhan:2012hf, Sherwin:2012nh, Schmittfull:2015mja, Seo:2015eyw}). By reversing the displacements of galaxies due to their bulk flow, the non-linear effects of structure formation and redshift-space distortions can be undone to a certain extent. Thanks to this method, it has now become possible to measure the frequency of the BAO spectrum~\cite{Beutler:2016ixs} and the location of the BAO peak~\cite{Ross:2016gvb, Vargas-Magana:2016imr} at the sub-percent level. Using the BAO signal as a standard ruler then allows to break degeneracies between cosmological parameters in the CMB (e.g.\ between $\Omega_m$ and~$\Omega_\Lambda$) resulting in tighter constraints.\footnote{Another common dataset that is used to break degeneracies are the local $H_0$~measurements from supernovae. Having said that, with both primordial, such as those from the~CMB and~BAO, and local measurements improving, the inferred values of~$H_0$ are currently statistically discrepant at the $3\sigma$~level~\cite{Riess:2016jrr}. It is however questionable whether we should be paying too much attention given the vast statistical power of the CMB in particular (see e.g.~\cite{Scott:2018adl}).}

Although the phase is absent for adiabatic initial conditions, as we will explicitly show in Section~\ref{sec:analytics}, it can be induced by free-streaming neutrinos. In general, the phase is also affected by the non-linearities induced by gravity. However, it has been proven that a constant phase, $\phi = \const$, is immune to these effects~\cite{Baumann:2017lmt}. This suggests that we can reliably extract the primordial phase from late-time observables. We will explore how to use the information encoded in the BAO~spectrum, in particular the phase~$\phi$, in Chapter~\ref{chap:bao-forecast} and establish a modified BAO analysis in Chapter~\ref{chap:bao-neutrinos}, which takes some of these considerations into account.
	\chapter{Light Species in Cosmology and Particle Physics}
\label{chap:background_species}
We saw in the last chapter that cosmological observables are very sensitive to the composition of the universe. We assumed that the Standard Model of particle physics accurately describes the baryonic matter content, which indeed it does to remarkable accuracy. In addition, we had to however invoke the presence of non-baryonic dark matter to explain some of the cosmological data, in particular the acoustic peaks of the~CMB. This is only one of the many reasons to believe that the~SM is incomplete. Other famous theoretical and experimental shortcomings of the~SM are the strong CP~problem, the origin of neutrino masses, the stability of the Higgs mass and the matter-antimatter asymmetry. Models of physics beyond the Standard Model usually invoke new degrees of freedom to address some or all of these problems. Many interesting SM~extensions contain new light species~\cite{Essig:2013lka}, such as axions~\cite{Peccei:1977hh, Weinberg:1977ma, Wilczek:1977pj}, axion-like particles~\cite{Arvanitaki:2009fg}, dark photons~\cite{Holdom:1985ag, Galison:1983pa} and light sterile neutrinos~\cite{Abazajian:2012ys}.\footnote{In addition, BSM~models often require new massive particles which can be too heavy to be produced at the energies available at colliders even in the distant future. Although it is also possible to constrain these types of particles using cosmology (see e.g.~\cite{Steigman:1984ac, Hu:1993gc, Kawasaki:1999na, Padmanabhan:2005es, Chluba:2008aw, Iocco:2008va, Galli:2009zc, Baumann:2011nk, Chluba:2013wsa, Dimastrogiovanni:2015pla, Lesgourgues:2015wza, Arkani-Hamed:2015bza, Flauger:2016idt, Lee:2016vti, Green:2018pmd}), in this thesis, we will focus on light weakly-coupled species.} These particles are often so weakly coupled to the~SM that they escape detection in terrestrial experiments. However, they may be efficiently produced in astrophysical systems and in the early universe, which therefore presents an alternative way of probing these elusive species.\medskip

In this chapter, we provide the connection between cosmological observables and the additional light relics predicted in some BSM~models. In Section~\ref{sec:estimates}, we give order-of-magnitude estimates for the constraining power of astrophysical systems and cosmology on weakly-interacting light particles. In Section~\ref{sec:BSMphysics}, we introduce the particle physics aspects of this thesis. We first review the current status of the Standard Model of particle physics and motivate its extensions with additional light particles. We will then present an effective field theory framework which offers a convenient parametrization of large classes of light species and their interactions with the SM. In Section~\ref{sec:neutrinosDarkRadiation}, we introduce the effective number of relativistic species,~$\Neff$, as the main cosmological parameter capturing neutrinos and any other light thermal relics, and determine the relation between the observable relic abundance of these particles and their decoupling temperatures. We show that a single species in thermal equilibrium produces a minimal non-zero contribution to~$\Neff$ which provides an interesting target for future cosmological measurements. In Section~\ref{sec:signaturesLightRelics}, we finally examine the possible signatures that these particles might leave in cosmological observables such as the CMB and LSS~power spectra. This will allow us to search for these signals and constrain the interactions of light relics with the Standard Model in the following chapters of this thesis.

\section{The Power of Astrophysics and Cosmology}
\label{sec:estimates}
Detecting new light species is challenging because their couplings to the Standard Model degrees of freedom are necessarily small (since we would have already detected them otherwise). In particular, since the scattering cross sections are tiny, it is difficult to probe these SM~extensions in terrestrial experiments on the intensity or energy frontier of particle physics, i.e.\ in the laboratory or at particle accelerators. In astrophysics and cosmology, however, we have access to high-density environments and/or the ability to follow the evolution over long time scales which can overcome the small cross sections and allow a significant production of the extra species. For example, new light particles can be produced in the interior of stars~\cite{Raffelt:1996wa}. Since these species are weakly interacting, they carry energy away from the stellar core similar to neutrinos. The absence of an anomalous extra cooling over the long lifetime of stars puts some of the best current constraints on weakly-coupled particles. 

To illustrate the origin of this sensitivity, we consider the fractional change in the number densities of the particles involved in the production process, which can schematically be written as
\beq
\frac{\Delta n}{n} \sim n\sigma \times \Delta t\, ,	\label{eq:schematicBoltzmannEquation}
\eeq
i.e.\ it is equal to the interaction rate, $\Gamma \sim n\sigma$, with thermally-averaged cross section~$\sigma$, times the interaction time $\Delta t$. This highlights how small cross sections can be compensated for by the high densities in the stellar interior, $n \sim (\SI{1}{\kilo\electronvolt})^3$, and especially the very long lifetime of stars which is typically of the order of $\Delta t \sim \SI{e8}{yrs}$. We therefore find significant changes in the stellar evolution, $\Delta n/n \gtrsim 1$, if 
\beq
\sigma > (n \Delta t)^{-1} \sim \left(\SI{e10}{\giga\electronvolt}\right)^{\!-2}\, .	\label{eq:stellarCoolingEstimate}
\eeq
These particles may also be produced in extreme astrophysical events, such as supernovae explosions, which happen on much shorter time scales, $\Delta t \sim \SI{10}{\second}$. Because the densities in this case are much higher, $n \sim (\SI{10}{\mega\electronvolt})^3$, the constraints which can be derived from the observed energy loss are at an order of magnitude comparable to~\eqref{eq:stellarCoolingEstimate}.

Since the early universe was dominated by radiation, constraints on light relics can also be inferred from cosmological measurements. In order to get a sense for the power of these possible bounds, a similar argument based on~\eqref{eq:schematicBoltzmannEquation} can be applied to cosmology. The high densities of the early universe, $n \sim T^3 \gg (\SI{1}{\mega\electronvolt})^3$, allow these light particles to have been in thermal equilibrium with the~SM (and therefore efficiently produced) for time scales of $\Delta t < \SI{1}{s}$. They can therefore make a significant contribution to the total radiation density of the universe and, hence, be possibly detected in CMB and LSS observables. The estimate~\eqref{eq:stellarCoolingEstimate} suggests that cosmological constraints will improve over astrophysical bounds for temperatures above \SI{e4}{\giga\electronvolt}. Moreover, cosmology may constrain all couplings to the Standard Model equally since thermal equilibrium in the early universe is democratic,\footnote{Any new light particle that was in thermal equilibrium in the past will have a number density which is comparable to that of photons. This is the reason why neutrinos have been detected with high significance in the CMB despite their weak coupling (cf.\ Chapter~\ref{chap:cmb-phases}; see also~\cite{Ade:2015xua, Follin:2015hya}).} whereas astrophysical systems and laboratory experiments are often only sensitive to a subset of these interactions, e.g.\ the coupling to photons. This universality of cosmological constraints is one of the reasons why the search for light thermal relics has been adopted as one of the main science targets of the next generation of CMB experiments, such as the CMB-S4 mission~\cite{Abazajian:2016yjj}. We will further quantify the constraining power of cosmology below and especially in Chapter~\ref{chap:cmb-axions}.

\section{Physics Beyond the Standard Model}
\label{sec:BSMphysics}
The Standard Model of particle physics is a great success. It provides a theoretically consistent description of all known particles and their interactions (except gravity) up to the Planck scale. On the experimental side, large improvements in detector technology and analysis techniques have led to all predicted particles being found, including the Higgs boson~\cite{Aad:2012tfa, Chatrchyan:2012xdj}, and their interactions being measured to exquisite precision. Having said that, the~SM is clearly incomplete and cannot be a fundamental theory since it does not address a number of open problems and several pieces of evidence for new physics. In the following, we briefly review the main aspects of the SM~(\textsection\ref{sec:SM}) and give a few pieces of evidence for dark BSM~sectors~(\textsection\ref{sec:theoryMotivation}). Finally, we introduce an effective field theory of light species as a convenient way of parametrizing the new dark sector and its interactions with the~SM for cosmological searches~(\textsection\ref{sec:EFTlightSpecies}).

\subsection{Standard Model of Particle Physics}
\label{sec:SM}
Theoretically speaking, the Standard Model of particle physics is a highly successful quantum field theory based on the local gauge group $SU(3)_C \otimes SU(2)_L \otimes U(1)_Y$. The elementary fields associated with these groups are the eight gluons $G^a$, which mediate the strong force, and the four gauge bosons $W^a$ and $B$, which mediate the electroweak force. The matter fields can be divided into quarks and leptons of three generations each. Quarks are charged under the entire SM~gauge group, whereas leptons only interact via the electroweak gauge group $SU(2)_L \otimes U(1)_Y$ as described by the Glashow-Weinberg-Salam model~\cite{Glashow:1961tr, Weinberg:1967tq, Salam:1968rm}. With the gauge bosons being spin-$1$ particles and matter consisting of elementary spin-$1/2$ fermions, the Higgs field is the only scalar (spin-$0$) quantity in the SM.

At high energies, all SM~particles are exactly massless and the Standard Model Lagrangian can schematically be written as
\beq
\L = - \frac{1}{4} X_{\mu\nu} X^{\mu\nu} + \ii \bar{\psi} \slashed{D} \psi + \left(y_{ij} H \psi_i  \psi_j + \mathrm{h.c.}\right) + |D_\mu H|^2 - V(H)\, .
\eeq
The first term captures the kinetic terms of the electroweak and strong gauge bosons, where $X_{\mu\nu} \equiv \{B_{\mu\nu}, W_{\mu\nu}^a, G_{\mu\nu}^a\}$ are the relevant field strength tensors. The second term is the kinetic term of the SM~Weyl fermions~$\psi$ and anti-fermions~$\bar{\psi} = \psi^\dagger \gamma^0$, with $\slashed{D} \equiv \gamma^\mu D_\mu$ and Dirac matrices~$\gamma^\mu$. The covariant derivative $D_\mu$ encodes the interaction of these fermions with the force carriers. The third term are the Yukawa couplings of the matter fields to the Higgs doublet~$H$ with Yukawa matrix $y_{ij}$. The last two terms finally characterise the kinetic term, the gauge boson interactions and the quartic potential of the Higgs field.

At low energies ($\lesssim\SI{100}{\giga\electronvolt}$), the Higgs field develops a non-zero vacuum expectation value, $v = \SI{246}{\giga\electronvolt}$, and the electroweak symmetry is spontaneously broken to $U(1)_\mathrm{em}$ with the photon being the associated gauge boson. In the process, the Higgs boson itself, the weak gauge bosons ($W^\pm$ and~$Z$) as well as the quarks and leptons\hskip1pt\footnote{It is currently unclear whether neutrinos are coupled to the SM~Higgs field as well. In the SM, they are exactly massless, i.e.\ the possible Yukawa couplings are taken to be zero.} receive their mass by means of the Higgs mechanism~\cite{Higgs:1964ia, Englert:1964et, Higgs:1964pj, Guralnik:1964eu}. The fermion masses are given by $m_i = y_{ii} v/\sqrt{2}$, for example. Moreover, the $B$ and $W^3$~bosons mix and become the photon and the $Z$~boson. The parameter governing this relationship is Weinberg's weak mixing angle~$\theta_w \approx \SI{30}{\degree}$ which also sets the difference between the $W$ and $Z$~boson masses, $m_W = m_Z \cos\theta_w$.

The Standard Model has 19 free parameters which have to be measured in experiments. Most of these parameters have been determined to exquisite precision and the~SM has passed most of its tests with flying colours~\cite{Patrignani:2016xqp}. For instance, the electromagnetic fine-structure constant, $\alpha \approx 1/137$, has now been measured with a relative uncertainty of less than one part in~\num{e9}~\cite{Bouchendira:2010es, Patrignani:2016xqp}. Remarkably, the value obtained from a measurement of the electron magnetic moment together with a quantum electrodynamics calculation to tenth-order in perturbation theory achieves the same level of precision and agrees~\cite{Aoyama:2012wj, Aoyama:2014sxa}.

\subsection{Motivations for New Physics}
\label{sec:theoryMotivation}
Apart from the obvious fact that the Standard Model of particle physics only describes three of the four fundamental forces of Nature, there are several experimental and theoretical pieces of evidence pointing towards physics beyond the Standard Model. In the following, we describe some of these puzzles with a particular focus on those problems which might be solved by dark sectors containing new light and weakly-interacting particles. For more in-depth reviews, we refer the reader to~\cite{Jaeckel:2010ni, Essig:2013lka, deGouvea:2013onf, Alexander:2016aln, Patrignani:2016xqp}.\medskip

It is well known that the weak interactions break charge-parity~(CP) symmetry through a complex phase in the CKM~matrix describing the mixing of quarks. A long-standing puzzle, however, is the non-observation of CP~violation in the strong interactions although the QCD~Lagrangian allows a CP-violating term, 
\beq
\L \supset \frac{\theta}{32\pi^2} G_{\mu\nu,a} \tilde{G}^{\mu\nu,a}\, ,
\eeq
where $\theta$ is the CP-violating phase and $\tilde{G}^{\mu\nu,a} \equiv \frac{1}{2} \epsilon^{\mu\nu\rho\sigma} G_{\rho\sigma}^a$ is the dual gluon field strength tensor. Among other implications, this term induces a neutron electric dipole moment. Current experimental bounds on this quantity imply that $|\theta| < \num{e-10}$~\cite{Baker:2006ts, Afach:2015sja} and not $\theta \sim \mathcal{O}(1)$ as one might expect. The absence of CP~violation in~QCD is therefore a fine-tuning problem with the number of proposed solutions being limited. The most popular suggestion is to promote this parameter to a dynamical field which can naturally make~$\theta$ small. This can be achieved by spontaneous breaking of the (approximate) Peccei-Quinn symmetry~\cite{Peccei:1977hh} at a scale~$f_a$ which is known as the decay constant. The spontaneous symmetry breaking gives rise to a pseudo-Nambu-Goldstone boson, the QCD axion, with a specific relation between the axion mass and its decay constant, $m_a \propto 1/f_a$, and generic couplings not only to the gluon, but also to quarks. Interestingly, axions which are non-thermally produced for example via the misalignment mechanism could make up all or part of the dark matter in the universe~\cite{Preskill:1982cy, Abbott:1982af, Dine:1982ah, Marsh:2015xka} and thereby solve two of the major problems in particle physics at once. In principle, axions can also be thermally produced~\cite{Turner:1986tb, Braaten:1991dd, Bolz:2000fu, Masso:2002np, Graf:2010tv, Salvio:2013iaa}. This is possible if the decay constant~$f_a$ is small enough and/or the reheating temperature of the universe large enough, so that the axion production rate $\Gamma \sim T^3/f_a^2$ was larger than the Hubble rate~$H$ at early times.

The neutrino sector is the least understood part of the SM. These very weakly interacting particles are famously predicted to be massless, but flavour oscillations\hskip1pt\footnote{A neutrino which is produced or emitted in a well-defined flavour eigenstate has a non-zero probability of being detected in a different flavour state.} have been observed for solar, atmospheric, reactor and accelerator neutrinos. This implies that these particles must have non-zero masses which are however constrained to be at the sub-\si{\electronvolt} level and, therefore, much smaller than the mass of any other known particle. From these neutrino oscillation experiments, we know the mass-squared differences, $\Delta m^2$, of two pairs of neutrinos, but their ordering is still unknown, i.e.\ we do not know whether there are two light neutrinos and one slightly heavier neutrino (normal hierarchy) or whether it is the other way around (inverse hierarchy). In fact, the overall mass scale and the mechanism by which neutrinos obtain their masses is still a mystery. It might be the standard Higgs mechanism with very small Yukawa couplings (if neutrinos are not their own anti-particle), it could be through a new Higgs-like field or the underlying process might be of an entirely different nature. Many attempts to incorporate neutrino masses in the Standard Model rely on new hidden sectors with new forces and/or particles, some of which might be light or even massless (see e.g.~\cite{Mohapatra:2005wg, King:2013eh, King:2015aea}). This also applies to solving additional problems related to neutrinos, such as the possibility of CP~violation and the question whether they are Majorana or Dirac fermions, i.e.\ their own anti-particle or not.

The anomalous magnetic moment of the muon can be measured accurately in the laboratory by studying the precession of~$\mu^+$ and~$\mu^-$ in a constant external magnetic field. Precise SM~calculations, however, are discrepant at the level of about~$3.5\hskip1pt\sigma$~\cite{Patrignani:2016xqp}. One possibility for addressing this tension is the presence of a new force mediated by an extra Abelian $U(1)$ gauge boson~$A_\mu'$~\cite{Jegerlehner:2009ry}, which is usually referred to as a dark photon. This new particle may be very weakly coupled to electrically charged particles (including muons) through kinetic mixing with the photon~\cite{Holdom:1985ag} and thereby alleviate the $g_\mu-2$ discrepancy. Further motivation for this type of BSM~model~\cite{ArkaniHamed:2008qn} comes from the unexpected energy-dependent rise in the ratio of positrons to electrons in cosmic rays, as observed for example by PAMELA~\cite{Adriani:2008zr}. Moreover, a substantial number of dark matter models also employ new millicharged species, i.e.\ particles with small un-quantized electric charge, as dark matter candidates. These new particles usually arise naturally in many SM~extensions which include dark photons or extra dimensions. As in the case of axions, light dark photons themselves may also constitute (part of) the dark matter.

One of the most famous theoretical problems of the~SM is the hierarchy problem. At the core, it raises the question why the weak force is \num{e24} times stronger than gravity or, in other words, why the Higgs boson mass, $m_H \approx \SI{125}{\giga\electronvolt}$, is so much smaller than the Planck mass, $\Mp \approx \SI{2.4e18}{\giga\electronvolt}$, despite of quantum corrections. For a long time, the most popular solution has been to introduce a new spacetime symmetry known as supersymmetry. One of its consequences would be that every SM~particle has a (heavier) superpartner. Apart from being theoretically appealing in a number of ways, supersymmetry could also, for instance, solve the problem of gauge coupling unification, provide a dark matter candidate and, imposed as a local symmetry, lead to a theory of supergravity. Having said that, no sign of any supersymmetric extension of the Standard Model has so far been seen at the LHC or elsewhere. This has led particle physicists to further explore alternative ways of solving the hierarchy problem. Examples include composite Higgs models in which the Higgs boson may arise as a pseudo-Nambu-Goldstone boson (pNGB)~\cite{Kaplan:1983fs, Kaplan:1983sm, Agashe:2004rs}, extra dimensions~\cite{ArkaniHamed:1998rs, Antoniadis:1998ig, Randall:1999ee}, a relaxation mechanism~\cite{Graham:2015cka} or a large number of hidden sectors~\cite{Arkani-Hamed:2016rle}. Instead of new heavy degrees of freedom, which have not been observed to date, light particles arise in these approaches for instance as mediators between the dark sector and the Standard Model or as a consequence of symmetry breaking patterns.

Ultimately, we would like to find a unified theory of the four fundamental forces or, in other words, of particle physics and gravity. Candidates for such a theory, e.g.\ string theory, generically predict (very rich) hidden sectors with a large number of moduli (scalar) fields, additional gauge bosons or even higher-spin particles. Some of these extra fields may have small or vanishing masses and may only be very weakly coupled to the SM~degrees of freedom. In fact, most of the scenarios discussed above, in particular axions and hidden photons, generically occur in string compactifications~\cite{Svrcek:2006yi, Jaeckel:2010ni, Baumann:2014nda, Marsh:2015xka}. For example, axions arise as Kaluza-Klein modes of higher-dimensional form fields when compactifying the $d>4$~spacetime dimensions to the usual four. These axions are exactly massless to all orders in perturbation theory, but receive their mass by non-perturbative effects, such as instantons. The presence of a large number of such axions with a wide range of masses is sometimes referred to as the string axiverse~\cite{Arvanitaki:2009fg}.\medskip

This comparably short and condensed list of current puzzles and their possible solutions only indicates the wealth of particle physics phenomenology these days. Having said that, many of the proposed SM~extensions have in common that they contain a dark sector which is only (very) weakly coupled to the Standard Model. Since the SM~symmetries restrict the kind of interactions or ``portals'' between the sectors, four of them are commonly used: the axion and vector portals, which we introduced above, as well as the Higgs and neutrino portals, which contain additional scalars or sterile neutrinos to mediate the interaction. There has been substantial experimental effort as well as progress in recent years at both the energy and intensity frontiers. This includes probing higher and higher energies especially at particle colliders, and using intense sources and ultra-sensitive detectors, for example in laser experiments~\cite{Jaeckel:2010ni, Essig:2013lka, deGouvea:2013onf}. In this thesis, we will contribute to the efforts on the cosmic frontier, i.e.\ the search for new physics in cosmology.

\subsection{Effective Field Theory of Light Species}
\label{sec:EFTlightSpecies}
We have just seen that additional light species arise in many well-motivated extensions of the Standard Model. In addition, cosmology tends to provide constraints on broad classes of models rather than very specific scenarios. Instead of working through these BSM~models one by one, it is therefore more efficient to study the interactions between the new species with the SM~degrees of freedom within the framework of effective field theory and thereby capture their main phenomenology. Generally speaking, this means parametrizing these interactions as
\beq
\L \supset \sum g\, \O_X \O_\mathrm{SM}\, ,	\label{eq:schematicEFT}
\eeq
where $\O_X$ and $\O_\mathrm{SM}$ are operators of light and SM~fields, respectively. Since the small masses of~$X$ could receive large quantum corrections, we employ (approximate) symmetries to prevent this. The allowed couplings in~\eqref{eq:schematicEFT} are then restricted by these protective symmetries. Ultimately, we will be able to put constraints on the interaction terms from cosmological measurements because the relic abundance of a new species is governed by its decoupling temperature, cf.\ Fig.~\ref{fig:deltaNeff}, which in turn depends on the coupling parameters~$g$.

\subsubsection{Introduction to effective field theory}
The effective field theory framework builds upon the realization that Nature comes with many separated scales and that we can usually analyse natural phenomena by considering one relevant scale at a time. For example, we are able to describe the formation of hydrogen during recombination in terms of protons and electrons, and do not have to take the dynamics of quarks and gluons inside the proton into account. Similarly, we can treat the primordial plasma prior to decoupling as a fluid with certain properties without knowing the exact trajectories of the baryons and photons. In the same spirit, we were able to use Fermi theory instead of the full Standard Model to study neutrino decoupling. This is due to the fact that Fermi theory well describes the weak interactions of neutrinos with the other leptons at energy scales far below the mass of the $W$~boson and can therefore be seen as an effective description of this part of the Standard Model.

An effective field theory is a quantum field theory which takes advantage of scale separation and only includes the appropriate degrees of freedom to describe certain phenomena occurring at a particular energy scale. This way, we capture the important aspects at the scale of interest and do not have to worry about the potentially rich spectrum of states in the underlying microscopic theory. In the case of Fermi theory, the full description in terms of the weak force within the~SM is known (and perturbative), but we can simplify the calculation of neutrino scattering by turning to the effective description. In the search for new physics, on the other hand, we usually do not have a full high-energy theory, but EFTs allow us to parametrize the unknown interactions, to estimate the magnitudes of these interactions and to classify their relative importance.\medskip

The guiding principle in the construction of EFTs are the symmetries obeyed by the relevant particle or field content. In this thesis, we are interested in EFTs which contain light fields in addition to the SM~degrees of freedom. Following~\cite{Brust:2013xpv}, we only consider models that are minimal and technically natural. Minimality here means that the additional particle content is as small as possible, i.e.\ it usually consists of only one additional elementary particle. The possible theory space is further reduced by imposing an (approximate) symmetry to protect the small mass from large quantum corrections.\footnote{The requirement of minimality essentially prevents us from employing strong dynamics in the new sector to generate large anomalous dimensions for the mass terms to render them small in the~EFT because these models generically have a rich spectrum.} We therefore require that our~EFT not only obeys the SM~symmetries, in particular Poincar\'e and gauge invariance, but also additional protective symmetries related to the specific scenario. Since the available symmetries depend on the spin of the new particle, it is convenient to organize the~EFT according to spin. In the following, we consequently study the effective field theories of light species with different spins~$s\leq2$. Since interacting particles with spin~$s>2$ necessarily have to be composites~\cite{Arkani-Hamed:2017jhn}, we therefore exhaust all possibilities for elementary particles.

\subsubsection{Spin-0: Goldstone bosons}
A particularly well-motivated example of a new light particle are Goldstone bosons which generically appear when global symmetries are spontaneously broken. Goldstone bosons are either massless (if the broken symmetry was exact) or naturally light (if it was approximate). Examples of light pseudo-Nambu-Goldstone bosons are axions~\cite{Peccei:1977hh, Weinberg:1977ma, Wilczek:1977pj}, familons~\cite{Wilczek:1982rv, Reiss:1982sq, Kim:1986ax}, and majorons~\cite{Chikashige:1980ui, Chikashige:1980qk}, associated with spontaneously broken Peccei-Quinn, family and lepton-number symmetry, respectively. Below the scale of the spontaneous symmetry breaking, the couplings of the Goldstone boson~$\phi$ to the SM~particles can be characterised through a set of effective interactions
\beq
\L \supset \sum\frac{\O_\phi\hskip1pt \O_\mathrm{SM}}{\Lambda^\Delta}\, ,	\label{eq:coupling}
\eeq
where $\Lambda$ is related to the symmetry breaking scale and the operators $\O_\phi$ are restricted by an approximate shift symmetry, $\phi \to \phi + \const$. 

Axion, familon and majoron models are characterised by different couplings in~\eqref{eq:coupling}.\footnote{We will follow the common practice of reserving the name axion or axion-like particle for pNGBs that couple to the gauge bosons of the~SM through operators like $\phi F_{\mu \nu} \tilde{F}^{\mu \nu}$. For simplicity, we will refer to all such particles simply as axions.} For example, below the electroweak symmetry breaking scale the axion couplings to the photon and gluon fields is given by
\beq
\L \supset -\frac{1}{4} \left(\frac{\phi}{\Lambda_\gamma} F_{\mu\nu}\tilde{F}^{\mu\nu} + \frac{\phi}{\Lambda_g} G_{\mu\nu,a}\tilde{G}^{\mu\nu,a}\right)\, ,	\label{eq:axionCoupling}
\eeq
with the field strength tensors~$F_{\mu\nu}$ and~$G_{\mu\nu}^a$, and their duals~$\tilde{F}_{\mu\nu}$ and~$\tilde{G}_{\mu\nu}^a$. The photon interaction term is somewhat model dependent, but typically arises together with the gluon coupling. Strictly speaking, only the latter has to be present in order to solve the strong CP~problem. The interaction of familons with the charged SM~fermions are generally governed by a current, e.g.\ the axial vector current $J_5^\mu$, derivatively coupled to the scalar field,
\beq
\L \supset -\frac{\partial_\mu \phi}{\Lambda_\psi}\, J_5^\mu = -\frac{\partial_\mu \phi}{\Lambda_\psi}\, \bar \psi \gamma^\mu \gamma^5 \psi\, , \label{eq:axialVectorCoupling}
\eeq
where $\psi$ is any charged SM~fermion. Majorons may arise as Goldstone bosons associated with the spontaneous breaking of the neutrino flavour symmetry which might be related to the existence of neutrino masses. Their interactions with the~SM may be of a similar form as for familons. For a more in-depth discussion of light spin-$0$ fields, we refer to Chapter~\ref{chap:cmb-axions}.

\subsubsection[Spin-\texorpdfstring{$\frac{1}{2}$}{1/2}: Light fermions]{Spin-$\mathbf{\frac{1}{2}}$: Light fermions}
Light spin-$\frac{1}{2}$ particles can either arise as Weyl or Dirac fermions. Both are natural possibilities since their mass terms can be protected by a chiral and axial symmetry, respectively. A hidden Dirac fermion~$\Psi$, which is invariant under $\Psi \to \e^{\ii\alpha\gamma^5}\Psi$, with arbitrary phase~$\alpha$ and $\gamma^5 \equiv \ii \gamma^0 \gamma^1 \gamma^2 \gamma^3$, can couple to the hypercharge gauge boson~$B_\mu$ through a dimension-5 dipole interaction,
\beq
\L \supset -\frac{1}{\Lambda_d}\, \bar{\Psi} \sigma^{\mu\nu} \Psi\, B_{\mu\nu}\, ,	\label{eq:dipoleInteraction}
\eeq
with $\sigma^{\mu\nu} \equiv \frac{\ii}{2}[\gamma^\mu,\gamma^\nu]$. This coupling may arise from loops in a high-energy theory that involve heavy charged particles. Another option is to couple the new particle~$\Psi$ to any SM~fermion~$\psi$ via dimension-6 four-fermion interactions,
\beq
\L \supset \frac{1}{\Lambda_f^2} \left( d_s\, \bar{\Psi} \Psi\, \bar\psi \psi+ d_p\, \bar{\Psi} \gamma^5 \Psi\, \bar\psi \gamma^5 \psi\ + d_a\, \bar{\Psi} \gamma^\mu \Psi\, \bar\psi \gamma_\mu \psi + d_v \bar{\Psi} \gamma^\mu \gamma^5 \Psi\, \bar\psi \gamma_\mu \gamma^5 \psi\right) ,	\label{eq:fourFermionInteraction}
\eeq
where the parameters $d_i$, $i=s,p,a,v$, are in principle $\mathcal{O}(1)$~numbers for the scalar, pseudo-scalar, axial and vector couplings, respectively. One possibility for inducing such an interaction is the exchange of a massive scalar or vector boson, such as those generated by spontaneously broken gauge symmetries. A well-studied example are light sterile neutrinos which may be coupled to the~SM via a new massive vector boson~$Z'$ from a spontaneously broken~$U(1)$ symmetry~\cite{Abazajian:2012ys, Anchordoqui:2012qu, Abazajian:2017tcc}.

A Weyl fermion~$\chi$, which is protected by the chiral symmetry $\chi \to \e^{\ii \alpha} \chi$, with arbitrary phase~$\alpha$, can be coupled to SM~fermions by four-fermion interactions as in~\eqref{eq:fourFermionInteraction}. Alternatively, we can couple $\chi$ to the hypercharge gauge boson $B_\mu$ through a dimension-6 anapole moment,
\beq
\L \supset -\frac{1}{\Lambda_a^2}\, \chi^\dagger \bar{\sigma}^\mu \chi\, \partial^\nu\! B_{\mu\nu}\, ,	\label{eq:anapoleInteraction}
\eeq
where $(\bar{\sigma}^\mu) = (\mathds{1}, -\sigma^i)$ with Pauli matrices $\sigma^i$.

\subsubsection{Spin-1: Vector bosons}
Massless spin-$1$ particles carry fewer degrees of freedom than their massive counterparts which means that a mass term cannot arise from perturbative quantum effects. This implies that massless vector bosons are technically natural. A small protected mass can nevertheless be generated, for example via the standard Higgs mechanism, a St{\"u}ckelberg mechanism or in LARGE volume compactifications of string theory~\cite{Jaeckel:2010ni}. The most minimal scenario is to couple one hidden~$U(1)$ gauge boson~$A_\mu'$, a dark photon, to the SM~fermions via a dipole interaction. Prior to electroweak symmetry breaking, this coupling can be written as
\beq
\L \supset - \frac{1}{\Lambda_{A'}^2} F'_{\mu\nu} \, H \bar{\psi} \sigma^{\mu\nu} \psi\, ,	\label{eq:spin1_dipole}
\eeq
where~$F'_{\mu\nu}$ is the field strength associated with the dark photon and the presence of the Higgs field~$H$ is required by gauge invariance. Once the Higgs has acquired its non-zero vacuum expectation value~$v$, this operator will be effectively of dimension-5 with an effective scale of $\tilde{\Lambda}_{A'} = \Lambda_{A'}^2/v$.

Another option to couple a dark photon to the Standard Model is by kinetically mixing the new boson $A'_\mu$ with the hypercharge gauge boson, $\L \supset -(\epsilon/2)\, F'^{\mu\nu} B_{\mu\nu}$~\cite{Holdom:1985ag}. If the extra particle~$A'_\mu$ is exactly massless, we can decouple it from the~SM by a field redefinition. This scenario therefore requires to enlarge the particle content by a new Dirac fermion~$\chi$ which is millicharged under the electroweak SM~gauge group, e.g.\
\beq
\L \supset -\epsilon g_{A'} \cos\theta_w\, \bar\chi \slashed{A} \chi -\epsilon g_{A'} \sin\theta_w\, \bar\chi \slashed{Z} \chi - g_{A'}\, \bar\chi \slashed{A}' \chi \, ,	\label{eq:spin1_millicharged}
\eeq
where~$g_{A'}$ is the coupling between~$\chi$ and the dark photon, and~$\epsilon$ is the kinetic mixing parameter. As mentioned above, the additional fermion~$\chi$ can be (a~fraction~of) the dark matter in many models.

\subsubsection[Spin-\texorpdfstring{$\frac{3}{2}$}{3/2}: Gravitino]{Spin-$\mathbf{\frac{3}{2}}$: Gravitino}
There is one unique elementary particle with spin-$3/2$, the gravitino. The existence of the gravitino as the superpartner to the graviton is a universal prediction of supergravity. Its mass is set by the supersymmetry breaking scale, $m_{3/2} \sim F/\Mp$, and can be very small in low-scale supersymmetry breaking scenarios. Although the gravitino typically interacts with gravitational strength, its longitudinal component couples to the~SM in the same way as the Goldstino,
\beq
\L \supset -\frac{1}{F^2}\, \chi^\dagger \sigma_\mu \partial_\nu \chi\, T^{\mu\nu} \, .	\label{eq:gravitino}
\eeq
The coupling parameter of this component of the gravitino is therefore enhanced compared to the Planck scale.

\subsubsection{Spin-2: Graviton}
The graviton is the unique elementary spin-$2$ particle and the force carrier of gravity. It is massless and only interacts with gravitational strength, i.e.\ all its interactions with the SM~fields are suppressed by~$\Mp\approx\SI{2.4e18}{\giga\electronvolt}$.\bigskip

\noindent
This concludes our discussion of the effective field theory of light species. We will return to these EFT couplings in~\textsection\ref{sec:eftConstraints} where we discuss some of the current constraints from particle accelerators, laboratory experiments, astrophysics and cosmology.

\section{Neutrinos and Dark Radiation}
\label{sec:neutrinosDarkRadiation}
In this thesis, we are particularly interested in measurements of the radiation density of the universe to probe particle physics. The contribution from photons,~$\rho_\gamma$, is fixed by the very well measured value of the CMB~temperature, 
\beq
\rho_{\gamma,0} = \frac{\pi^2}{15} T_0^4 \approx \SI{2.0e-15}{\electronvolt\tothe{4}}\, ,
\eeq
which corresponds to a physical photon density of $\omega_\gamma = \Omega_\gamma h^2 \approx \num{2.5e-5}$. As we discussed in~\textsection\ref{sec:cnub}, the standard models of cosmology and particle physics also predict a contribution from neutrinos. According to equations~\eqref{eq:gStar} and~\eqref{eq:neutrinoTemperature}, the expected radiation density from each neutrino species in the instantaneous decoupling limit is\footnote{In principle, the decoupling temperatures of the three different neutrino species slightly differ from each other (see e.g.~\cite{Dolgov:2002wy, Lesgourgues:2013ncb}). We neglect this and effectively take~$T_\nu$ to be the average neutrino temperature.}
\beq
\rho_{\nu_i} = \frac{7}{8} \left(\frac{T_{\nu,0}}{T_0}\right)^{\!4/3} \rho_\gamma = \frac{7}{8} \left(\frac{4}{11}\right)^{\!4/3} \rho_\gamma \equiv a_\nu^{-1} \rho_\gamma\, ,	\label{eq:anu}
\eeq
where we defined $a_\nu \approx 4.40$ instead of its inverse for later convenience. The three neutrino species of the Standard Model consequently contribute a significant amount to the total radiation density in the early universe: $\rho_\nu/\rho_r = \sum_i \rho_{\nu_i}/\rho_r \approx \SI{41}{\percent}$. The gravitational effects of neutrinos are therefore significant at early times which is why we can observe their imprints in the CMB and BAO~spectra (cf.\ Chapters~\ref{chap:cmb-phases} and~\ref{chap:bao-neutrinos}; see also~\cite{Follin:2015hya}) although their contribution to the total energy density today is very small.

\subsection{Effective Number of Relativistic Species}
\label{sec:Neff}
We have assumed so far that neutrinos decoupled instantaneously and, therefore, did not receive any of the entropy that was released when electrons and positrons annihilated. However, a small amount of the entropy is actually transferred to the neutrino sector as well. Instead of changing the prefactor $a_\nu$ in~\eqref{eq:anu}, it is common to introduce the effective number of neutrinos $N_\nu$ as
\beq
\rho_r = \rho_\gamma + \rho_\nu = \left[ 1 + a_\nu^{-1} N_\nu \right] \rho_\gamma\, .
\eeq
In the instantaneous decoupling limit, we have $N_\nu = 3$. Accounting for plasma corrections of quantum electrodynamics, flavour oscillations and, in particular, the fact that neutrinos have not fully decoupled when electrons and positrons annihilated, one finds $N_\nu = 3.046$ in the SM~\cite{Mangano:2005cc}.\footnote{Recently, a more accurate calculation including neutrino oscillations with the present values of the mixing parameters found $N_\nu = 3.045$~\cite{deSalas:2016ztq}. At the level of precision anticipated in the near future and under consideration in this thesis, the difference is irrelevant which is why we keep the standard value of $N_\nu = 3.046$ as our baseline assumption.} In this sense, measurements of $N_\nu > 0$ probe the energy density of the C$\nu$B and $N_\nu \neq 3.046$ could be a sign for non-standard properties of neutrinos or changes to the standard thermal history. 

The introduction of this new parameter also provides a convenient way to include the possibility of radiation in excess of the SM~expectations. BSM~physics may add extra radiation density~$\rho_X$ to the early universe which is often referred to as dark radiation. It is conventional to measure this radiation density relative to the density $\rho_{\nu_i}$ of a single SM~neutrino,
\beq
\Delta\Neff \equiv \frac{\rho_X}{\rho_{\nu_i}} = a_\nu \frac{\rho_X}{\rho_\gamma}\, ,
\eeq
and define the \textit{effective number of relativistic species}
\beq
\Neff \equiv N_\nu + \Delta\Neff = 3.046 + \Delta\Neff\, . \label{eq:Neff}
\eeq
The parameter $\Neff$ therefore captures the difference between the total radiation density of the universe, $\rho_r$, and the CMB~photon energy density, $\rho_\gamma$, normalized to the energy density of a single neutrino species, $\rho_{\nu_i}$. This is the central cosmological parameter under investigation in this thesis. Slightly generalizing the statement from above, measurements of $\Neff > 0$ therefore probe the energy density of the C$\nu$B, and $\Neff \neq 3.046$ would be a signature of physics beyond the standard models of particle physics and/or cosmology.

Current observations of the CMB~anisotropies and the light element abundances find~\cite{Ade:2015xua, Cyburt:2015mya}\footnote{Stated imprecisely, colliders also constrain the number of neutrino species through precision measurements of the width of the $Z$~decay. Yet, when stated more carefully, collider measurements only tell us how many fermions with mass below~$\frac{1}{2}m_Z$ couple to the~$Z$ boson~\cite{ALEPH:2005ab}. These experiments find very close agreement with three families of active neutrinos.}\footnote{The quoted CMB~constraint includes high-$\ell$~polarization data which has been labelled as preliminary by the Planck collaboration. Only considering temperature and low-$\ell$~polarization data results in $\Neff=3.13^{+0.30}_{-0.34}$~\cite{Ade:2015xua}.}
\begin{align}
\Neff &= 3.04 \pm 0.18 \quad \text{(CMB)}\, ,	\label{eq:NCMB}		\\
\Neff &= 2.85 \pm 0.28 \quad \text{(BBN)}\, ,
\end{align}
which are consistent with the SM~prediction of $\Neff = N_\nu = 3.046$. These measurements represent a highly significant detection of the energy density associated with the cosmic neutrino background and a non-trivial confirmation of the thermal history back to about one second after the big bang when neutrinos decoupled. The consistency of the measurements is remarkable, although the interpretation is somewhat sensitive to assumptions about the cosmological model and constraints weaken considerably in some extensions of the $\Lambda\mathrm{CDM}$~model (cf.\ Section~\ref{sec:analysis}). Furthermore, these measurements put interesting limits on many extensions of the~SM containing additional light fields and/or thermal histories that enhance or dilute the radiation density (see e.g.~\cite{Jungman:1995bz, Cadamuro:2010cz, Menestrina:2011mz, Boehm:2012gr, Brust:2013xpv, Weinberg:2013kea, Cyr-Racine:2013fsa, Vogel:2013raa, Millea:2015qra, Chacko:2015noa}).\medskip

\begin{figure}
	\includegraphics[scale=0.65]{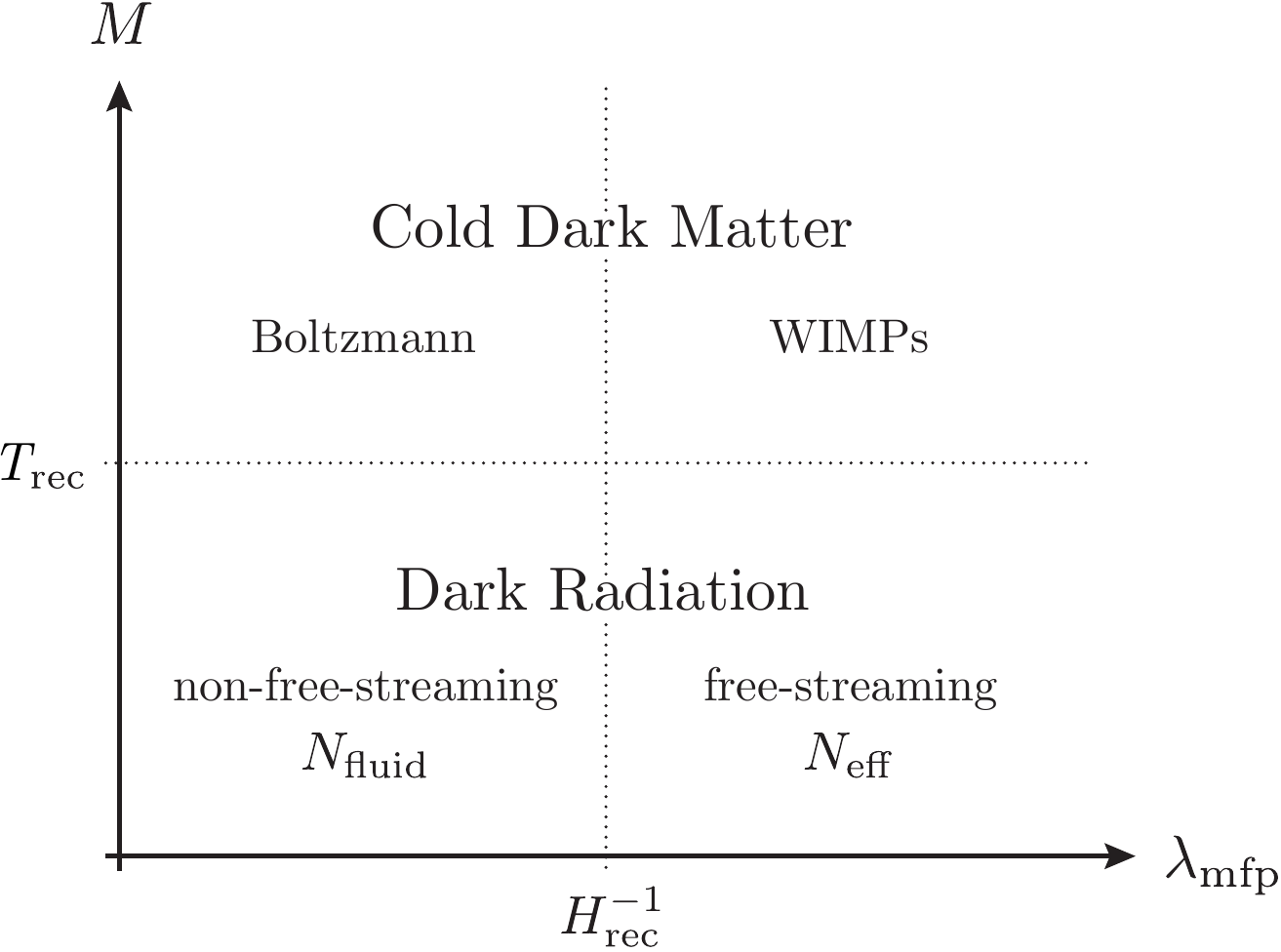}
	\caption{Particles beyond the Standard Model can be classified according to their masses~$M$ and their mean free paths~$\lambda_\mathrm{mfp}$ (both normalized relative to the Hubble rate at recombination,~$H_\rec$, relevant for CMB~observations). Particles with $M > T_\rec$ contribute to the cold dark matter of the universe at recombination, while particles with $M < T_\rec$ are relativistic at recombination. Massive, strongly interacting particles are Boltzmann-suppressed and, therefore, do not contribute a cosmologically interesting radiation density. Dark radiation separates into free-streaming and non-free-streaming particles. Note that non-thermal relics, such as non-thermally produced axions, escape the simple characterisation of this figure.}
	\label{fig:classification}
\end{figure}

Strictly speaking, the parameter~$\Neff$ is usually taken to capture neutrinos and neutrino-like species, i.e.\ it refers to free-streaming radiation. As we illustrate in Fig.~\ref{fig:classification},%
not all relativistic BSM~particles have to fall into this category. We will hence also allow for a contribution from non-free-streaming radiation and capture this by the following parameter:\footnote{Another attempt to parametrize (non-)free-streaming radiation is in terms of a viscosity parameter~$c_\mathrm{vis}$~\cite{Hu:1998kj}. This parameter has recently been detected by Planck, $c_\mathrm{vis}^2= 0.331\pm 0.037$~\cite{Ade:2015xua} (see also~\cite{Archidiacono:2013lva, Audren:2014lsa}). However, as discussed in~\cite{Sellentin:2014gaa}, $c_\mathrm{vis}^2 =\frac{1}{3}$ is not equivalent to free-streaming radiation and differs from $\Lambda\mathrm{CDM}$ by $\Delta\chi^2 = 20$. Our parametrization has the advantage that it reproduces $\Lambda\mathrm{CDM}$ when $\Nf = N_\nu = 3.046$ and $\Nn = 0$. A similar parametrization has appeared in~\cite{Bell:2005dr, Friedland:2007vv, Chacko:2015noa, Brust:2017nmv} and was analysed with WMAP~data in~\cite{Bell:2005dr, Friedland:2007vv}. However, it has only recently become possible to distinguish these parameters with high significance (see Section~\ref{sec:analysis}).}\footnote{This approach of reducing large numbers of models to a single model which captures their essential features is analogous to the use of simplified models to search for new physics at the LHC~\cite{Alves:2011wf}.}
\beq
\Nn \equiv a_\nu \, \frac{\rho_Y}{\rho_\gamma}\, ,	\label{eq:Nfluid}
\eeq
with non-free-streaming radiation density $\rho_Y$. In Chapter~\ref{chap:cmb-phases}, we will characterise the different effects of~$\Nf$ and~$\Nn$ on the photon-baryon fluid, and study their distinct cosmological imprints in detail. In particular, we will keep $\Nn \neq 0$ when analysing Planck data and forecasting future constraints. In Chapters~\ref{chap:bao-forecast} and~\ref{chap:bao-neutrinos}, we will restrict ourselves to~$\Neff$ and implicitly set $\Nn \equiv 0$ because many BSM~models, in particular simple extensions of the SM, only predict free-streaming species.

\subsection{Thermal History with Additional Species}
\label{sec:thermalHistoryAdditional}
\begin{figure}[t!]
	\centering
	\includegraphics{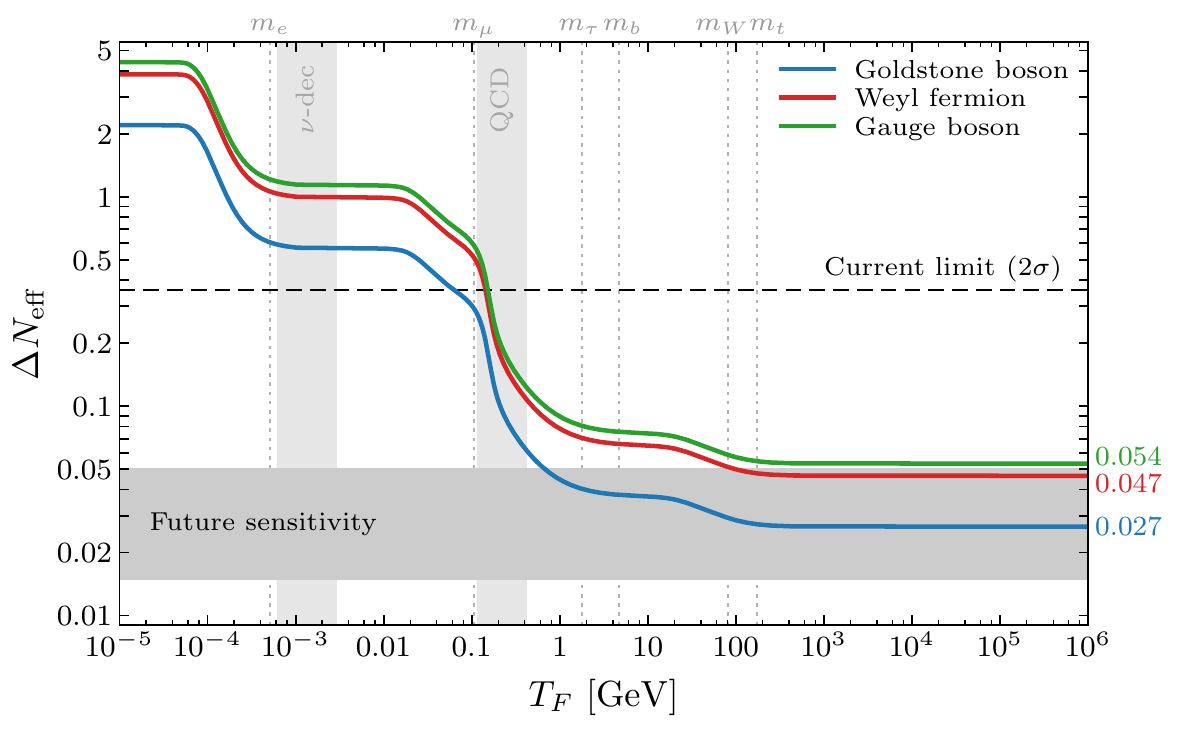}
	\caption{Contributions of a single thermally-decoupled Goldstone boson, Weyl fermion or massless gauge boson to the effective number of relativistic species,~$\Delta\Neff$, as a function of its decoupling temperature~$\Tf$. The current Planck limit at \SI{95}{\percent}~c.l.\ from~\cite{Ade:2015xua} and the possible future sensitivity (cf.~Chapter~\ref{chap:bao-forecast}) illustrate the current and future power of cosmological surveys to constrain light thermal relics. The drop in $\Delta\Neff$ by about one order of magnitude around $\Tf \sim \SI{150}{\mega\electronvolt}$ is due to the QCD phase transition, which is denoted by a vertical gray band as is neutrino decoupling. The dotted lines denote some of the mass scales at which SM~particles and anti-particles annihilate.}
	\label{fig:deltaNeff}
\end{figure}

A natural source for~$\Delta\Neff \neq 0$ are extra relativistic particles.\footnote{For simplicity, we set $\Nn=0$ in this section, but note that the thermal history is unchanged if an additional species contributes to $\Nn$ instead of $\Delta\Neff$ because we will only be concerned with the relic energy density. As we will see later, distinguishing $\Neff$ and $\Nn$ relies on more subtle effects.} Let us therefore consider a light species~$X$ as the only additional particle in some BSM~theory. Assuming this species was in thermal equilibrium with the SM~particles at some point in the history of the universe, we can compute its contribution to~$\Delta\Neff$ in a similar way to our calculation of the relic density of neutrinos. Furthermore, assuming that the species~$X$ freezes out well before neutrino decoupling,\footnote{If the particle~$X$ froze out after electron-positron annihilation, its temperature would be the same as for photons, $T_X = T$, for all times. On the other hand, we would have $T > T_X > T_\nu$ if the new species decoupled in the relatively small window between the onset of neutrino decoupling and the conclusion of $e^+ e^-$ annihilation. As illustrated in Fig.~\ref{fig:deltaNeff}, both scenarios are ruled out by current measurements at more than $5\sigma$ ($3\sigma$) for new particles with (without) spin. We therefore do not discuss them further.} $T_{F,X} \gtrsim \SI{10}{\mega\electronvolt}$, the temperature associated with~$X$ in relation to the neutrino temperature~$T_\nu$ is
\beq
T_X = \left(\frac{g_*(T_{F,\nu})}{g_*(T_{F,X})}\right)^{\!1/3} T_\nu = \left(\frac{10.75}{g_{*,SM}}\right)^{\!1/3} \left(\frac{g_{*,SM}}{g_*(T_{F,X})}\right)^{\!1/3} T_\nu \approx 0.465 \left(\frac{g_{*,SM}}{g_*(T_{F,X})}\right)^{\!1/3} T_\nu\, ,
\eeq
where we employed entropy conservation among the particles in thermal equilibrium, i.e.\ we assumed no significant entropy production after the decoupling of~$X$. We also inserted $g_{*,SM} = 106.75$, and $g_*(T_{F,\nu}) = 10.75$ due to photons, electrons and neutrinos. (We omitted the additional subscript for entropy since $g_*(T) = g_{*S}(T)$ in thermal equilibrium.) Note that neither~$g_*(T_{F,\nu})$ nor~$g_*(T_{F,X})$ include a contribution from the particle~$X$ as it has decoupled and does not receive any of the released entropy of later annihilation processes.

After neutrinos decouple, the evolution of~$T_X$ and~$T_\nu$ is the same since both scale as $T_i \propto a^{-1}$. Provided that the extra species are relativistic when they freeze out and remain decoupled, their energy ratio stays constant and we get
\beq
\Delta\Neff = \frac{\rho_X}{\rho_{\nu_i}} = \frac{g_{*,X}\, T_X^4}{g_{*,{\nu_i}}\, T_\nu^4} \approx 0.027\, g_{*,X} \left(\frac{g_{*,SM}}{g_*(T_{F,X})}\right)^{\!4/3}\, ,	\label{eq:DeltaNeffCalculation}
\eeq
where we used $g_{*,{\nu_i}} = 7/4$ and $g_{*,X}$ depends on the spin of the particle (cf.~\textsection\ref{sec:equilibrium}). Figure~\ref{fig:deltaNeff}%
shows the contribution to~$\Delta\Neff$ from a single thermally-decoupled species as a function of the decoupling temperature $\Tf$ and the spin of the particle based on~\eqref{eq:DeltaNeffCalculation}. We see that decoupling after the QCD phase transition produces a contribution to~$\Neff$ that is comparable to that of a single neutrino species, which is ruled out (or at least strongly disfavoured) by current observations. On the other hand, decoupling before the QCD phase transition creates an abundance that is smaller by an order of magnitude due to the much larger number of available degrees of freedom~$g_*$. Since Planck is blind to these particles, such scenarios are still consistent with current limits. Future observations will therefore give us access to particles that are more weakly coupled than neutrinos and decoupled before the end of the QCD phase transition.

Employing $g_*(T_{F,X}) \leq g_{*,SM} = 106.75$ in~\eqref{eq:DeltaNeffCalculation}, we get the asymptotic values displayed in Fig.~\ref{fig:deltaNeff}. We find that any additional species~$X$ which has been in thermal equilibrium with the Standard Model at any point in the history of the universe contributes the following minimal amount to the radiation density of the universe:
\beq
\Delta \Neff \geq 0.027 g_{*,X} = 
		\begin{dcases}
			\, 0.027	& \text{Goldstone boson (spin-0)},		\\
			\, 0.047	& \text{Weyl fermion (spin-1/2)},		\\
			\, 0.054	& \text{massless gauge boson (spin-1)}.	\\
		\end{dcases}
\label{eq:DeltaNeffValues}
\eeq
This is an important result. Provided that future cosmological surveys are sensitive to $\Delta\Neff = 0.027$, we can detect any particle which has ever been in thermal equilibrium with the SM. Reaching this minimal thermal abundance is therefore a very interesting target for upcoming measurements with important consequences for BSM~physics (see Chapter~\ref{chap:cmb-axions} and~\cite{Brust:2013xpv, Chacko:2015noa}, for instance). We find it intriguing that this threshold seems to be within reach of future observations. In Chapters~\ref{chap:cmb-phases} and~\ref{chap:bao-forecast}, we will quantify this expectation and see that future CMB and LSS~experiments indeed have the potential to achieve such measurements. In this case, these cosmological surveys either have to detect extra light relics or we can put strong constraints on their couplings to the~SM, as we will show in Chapter~\ref{chap:cmb-axions} for the case of Goldstone bosons.\medskip

When deriving equation~\eqref{eq:DeltaNeffCalculation}, we assumed an extension of the~SM in which there is no significant entropy production and the species~$X$ is the only addition to the SM~particle content. We briefly revisit these assumptions in the following. Arguably, many BSM~models come with additional massive particles which increase the effective number of relativistic degrees of freedom~$g_*$ at high temperatures, $T \gg \SI{100}{\giga\electronvolt}$. Moreover, their annihilation might subsequently reduce the relic abundance of~$X$ according to~\eqref{eq:DeltaNeffCalculation}, since $g_*(T_{F,X}) > g_{*,SM}$, allowing for $\Delta\Neff<0.027g_{*,X}$. This is however degenerate with the uncertainty on the reheating temperature, i.e.\ the question whether such an additional particle has ever been in thermal equilibrium, because the energy density is only diluted for decoupling temperatures above the masses of these new particles. Furthermore, a dilution of the minimal contribution to $\Neff$ by a factor of two requires $g_{*,BSM} \approx 1.7 g_{*,SM}$, i.e.\ almost a doubling of the SM degrees of freedom. Although this is possible in SM~extensions, these models commonly contain many additional light particles as well (see e.g.~\cite{Arvanitaki:2009fg, Arkani-Hamed:2016rle}) which in turn enhance the expected value of $\Delta\Neff$. For instance, just three light degrees of freedom can compensate for the large increase in field content in the Minimal Supersymmetric Standard Model~(MSSM). In this sense, our assumption of one new particle~$X$ is to be considered as a conservative choice.

In certain BSM~models, the minimal contribution to~$\Delta\Neff$ might also be diluted because entropy is produced, e.g.\ by out-of-equilibrium decays of massive BSM~particles or a phase transition. Introducing the parameter $\gamma = s(T_{F,X})/s(T_{F,\nu})$ to capture the amount of entropy production after the species~$X$ decoupled, equation~\eqref{eq:DeltaNeffCalculation} becomes
\beq
\Delta\Neff = 0.027\, g_{*,X} \left(\frac{g_{*,SM}}{g_*(T_{F,X})}\right)^{\!4/3} \times \gamma^{-4/3}\, .
\eeq
To illustrate the possible size of~$\gamma$, we consider a massive particle~$\chi$ which decays out of equilibrium, i.e.\ after it has decoupled and acquired an abundance $Y_i = n_\chi/s$. If we assume that this particle dominates the energy density of the universe when it decays, the entropy ratio is given by $\gamma \sim g_{*,d}^{1/4} Y_i\, (m_\chi^2 \tau_\chi/\Mp)^{1/2}$~\cite{Kolb:1990vq}, with particle mass~$m_\chi$ and lifetime~$\tau_\chi$. This might potentially be large and, for example, heavy gravitinos are constrained to have never been in thermal equilibrium during the hot big bang evolution in this way. For concreteness, let us consider a light scalar particle, which froze out at high temperatures, and an additional massive species with mass $m_\chi = \SI{1}{\tera\electronvolt}$, which decoupled at $T>m_\chi$ and has a lifetime corresponding to a decay temperature $\Td \sim \SI{10}{\giga\electronvolt}$. Such a particle would dominate the energy density of the universe at $T \sim \SI{1}{\tera\electronvolt}$ and easily dilute the contribution of the light particle to~$\Delta\Neff$. Having said that, this scenario is no different than assuming the reheating temperature to be $\Tr \sim \SI{10}{\giga\electronvolt}$, in which case the scalar particle would have never been in thermal equilibrium. In a sense, we are therefore expanding the definition of the reheating temperature~$\Tr$ to include a possible second reheating phase caused by an out-of-equilibrium decay process after inflationary reheating. This is therefore degenerate with our general uncertainty on the reheating temperature. Depending on the involved time scales, the combined constraints on~$\Neff$ from~BBN and the CMB/LSS, together with other cosmological signatures of particle decays, might be able to put some constraints on the entropy factor~$\gamma$ and thereby limit the possible dilution of~$\Delta\Neff$ due to entropy production at correspondingly late times (see e.g.~\cite{Kawasaki:1999na, Hannestad:2004px, Iocco:2008va}).\vspace{-1pt}\medskip

So far, we have implicitly assumed that $\Delta\Neff>0$. The annihilation or decay of particles into neutrinos after their decoupling could however produce $\Delta\Neff<0$, for instance. This type of signature would therefore hint towards certain types of BSM models. One such example may be a light particle which is unstable and may decay to neutrinos. Interestingly, its out-of-equilibrium decay, i.e.\ after it decoupled at early times, would lead to a larger suppression of~$\Neff$ than its equilibrium decay and would therefore be easier to detect. Alternatively, non-standard neutrino properties or particles decaying to photons could effectively lead to $\Delta\Neff<0$ as well.\vspace{-1pt}

\subsection{Current Constraints on EFT Parameters}
\label{sec:eftConstraints}
As we have already argued in Section~\ref{sec:estimates}, astrophysics and cosmology can put strong constraints on light and weakly-coupled particles. In the following, we reconsider the EFT of light species and discuss some of the constraints on the effective interactions from current measurements. We will also connect these couplings to the thresholds~\eqref{eq:DeltaNeffValues} for~$\Delta\Neff$. As in~\textsection\ref{sec:EFTlightSpecies}, we will study the new species separately according to their spin.\vspace{-1pt}

\subsubsection{Spin-0: Goldstone bosons}\vspace{-1pt}
The Goldstone couplings~\eqref{eq:axionCoupling} and~\eqref{eq:axialVectorCoupling} are constrained by laboratory experiments~\cite{Essig:2013lka, Graham:2015ouw}, by astrophysics~\cite{Raffelt:1996wa, Raffelt:2012kt} and by cosmology~\cite{Brust:2013xpv, Marsh:2015xka} (see~\cite{Irastorza:2018dyq} for a recent review).\footnote{With the advent of gravitational-wave astronomy, very light~pNGBs could potentially also be probed around rapidly-rotating black holes (see e.g.~\cite{Arvanitaki:2010sy, Yoshino:2013ofa, Brito:2014wla, Brito:2017zvb, Baumann:2018vus}).} While laboratory constraints, such as those from light-shining-through-walls experiments, helioscopes or haloscopes, have the advantage of being direct measurements, their main drawback is that they are usually rather model-specific and sensitive only to narrow windows of pNGB masses. Astrophysical and cosmological constraints are complementary since they are relatively insensitive~to the detailed form of the couplings to the~SM and span a wide range of masses. The main astrophysical~constraints on these new light particles come from stellar cooling~\cite{Raffelt:1996wa}. In order not to disrupt successful models of stellar evolution, any new light particles must be more weakly coupled than neutrinos. The axion-photon coupling, for example, is bounded by $\Lambda_\gamma \gtrsim \SI{1.5e10}{\giga\electronvolt}$ (\SI{95}{\percent}~c.l.)\ in this way~\cite{Ayala:2014pea}, in agreement with our estimate~\eqref{eq:stellarCoolingEstimate}. Moreover, since neutrinos couple to the rest of the~SM through a dimension-six operator (suppressed by the electroweak scale), the constraints on extra particles are particularly severe for dimension-four and dimension-five couplings to the~SM.

We refer to Chapter~\ref{chap:cmb-axions} for constraints on the SM~interactions of Goldstone bosons from precise cosmological measurements. We will see that current bounds may be improved by many orders of magnitude if the threshold value $\Delta\Neff = 0.027$ can be observationally excluded. In addition, most bounds on familon couplings can already be substantially improved by cosmological experiments that are only sensitive to $\Delta\Neff \sim 0.05$.

\subsubsection[Spin-\texorpdfstring{$\frac{1}{2}$}{1/2}: Light fermions]{Spin-$\mathbf{\frac{1}{2}}$: Light fermions}
The best constraint on the dipole interaction~\eqref{eq:dipoleInteraction} between new light Dirac fermions and the Standard Model comes from stellar cooling with $\Lambda_d \gtrsim \SI{e9}{\giga\electronvolt}$~\cite{Raffelt:2012kt}. On the other hand, collider searches provide the most competitive bounds on the four-fermion couplings of~\eqref{eq:fourFermionInteraction}, $\Lambda_f \gtrsim \O(\SI{1}{\tera\electronvolt})$, with the precise numbers depending somewhat on the type of SM~fermion~\cite{Brust:2013xpv}. The anapole coupling~\eqref{eq:anapoleInteraction} of extra Weyl fermions has similar constraints from particle accelerators.

Figure~\ref{fig:deltaNeff} shows that any dark Dirac or Weyl fermion must have decoupled before the end of the QCD phase transition in order to be compatible with the current Planck limits on the radiation density. Cosmological measurements will be able to put strong constraints on these couplings since Dirac fermions minimally contribute $\Delta\Neff = 0.094$ to the radiation density, which will be well within the sensitivity of future surveys. For interactions including light Weyl fermions, the cosmological threshold is smaller by a factor of 2, but the combination of upcoming CMB and LSS~experiments still has the potential to improve these bounds by many orders of magnitude.

\subsubsection{Spin-1: Vector bosons}
The constraints on the dipole coupling~\eqref{eq:spin1_dipole} of dark photons from stellar and supernova cooling depend on the type of SM~fermions that are involved. For instance, the coupling to electrons is bounded by $\Lambda_{A'} \gtrsim \SI{e7}{\giga\electronvolt}$ and the muon interaction by $\Lambda_{A'} \gtrsim \SI{e4}{\giga\electronvolt}$~\cite{Brust:2013xpv}. Models with extra millicharged particles, such as those of~\eqref{eq:spin1_millicharged}, also have strong stellar and supernova bounds. These constraints are complemented by a number of collider searches and laboratory experiments, for instance in beam dump experiments and using helioscopes (see e.g.~\cite{Jaeckel:2010ni, Essig:2013lka}). For hidden millicharged fermion masses of $m_\chi \lesssim \SI{100}{\kilo\electronvolt}$ for example, the current bound on the kinetic mixing parameter is $\epsilon \lesssim \num{e-13}$~\cite{Jaeckel:2010ni}. These constraints are again supplemented by current measurements of~$\Neff$ from~BBN and the~CMB as well~\cite{Brust:2013xpv}.

Since massless gauge bosons in thermal equilibrium contribute at least~$0.054$ to~$\Neff$, future cosmological measurements will be able to severely restrict the allowed parameter space of these models, including scenarios where millicharged fermions make up the dark matter. It goes without saying that bounds on non-Abelian gauge fields will be much stronger because they would carry a radiation density equivalent to $\Nn \sim 0.07(N^2-1)$ for a dark~$SU(N)$ gauge group~\cite{Buen-Abad:2015ova}.

\subsubsection[Spin-\texorpdfstring{$\frac{3}{2}$}{3/2}: Gravitino]{Spin-$\mathbf{\frac{3}{2}}$: Gravitino}
Since the gravitino coupling~\eqref{eq:gravitino} is suppressed by the supersymmetry breaking scale and not the Planck scale, the gravitino may have been in thermal equilibrium with the Standard Model at early times. It minimally contributes $\Delta\Neff=0.047$ to the radiation density in the early universe because it behaves like a Weyl fermion as far as its contribution to~$\Delta\Neff$ is concerned. In fact, current cosmological constraints on $m_{3/2}$ could indicate that the gravitino may be so light (and hence~$F$ so small) that it only decouples at $T \lesssim \SI{100}{\giga\electronvolt}$~\cite{Abazajian:2016yjj}, i.e.\ the remaining minimal contribution to~$\Neff$ gets enhanced compared to the value for a Weyl fermion, $\Delta\Neff \gtrsim 0.057$, and therefore more easily detected in upcoming observations.

\subsubsection{Spin-2: Graviton}
It is unlikely that the graviton has ever been in thermal equilibrium because it only interacts with gravitational strength. Its thermal abundance is therefore negligible. Having said that, since gravitational waves are massless and free-streaming, the energy density of a stochastic background in the early universe may contribute a small amount to the radiation density and therefore to the effective number of relativistic species~$\Neff$~\cite{Smith:2006nka, Boyle:2007zx, Meerburg:2015zua, Lasky:2015lej}.

\section{Cosmological Signatures of Light Relics}
\label{sec:signaturesLightRelics}
Changing the radiation density by adding additional light species\hskip1pt\footnote{A change in~$\Neff$ might not be related to new light, but new heavy particles which could decay or otherwise mimic the effects of light degrees of freedom~\cite{Fischler:2010xz, Menestrina:2011mz, Hooper:2011aj, Hasenkamp:2012ii}. However, these scenarios are usually accompanied by additional signatures. The late decay or annihilation to photons led to $\Delta\Neff>0$, for example, but could also change~BBN, prompt spectral distortions of the CMB~frequency spectrum or leave distinct signals in the CMB~anisotropy spectrum, depending on when this additional energy injection into the thermal bath of photons occurred (see e.g.~\cite{Hu:1993gc, Chluba:2008aw, Chluba:2013wsa} and references therein).} to the universe will impact a number of cosmological observables in more or less subtle ways. In the following, we will describe the influence of light relics on the CMB~anisotropies, the matter power spectrum, the BAO~spectrum and big bang nucleosynthesis. We will see that precise measurements of the various observables and reliable extraction of several signatures turns cosmology into an accurate tool for probing cosmic neutrinos as well as other light thermal relics and, hence, this type of SM~and BSM~physics.

\subsection{Diffusion Damping}
\label{sec:species_diffusionDamping}
\begin{figure}
	\centering
	\includegraphics{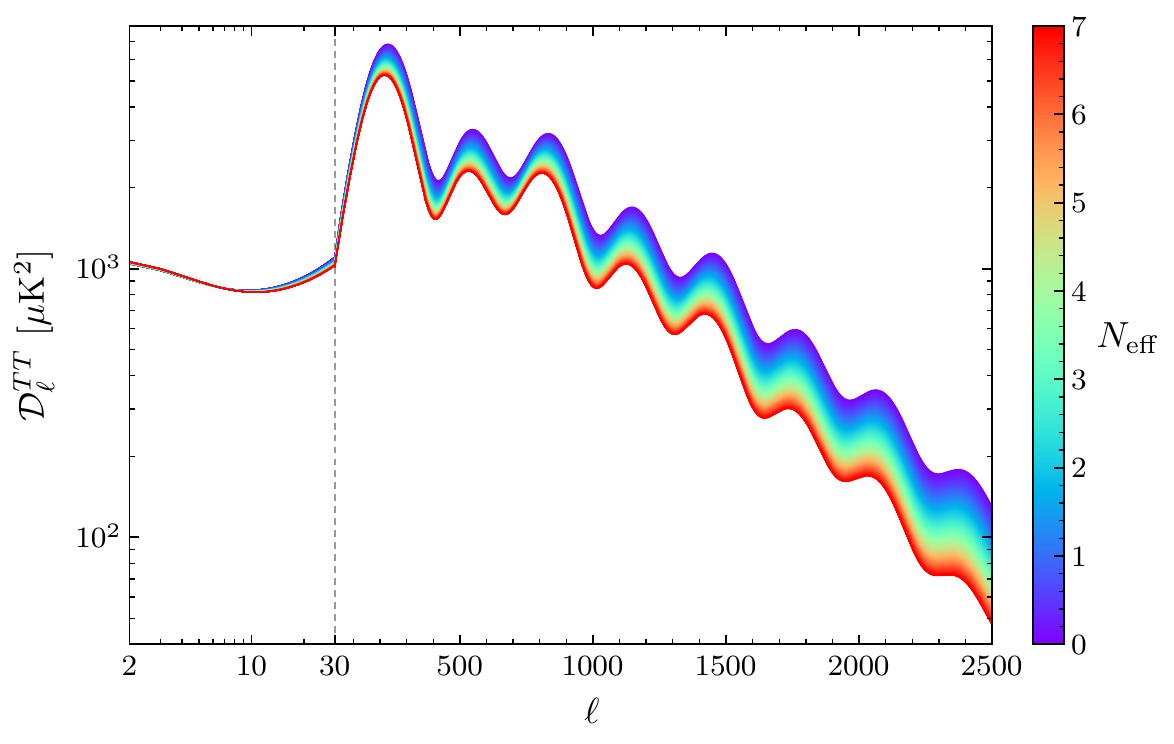}
	\caption{Variation of the CMB~temperature power spectrum $\mathcal{D}_\ell^{TT}$ as a function of $\Neff$ for fixed angular size of the sound horizon~$\theta_s$.}
	\label{fig:cmbPowerSpectrumNeffSequence}
\end{figure}

A change in the radiation density impacts the CMB~power spectrum in several ways (see~\cite{Hou:2011ec} for an excellent discussion). At the level of the homogeneous cosmology, the largest effect of relativistic particles is a change in the expansion rate during radiation domination, which according to the first Friedmann equation~\eqref{eq:FriedmannEquations} is given by
\beq
3 M_\mathrm{pl}^2 H^2 = \rho_\gamma \left(1+\frac{\Nf + \Nn}{a_\nu}\right) ,
\eeq
where free-streaming particles~($\Nf$) and non-free-streaming particles~($\Nn$) contribute equally, and $a_\nu \approx 4.40$ was introduced in~\eqref{eq:anu}. As we anticipated in~\textsection\ref{sec:cosmicSoundWaves}, the change in the Hubble rate manifests itself in a modification of the damping tail of the CMB (see Fig.~\ref{fig:cmbPowerSpectrumNeffSequence}).%
However, understanding the precise impact of a change in the radiation density is non-trivial~\cite{Hou:2011ec}, since changing $H$ will also affect the location of the first acoustic peak, which is extremely well measured. To study the effects of~$\Nf+\Nn$, it is instructive to consider the ratio of the angular sizes of the damping scale and the sound horizon, $\theta_d = 1/(k_d D_A)$ and $\theta_s = r_s/D_A$, in order to eliminate the a priori unknown angular diameter distance to the last-scattering surface, $D_A$, which also depends on the Hubble rate. Around the time of recombination, the wavelength associated with the mean squared diffusion distance is proportional to the Hubble rate, $k_d^2 \propto H_\rec$, according to~\eqref{eq:damping}. Since the sound horizon is inversely proportional to $H_\rec$, the ratio scales as
\beq
\frac{\theta_d}{\theta_s} = \frac{1}{k_d\,r_s} \propto \frac{1}{H_\rec^{1/2} H_\rec^{-1}} = H_\rec^{1/2}\, .	\label{eq:thetadOverThetas}
\eeq
Increasing~$\Neff$ (and hence~$H_\rec$) therefore leads to a larger~$\theta_d$ if we keep the angular scale of the first acoustic peak,~$\theta_s$, fixed, e.g.\ by simultaneously varying the Hubble constant~$H_0$. This implies that the damping kicks in at larger angular scales (smaller multipoles) and reduces the power in the damping tail when increasing the radiation density. This is exactly the behaviour that we observe for the CMB~power spectrum in Fig.~\ref{fig:cmbPowerSpectrumNeffSequence}. The constraint in~\eqref{eq:NCMB} is, in fact, mostly derived from measurements of the CMB damping tail~\cite{Keisler:2011aw, Ade:2015xua}.

It is useful to anticipate the possible degeneracies between the effects of extra relativistic species and changes in the cosmological parameters as these may limit the constraining power. The physical origin of the effect on the damping tail is given by~\eqref{eq:damping}, so we can understand the most severe degeneracy analytically. As pointed out in~\cite{Bashinsky:2003tk}, there is an important degeneracy between the expansion rate~$H$ and the primordial helium fraction~$Y_p$. Since helium has a much larger binding energy than hydrogen, increasing the helium fraction for fixed~$\omega_b$ (which is well determined by the relative peak heights) will decrease the number of free electrons at the time of hydrogen recombination, $n_e \propto (1-Y_p)$, resulting in a larger photon mean free path and, in consequence, more damping. It is then possible to change~$Y_p$ and~$H$ simultaneously in a way that keeps the damping scale fixed, cf.~$k_d^{-2} \propto (n_e H)^{-1}$. Including the dependence on~$n_e$ in~\eqref{eq:thetadOverThetas}, we therefore have $\theta_d/\theta_s \propto (H_\rec/n_e)^{1/2}$. For fixed~$\theta_s$, the damping scale consequently remains unchanged if $Y_p$ is reduced while simultaneously increasing~$\Neff$, i.e.\ the parameters~$\Neff$ and~$Y_p$ are anti-correlated. If we want to measure~$Y_p$ independently from~BBN, this degeneracy will be one of the main limiting factors for constraints on extra relativistic species in the future.\medskip

Given that the physics underlying the acoustic peaks in the CMB and BAO~spectra is the same, we expect to see similar effects of relativistic species. In particular, the baryon acoustic oscillations are also damped due to photon diffusion and hence sensitive to~$\Neff$ in the same way as the CMB~anisotropies. However, there is an additional exponential damping in the BAO~spectrum because it is non-linearly processed by gravitational evolution. Since a change in diffusion damping is therefore degenerate with theoretical uncertainties in the amount of non-linear damping, the constraining power of the damping tail is restricted in this observable.

\subsection{Phase Shift}
\label{sec:species_phaseShift}
Before Planck, the CMB~constraints on~$\Neff$ were mainly provided by the damping tail and therefore just probed the homogeneous radiation density. Recently, the experimental sensitivity has however improved to such a level that the measurements have started to become susceptible to neutrino perturbations (and those of other free-streaming relics). These affect the photon-baryon fluid through their gravitational influence and lead to a distinct imprint in the acoustic oscillations~\cite{Bashinsky:2003tk}. In the following, we present a heuristic description of this effect and refer to Section~\ref{sec:analytics} for the proper treatment.

\begin{figure}[t!]
	\includegraphics{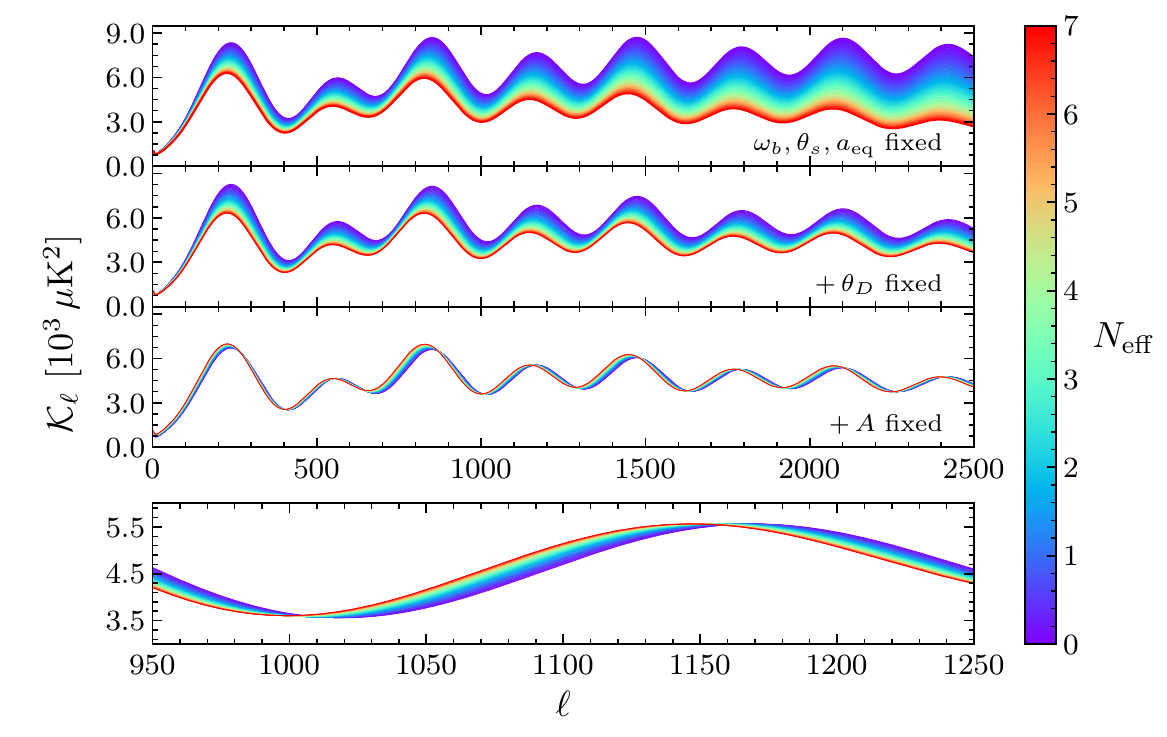}
	\caption{Variation of the CMB~temperature power spectrum as a function of $\Neff$. The spectra have been rescaled, so that the fiducial spectrum for $\Neff=3.046$ is undamped, $\mathcal{K}_\ell = \mathcal{D}_\ell^{TT} \exp\!\left\{a (\ell \theta_d)^\kappa\right\}$, with $\theta_d \approx \num{1.6e-3}$ and the fitting parameters $a \approx 0.68$, $\kappa \approx 1.3$, i.e.~the exponential diffusion damping was removed. Following~\cite{Follin:2015hya}, the physical baryon density~$\omega_b$, the scale factor at matter-radiation equality $\aeq \equiv \omega_m/\omega_r$ and the angular size of the sound horizon~$\theta_s$ are held fixed in all panels. The dominant effect in the first panel is the variation of the damping scale $\theta_D$. In the second panel, we fixed $\theta_D$ by adjusting the helium fraction $Y_p$. The dominant variation is now the amplitude perturbation $\delta A$. In the third panel, the spectra are normalized at the fourth peak. The remaining variation is the phase shift $\phi$ (see the zoom-in in the fourth panel).}
	\label{fig:CMBPhaseShift}
\end{figure}

For this purpose, let us reconsider the single point-like overdensities that we discussed in~\textsection\ref{sec:cosmicSoundWaves} and whose evolution is visualized in Fig.~\ref{fig:overdensities}. Since neutrinos freely stream through the universe close to the speed of light after their decoupling, they induce metric perturbations ahead of the sound horizon. The baryons and photons experience this change in the gravitational potential and the peak of these perturbations is slightly displaced to a larger radius. In Fourier space, this results in a small change in the temporal phase of the acoustic oscillations which is imprinted in both the temperature and polarization spectra of the~CMB as a coherent shift in the peak locations (see Fig.~\ref{fig:CMBPhaseShift}).%
A precise determination of the acoustic peaks therefore allows a measurement of free-streaming radiation independent of the Hubble rate. As a consequence, the degeneracy between $\Neff$ and $\Nn$, as well as between $\Neff$ and $Y_p$ can be broken by this subtle effect. We will study this \textit{phase shift} in the acoustic oscillations and its measurable implications in great detail throughout this thesis.\medskip

\begin{figure}[t!]
	\includegraphics{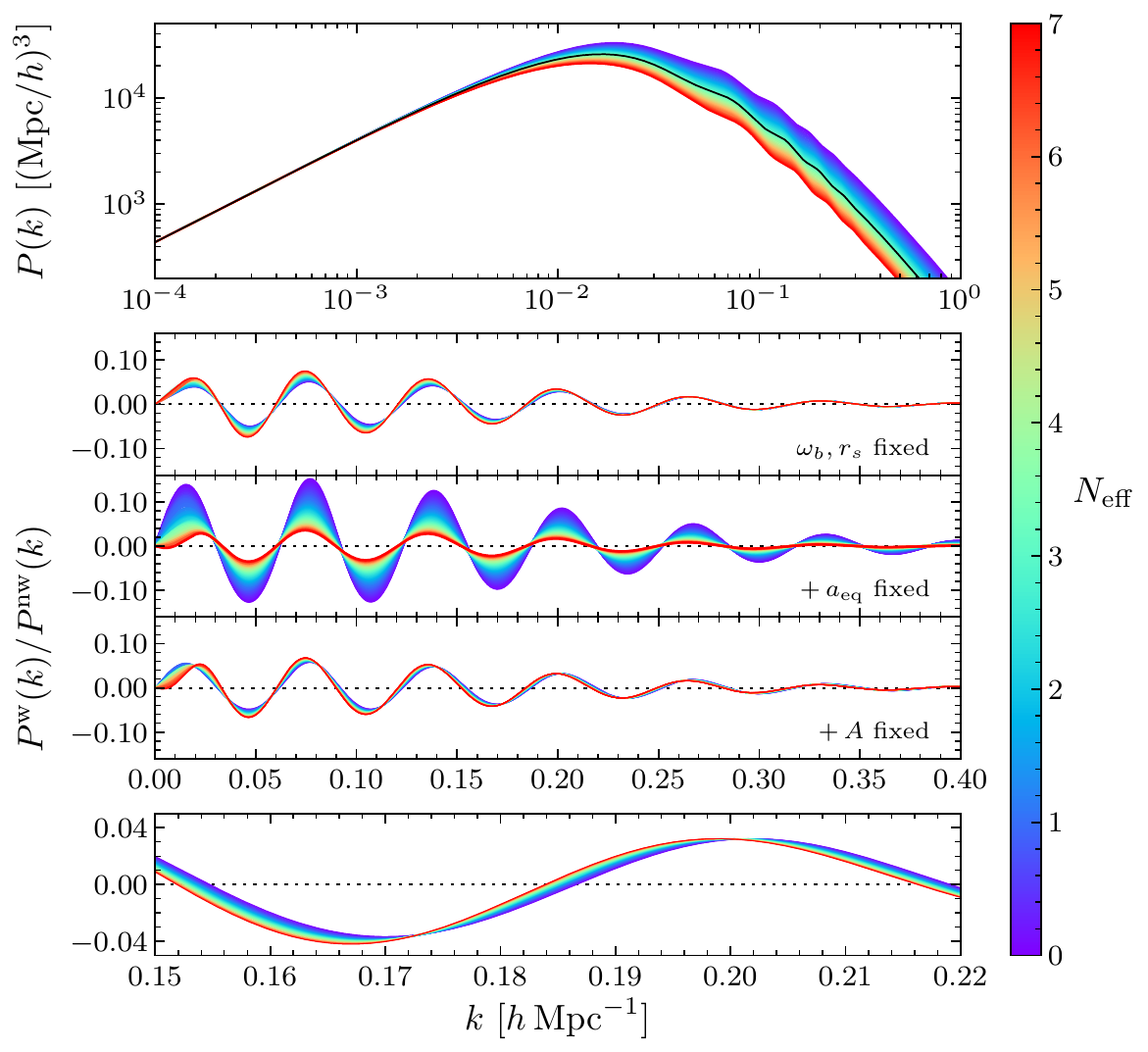}
	\caption{Variation of the matter power spectrum~$P(k)$~(\textit{top}) and the BAO spectrum~$\Pw(k)/\Pnw(k)$~(\textit{bottom}) as a function of~$\Neff$. The physical baryon density~$\omega_b$ and the physical sound horizon at the drag epoch, $r_s$, are held fixed in all panels of the BAO spectrum. In the second BAO panel, we fixed the scale factor at matter-radiation equality, $\aeq \equiv \omega_m/\omega_r$. Note that we effectively fix~$r_s$ by appropriately rescaling the wavenumbers for each spectrum, whereas we can vary~$\omega_c$ to fix~$\aeq$. The variation in the BAO amplitude $\delta A$ is then the dominant contribution. In the third BAO panel, the spectra are normalized at the fourth peak and the bottom panel shows a zoom-in illustrating the remaining phase shift.}
	\label{fig:BAOPhaseShift}
\end{figure}

Since the baryon acoustic oscillations in the matter power spectrum originate from the same mechanism as those in the CMB~spectra, the BAO~spectrum exhibits the same phase shift induced by the supersonic propagation of free-streaming species, cf.\ Fig.~\ref{fig:BAOPhaseShift}.%
Importantly, it was recently pointed out that this shift should also be robust to non-linear gravitational evolution in the late universe~\cite{Baumann:2017lmt}. This means that we can circumvent the usual LSS~complications related to non-linearities and use linear perturbation theory to predict and measure this shift in late-time observables such as the BAO~spectrum of galaxy clustering. This will be the basis of the modified BAO~analysis that we propose in Chapter~\ref{chap:bao-forecast} and will ultimately lead to the first measurement of this imprint of neutrinos in the distribution of galaxies in Chapter~\ref{chap:bao-neutrinos}.

The largest impact on the BAO~spectrum is actually a change in the sound horizon coming from a difference in the expansion history. By increasing the expansion rate during the radiation era and, hence, reducing the time over which the sound waves can propagate and diffuse, the acoustic scale decreases, $r_s \propto H^{-1}$. This of course appears as a variation in the frequency of the acoustic oscillations that BAO~surveys are very sensitive to. The influence of light species on the BAO~amplitude should be taken with care when constraining $\Neff$ because it is degenerate with theoretical uncertainties on the non-linear damping.\medskip

\subsection{Matter-Radiation Equality}
\label{sec:species_equality}
The two main consequences of an enhanced (diminished) radiation density on the matter power spectrum are a change in the location of the turn-over of the spectrum towards larger (smaller) scales and a decrease (increase) in power on small scales (see the top panel of Fig.~\ref{fig:BAOPhaseShift}). These two effects are related as they are both linked to a change in the time of matter-radiation equality. For fixed matter density, $\omega_m = \const$, an increase in the radiation density leads to a longer epoch of radiation domination. The maximum of the matter power spectrum therefore shifts to larger scales as its location corresponds to the wavenumber which enters the horizon at~$a_\mathrm{eq}$. At the same time, the amplitude of those modes that crossed the horizon before matter-radiation equality is suppressed since the growth of structures is only logarithmic during radiation domination (compared to linear during matter domination). Larger scales remain unchanged as they evolve deep in the matter-dominated era. Although these effects are clearly visible in the linear matter power spectrum, non-linearities make the matter power spectrum a less robust probe of~$\Neff$ than the phase shift.

In the~CMB, the power of more modes gets increased by radiation driving if the radiation density is larger. Moreover, a later matter-radiation equality also results in an enhanced ISW~effect because the gravitational potentials evolve by a slightly larger amount after recombination. This effect is however sub-dominant, unless we marginalize over $Y_p$~\cite{Hou:2011ec}. In this case, the ISW~effect contributes some constraining power as the sensitivity of the damping tail to~$\Neff$ is greatly reduced.

\subsection{Light Element Abundances}
\label{sec:species_bbn}
Before the advent of precision cosmology, in particular before the leap in CMB~detector sensitivity, the most important model-independent constraints on light relics came from big bang nucleosynthesis. Still today, the measurements of the primordial relic abundance of light elements are an important and independent probe of additional light species. The sensitivity of the predicted abundances from~BBN on~$\Neff$ is mainly through its impact on the expansion rate between about one second and a few minutes after the big bang. This is due to the fact that a larger expansion rate implies an earlier freeze-out of neutrons which leads to an increase in the neutron-to-proton ratio. As a consequence, more neutrons are available for the synthesis of helium and the other elements which results in an enhancement of the primordial abundances (see e.g.~\cite{Sarkar:1995dd, Iocco:2008va, Pospelov:2010hj, Steigman:2012ve, Cyburt:2015mya, Patrignani:2016xqp} for reviews of~BBN and its implications on BSM~physics).

With the sensitivity of the CMB and BBN to~$\Neff$ and~$Y_p$ reaching similar levels, there are a number of exciting possibilities to use these measurements in conjunction. First, we can independently check the predictions of~BBN by measuring the two BBN~input parameters~$\Neff$ and~$\omega_b$ as well as the output $Y_p$. Moreover, since the CMB and BBN provide snapshots of the universe at different times, we can also compare the measured values and investigate whether the radiation density might have evolved over time, for example due to the decay or presence of additional non-relativistic particles~\cite{Fischler:2010xz, Boehm:2012gr}. Having said that, for most parts of this thesis, we will assume that the physics underlying the cosmological observables and BBN are consistent which will result in the tightest possible constraints on the radiation density in the early universe.
	\chapter{New Target for Cosmic Axion Searches}
\label{chap:cmb-axions}
We have seen that many interesting extensions of the Standard Model of particle physics predict new light species. These particles have to be more weakly coupled than neutrinos, but may have been in thermal equilibrium in the early universe. In addition, we established that cosmological observations are, in principle, very sensitive to these types of particles. Given the Moore's law-like improvements in CMB~detector sensitivity~\cite{Abazajian:2013oma, Wu:2014hta}, cosmology will push the sensitivity to new light particles beyond the strength of weak-scale interactions and has the potential to explore a fundamentally new territory of BSM~physics. Specifically, observations may become sensitive to any light relics that have ever been in thermal equilibrium with the Standard Model. In this chapter, we demonstrate that even the absence of a detection would be informative since it would allow us to put constraints on the Standard Model interactions of light BSM~particles, such as axions. In many cases, the constraints achievable from cosmology will surpass existing bounds from laboratory experiments and astrophysical observations by orders of magnitude.\medskip

The outline of this chapter is as follows. In Section~\ref{sec:cmb-axions_introduction}, we lay out the general philosophy underlying these new constraints on light relics. In principle, this is applicable to all effective field theories introduced in~\textsection\ref{sec:EFTlightSpecies}. As an example, we then focus on the EFT of pseudo-Nambu-Goldstone bosons and consider the couplings to gauge bosons, to fermions and to neutrinos in turn. In Section~\ref{sec:axions}, we study the axion interactions focussing on the photon and gluon couplings, which are of particular phenomenological interest. In Section~\ref{sec:fermions}, we derive two types of bounds on familon interactions. Apart from constraints arising from the possibility that these pNGBs could have frozen out in the early universe, we also consider the case in which these particles might have frozen in, i.e.\ came back into thermal contact at later times. In Section~\ref{sec:neutrinos}, we finally discuss the same two classes of constraints for majorons. We conclude this chapter in Section~\ref{sec:conclusions_cmb-axions} with a summary of the derived constraints. Some technical details underlying these bounds are relegated to Appendix~\ref{app:cmb-axions_appendices}.

\section{Novel Constraints on Light Relics}
\label{sec:cmb-axions_introduction}
In~\textsection\ref{sec:thermalHistoryAdditional}, we derived the minimal contribution of any light thermal relic to the radiation density of the universe to be $\Delta\Neff = 0.027$. The fact that this contribution may be detectable in future cosmological observations has interesting consequences. First, the level $\Delta\Neff = 0.027$ provides a natural observational target (see e.g.~\cite{Brust:2013xpv, Salvio:2013iaa, Kawasaki:2015ofa, Chacko:2015noa, Ferreira:2018vjj} for related discussions). Second, even the absence of a detection would be very informative, because it would strongly constrain the EFT~couplings between the extra light relics and the SM~degrees of freedom, which can schematically be written as
\beq
\L \supset \sum \frac{\O_\phi\hskip1pt \O_\mathrm{SM}}{\Lambda^\Delta}\, .	\label{eq:coupling2}
\eeq
These new bounds arise since a thermal abundance can be avoided\hskip1pt\footnote{We remind ourselves that a thermal abundance may be diluted below the level of Fig.~\ref{fig:deltaNeff} if extra massive particles are added to the~SM. As we have discussed in~\textsection\ref{sec:thermalHistoryAdditional}, however, a significant change to our conclusions would require a very large number of new particles or a significant amount of non-equilibrium photon production. In addition, the possibility that dark sectors never reach thermal equilibrium with the~SM (see e.g.~\cite{Berezhiani:1995am, Hodges:1993yb, Feng:2008mu, Foot:2014uba, Reece:2015lch}) is strongly constrained by the physics of reheating~\cite{Adshead:2016xxj}.} if the reheating temperature of the universe,~$\Tr$, is below the would-be freeze-out temperature, i.e.~$\Tr < \Tf$. In that case, the extra particles have never been in thermal equilibrium and their densities therefore do not have to be detectable. In the absence of a detection, requiring $\Tf(\Lambda) > \Tr$ would place very strong bounds on the scale(s) in~\eqref{eq:coupling2}, i.e.~$\Lambda > \Tf^{-1}(\Tr)$. As we will see, in many cases the cosmological bounds will be much stronger than existing bounds from laboratory experiments and astrophysical observations. We note that these constraints make no assumption about the nature of dark matter because the thermal population of axions arises independently of a possible cold population. On the other hand, we have to assume that the effective description of the pNGBs with interactions of the form of~\eqref{eq:coupling2} holds up to $\Tf\ll\Lambda$. This is equivalent to assuming that the ultraviolet completion of the effective theory is not too weakly coupled. Moreover, we also require the absence of any significant dilution of~$\Delta\Neff$ after freeze-out. In practice, this means that we are restricting to scenarios in which the number of additional relativistic degrees of freedom at the freeze-out temperature is bounded by $\Delta{g_*(\Tf)}\lesssim{g_{*}^\mathrm{SM}(\Tf)} \approx \num{e2}$ (cf.\ our discussion in~\textsection\ref{sec:thermalHistoryAdditional}).\medskip

The couplings of~pNGBs to SM~fermions~$\psi$ can lead to a more complicated thermal evolution than the simple freeze-out scenario. Below the scale of electroweak symmetry breaking, the approximate chiral symmetry of the fermions makes the interactions with the~pNGBs effectively marginal. The temperature dependence of the interaction rate is then weaker than that of the Hubble expansion rate, leading to a recoupling (i.e.~freeze-in) of the~pNGBs at low temperatures. To avoid a large density of~pNGBs requires that the freeze-in temperature~$\Tff$ is smaller than the mass of the fermions participating in the interactions, $\Tff < m_\psi$, so that the interaction rate becomes Boltzmann suppressed before freeze-in can occur. Again, this constraint can be expressed as a bound on the scale(s)~$\Lambda$ that couple the~pNGBs to the SM~fermions. Although the freeze-in constraints are usually weaker than the freeze-out constraints, they have the advantage that they do not make any assumptions about the reheating temperature (as long as reheating occurs above~$T \sim m_\psi$). Furthermore, freeze-in produces larger contributions to~$\Delta\Neff$ which are detectable with a less sensitive experiment.\medskip\smallskip

In the rest of this chapter, we will show that cosmology is highly sensitive to axions, and other~pNGBs, when $\Delta\Neff = 0.027$ is detectable. To simplify the narrative, we will assume that this sensitivity will be reached with CMB-S4, either on its own or in conjunction with other data (cf.\ Chapters~\ref{chap:cmb-phases} and~\ref{chap:bao-forecast}; see also~\cite{Font-Ribera:2013rwa, Manzotti:2015ozr}). Alternatively, our arguments could be viewed as strong motivation for reaching this critical level of sensitivity in future experiments. In the following, we will derive bounds on the couplings of~pNGBs to the~SM arising from the absence of a detection. We will assume the mass range $0 \leq m_\phi < \SI{1}{MeV}$, so that the only possible decays of the~pNGBs are to photons or neutrinos. This regime is probed by measurements of~$\Neff$ for $m_\phi \leq \Trec$ and by warm dark matter constraints for $m_\phi > \Trec$ (see e.g.~\cite{Archidiacono:2015mda, DiValentino:2015wba}), where $\Trec \approx \SI{0.26}{eV}$ is the temperature at recombination.

\section{Constraints on Axions}
\label{sec:axions}
Axions arise naturally in many areas of high-energy physics, the QCD~axion being a particularly well-motivated example. Besides providing a solution to the strong CP~problem~\cite{Peccei:1977hh, Weinberg:1977ma, Wilczek:1977pj}, the QCD~axion also serves as a natural dark matter candidate~\cite{Preskill:1982cy, Abbott:1982af, Dine:1982ah}. Moreover, light axions appear prolifically in string theory~\cite{Svrcek:2006yi, Arvanitaki:2009fg, Baumann:2014nda} and have been proposed to explain the small mass of the inflaton~\cite{Freese:1990rb} as well as to solve the hierarchy problem~\cite{Graham:2015cka}. Finally, axions are a compelling example of a new particle that is experimentally elusive~\cite{Essig:2013lka, Graham:2015ouw} because of its weak coupling rather than due to kinematic constraints.\medskip 
 
What typically distinguishes axions from other~pNGBs are their unique couplings to the SM~gauge fields. Prior to~EWSB, we consider the following effective theory with shift-symmetric couplings of the axion to the SM~gauge sector:
\beq
\L_{\phi\mathrm{EW}} = - \frac{1}{4}\frac{\phi}{\Lambda} \left( c_1 \hskip1pt B_{\mu\nu} \tilde{B}^{\mu\nu} + c_2 \hskip1pt W_{\mu\nu}^a \tilde{W}^{\mu\nu,a} +c_3 \hskip1pt G_{\mu\nu}^a \tilde{G}^{\mu\nu,a} \right) ,	\label{eq:LEW0}
\eeq
where $X_{\mu\nu} \equiv \{B_{\mu\nu}, W_{\mu\nu}^a, G_{\mu\nu}^a\}$ are the field strengths associated with the gauge groups $\{U(1)_Y, SU(2)_L, SU(3)_C\}$, and $\tilde{X}^{\mu\nu} \equiv \frac{1}{2} \epsilon^{\mu\nu\rho\sigma} X_{\rho\sigma}$ are their duals. Axion models will typically include couplings to all SM~gauge fields, but only the coupling to gluons is strictly necessary to solve the strong CP~problem.\medskip

At high energies, the rate of axion production through the gauge field interactions in~\eqref{eq:LEW0} can be expressed as~\cite{Salvio:2013iaa} (see also~\cite{Braaten:1991dd, Bolz:2000fu, Masso:2002np, Graf:2010tv})
\beq
\Gamma(T,\Lambda_n) = \sum_{n=1}^3\gamma_n(T)\, \frac{T^3}{\Lambda_n^2}\, ,	\label{eqn:Gamma}
\eeq
where $\Lambda_n \equiv \Lambda/c_n$. The prefactors~$\gamma_n(T)$ have their origin in the running of the couplings and are only weakly dependent on temperature. For simplicity of presentation, we will treat these functions as constants in the main text, but take them into account in Appendix~\ref{app:cmb-axions_appendices}. We see that the production rate, $\Gamma \propto T^3$, decreases faster than the expansion rate during the radiation era, $H \propto T^2$. The axions therefore freeze out when the production rate drops below the expansion rate, with the freeze-out temperature~$\Tf$ determined by $\Gamma(\Tf) = H(\Tf)$. This thermal abundance can be avoided if the reheating temperature of the universe~$\Tr$ was below the would-be freeze-out temperature, i.e.~$\Tr < \Tf$. In that case, the temperature of the universe was simply never high enough to bring the axions into thermal equilibrium. We can express this condition as
\beq
\Gamma(\Tr,\Lambda_n) < H(\Tr) = \frac{\pi}{\sqrt{90}} \sqrt{g_{*,R}}\,\frac{\Tr^2}{\Mp}\, ,
\eeq
where $g_{*,R} \equiv g_*(\Tr)$ denotes the effective number of relativistic species at~$\Tr$. For a given reheating temperature, this is a constraint on the couplings~$\Lambda_n$ in~\eqref{eqn:Gamma}. Treating the different axion couplings separately, we can write
\beq
\Lambda_n \, >\, \left(\frac{\pi^2}{90} g_{*,R}\right)^{\!-1/4} \sqrt{\gamma_{n,R} \hskip1pt \Tr\hskip1pt \Mp}\, ,	\label{eq:freezeoutconstraint}
\eeq
where $\gamma_{n,R} \equiv \gamma_n(\Tr)$. In the following, we will evaluate these bounds for the couplings to photons~(\textsection\ref{sec:photons}) and gluons~(\textsection\ref{sec:gluons}), and compare them to existing laboratory and astrophysical constraints.

\subsection{Coupling to Photons}
\label{sec:photons}
The operator that has been most actively investigated experimentally is the coupling to photons,
\beq
\L_{\phi\mathrm{EW}} \supset \L_{\phi \gamma} = - \frac{1}{4}\frac{\phi}{\Lambda_\gamma} F_{\mu \nu} \tilde F^{\mu \nu}\, . \label{eq:phiF}
\eeq
The photon coupling~$\Lambda_\gamma$ is related to the electroweak couplings~$\Lambda_1$ and~$\Lambda_2$ via $\Lambda_\gamma^{-1} = \cos^2\theta_w \hskip1pt\Lambda_1^{-1} + \sin^2\theta_w \hskip1pt\Lambda_2^{-1}$, where $\theta_w \approx \ang{30}$ is Weinberg's mixing angle. Photons are easily produced in large numbers in both the laboratory and in many astrophysical settings which makes this coupling a particularly fruitful target for axion searches.\medskip

\begin{figure}[t!]
	\includegraphics{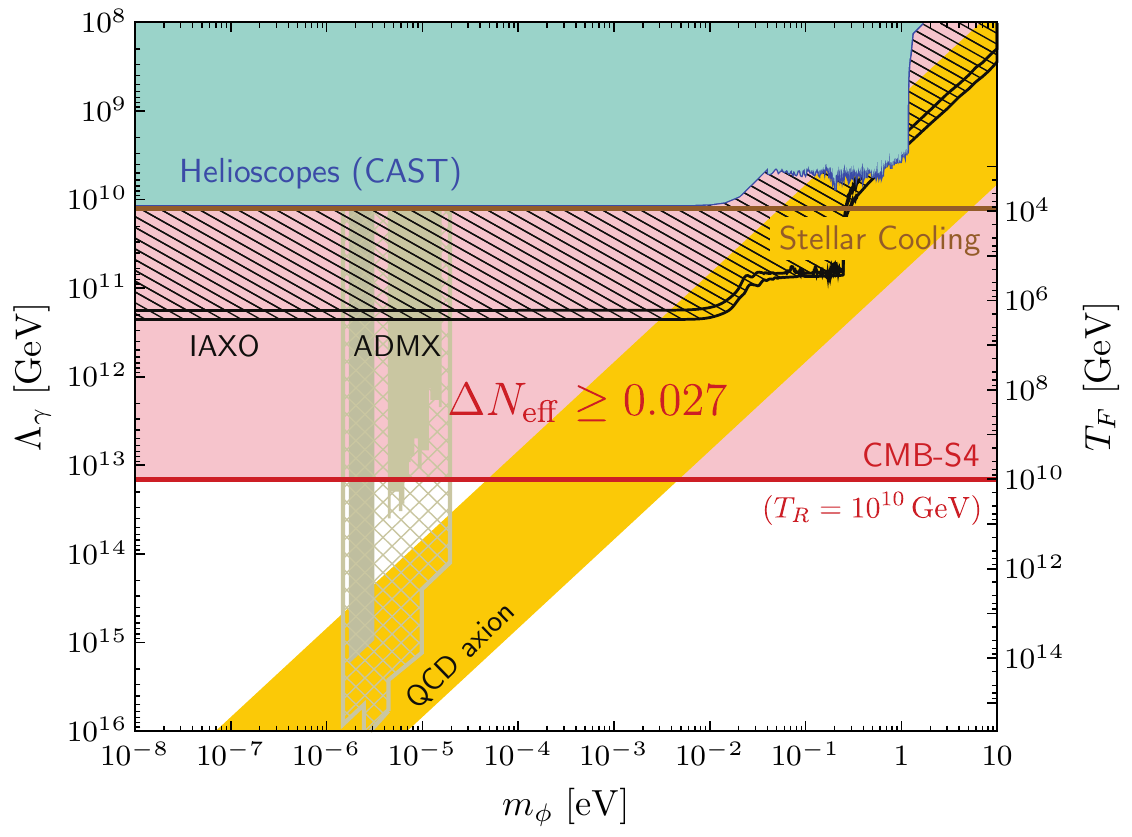}
	\caption{Comparison between current constraints on the axion-photon coupling and the sensitivity of a future CMB-S4 experiment (figure adapted from~\cite{Carosi:2013rla}). Future laboratory constraints (IAXO and ADMX) are shown as shaded regions. The yellow band indicates a range of representative models for the QCD axion (not assuming that it provides all of the dark matter). The future CMB~bound is a function of the reheating temperature~$\Tr$ and the displayed constraint conservatively assumes that the photon coupling derives only from the coupling~$U(1)_Y$ above the electroweak scale. Specific axion models typically also involve a coupling to~$SU(2)_L$ in which case the bound would strengthen by an order of magnitude or more (see Appendix~\ref{app:cmb-axions_appendices}). We note that~ADMX assumes that the axion is all of the dark matter, while all other constraints do not have this restriction.}
	\label{fig:S4axion}
\end{figure}

In Appendix~\ref{app:cmb-axions_appendices}, we show in detail how the constraints~\eqref{eq:freezeoutconstraint} on the couplings to the electroweak gauge bosons map into a constraint on the coupling to photons. This constraint is a function of the relative size of the couplings to the $SU(2)_L$ and~$U(1)_Y$ sectors, as measured by the ratio~$c_2/c_1$ in~\eqref{eq:LEW0}. To be conservative, we will here present the weakest constraint which arises for $c_2=0$ when the axion only couples to the $U(1)_Y$ gauge field. A specific axion model is likely to also couple to the $SU(2)_L$ sector, i.e.~have $c_2\ne 0$, and the constraint on~$\Lambda_\gamma$ would then be stronger (as can be seen explicitly in Appendix~\ref{app:cmb-axions_appendices}). Using $\gamma_{1,R} \approx \gamma_1(\SI{e10}{GeV}) = 0.017$ and $g_{*,R} = 106.75+1$, we find
\beq
\Lambda_\gamma > \SI{1.4e13}{GeV} \left(\frac{\Tr}{\SI{e10}{GeV}}\right)^{\!1/2}\, .	\label{eq:PhotonBound}
\eeq
For a reheating temperature of about \SI{e10}{GeV}, the bound in~\eqref{eq:PhotonBound} is three orders of magnitude stronger than the best current constraints (cf.~Fig.~\ref{fig:S4axion}).%
Even for a reheating temperature as low as $\SI{e4}{GeV}$ the bound from the~CMB would still marginally improve over existing constraints.\medskip

Massive axions are unstable to decay mediated by the operator~$\phi F \tilde F$. However, for the range of parameters of interest, these decays occur after recombination and, hence, do not affect the~CMB. To see this, we consider the decay rate for $m_\phi \gtrsim T$~\cite{Peskin:1995ev}, 
\beq
\Gamma_{D,\gamma} = \frac{1}{64\pi} \frac{m_\phi^3}{\Lambda_\gamma^2} \, .
\eeq
The decay time is $\tau_D = \Gamma_{D,\gamma}^{-1}$ and the temperature at decay is determined by $H(\Td) \approx \tau_D^{-1} = \Gamma_{D,\gamma}$. We will not consider the regime $m_\phi < \Td$ as it does not arise in the range of parameters of interest. Assuming that the universe is matter dominated at the time of the decay, we get
\beq
\frac{\Td}{\Trec} \,\approx\, \num{9.5e-10} \left(\frac{\Lambda_\gamma}{\SI{e10}{GeV}} \right)^{\!-4/3} \left(\frac{m_\phi}{\Trec}\right)^{\!2} \, .
\eeq
Using the stellar cooling constraint, $\Lambda_\gamma > \SI{1.3e10}{GeV}$~\cite{Friedland:2012hj}, we therefore infer that $\Td < \num{7.1e-10} \,\Trec \left({m_\phi/\Trec}\right)^{2}$, so that the axions are stable on the time-scale of recombination as long as $m_\phi \lesssim \SI{10}{keV}$. CMB-S4 will probe this regime through sensitivity to~$\Neff$ for $m_\phi \lesssim \Trec$ and through sensitivity to warm dark matter for larger masses. Warm dark matter is already highly constrained by cosmology, with current CMB~data limiting the mass of the QCD~axion to $m_\phi < \SI{0.53}{eV}$ (\SI{95}{\percent}~c.l.)~\cite{DiValentino:2015wba}. The regime $\SI{10}{keV} < m_\phi < \SI{1}{MeV}$ (where the axion decays between neutrino decoupling and recombination) is constrained by effects on the~CMB and on big bang nucleosynthesis~\cite{Cadamuro:2010cz,Cadamuro:2011fd, Millea:2015qra}.

\subsection{Coupling to Gluons}
\label{sec:gluons}
\begin{figure}[t!]
	\includegraphics{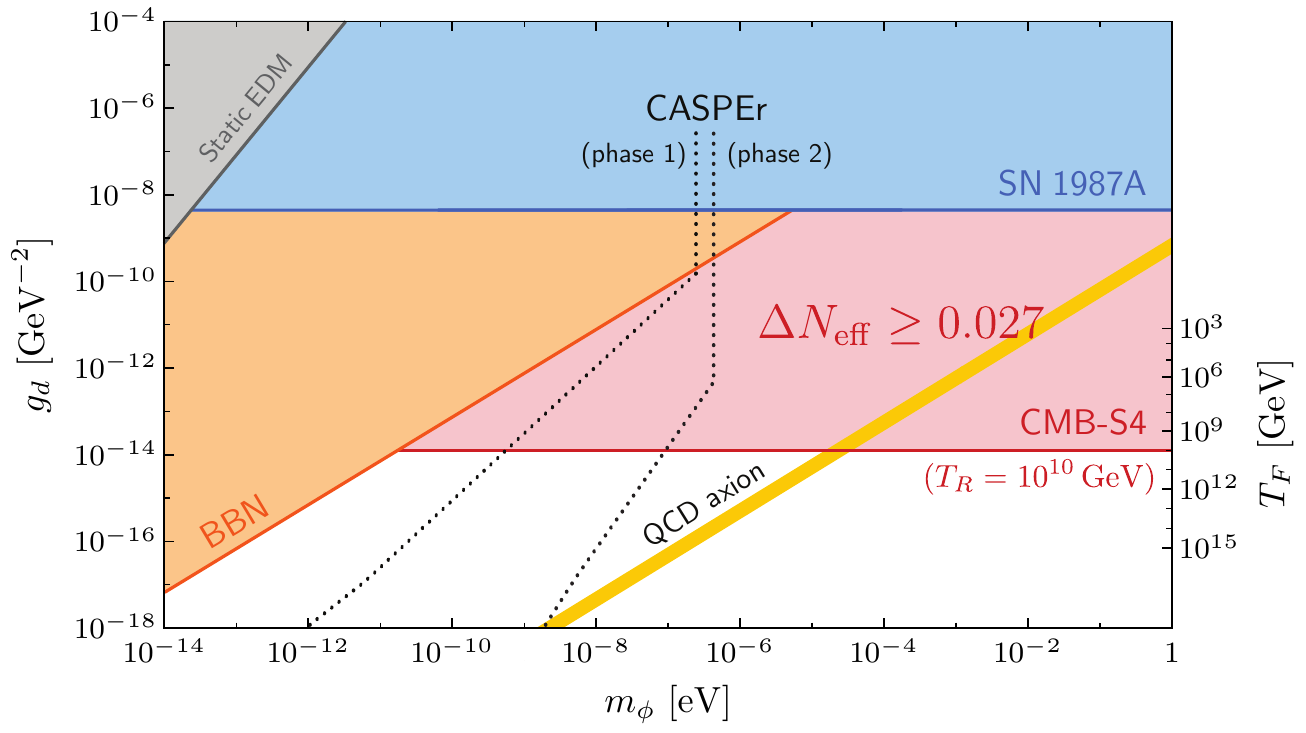}\vspace{-3pt}
	\caption{Comparison between current constraints on the axion-gluon coupling and the sensitivity of a future CMB-S4~experiment (figure adapted from~\cite{Graham:2013gfa,Blum:2014vsa}). The dotted lines are the projected sensitivities of the NMR~experiment CASPEr~\cite{Budker:2013hfa}. We note that CASPEr, the static EDM~\cite{Graham:2013gfa} and BBN constraints~\cite{Blum:2014vsa} assume that the axion is all of the dark matter, while SN~1987A~\cite{Raffelt:1996wa} and the future CMB~constraint do not have this restriction.\vspace{-2pt}}
	\label{fig:S4dipole}
\end{figure}

The coupling to gluons is especially interesting for the~QCD axion since it has to be present in order to solve the strong CP problem. The axion production rate associated with the interaction~$\phi \hskip1pt G \tilde G$ is~\cite{Salvio:2013iaa}
\beq
\Gamma_g \simeq 0.41\hskip1pt \frac{T^3}{\Lambda_g^2}\, ,
\eeq
where $\Lambda_g \equiv \Lambda/c_3$. As before, we have dropped a weakly temperature-dependent prefactor, but account for it in Appendix~\ref{app:cmb-axions_appendices}. The bound~\eqref{eq:freezeoutconstraint} then implies
\beq
\Lambda_g > \SI{5.4e13}{GeV} \left(\frac{\Tr}{\SI{e10}{GeV}}\right)^{\!1/2}\, .
\eeq
Laboratory constraints on the axion-gluon coupling are usually phrased in terms of the induced electric dipole moment~(EDM) of nucleons: $d_n = g_d \phi_0$, where~$\phi_0$ is the value of the local axion field. The coupling $g_d$ is given for the QCD axion with an uncertainty of about~\SI{40}{\percent} by~\cite{Pospelov:1999ha, Graham:2013gfa}
\beq
g_d \,\approx\, \frac{2\pi}{\alpha_s} \times \frac{\SI{3.8e-3}{\per\GeV}}{\Lambda_g} \ <\ \SI{1.3e-14}{\per\GeV\squared} \left(\frac{\Tr}{\SI{e10}{GeV}}\right)^{\!-1/2}\, .
\eeq
Constraints on~$g_d$ (and hence~$\Lambda_g$) are shown in Fig.~\ref{fig:S4dipole}. We see that future cosmological observations will improve over existing constraints on~$\Lambda_g$ by up to six orders of magnitude if $\Tr = \O(\SI{e10}{GeV})$. Even if the reheating temperature is as low as~\SI{e4}{GeV}, the future CMB constraints will be tighter by three orders of magnitude. In Figure~\ref{fig:S4dipole},%
we also show the projected sensitivities of the proposed EDM~experiment CASPEr~\cite{Budker:2013hfa}. We see that CASPEr and CMB-S4 probe complementary ranges of axion masses. It should be noted that CASPEr is only sensitive to axion dark matter, while the CMB~constrains a separate thermal population of axions which does not require assumptions about the dark matter.

\section{Constraints on Familons}
\label{sec:fermions}
Spontaneously broken global symmetries have also been envoked to explain the approximate $U(3)^5$~flavour symmetry of the Standard Model. As we discussed in~\textsection\ref{sec:EFTlightSpecies}, the associated pNGBs---called \textit{familons}~\cite{Wilczek:1982rv, Reiss:1982sq, Kim:1986ax}---couple to the~SM through Yukawa couplings,\vspace{-1pt}
\begin{align}
\L_{\phi\psi}	&= - \frac{\partial_\mu \phi}{\Lambda_\psi}\, \bar \psi_i \gamma^\mu \big(g^{ij}_V+ g^{ij}_A \gamma^5\big) \psi_j \nonumber \\[4pt]
				&\to \frac{\phi}{\Lambda_\psi}\bigg( \ii H \,\bar\psi_{L,i} \!\left[ (\lambda_i - \lambda_j) g_V^{ij} + (\lambda_i + \lambda_j) g_A^{ij} \right]\! \psi_{R,j} + \mathrm{h.c.} \bigg) \,+\, \O(\phi^2)\, ,\vspace{-1pt}	\label{eq:Lfamilon}
\end{align}
where~$H$ is the Higgs doublet and $\psi_{L,R} \equiv \tfrac{1}{2}(1\mp \gamma^5) \psi$. The~$SU(2)_L$ and $SU(3)_C$~structures in~\eqref{eq:Lfamilon} take the same form as for the SM~Yukawa couplings~\cite{Peskin:1995ev}, but this has been left implicit to avoid clutter. In the second line we have integrated by parts and used the equations of motion. The subscripts~$V$ and~$A$ denote the couplings to the vector and axial-vector currents, respectively, and $\lambda_i \equiv \sqrt{2} m_i /v$ are the Yukawa couplings, with $v = \SI{246}{GeV}$ being the Higgs vacuum expectation value. We note that the diagonal couplings,~$i=j$, are only to the axial part, as expected from vector current conservation. Due to the chiral anomaly, a linear combination of the axial couplings is equivalent to the coupling of axions to gauge bosons. In this section, we only consider the effects of the couplings to matter with no contribution from anomalies.\medskip

In Table~\ref{tab:bounds},%
\begin{table}[t!]
	\begin{tabular}{c c r c c c c} 
			\toprule
							& \multicolumn{2}{c}{Current Constraints}				& & \multicolumn{3}{c}{Future CMB Constraints}								\\
			\cmidrule(lr){2-3} \cmidrule(lr){5-7}
		Coupling 			& Bound [GeV]		& Origin 							& & Freeze-Out [GeV]	& Freeze-In [GeV]	& $\Delta\tilde{N}_\mathrm{eff}$\\
			\midrule[0.065em]
		$\Lambda_{ee}$		& \ \num{1.2e10}	& White dwarfs 						& & \num{6.0e7}			& \num{2.7e6}		& 1.3\phantom{0}				\\
		$\Lambda_{\mu\mu}$	& \num{2.0e6}		& Stellar cooling					& & \ \num{1.2e10}		& \num{3.4e7}		& 0.5\phantom{0}				\\
		$\Lambda_{\tau\tau}$& \num{2.5e4}		& Stellar cooling					& & \ \num{2.1e11}		& \num{9.5e7}		& 0.05							\\
		$\Lambda_{bb}$		& \num{6.1e5}		& Stellar cooling					& & \ \num{9.5e11}		& {--}				& 0.04							\\
		$\Lambda_{tt}$		& \num{1.2e9}		& Stellar cooling					& & \ \num{3.5e13}		& {--}				& 0.03							\\
			\midrule
		$\Lambda_{\mu e}^V$	& \num{5.5e9}		& $\mu^+ \to e^+ \,\phi $			& & \num{6.2e9}			& \num{4.8e7}		& 0.5\phantom{0}				\\
		$\Lambda_{\mu e}$	& \num{3.1e9}		& $\mu^+ \to e^+ \,\phi\,\gamma$	& & \num{6.2e9}			& \num{4.8e7}		& 0.5\phantom{0}				\\
		$\Lambda_{\tau e}$	& \num{4.4e6}		& $\tau^- \to e^- \phi$ 			& & \ \num{1.0e11}		& \num{1.3e8}		& 0.05							\\
		$\Lambda_{\tau \mu}$& \num{3.2e6}		& $\tau^- \to \mu^- \phi$ 			& & \ \num{1.0e11}		& \num{1.3e8}		& 0.05							\\
		$\Lambda_{cu}^A$	& \num{6.9e5}		& $D^{0}$-$\bar D^{0}$ 				& & \ \num{1.3e11}		& \num{2.0e8}		& 0.05							\\
		$\Lambda_{bd}^A$	& \num{6.4e5}		& $B^{0}$-$\bar B^{0}$ 				& & \ \num{4.8e11}		& \num{3.7e8}		& 0.04							\\
		$\Lambda_{bs}$		& \num{6.1e7}		& $b \to s \phi$					& & \ \num{4.8e11}		& \num{3.7e8}		& 0.04							\\
		$\Lambda_{tu}$		& \num{6.6e9}		& Mixing							& & \ \num{1.8e13}		& \num{2.1e9}		& 0.03							\\
		$\Lambda_{tc}$		& \num{2.2e9}		& Mixing							& & \ \num{1.8e13}		& \num{2.1e9}		& 0.03							\\
			\bottomrule 
	\end{tabular}
	\caption{Current experimental constraints on Goldstone-fermion couplings (taken from~\cite{Brust:2013xpv, Feng:1997tn, Hansen:2015lqa}) and future CMB~constraints. In some cases, the current constraints are only on the coupling to right-handed particles (namely for~$\Lambda_{\tau\tau}$, $\Lambda_{bb}$, $\Lambda_{tt}$) and to left-handed particles (namely for~$\Lambda_{tu}, \Lambda_{tc}$). The quoted freeze-out bounds are for $\Tr = \SI{e10}{GeV}$ and require that a future CMB~experiment excludes $\Delta\Neff=0.027$. In contrast, the freeze-in bounds from avoiding recoupling of the familons to the~SM at low temperatures do not depend on $T_R$ and assume weaker exclusions $\Delta \tilde N_\mathrm{eff}$ [see the last column for estimates of the freeze-in contributions associated with the different couplings, $\Delta \tilde N_\mathrm{eff} \simeq \Delta N_\mathrm{eff}(\frac{1}{4}m_i)$]. Hence, they may be detectable with a less sensitive experiment. Qualitatively, the bounds from the~CMB are stronger for the second and third generations, while laboratory and stellar constraints are strongest for the first generation (with the exception of the constraint on~$\Lambda_{tt}$).}
	\label{tab:bounds}
\end{table}
we have collected accelerator and astrophysics constraints on the effective couplings $\Lambda_{ij}^I \equiv \Lambda_\psi/g^{ij}_I$ and $\Lambda_{ij} \equiv {\Lambda_\psi/[(g_V^{ij})^2 + (g_A^{ij})^2}]^{1/2}$. We see that current data typically constrain the couplings to the first generation fermions much more than those to the second and third generations. We wish to compare these constraints to the reach of future CMB~observations. We will find distinct behaviour above and below the EWSB~scale, due to the presence of the Higgs. The effective scaling of the operator~\eqref{eq:Lfamilon} changes from irrelevant to marginal and we therefore have both freeze-out and freeze-in contributions.

\subsection{Freeze-Out}
At high energies, the flavour structure of~\eqref{eq:Lfamilon} is unimportant since all SM~particles are effectively massless. The role of the flavour is only to establish the strength of the interaction by the size of the Yukawa coupling. Above the EWSB~scale, the production of the familon $\phi$ is determined by a four-point interaction. This allows the following processes: $\bar\psi_i + \psi_j \rightarrow H +\phi$ and $\psi_i + H \rightarrow \psi_j + \phi$. The total production rate is derived in Appendix~\ref{app:cmb-axions_appendices},
\beq
\Gamma_{ij}^{I} \simeq 0.37 \hskip1pt N_\psi\, \frac{(\lambda_i \mp \lambda_j)^2}{8\pi} \frac{T^3}{(\Lambda_{ij}^{I})^2} \, ,	\label{eq:GijI}
\eeq
where $N_\psi=1$ for charged leptons and $N_\psi=3$ for quarks. The `$-$' and `$+$'~signs in~\eqref{eq:GijI} apply to $I=V$ and $I=A$, respectively. We see that the rate vanishes for the diagonal vector coupling, as it should by current conservation. Deriving the freeze-out temperature and imposing $\Tf > \Tr$, we find
\beq
\Lambda_{ij}^{I} \ >\ 
	\begin{dcases}
		\, \SI{1.0e11}{GeV} \ \frac{m_i \mp m_j}{m_\tau} \left(\frac{\Tr}{\SI{e10}{GeV}}\right)^{\!1/2} & \quad i,j=\text{leptons}, \\
		\, \SI{1.8e13}{GeV} \ \frac{m_i \mp m_j}{\mt} \left(\frac{\Tr}{\SI{e10}{GeV}}\right)^{\!1/2} 	& \quad i,j=\text{quarks},
	\end{dcases}
\eeq
where $m_\tau \approx \SI{1.8}{GeV}$ and $\mt \approx \SI{173}{GeV}$. In Table~\ref{tab:bounds}, we show how these bounds compare to current laboratory and astrophysics constraints for a fiducial reheating temperature of $\SI{e10}{GeV}$. Except for the coupling to electrons, the constraints from future CMB~experiments are orders of magnitude stronger than existing constraints. For lower reheating temperatures the constraints would weaken proportional to~$\sqrt{\Tr}$. We note that while laboratory and astrophysical constraints are considerably weaker for second and third generation particles because of kinematics, the cosmological constraints are strengthened for the higher mass fermions due to the larger effective strength of the interactions. The exception to this pattern is the top quark which is strongly constrained by stellar cooling due to a loop correction to the coupling of $W^\pm$ and $Z$ to~$\phi$, with the loop factor suppression being offset by the large Yukawa coupling of the top quark.

\subsection{Freeze-In}
Below the EWSB~scale, the leading coupling of the familon to fermions becomes marginal after replacing the Higgs in~\eqref{eq:Lfamilon} with its vacuum expectation value. As the temperature decreases, the production rate will therefore grow relative to the expansion rate and we may get a thermal freeze-in abundance. By observationally excluding such a relic density, we can then put constraints on the familon interactions (cf.\ Section~\ref{sec:cmb-axions_introduction}). The leading familon production mechanism will depend on whether the coupling is diagonal or off-diagonal in the mass eigenbasis.\medskip

\noindent
\textit{Diagonal couplings.}---For the diagonal couplings in~\eqref{eq:Lfamilon}, the production rate is dominated by a Compton-like process, $\{\gamma,g\} + \psi_i \to \psi_i + \phi$, and by fermion/anti-fermion annihilation, $\bar \psi_i + \psi_i \to \{\gamma,g\} + \phi$, where $\{\gamma,g\}$ is either a photon or gluon depending on whether the fermion is a lepton or quark. The full expression for the corresponding production rate is given in Appendix~\ref{app:cmb-axions_appendices}. Since freeze-in occurs at low temperatures, the quark production becomes sensitive to strong coupling effects. Although qualitative bounds could still be derived for the quark couplings, we choose not to present them and instead focus on the quantitative bounds for the lepton couplings. Below the scale of~EWSB, but above the lepton mass, the production rate is
\beq
\tilde \Gamma_{ii} \simeq 5.3 \hskip1pt \alpha \, \frac{|\tilde \epsilon_{ii}|^2}{8 \pi} T\, , 
\eeq
where $\tilde \epsilon_{ii} \equiv 2 m_i/\Lambda_{ii}$. The freeze-in temperature~$\Tff$ follows from $\tilde \Gamma_{ii}(\Tff) = H(\Tff)$. To avoid producing a large familon abundance requires that the fermion abundance becomes Boltzmann suppressed before freeze-in could occur. This implies $\Tff < m_i$, or 
\beq
 \Lambda_{ii} \ >\ \SI{9.5e7}{GeV} \left(\frac{g_{*,i}}{g_{*,\tau}}\right)^{\!-1/4} \left(\frac{\alpha_{i}}{\alpha_\tau}\right)^{\!1/2} \left(\frac{m_i}{m_\tau}\right)^{\!1/2}\, , \quad i=\text{lepton},	\label{eq:Lii}
\eeq
where~$g_{*,i}$ and~$\alpha_i$ are the effective number of relativistic species and the fine-structure constant at $T=m_i$. The scalings in~\eqref{eq:Lii} have been normalized with respect to~$g_*$ and~$\alpha$ at $T= m_\tau$, i.e.~we use $g_{*,\tau} = 81.0$ and $\alpha_\tau = 134^{-1}$. In Table~\ref{tab:bounds}, these bounds are compared to the current astrophysical constraints. Except for the coupling to electrons, these new bounds are significantly stronger than the existing constraints.\medskip

\noindent
\textit{Off-diagonal couplings.}---For the off-diagonal couplings in~\eqref{eq:Lfamilon}, we have the possibility of a freeze-in population of the familon from the decay of the heavy fermion, $\psi_i \to \psi_j +\phi$. For $m_i \gg m_j$, the production rate associated with this process is
\beq
\tilde\Gamma_{ij} \simeq 0.31\hskip1pt N_\psi\,\frac{|\tilde \epsilon_{ij}|^2}{8\pi} \frac{m_i^2}{T} \, ,
\eeq
where $\tilde \epsilon_{ij} \approx m_i/\Lambda_{ij}$. Requiring the corresponding freeze-in temperature to be below the mass of the heavier fermion, $\Tff < m_i$, we get 
\beq
\Lambda_{ij} \ >\ 
	\begin{dcases}
		\, \SI{1.3e8}{GeV} \left(\frac{g_{*,i}}{g_{*,\tau}}\right)^{\!-1/4} \left(\frac{m_i}{m_\tau}\right)^{\!1/2} 	& \quad i,j=\text{leptons}, \\
		\, \SI{2.1e9}{GeV} \left(\frac{g_{*,i}}{g_{*,t}}\right)^{\!-1/4} \left(\frac{m_i}{m_t}\right)^{\!1/2}			& \quad i,j=\text{quarks}.
	\end{dcases}
\eeq
We see that this improves over existing constraints for the third generation leptons and for the second and third generation quarks (except the top).\medskip

The freeze-in abundance is created \textit{after} the annihilation of the most massive fermion in the coupling. In the presence of a single massive fermion, the prediction for a freeze-in scenario is the same as that for a freeze-out scenario with $\Tf \ll m_i$ since decoupling occurs after most of the fermions~$\psi_i$ have annihilated and their abundance is exponentially suppressed. This then results in a relatively large contribution to $\Neff$. Of course, the~SM contains fermions with different masses. To capture the energy injection from the relevant fermion annihilation without incorrectly including the effects from the annihilation of much lighter fermions, we take the decoupling temperature to be~$\tfrac{1}{4}m_i$. This choice of decoupling temperature gives good agreement with numerical solutions to the Boltzmann equations and leads to the following estimate for the freeze-in contributions:
\beq
\Delta \tilde N_\mathrm{eff} \simeq \Delta\Neff(\tfrac{1}{4}m_i)= \frac{4}{7} \left(\frac{43}{4\,g_*(\tfrac{1}{4}m_i)}\right)^{4/3}\, .
\eeq
When the heaviest fermion is a muon (electron), one finds $\Delta \tilde N_\mathrm{eff}\simeq 0.5$ ($1.3$) which is excluded by Planck at about~$3\sigma$~($7\sigma$). It is worth noting that the Planck constraint on the diagonal muon coupling, $\Lambda_{\mu\mu}>\SI{3.4e7}{GeV}$, improves on the current experimental bound by more than an order of magnitude. Couplings involving the tau and the charm or bottom quark produce $\Delta \tilde N_\mathrm{eff} \sim 0.05$ which will become accessible when the sensitivity of CMB experiments reaches $\sigma(\Neff) \lesssim 0.025$.

\section{Constraints on Majorons}
\label{sec:neutrinos}
In the Standard Model, the masses of Majorana neutrinos do not arise from renormalizable couplings to the Higgs, but instead must be written as irrelevant operators suppressed by a scale of about~\SI{e15}{GeV}. Moreover, the existence of neutrino masses and mixings point to structure in the flavour physics of neutrinos. Much like in the case of familons, it is plausible that this structure could arise from the spontaneous breaking of the neutrino flavour symmetry. The associated Goldstone bosons are often referred to as \textit{majorons}~\cite{Chikashige:1980ui, Chikashige:1980qk}.\medskip 

Assuming that neutrinos are indeed Majorana fermions, the leading coupling of the majoron~is
\begin{align}
\L_{\phi\nu}	&= - \frac{1}{2} \left(e^{\ii \phi T_{ik}/ (2\Lambda_\nu)} \hskip1pt m_{kl} \hskip1pt e^{\ii \phi T_{l j}/(2 \Lambda_\nu)} \nu_i \nu_j +\mathrm{h.c.}\right) \nonumber \\[4pt]
				&= - \frac{1}{2}\left[ \left(m_{ij} \nu_i \nu_j + \ii \hskip1pt \tilde \epsilon_{ij} \phi \nu_i \nu_j - \frac{1}{2\Lambda_\nu} \epsilon_{ij} \phi^2 \nu_i \nu_j + \cdots \right) + \mathrm{h.c.}\right] ,	\label{eq:Lmajoron} 
\end{align}
where~$\nu_{i}$ are the two-component Majorana neutrinos in the mass eigenbasis,~$m_{ij}$ is the neutrino mass matrix and~$T_{ij}$ are generators of the neutrino flavour symmetry. After expanding the exponentials, we have defined the dimensionless couplings $\tilde \epsilon_{ij} \equiv (T_{i k} m_{k j} +m_{i k} T_{k j})/(2\Lambda_\nu)$ and $\epsilon_{ij} \equiv (m_{ik} T_{kl} T_{lj} + 2 T_{ik} m_{kl} T_{lj} +T_{ik} T_{kl} m_{lj})/(4\Lambda_\nu)$. For numerical estimates, we will use the cosmological upper limit on the sum of the neutrino masses~\cite{Ade:2015xua}, $\sum m_i < \SI{0.23}{eV}$, and the mass splittings $m_2^2-m_1^2 \approx \SI{7.5e-5}{eV\squared}$ and $|m_3^2 - m_1^2| \approx \SI{2.4e-3}{eV\squared}$ from neutrino oscillation measurements~\cite{Patrignani:2016xqp}. The couplings in~$\L_{\phi\nu}$ are identical to the familon couplings after a chiral rotation, except that there is no analogue of the vector current in the case of Majorana neutrinos. The representation of the coupling in~\eqref{eq:Lmajoron} is particularly useful as it makes manifest both the marginal and irrelevant couplings between~$\phi$ and~$\nu$. As a result, we will get both a freeze-out\hskip1pt\footnote{Technically speaking the operator in~\eqref{eq:Lmajoron} is only well-defined below the EWSB~scale. However, in~\textsection\ref{sec:nufreeze} we will find that in order for freeze-out to occur in the regime of a consistent effective field theory description ($T< \Lambda_\nu$), we require $\Tf \lesssim \SI{33}{MeV}$ and, therefore, the operator as written will be sufficient for our purposes.} and a freeze-in production of the majorons.

\subsection{Freeze-Out}
\label{sec:nufreeze}
Thermalization at high energies is dominated by the dimension-five operator~$\phi^2 \nu_i \nu_j$ in~\eqref{eq:Lmajoron}. In Appendix~\ref{app:cmb-axions_appendices}, we show that the corresponding production rate is 
\beq
\Gamma_{ij} \simeq 0.047 \hskip1pt s_{ij} \, \frac{| \epsilon_{ij}|^{2}}{8\pi} \,\frac{T^3}{\Lambda_\nu^2} \, , 
\eeq
where $s_{ij} \equiv 1-\frac{1}{2} \delta_{ij}$ is the symmetry factor for identical particles in the initial state. This leads to a freeze-out temperature of
\beq
\Tf \simeq \SI{0.23}{MeV} \, s_{ij}^{-1} \, \bigg(\frac{g_{*,F}}{10}\bigg)^{\!1/2} \left(\frac{\mu_{ij}}{\SI{0.1}{eV}}\right)^{\!-2} \left(\frac{\Lambda_\nu}{\SI{10}{MeV}}\right)^{\!4} \, ,	\label{eq:Tffam}
\eeq
where $\mu_{ij} \equiv |\epsilon_{ij}| \Lambda_\nu$. Consistency of the effective field theory description requires~$\Tf$ to be below the cutoff~$\Lambda_\nu$ associated with the interactions in~\eqref{eq:Lmajoron}. Using~\eqref{eq:Tffam}, this implies 
\beq
\Tf < \Lambda_\nu <\, \SI{35}{MeV} \, s_{ij}^{1/3} \left(\frac{g_{*,F}}{10}\right)^{\!-1/6} \left(\frac{\mu_{ij}}{\SI{0.1}{eV}}\right)^{\!2/3}\, .
\eeq
Taking $\mu_{ij} \lesssim m_3<\SI{0.1}{eV}$ from both the mass splittings and the bound on the sum of neutrino masses, and~$g_* \approx 14$, we obtain $\Tf\lesssim \SI{33}{MeV}$. Such a low freeze-out temperature would lead to $\Delta\Neff\gtrsim0.44$ (cf.~Fig.~\ref{fig:deltaNeff}) which is ruled out by \textit{current} CMB~measurements at more than $2\sigma$. To avoid this conclusion, we require $\Lambda_\nu > \SI{33}{MeV}$, so that the would-be freeze-out is pushed outside the regime of validity of the EFT. Moreover, we have to assume that the production of majorons is suppressed in this regime. This logic leads to the following constraint: 
\beq
\Lambda_\nu > \SI{33}{MeV} \quad \xrightarrow{\ \mu_{ij} \,\lesssim\, \SI{0.1}{eV} \ } \quad |\epsilon_{ij}| < \num{3e-9} \, .	\label{eqn:nufreezeout}
\eeq
Somewhat stronger bounds can be derived for individual elements of~$\epsilon_{ij}$. This simple bound is much stronger than existing constraints from neutrinoless double beta decay~\cite{Gando:2012pj, Albert:2014fya} and supernova cooling~\cite{Farzan:2002wx}, $\epsilon_{ij} \lesssim 10^{-7}$. Note also that the constraints on~$\epsilon_{ij}$ are stronger for smaller values of~$\mu_{ij}$.

\subsection{Freeze-In}
\label{sec:nufreezein}
At low energies, the linear coupling $\phi \nu_i \nu_j$ in~\eqref{eq:Lmajoron} will dominate. The corresponding two-to-one process is kinematically constrained and we therefore get qualitatively different behaviour depending on whether the majoron mass is larger or smaller than that of the neutrinos.\medskip

\noindent
\textit{Low-mass regime.}---For $m_\phi \ll m_i -m_j$, with $m_i > m_j$, the off-diagonal couplings allow the decay $\nu_i \to \nu_j + \phi$, while other decays are kinematically forbidden. As a result, only the off-diagonal couplings are constrained by freeze-in. Including the effect of time dilation at finite temperature, the rate is
\beq
\tilde\Gamma_{ij} \simeq 0.31\, \frac{|\tilde \epsilon_{ij}|^2}{8\pi}\, \frac{m_i^2}{T} \, ,	\label{eqn:nudecay}
\eeq
where we have assumed $m_i \gg m_j$, which is guaranteed for the minimal mass normal hierarchy (for the general result see~Appendix~\ref{app:cmb-axions_appendices}). When the freeze-in occurs at $\Tff>m_i$, then the majorons and neutrinos are brought into thermal equilibrium, while the comoving energy density is conserved. However, since the momentum exchange at each collision is only $\Delta p^2 \simeq m_i^2 \ll T^2$, the neutrino-majoron radiation is free-streaming at the onset of the freeze-in and is difficult\footnote{Since neutrinos have been converted to majorons with $m_\phi \ll m_i$, this scenario predicts that the cosmological measurement of the sum of the neutrino masses would be significantly lower than what would be inferred from laboratory measurements.} to distinguish from conventional neutrinos. As the temperature drops below~$\Tfl$, with $\tilde\Gamma_{ij}(\Tfl) = (\Tfl/m_i)^2\, H(\Tfl)$, enough momentum is exchanged between the neutrinos and they will behave as a relativistic fluid rather than free-streaming particles~\cite{Chacko:2003dt, Hannestad:2005ex, Friedland:2007vv}. From the rate~\eqref{eqn:nudecay}, we find
\beq
\Tfl \simeq \num{0.10} \, \Teq \times \left(\frac{\tilde \epsilon_{ij}}{10^{-13}} \right)^{\!2/5} \, \left(\frac{m_i}{\SI{0.05}{eV}} \right)^{\!4/5} \, , 
\eeq
where we used $g_{*,\tilde F} \approx 3.4$ and $\Teq\approx\SI{0.79}{eV}$ for the temperature at matter-radiation equality. In this regime, the majoron scenario predicts both free-streaming and non-free-streaming radiation with $\Neff \leq 2$ and $\Nn \geq 1$ (with equality when the majoron couples to one single neutrino species), which is inconsistent with the constraints from Planck data that we will find in the next chapter. To avoid this conclusion requires $\Tfl < \Teq$,\footnote{The imprint of dark radiation is suppressed during matter domination since its contribution to the total energy density is sub-dominant. As a result, constraints on~$\Neff$ are driven by the high-$\ell$ modes of the~CMB which are primarily affected by the evolution of fluctuations during radiation domination (see Section~\ref{sec:signaturesLightRelics}).} which puts a bound on the neutrino-majoron coupling\hskip1pt\footnote{The effect of the linear coupling between a massless majoron and neutrinos on the~CMB was also studied in~\cite{Forastieri:2015paa} and a flavour-independent constraint of $\tilde\epsilon_{ij} < \num{8.2e-7}$ was obtained. This constraint is substantially weaker than our bound~\eqref{eqn:lowmassbound} because it only accounted for the scattering of neutrinos through the exchange of a \textit{virtual} Goldstone boson. The neutrino cross section in that case is suppressed by a factor of~$|\tilde \epsilon_{ij}|^4$ which is much smaller than the rate for the production of real Goldstone bosons in~\eqref{eqn:nudecay}.}
\beq
\tilde\epsilon_{ij} \,<\, \num{3.2e-11} \times \left(\frac{m_i}{\SI{0.05}{eV}}\right)^{\!-2} \, .	\label{eqn:lowmassbound}
\eeq
This constraint has been pointed out previously in~\cite{Chacko:2003dt, Hannestad:2005ex, Friedland:2007vv, Archidiacono:2013dua}.\medskip

\noindent
\textit{High-mass regime.}---For $m_\phi \gg m_i \ge m_j$, the majoron decays into neutrinos, $\phi \to \nu_i+ \nu_j$, and is produced by the inverse decay. For $T \gg m_\phi$, the production rate of the majoron is identical to the rate in~\eqref{eqn:nudecay} after making the replacement $m_i \to m_\phi/\sqrt{1- 4/\pi^2}$ and the corresponding freeze-in temperature is
\beq
\Tff \simeq 1.0 \,\Teq \times \, s_{ij}^{1/3} \left(\frac{\tilde \epsilon_{ij}}{10^{-13}} \, \frac{m_\phi}{\Teq} \right)^{\!2/3} \, .
\eeq
If $\Tff>m_\phi$, then freeze-in occurs while the majorons are relativistic, and the neutrinos and the majorons are brought into thermal equilibrium. How this affects the CMB will depend on whether~$m_\phi$ is greater or smaller than~$\Teq$. For $m_\phi > \Teq$, the majorons decay to neutrinos before matter-radiation equality. To compute the effect on the CMB, we note that the initial (relativistic) freeze-in process conserves the comoving energy density and, once in equilibrium, the decay will conserve the comoving entropy density. This information allows us to derive the final neutrino temperature analytically (see Appendix~\ref{app:cmb-axions_appendices}) and to determine the extra contribution to the radiation density,
\beq
\Delta\Neff \,\geq\, \left(1 + \frac{4}{7}\right)^{\!1/3} - 1 = 0.16\, .
\eeq 
This extra radiation density is easily falsifiable (or detectable) with future CMB~experiments. If $m_\phi \ll \Teq$, on the other hand, the neutrinos and the majorons could effectively form a fluid at matter-radiation equality leading to a similar constraint as~\eqref{eqn:lowmassbound} with $m_i \to m_\phi$.

Assuming that future experiments do not detect the above effects would require either that the would-be freeze-in temperature is below the mass of the majoron, $\Tff < m_\phi$, or that freeze-in occurs after matter-radiation equality, $\Tff < \Teq$. Converting these constraints into a bound on the coupling, we find
\beq
\tilde \epsilon_{ij} \,<\, \num{9.9e-14} \, s_{ij}^{-1/2} \left(\frac{m_\phi}{\Teq} \right)^{\!1/2}\, , \quad \mathrm{for} \quad m_\phi>\Teq\, .
\eeq 
A similar constraint, of the same order of magnitude, applies in the narrow range $m_i \ll m_\phi<\Teq$. This bound is stronger than the freeze-out constraint~\eqref{eqn:nufreezeout} over the full range of allowed masses up to the neutrino decoupling temperature $T_{F,\hskip1pt\nu} \simeq \SI{1}{MeV}$ (note that although in general $\epsilon_{ij} \neq \tilde \epsilon_{ij}$, the two parameters are related by the symmetry under which the majoron transforms). For $m_\phi > T_{F,\hskip1pt\nu}$, the decay of the majorons occurs while the neutrinos are still in equilibrium with the~SM and, therefore, it has no impact on~$\Neff$.

\section{Summary}
\label{sec:conclusions_cmb-axions}
Light pseudo-Nambu-Goldstone bosons arise naturally in many proposals for physics beyond the Standard Model and are an exciting window into the early universe. In this chapter, we showed that future cosmological surveys will either detect these new particles, or place very strong constraints on their couplings to the~SM. These constraints arise because the couplings to the~SM can bring the Goldstone bosons into thermal equilibrium in the early universe. At the same time, cosmological experiments are becoming sensitive enough to detect thermal relics up to arbitrarily high freeze-out temperatures (see Fig.~\ref{fig:deltaNeff}). To avoid producing this detectable relic abundance requires that the reheating temperature of the universe was below the would-be freeze-out temperature. In that case, the temperature in the universe simply was never high enough to bring the extra particles into thermal equilibrium with the SM. For a given reheating temperature~$\Tr$, this puts bounds on the scales~$\Lambda_i$ in the effective interactions between the Goldstone boson~$\phi$ and the SM~fields,
\beq
\L_{\phi\mathrm{SM}} = - \frac{1}{4}\frac{\phi}{\Lambda_\gamma} F \tilde F - \frac{1}{4}\frac{\phi}{\Lambda_g} \mathrm{Tr}(G \tilde G) - \frac{\partial_\mu \phi}{\Lambda_\psi}\, \bar \psi \gamma^\mu \gamma^5 \psi \, +\, \cdots\, .
\eeq 
The bounds on the couplings to photons and gluons are
\begin{align}
\Lambda_\gamma	&> \SI{1.4e13}{GeV}\,\sqrt{T_{R,10}} \, ,	\label{eq:conclusionAxion}	\\
\Lambda_g 		&> \SI{5.4e13}{GeV}\,\sqrt{T_{R,10}} \, , 
\end{align}
where $T_{R,10} \equiv \Tr/\SI{e10}{GeV}$. When considering the interactions with fermions, we distinguish between the couplings to charged leptons and quarks. The resulting bounds are 
\beq
\Lambda_\psi \ > \ 
	\begin{dcases}
		\, \SI{2.1e11}{GeV} \,\,m_{\psi,\tau}\,\sqrt{T_{R,10}}	& \quad \psi=\text{lepton}, \\[4pt]
		\, \SI{3.5e13}{GeV} \,\,m_{\psi,t}\,\sqrt{T_{R,10}} 	& \quad \psi=\text{quark},
	\end{dcases}
\label{eq:conclusionFamilon}
\eeq
where $m_{\psi,\tau} \equiv m_\psi/\SI{1.8}{GeV}$ and $m_{\psi,t} \equiv m_\psi/\SI{173}{GeV}$. For all reasonable reheating temperatures these bounds improve significantly over existing constraints, sometimes by many orders of magnitude. Moreover, while some of the current constraints only apply if the new particles are identified with the dark matter, our bounds do not have this restriction.\medskip

Below the scale of electroweak symmetry breaking, the couplings to the SM~fermions become effectively marginal which can bring the decoupled Goldstone bosons back into thermal equilibrium leading to a detectable freeze-in abundance. Furthermore, the coupling to the light Goldstone boson can lead to a new force between the fermions which becomes relevant at low temperatures~\cite{Chacko:2003dt, Hannestad:2005ex, Friedland:2007vv, Archidiacono:2013dua}. As we will show in the next chapter, both of these effects are highly constrained, even with current data. These arguments are particularly relevant for the couplings to neutrinos, 
\beq
\L_{\phi\nu} = -\frac{1}{2} \big(\,\ii \hskip1pt \tilde\epsilon_{ij} \phi \nu_i \nu_j + \mathrm{h.c.}\,\big)+ \cdots\, . 
\eeq
For the off-diagonal couplings, the following constraints apply:
\beq
\tilde \epsilon_{ij} \ <\ 
	\begin{dcases}
		\,\num{3.2e-11} \times \left(\frac{m_i}{\SI{0.05}{eV}}\right)^{\!-2}	& \quad m_\phi \ll m_i\, , \\[2pt]
		\,\num{9.9e-14} \times \left(\frac{m_\phi}{\Teq} \right)^{\!1/2}		& \quad m_\phi > \Teq\, ,
	\end{dcases}
\eeq
where~$m_i$ is the mass of the heavier neutrino in the off-diagonal interaction. A combination of freeze-in and freeze-out also constrain the diagonal couplings~$\tilde \epsilon_{ii}$. These constraints are orders of magnitude stronger than existing laboratory and astrophysics constraints.\medskip\smallskip

It is also interesting to consider a scenario in which one of the many ongoing searches directly detects axions, familons or majorons. This would determine the coupling strength to at least one of the SM~fields (depending on the detection channel) and would predict the freeze-out temperature of these particles; cf.~Figs.~\ref{fig:S4axion} and~\ref{fig:S4dipole}. Excitingly, the cosmological estimation of $\Delta\Neff$ would then provide information about the reheating temperature of the universe: the absence of a detection of $\Neff \ne 3.046$ would put an upper bound on~$\Tr$ [see e.g.~\eqref{eq:conclusionAxion}--\eqref{eq:conclusionFamilon}], while a measurement of $\Delta\Neff \geq 0.027$ would imply a lower bound on~$\Tr$. The combination of a cosmological measurement of~$\Neff$ and a direct detection could therefore be used to probe the energy scale of the beginning of the hot big bang.\medskip\smallskip

In closing, we would like to re-emphasize that $\Delta\Neff = 0.027$ is an important theoretical threshold. Remarkably, we will show in the next two chapters that this target is within reach of future cosmological surveys. These observations therefore have the potential to probe for light thermal relics up to arbitrarily high decoupling temperatures. We consider this to be a unique opportunity to detect new particles, or place very strong constraints on their couplings to the Standard Model.
	\chapter{Searching for Light Relics with the CMB}
\label{chap:cmb-phases}
Fluctuations in the cosmic neutrino background are known to produce a phase shift in the acoustic peaks of the~CMB. It is through the sensitivity to this effect that the recent CMB~data has provided a robust detection of free-streaming neutrinos~\cite{Follin:2015hya}. In this chapter, we revisit the phase shift of the CMB~anisotropy spectrum as a probe of new physics. The phase shift is particularly interesting because its physical origin is strongly constrained by the analytic properties of the Green's function of the gravitational potential. For adiabatic fluctuations, a phase shift requires modes that propagate faster than the speed of fluctuations in the photon-baryon plasma. This possibility is realized by free-streaming relativistic particles, such as neutrinos or other forms of dark radiation. Alternatively, a phase shift can arise from isocurvature fluctuations. We present simple models to illustrate each of these effects and provide observational constraints from the Planck temperature and polarization data on additional forms of radiation. We also forecast the capabilities of future CMB~Stage-4 experiments. Whenever possible, we give analytic interpretations of our results.\medskip

The outline of this chapter is as follows. In Section~\ref{sec:cmb-phases_introduction}, we briefly review the phase shift in the acoustic oscillations and preview some of this chapter's results. In Section~\ref{sec:analytics}, we analytically derive the effects of new relativistic particles on the perturbations of the photon density. We identify the precise physical conditions that produce a phase shift in the CMB~anisotropy spectrum and illustrate these effects through an exactly solvable toy model. In addition, we compute the phase shift for a simple model with isocurvature fluctuations and for free-streaming relativistic particles. In Section~\ref{sec:analysis}, we confirm some of these pen-and-paper results through a numerical analysis. We present new constraints on dark radiation from the Planck~2015 data~\cite{Aghanim:2015xee} and forecast the capabilities of future CMB-S4~experiments~\cite{Abazajian:2016yjj}. In Section~\ref{sec:consequences}, we consider possible implications of this future sensitivity of the~CMB on different scenarios beyond the Standard Model. Section~\ref{sec:conclusions_cmb-phases} contains a summary. In Appendix~\ref{app:cmb-phases_appendices}, we comment on the inclusion of matter and polarization in our analytic treatment.

\section{Phases of New Physics in the CMB Spectrum}
\label{sec:cmb-phases_introduction}
\begin{figure}[h!t]
	\includegraphics{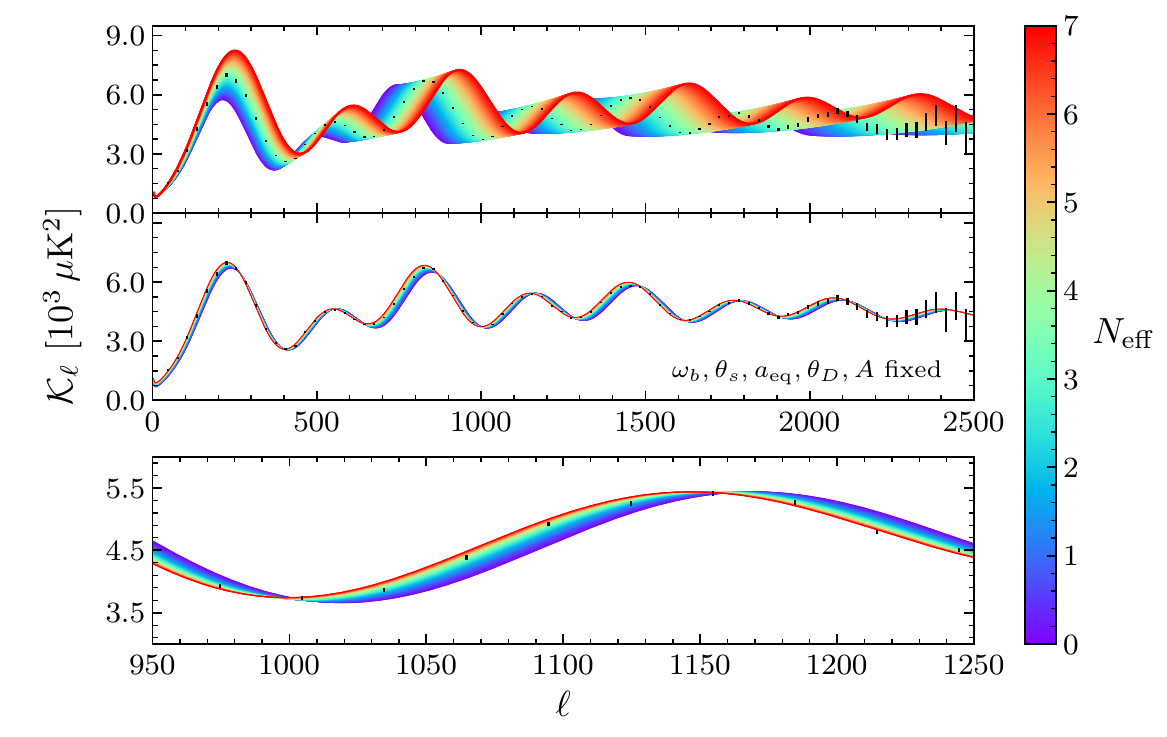}
	\caption{Variation of the CMB~power spectrum as a function of~$\Neff$. The spectra have been undamped, i.e.~the exponential diffusion damping was removed. Following~\cite{Follin:2015hya}, the physical baryon density~$\omega_b$, the scale factor at matter-radiation equality $\aeq \equiv \omega_m/\omega_r$, the angular size of the sound horizon~$\theta_s$ and the angular size of the damping scale~$\theta_D$ are held fixed in the second panel. In addition, the spectra are normalized at the fourth peak. The remaining variation is the phase shift~$\phi$ with a zoom-in shown in the bottom panel. To illustrate the sensitivity of the Planck~2015 high-$\ell$ temperature data, we also display their $1\sigma$~error bars.}
	\label{fig:CMBPhaseShift2}
\end{figure}

After accounting for the difference in the sound horizon, the main effect of increasing the radiation density of the early universe on the CMB~anisotropy spectrum is to enhance the damping of the high-$\ell$~multipoles. As we discussed in~\textsection\ref{sec:species_phaseShift}, neutrinos and other free-streaming particles also induce a distinct shift in the temporal phase of sound waves in the primordial plasma~\cite{Bashinsky:2003tk}. This subtle effect manifests itself in the CMB~spectrum as a coherent shift in the locations of the acoustic peaks (see Fig.~\ref{fig:CMBPhaseShift2}).\footnote{This phase shift refers to a coherent shift in the locations of the high-$\ell$~acoustic peaks. We emphasize that this is a distinct effect from the locations of the first few acoustic peaks which are sensitive to many cosmological parameters, as studied e.g.\ in~\cite{Hu:2000ti, Doran:2001yw, Corasaniti:2007rf}. A detailed study of the CMB~peak locations recently appeared in~\cite{Pan:2016zla}.}%
\medskip

In this chapter, we will revisit the analytic treatment of the CMB~anisotropies with an eye towards BSM~applications. While numerical codes are ultimately necessary in order to make precise predictions for any particular model, analytic results still play a vital role. It is through the physical understanding of the data that we can devise new tests and motivate new models. For example, the use of B-modes in the search for primordial gravitational waves arose from a clear analytic understanding of CMB~polarization~\cite{Seljak:1996ti, Seljak:1996gy, Zaldarriaga:1996xe, Kamionkowski:1996ks}. Similarly, we wish to identify CMB~observables that are sensitive to well-motivated forms of BSM~physics, but are not strongly degenerate with other cosmological parameters. We will advocate the phase shift of the acoustic peaks of the CMB spectrum as an observable with the desired characteristics. As we will see below, the physical conditions that lead to a phase shift are rather restrictive and determined by the analytic properties of the Green's function of the gravitational potential. For adiabatic fluctuations, a phase shift requires fluctuations that travel faster than the sound speed of the photon-baryon fluid, which arises naturally for free-streaming relativistic particles. Alternatively, a shift in the temporal phase of the cosmic sound waves can also arise from isocurvature fluctuations. The phase shift therefore probes an interesting regime in the parameter space of BSM~models.\medskip

We have previously seen that it is useful to characterise the effects of extra light species in terms of free-streaming species~($X$) and non-free-streaming species~($Y$\hskip-0.5pt). We parametrize their respective energy density in terms of the parameters~$\Nf$ and~$\Nn$, i.e.\ with respect to the energy density of a single SM neutrino species (cf.\ Section~\ref{sec:neutrinosDarkRadiation}). Until recently, CMB observations were not sensitive enough to distinguish between these types of relativistic species since they contribute equally to the background density of the universe and, therefore, affect the CMB~damping tail in the same way~\cite{Hou:2011ec}. To separate~$\Nf$ and~$\Nn$ requires measuring subtle differences in the evolution of perturbations. Free-streaming particles (like neutrinos) create significant anisotropic stress which induces a characteristic phase shift in the CMB~anisotropies~\cite{Bashinsky:2003tk}. This phase shift has recently been detected for the first time~\cite{Follin:2015hya}. As we will show, non-free-streaming particles (e.g.~\cite{Cyr-Racine:2013jua, Archidiacono:2013dua, Oldengott:2014qra, Buen-Abad:2015ova, Chacko:2015noa}), in general, do not produce a phase shift (at least as long as the fluctuations are adiabatic and their sound speed is not larger than that of the photons).

Guided by our analytic understanding, we will explore the sensitivity to these effects with the Planck satellite and with a future CMB-S4~experiment, focussing on the ability to distinguish the parameters~$\Nf$ and~$\Nn$. Our analysis of the Planck temperature and polarization data leads to the following constraints:\footnote{These constraints assume that the helium fraction~$Y_p$ is fixed by consistency with~BBN. Results that marginalize over~$Y_p$ are presented in~\textsection\ref{sec:PlanckResults}.}
\beq
\Nf = 2.80^{+0.24}_{-0.23}\, \ (1\sigma)\, , \qquad \Nn < 0.67 \, \ (2\sigma) \, .
\eeq
We see that the current data is already sensitive to the free-streaming nature of the fluctuations. We will explain the important role played by the polarization data in breaking the degeneracy between~$\Nf$ and~$\Nn$, as well as that with the helium fraction~$Y_p$. We will also show that a CMB-S4~experiment would improve these constraints by up to an order of magnitude under a number of experimental configurations. We will highlight how much present and future constraints are driven by measurements of the phase shift. In addition, we will explore how these measurements may be optimized, including through the use of delensing to sharpen the acoustic peaks.

\section{Physical Origin of the Phase Shift}
\label{sec:analytics}
\begin{figure}[h!t]
	\includegraphics{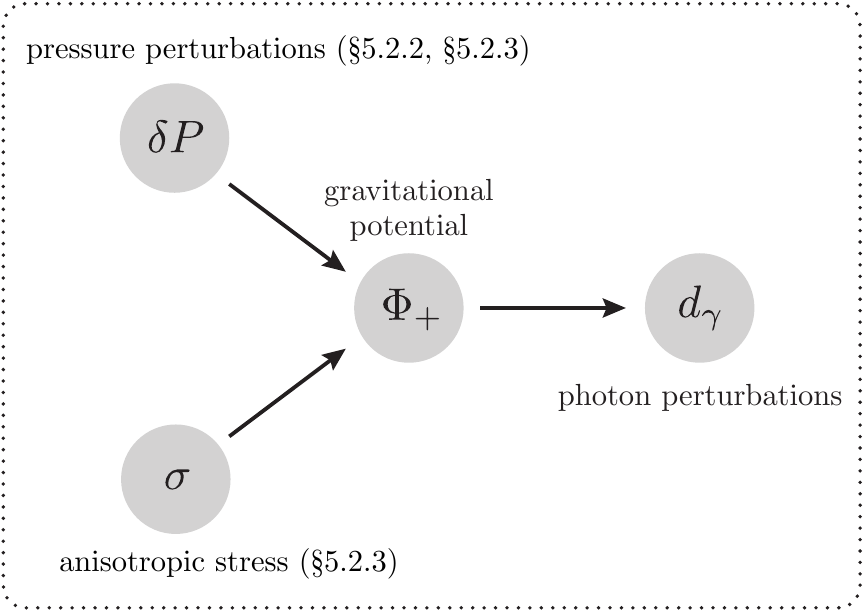}
	\caption{Illustration of the coupled perturbations in the primordial plasma.}
	\label{fig:illustration}
\end{figure}

We will now study under which circumstances a phase shift is imprinted in the CMB~anisotropies. The structure of the acoustic peaks in the~CMB is largely determined by the propagation of fluctuations in the photon-baryon plasma. The physics of the cosmic sound waves is that of a harmonic oscillator with a time-dependent gravitational forcing, 
\beq
\ddot{d}_\gamma - c_\gamma^2\hskip2pt \nabla^2 d_\gamma = \nabla^2 \Phi_+\, ,	\label{eq:dgamma}
\eeq
where $c_\gamma^2 \approx \frac{1}{3}$. Following~\cite{Bashinsky:2003tk}, we wrote this equation in terms of the overdensity in the particle number with respect to the coordinate volume, $d_a \equiv \delta\rho_a/(\bar \rho_a+ \bar P_a) - 3\Psi = \delta_a/(1+w_a) - 3 \Psi$. Moreover, it will be convenient to work with the sum and the difference of the two metric potentials,
\beq
\Phi_\pm \equiv \Phi \pm \Psi \, .
\eeq
A non-trivial evolution of~$\Phi_+$ is sourced either by anisotropic stress~$\sigma$ or by pressure perturbations~$\delta P$ (see Fig.~\ref{fig:illustration}). Under certain conditions, which we will identify, this induces a contribution to~$d_\gamma$ which is out of phase with its freely oscillating part.
\medskip

In this section, we will give an analytic description of these effects, building on the pioneering work of Bashinsky\,\&\,Seljak~\cite{Bashinsky:2003tk}. We first extract the two possible origins of a phase in the acoustic oscillations~(\textsection\ref{sec:originPhaseShift}). We then study a simple toy model to illustrate these abstract conditions~(\textsection\ref{sec:toyModel}). Finally, we explicitly derive the phase shift in a curvaton-like model with isocurvature fluctuations and for light free-streaming particles, such as neutrinos~(\textsection\ref{sec:examples}).

\subsection{Conditions for a Phase Shift}
\label{sec:originPhaseShift}
In the following, we will analyse the evolution of perturbations in the photon-baryon plasma. We have seen in Section~\ref{sec:inhomogeneousCosmology} that the~CMB couples gravitationally to fluctuations in the matter densities. In terms of the variable~$d_\gamma$, the evolution equation~\eqref{eq:combinedConservationEquations} for the photon perturbations is given by
\beq
\ddot d_\gamma + \chi_\gamma \dot d_\gamma - c_\gamma^2 \nabla^2 d_\gamma = \nabla^2(\Phi + 3c_\gamma^2 \Psi) \, ,	\label{eq:perturbationsEoMphotons}
\eeq
where we omitted the term~$\nabla^4 \sigma_\gamma$, since the anisotropic stress is absent before decoupling, $\sigma_\gamma \approx 0$. For simplicity, we will ignore the small effect due to the baryons,\footnote{We ignore the contributions of baryons and dark matter to the energy density, but we are implicitly including the baryons when we assume that the photons are not free-streaming particles.} so that the speed of the perturbations in the photon density is $c_\gamma^2 \approx \frac{1}{3}$. The Hubble drag rate in~\eqref{eq:perturbationsEoMphotons} therefore vanishes, $\chi_\gamma \approx 0$, and we get the evolution equation~\eqref{eq:dgamma}. The solution for~$d_\gamma$ can then be written as
\beq
d_\gamma(y) = d_{\gamma,\mathrm{in}} \cos y - c_\gamma^{-2} \int^y_{0} \d y'\, \Phi_+(y') \sin(y- y')\, ,	\label{eq:dgammaSol}
\eeq
where $y \equiv c_\gamma k \tau$ and the first term is the homogeneous solution with constant superhorizon adiabatic initial condition~$d_{\gamma,\mathrm{in}} \equiv d_\gamma(y_\mathrm{in} \ll 1)$. Primordial isocurvature modes would be straightforward to include here and in the following by an additional sine contribution. The second term is the inhomogeneous correction induced by the evolution of metric fluctuations. Since $\sin(y- y') = \sin y \cos y' - \cos y \sin y'$, we can write~\eqref{eq:dgammaSol}~as
\beq
d_\gamma(y) = \Big[d_{\gamma,\mathrm{in}} + c_\gamma^{-2} A(y) \Big] \cos y - c_\gamma^{-2} B(y) \sin y \, ,	\label{eq:dg2}
\eeq
where we defined
\begin{align}
A(y) &\equiv \int^y_0 \d y'\, \Phi_+(y')\hskip1pt\sin y'\, ,	\label{eq:A}	\\[4pt]
B(y) &\equiv \int^y_0 \d y'\, \Phi_+(y')\hskip1pt\cos y'\, . 	\label{eq:B}
\end{align}
We wish to evaluate~\eqref{eq:dg2} at recombination, $y \to y_\rec$. For the high-$\ell$ modes of the CMB, it is a good approximation to take the limit $y\to \infty$ and assume that the background is radiation dominated (see Appendix~\ref{app:cmb-phases_appendices} for further discussion). If the integral in~\eqref{eq:B} converges in this limit, then a non-zero value of $B \equiv \lim_{y \to \infty} B(y)$ will produce a constant phase shift~$\phi$ relative to the homogeneous solution,
\beq
\sin\phi = \frac{B}{\sqrt{\left(A + c_\gamma^2\hskip1pt d_{\gamma,\mathrm{in}}\right)^2+B^2}}\, .	\label{eq:theta}
\eeq
This phase shift will be reflected in a shift of the acoustic peaks of the CMB anisotropy spectrum. In the following, we will identify the precise physical conditions for which such a phase shift is generated.\medskip

It will be convenient to combine~$B$ and~$A$ into a complex field 
\beq
B+iA \,=\, \int_0^\infty \d y\, e^{i y}\, \Phi_+(y) \,=\, \frac{1}{2} \int_{-\infty}^\infty \d y\, e^{i y} \left[ \Phi^{(s)}_+(y) + \Phi^{(a)}_+(y)\right] ,	\label{eq:AB}
\eeq
where~$\Phi^{(s)}_+(y)$ is an even function of~$y$, while~$\Phi^{(a)}_+(y)$ is an odd function. It is easy to see that the even part of~$\Phi_+$ determines~$B$ and the odd part determines~$A$:
\beq
B = \frac{1}{2} \int_{-\infty}^\infty \d y\, e^{i y} \, \Phi^{(s)}_+(y)\, , \qquad i A = \frac{1}{2} \int_{-\infty}^\infty \d y\, e^{i y} \, \Phi^{(a)}_+(y) \, .	\label{eq:Bdef}
\eeq
We will get $B=0$ as long as~$\Phi^{(s)}_+(y)$ is an analytic function and $e^{i y} \Phi^{(s)}_+(y)$ vanishes faster than~$y^{-1}$ for $|y| \to \infty$.\footnote{Since the equations are symmetric in $y \to -y$, the odd part~$ \Phi^{(a)}_+(y)$ is not analytic around $y= 0$. This is why we always find contributions to~$A$.} This suggests two ways of generating a non-zero~$B$ and, hence, a phase shift in the solution for the photon density:
\vskip8pt
\begin{center}
\begin{tabular}{r l c l}
\textit{i.}		& rapid growth of~$\Phi_+^{(s)}(\pm iy)$ 		& $\longrightarrow$	& mode traveling faster than~$c_\gamma$,	\\[4pt]
\textit{ii.}	& non-analytic behaviour of~$\Phi^{(s)}_+(y)$ 	& $\longrightarrow$	& non-adiabatic fluctuations.
\end{tabular}
\end{center}
\vskip2pt
The mathematical requirements listed on the left are mapped directly into physical conditions, shown on the right. 

\begin{itemize}
\item The first condition is easy to understand physically: in~\eqref{eq:AB}, the Green's function of~$d_\gamma$, i.e.~$\sin(y-y')$, leads to exponential suppression for $y \to i \infty$. To have a growing solution at $y = i \infty$, we therefore need a term in~$\Phi_+$ of the form $e^{-i c_s k \tau} = e^{-i(c_s/c_\gamma)y}$ with $c_s > c_\gamma$.\footnote{Note that~$c_s$ is just a parameter of the wave-like solution and is not necessarily the sound speed of a fluid. Indeed, in the case of free-streaming radiation, it corresponds to the propagation speed of the individual particles.}
\item The second possibility, non-analyticity, is easy to understand mathematically, but the physical requirements are less transparent. First of all, the equations of motion for any mode should be analytic around any finite value of~$k\tau$ in the radiation-dominated era since there is no preferred time. Hence, the only moment at which non-analytic behaviour is possible is around $k\tau = 0$, i.e.~where the initial conditions are defined. Let us first show that adiabatic initial conditions are analytic at $k\tau = 0$. By definition, for adiabatic initial conditions, any long-wavelength mode is locally generated by a diffeomorphism~\cite{Weinberg:2003sw}. In the limit $k \tau \to 0$, we then have $\Phi_+ = \Phi_{+,\mathrm{in}} + \mathcal{O}(k^2 \tau^2)$. This expansion is necessarily analytic in~$k^2$ (by locality and rotational invariance), but also in~$k^2\tau^2$, because the scaling $k \to \lambda k$ and $ \tau \to \lambda^{-1} \tau$ can be absorbed into the overall normalization of the scale factor $a$ which has no physical effect.\footnote{In a universe with a preferred time, this rescaling would also require a shift in this preferred time to keep the density fluctuations fixed. For adiabatic modes, the curvature perturbation~$\zeta$ is conserved outside the horizon even in the presence of such a preferred time and so this is unlikely to have an impact on gauge-invariant observables.} Hence, $\Phi^{(s)}_+(y)$ must be analytic around $y = c_\gamma k \tau = 0$, as long as the modes are adiabatic. Conversely, any violation of analyticity requires a source of non-adiabaticity.
\end{itemize}
In the following sections, we will illustrate the different physical origins of the CMB phase shift through a number of simple examples.

\subsection{Intuition from a Toy Model}
\label{sec:toyModel}
To gain more intuition for the system of equations discussed in the previous section, let us solve them exactly in a simple toy model. In particular, we will study an example in which the metric fluctuations~$\Phi_+$ propagate with a different speed than the photons, $c_s \ne c_\gamma$. We wish to understand under which conditions this mismatch leads to a phase shift in the photon oscillations.\medskip

The Einstein equations for the metric potentials, eq.~\eqref{eq:evolution}, can be rewritten in terms of the fields~$\Phi_\pm$ as
\beq
\ddot \Phi_+ + 3 \H \dot \Phi_+ + (2\dot \H + \H^2) \Phi_+ \,=\, 8\pi G a^2\, \delta P \,+\, \mathcal{S}[\Phi_-] \, ,	\label{eq:Einstein0}
\eeq
where the source term on the right-hand side is defined as
\beq
\mathcal{S}[\Phi_-] \equiv \ddot\Phi_- + \H\dot\Phi_- - \left(2\dot\H+\H^2 + \frac{2}{3}\nabla^2\right)\Phi_-\, . 
\eeq
The field~$\Phi_-$ is related to the total anisotropic stress~$\sigma$ via the constraint equation~\eqref{eq:EinsteinC}. In the standard model, both photons and neutrinos contribute to~$\delta P$, but only neutrinos provide a source for~$\sigma$ (and hence~$\Phi_-$). BSM~particles may lead to additional pressure and/or anisotropic stress. In general, we can write the pressure perturbation as~\cite{Peter:2013pri}
\beq
\delta P = c_s^2\hskip1pt \delta \rho + \delta P_\mathrm{en}\, ,
\eeq
where $c_s$ is the speed controlling the propagation of the total density perturbation~$\delta \rho$ and~$\delta P_\mathrm{en}$ denotes the non-adiabatic entropy perturbation. For adiabatic fluctuations, one has $\delta P_\mathrm{en} = 0$ and $c_s^2 = w- [3\mathcal{H}(1+w)]^{-1}\hskip2pt \dot w$, where we emphasize that we are \textit{not} assuming that $P=w\rho$, but $\bar{P} = w\bar{\rho}$. We eliminate the density perturbation~$\delta \rho$ using the relativistic generalization of the Poisson equation~\eqref{eq:Poisson}. Equation~\eqref{eq:Einstein0} can then be written as
\beq
\ddot \Phi_+ + 3 \H (1+c_s^2) \dot \Phi_+ - 3 \H^2 (w- c_s^2) \Phi_+ + c_s^2 k^2 \Phi_+ \,=\, 8\pi G a^2\, \delta P_\mathrm{en}\, ,	\label{eq:Phi++0}
\eeq
where we further assumed the absence of anisotropic stress, in which case the source term~$\mathcal{S}[\Phi_-]$ vanishes.

So far, this is fairly general and has only assumed vanishing anisotropic stress. In particular, at this point $w$~and~$c_s$ are still general, possibly time-dependent parameters. To be able to derive an analytic solution for the evolution of~$\Phi_+(\tau)$, we will now make a few simplifying assumptions. First, we assume that the equation of state~$w$ is nearly constant, so that we can integrate the combined Friedmann equations~\eqref{eq:FriedmannEquations}
to get
\beq
\H = \frac{2}{1+3w} \frac{1}{\tau}\, .
\eeq
Second, we take $c_s^2 \approx \const$ and $\delta P_\mathrm{en} \approx 0$. This allows us to solve~\eqref{eq:Phi++0} analytically. For arbitrary $w$~and~$c_s$, these assumptions are not guaranteed to be easily realizable in a physical model. Our analysis only serves as a simple illustration of some of the effects that give rise to phase shifts in the CMB. More concrete examples of these effects will be discussed in~\textsection\ref{sec:examples}.\medskip

It is convenient to define $z \equiv \c k \tau$ and write~\eqref{eq:Phi++0} as
\beq
\frac{\d^2}{\d z^2}\Phi_+ + \frac{1-2\alpha}{z} \, \frac{\d}{\d z}\Phi_+ +\left[1- \frac{\beta}{z^2} \right]\Phi_+ = 0\, ,	\label{eq:Phi++}
\eeq
with
\beq
\alpha \equiv \frac{1}{2}-\frac{3(1+\c^2)}{1+3w}\ , \qquad\ \beta \equiv \frac{12 (w-\c^2)}{(1+3w)^2}\, .	\label{eq:abG}
\eeq	
In the physically interesting parameter regime, $0 \leq (c_s^2, w) \leq 1$, we have $-\frac{11}{2} \leq \alpha \leq -\frac{1}{4}$, with equality for $(c_s^2,w) = (1,0)$ and $(0,1)$, respectively. The general solution of~\eqref{eq:Phi++} is 
\beq
\Phi_+(z) = z^{\alpha}\left(c_1J _{\kappa}(z)+ c_2Y _{\kappa}(z)\right) , \qquad \kappa \equiv \sqrt{\alpha^2+\beta}\, ,	\label{eq:Phi+nfs}
\eeq
where~$J _{\kappa}(z)$ and~$Y _{\kappa}(z)$ are Bessel functions of the first and second kind, respectively. Note that~$\kappa$ is strictly positive and real-valued for the physically relevant parameter range, $ \frac{1}{2}\sqrt{3} \leq \kappa \leq \frac{1}{2} \sqrt{73} $, with the minimum at $(c_s^2,w) = (\frac{1}{3},1)$ and the maximum at $(c_s^2,w) = (1,0)$.

\begin{figure}[t]
	\includegraphics{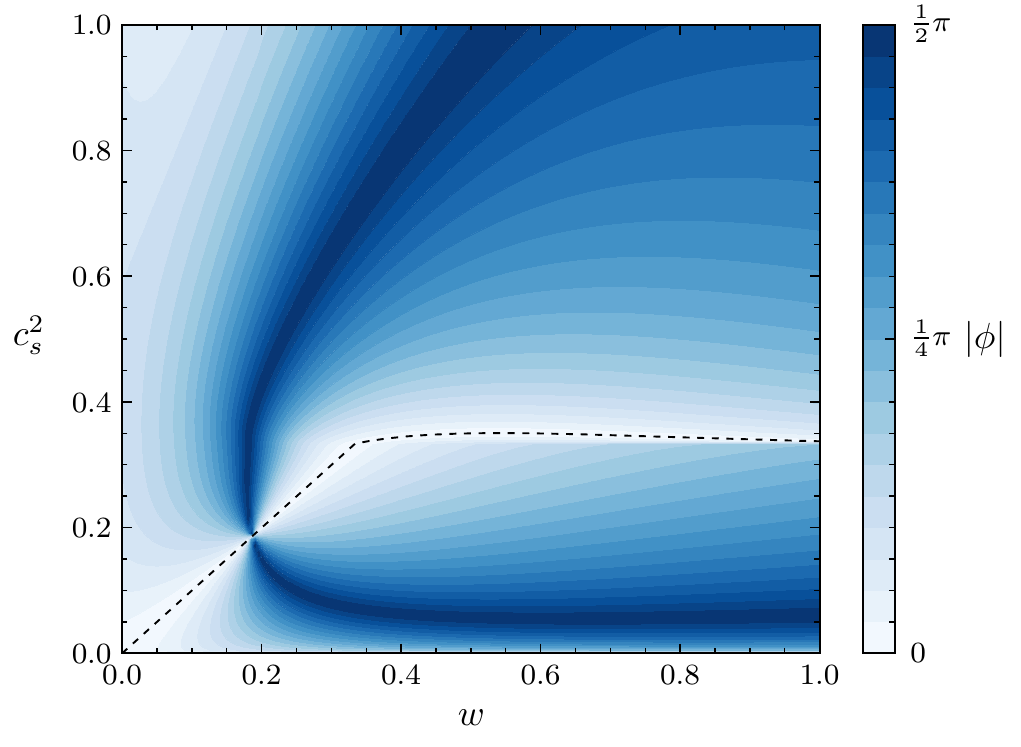}
	\caption{Phase shift~$\phi$ for varying speed of sound~($\c$) and equation of state~($w$). The dashed line denotes~$\phi=0$. Below this line, the phase shift is negative, while above it is positive.}
	\label{fig:toyModel_phi_contour}
\end{figure}

To impose initial conditions, we consider the superhorizon limit, $z \ll 1$, 
\beq
\Phi_+(z) \simeq \frac{2^{-\kappa} \left(c_1+c_2 \cot{(\pi\kappa)}\right)}{\Gamma(1+\kappa)} z^{\alpha +\kappa} - \frac{2^{\kappa} c_2 \Gamma(\kappa)}{\pi} z^{\alpha-\kappa} + \cdots\, .
\eeq
Since $\alpha + \kappa > \alpha-\kappa$, the ``growing mode'' solution corresponds to $c_2 \equiv 0$. Hence, we have
\beq
\Phi_+(z) = c_1 z^{\alpha} J _{\kappa}(z)\, .	\label{eq:Phi+nfs2}
\eeq
The overall normalization in~\eqref{eq:Phi+nfs2} will depend on the nature of the initial conditions (adiabatic or entropic). For $c_s^2 = w$, the superhorizon limit of~$\Phi_+$ is a constant, which we match to the superhorizon value of the primordial curvature perturbation~$\zeta$. This leads to the normalization $c_1 =2 \sqrt{2 \pi}\,\zeta$. We will maintain this normalization even for $c_s^2 \ne w$, although, in principle, the normalization of non-adiabatic modes is model-dependent. The $y \to \infty$ limit of~\eqref{eq:B} then becomes
\beq
B = 2 \sqrt{2 \pi}\,\zeta \int^\infty_0 \d y\, \left(\frac{\c}{c_\gamma} y\right)^{\!\alpha} J_\kappa\!\left(\frac{\c}{c_\gamma} y\right)\cos y\, ,
\eeq
and similarly for~$A$. Together with $d_{\gamma,\mathrm{in}}=-3\zeta$, this allows us to compute the phase shift via~\eqref{eq:theta}. A graphical illustration of the dependence of the phase shift~$\phi$ on the parameters~$\c^2$ and~$w$ is given in Fig.~\ref{fig:toyModel_phi_contour}.%
Let us emphasize again that we do not imagine that all of the combinations of~$\c^2$ and~$w$ that we show in the figure can be realized in a physically realistic model. In the following, we take slices through the parameter space to show that the most important features of the figure can be understood analytically.

\begin{figure}[b]
	\includegraphics{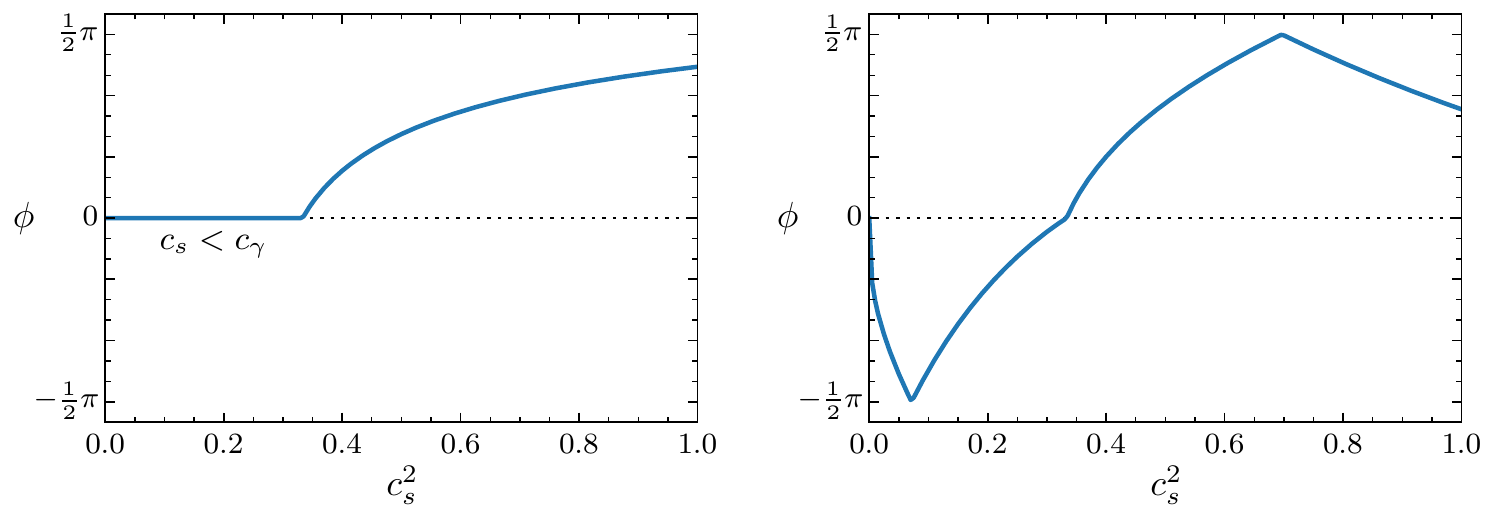}
	\caption{Phase shift~$\phi$ for varying $c_s^2=w$~(\textit{left}), and for varying $c_s^2$ at fixed $w=\frac{1}{3}$~(\textit{right}).}
	\label{fig:toyModel_phi_slices}
\end{figure}

Consider first the special case $c_s^2 = w$, which corresponds to adiabatic fluctuations. The parameters in~\eqref{eq:abG} and~\eqref{eq:Phi+nfs} then reduce to
\beq
\alpha = - \frac{5+3\c^2}{2(1+3\c^2)}\, , \qquad \beta =0\, , \qquad \kappa = |\alpha| \, .
\eeq		
As shown in the left panel of Fig.~\ref{fig:toyModel_phi_slices},%
the phase shift vanishes in this case for $\c \leq c_\gamma = \frac{1}{\sqrt{3}}$ and is positive for $\c > c_\gamma$. This is consistent with our abstract reasoning of the previous section. At large $z= c_s k\tau$, the solution~\eqref{eq:Phi+nfs2} behaves as $z^{\alpha - \frac{1}{2}} \cos(z) \propto \cos(\c/c_\gamma\, y)$, with $y=c_\gamma k \tau$. Since the contour at infinity in~\eqref{eq:Bdef} will not vanish when $\c > c_\gamma$, we cannot conclude that $\phi = 0$ (cf.~condition~\textit{i.}\ in~\textsection\ref{sec:originPhaseShift}). To find $\phi \neq 0$ was still not guaranteed, but there was no reason to expect otherwise. In contrast, $\phi$ vanishes for $\c \leq c_\gamma $ for exactly the reasons discussed before. In particular, the solution~\eqref{eq:Phi+nfs2} now takes the form $z^\alpha J_{|\alpha|} (z)$ with $\alpha < 0$. Near~$z=0$, the solution is analytic (cf.~condition~\textit{ii.}\ in~\textsection\ref{sec:originPhaseShift}) since the non-analytic behaviour of $z^\alpha$ cancels that of the Bessel function. Of course, this is precisely what we expected for adiabatic modes.

Taking $\c^2 \neq w$ corresponds to non-adiabatic fluctuations, i.e.~fluctuations which contain an isocurvature component. In this case, we expect a phase shift to arise for any values of~$\c^2$ and~$w$. To illustrate this, let us take $w = \frac{1}{3}$ and only allow~$\c^2$ to vary. We then have
\beq
\alpha = -1 -\frac{3\c^2}{2}\, , \qquad \beta = 1-3\c^2\, , \qquad \kappa = \frac{1}{2}\sqrt{8+9\c^4}\, .
\eeq
The corresponding phase shift is shown in the right panel of~Fig.~\ref{fig:toyModel_phi_slices}. We see that the phase shift now only vanishes at the special point $\c^2 = w = \frac{1}{3}$, where the fluctuations are adiabatic. This is also where the parameter $\beta$ changes sign, which is the origin of the change in the sign of the phase shift, $\phi \lessgtr 0$ for $\c^2 \lessgtr w$. This time the phase shift is associated with the non-analytic behaviour of~$\Phi_+(z)$ near the origin. To see this explicitly, consider the $z \to 0$ limit of~\eqref{eq:Phi+nfs2}:
\beq
\Phi_+(z) = \frac{c_1}{2^\kappa \Gamma(1+\kappa)} \, z^{\alpha + \kappa} \left[1+\mathcal{O}(z^2)\right] .
\eeq
For $0\leq c_s^2 \leq 1$, we have $\alpha+\kappa < 2$ and, hence, $\Phi_+(z)$ is non-analytic at $z=0$ (cf.~condition~\textit{ii.}\ in~\textsection\ref{sec:originPhaseShift}), except for the special case of the adiabatic limit where $\alpha+\kappa = 0$. This corresponds to the non-trivial superhorizon evolution of $\Phi_+$ in the presence of isocurvature modes.

\subsection{Simple Examples}
\label{sec:examples}
The toy model of the previous section suggests that isocurvature perturbations and free-streaming particles produce a phase shift. To study this further, it is useful to consider more realistic models. In the following, we will therefore compute the phase shift in a curvaton-like model, and for neutrinos and neutrino-like species.

\subsubsection{Isocurvature perturbations}
To simplify the calculations as much as possible, our curvaton-like model will include three species: photons~($\gamma$), a dark radiation fluid~($Y$) and a matter component~($m$) that decays into the dark radiation. The matter will carry the isocurvature fluctuations. We include the dark radiation because we are only interested in the gravitational effects on the photons, i.e.~we want to avoid the direct heating of the photons due to the decaying matter. The model will allow us to explore small deviations from the adiabatic limit $c_s^2=w$.\medskip

The coupled equations for the background densities of~$m$ and~$Y$ are 
\begin{align}
\frac{1}{a^3} \frac{\d}{\d\tau}(a^3 \bar{\rho}_m)	&= - \Gamma \hskip1pt a \hskip2pt \bar{\rho}_m \, ,	\label{eq:rhom}\\[4pt]
\frac{1}{a^4} \frac{\d}{\d\tau}(a^4 \bar{\rho}_Y)	&= + \Gamma \hskip1pt a \hskip2pt \bar{\rho}_m \, ,	\label{eq:rhoY}
\end{align}
where~$\Gamma$ is a constant decay rate. To simplify the calculations, we will work perturbatively in $\epsilon_m \equiv \bar \rho_m / \bar \rho$. At zeroth order in~$\epsilon_m$, the universe is radiation dominated and hence $a = \tau / \tau_\mathrm{in}$. Integrating~\eqref{eq:rhom}, we get
\beq
\bar\rho_m(a) = \frac{\bar\rho_{m,\mathrm{in}}}{a^3} \,e^{-\frac{1}{2}\Gamma \tau_\mathrm{in}(a^2-1)} \, ,	\label{eq:rhom2}
\eeq
where we set the initial value $\bar\rho_{m,\mathrm{in}} \equiv \bar\rho_m(\tau_\mathrm{in})$. Substituting~\eqref{eq:rhom2} into~\eqref{eq:rhoY}, we would get the solution for~$\bar\rho_Y(a)$, but this will not be needed for our purposes.

We now wish to determine how the decaying matter affects the evolution of the metric perturbation~$\Phi_+$ as given by~\eqref{eq:Einstein0}. The pressure perturbations only receive contributions from~$\gamma$ and~$Y$, so we have 
\beq
\delta P = c_\gamma^2(\delta \rho_\gamma + \delta \rho_Y) \,=\, \frac{1}{3}\delta \rho - \frac{1}{3}\delta \rho_m\, ,
\eeq
where we have used $c_Y^2= c_\gamma^2 = \frac{1}{3}$. Using the Poisson equation~\eqref{eq:Poisson}, we can write~\eqref{eq:Einstein0} in the absence of anisotropic stress as
\beq
\ddot \Phi_+ + 4 \H \dot \Phi_+ - \frac{1}{3} \nabla^2 \Phi_+ \,=\, (3w - 1) \H^2 \Phi_+ - \H^2 \epsilon_m \delta_m \, .	\label{eq:PhiPlus}
\eeq
We wish to solve this at linear order in~$\epsilon_m$.

We will shortcut the computation by isolating the isocurvature contribution. Suppose we write $\Phi_+ = \Phi_+^\mathrm{ad} + \Phi_+^\mathrm{iso}$ and similarly for~$\delta_m$. Equation~\eqref{eq:PhiPlus} then implies
\beq
\Phi_+^\mathrm{iso}\hskip1pt{}'{}' + \frac{4}{y} \Phi_+^\mathrm{iso}\hskip1pt{}' + \Phi_+^\mathrm{iso} \,=\,-\frac{\epsilon_m}{y^2}\hskip1pt \delta_m^\mathrm{iso} + \mathcal{O}(\epsilon_m^2) \, ,	\label{eq:Phi+2}
\eeq
where the primes denote derivatives with respect to $y \equiv c_\gamma k \tau$ and we have used that $\Phi_+^\mathrm{iso} \sim \mathcal{O}(\epsilon_m)$, so that all terms multiplying $\Phi_+^\mathrm{iso}$ can be evaluated at zeroth order in~$\epsilon_m$. The evolution of~$\delta_m^\mathrm{iso}$ is governed by $\delta_m^\mathrm{iso}\hskip1pt{}'{}' + \frac{1}{y}\delta_m^\mathrm{iso}\hskip1pt{}' = 0 + \mathcal{O}(\epsilon_m)$ according to~\eqref{eq:combinedConservationEquations}. Since the right-hand side of~\eqref{eq:Phi+2} is proportional to~$\epsilon_m$, we only need the homogeneous solution, which is
\beq
\delta_m^\mathrm{iso}(y) = c_1 + c_2 \ln y\, ,	\label{eq:deltaIso}
\eeq
where~$c_{1,2}$ are constants that may depend on~$k$. We solve~\eqref{eq:Phi+2} using the Green's function~$G_{\Phi_+}(y,y')$. Substituting~\eqref{eq:rhom2} and~\eqref{eq:deltaIso}, we get	
\beq
\Phi_+^\mathrm{iso}(y) \,=\, \frac{1}{c_\gamma k \tau_\mathrm{in}}\, \tilde\epsilon_{m,\mathrm{in}}\, \underbrace{\int_{y_\mathrm{in}}^y \d y'\, G_{\Phi_+}(y,y') \, \left(-\exp\left[-\frac{1}{2}\frac{(y')^2}{(c_\gamma k \tau_\mathrm{dec})^2}\right] \, \frac{c_1 + c_2 \ln(y')}{y'} \right)}_{\displaystyle \equiv \mathcal{I}(y)}\,\hskip1pt ,	\label{eq:PhiPlusF}
\eeq
where $\tilde\epsilon_{m,\mathrm{in}} \equiv \epsilon_{m,\mathrm{in}} \hskip1pt e^{\frac{1}{2}(\tau_\mathrm{in}/\tau_\mathrm{dec})^2}$ and we have introduced the ``decay time scale'' $\tau_\mathrm{dec}^2 \equiv \tau_\mathrm{in}/\Gamma$.

Let us comment on a few features of this solution. First of all, we notice that the integral is highly suppressed when $k \tau_\mathrm{dec} \ll 1$. The reason is easy to understand: the integral would have been dominated by contributions around the time of horizon crossing, $y \sim \mathcal{O}(1)$, but, for $k \tau_\mathrm{dec} \ll 1$, this is long after~$\rho_m$ has decayed. Second, we see that the solution has an overall factor of~$(c_\gamma k \tau_\mathrm{in})^{-1}$. This reflects the growth of~$\epsilon_m$ from the initial time,~$\tau_\mathrm{in}$, to the time of horizon crossing,~$(c_\gamma k)^{-1}$.

\begin{figure}[t]
	\includegraphics{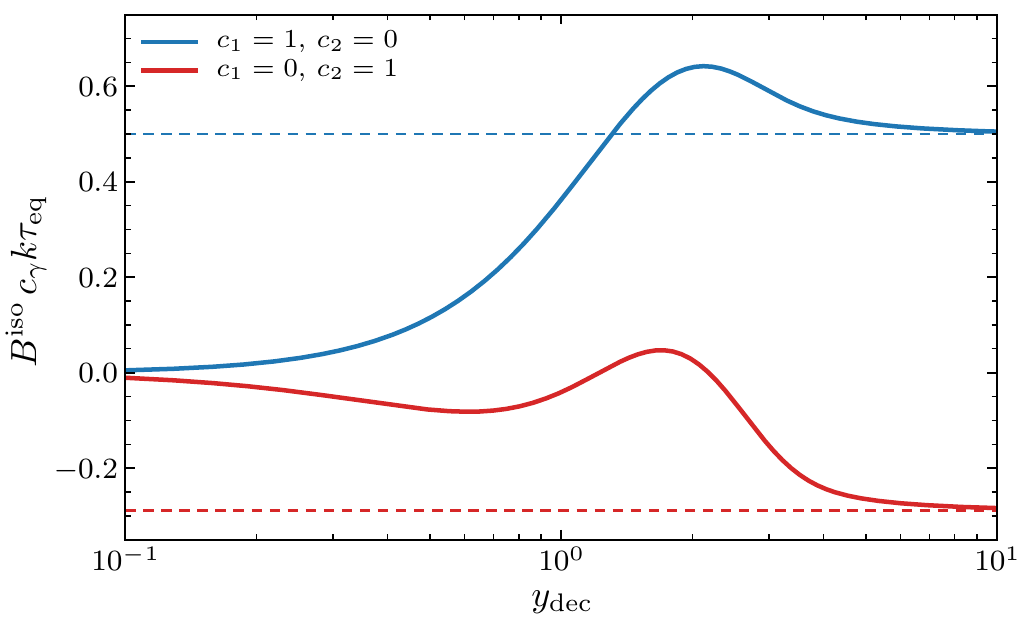}
	\caption{Numerical value of $B^\mathrm{iso} \, c_\gamma k \tau_\mathrm{eq}$ as a function of $y_\mathrm{dec}$. The \textcolor{pyBlue}{blue} and \textcolor{pyRed}{red} solid lines show the effect from $c_1$ and $c_2$, respectively. The dashed lines are the asymptotic values calculated in~\eqref{eq:Bisofinal}.}
	\label{fig:Isocurvature}
\end{figure}

It is convenient to define~$\tau_\mathrm{eq}$ as the time at which~$\bar \rho_m$ and~$\bar \rho_\gamma + \bar \rho_Y$ would be equal \textit{if} there was no decay. This is given by $\tau_\mathrm{eq} \simeq \tau_\mathrm{in}/\tilde\epsilon_{m,\mathrm{in}}$. Equation~\eqref{eq:PhiPlusF} then becomes 
\beq
\Phi_+^\mathrm{iso}(y) = \frac{1}{c_\gamma k \tau_\mathrm{eq}}\, \mathcal{I}(y) \, .	\label{eq:PhiPlusF2}
\eeq
We compute the phase shift by substituting~\eqref{eq:PhiPlusF2} into~\eqref{eq:A} and~\eqref{eq:B}, and taking the limits $y_\mathrm{in} \to 0$ and $y \to \infty$,
\begin{align}
A^\mathrm{iso} &\equiv \frac{1}{c_\gamma k \tau_\mathrm{eq}} \int^\infty_0 \d y'\, \mathcal{I}(y')\hskip1pt\sin y'\, ,	\\[4pt]
B^\mathrm{iso} &\equiv \frac{1}{c_\gamma k \tau_\mathrm{eq}} \int^\infty_0 \d y'\, \mathcal{I}(y')\hskip1pt\cos y'\, .
\end{align}
In Figure~\ref{fig:Isocurvature}, we display the numerical result for~$B^\mathrm{iso}$ as a function of $y_\mathrm{dec} \equiv c_\gamma k \tau_\mathrm{dec}$. In the limit $y_\mathrm{dec} \gg 1$, we can simplify the calculation by dropping the exponential in~\eqref{eq:PhiPlusF}. We then get
\begin{align}
A^\mathrm{iso} &\,\xrightarrow{\ y_\mathrm{dec} \gg 1 \ }\, \frac{\pi}{4} c_2\, \frac{1}{c_\gamma k \tau_\mathrm{eq}}\, , \\
B^\mathrm{iso} &\,\xrightarrow{\ y_\mathrm{dec} \gg 1 \ }\, \frac{1}{2}( c_1 - c_2 \gamma_\mathrm{E} )\, \frac{1}{c_\gamma k \tau_\mathrm{eq}} \, ,	\label{eq:Bisofinal}
\end{align}
where $\gamma_\mathrm{E} \approx 0.5772$ is the Euler-Mascheroni constant. We see from Fig.~\ref{fig:Isocurvature}%
that the analytic result~\eqref{eq:Bisofinal} becomes a good approximation for $y_\mathrm{dec} \gtrsim 5$.\medskip

To summarize, we have demonstrated in a simple model that isocurvature perturbations give rise to a phase shift, as we expected from condition~\textit{ii.}\ of \textsection\ref{sec:originPhaseShift}. As suggested by Fig.~\ref{fig:Isocurvature}, this phase shift has a nontrivial scale dependence which probably allows it to be distinguished from other sources for a phase shift. This scale dependence is likely to be a general feature of isocurvature models.

\subsubsection{Free-streaming particles}
\label{sec:fs}
Above, we have seen that a phase shift is also generated if fluctuations in the gravitational potential travel faster than the speed of sound in the photon-baryon fluid. A simple way to realize this is through free-streaming relativistic particles,\hskip1pt\footnote{While it should be physically clear that free-streaming radiation travels at the speed of light, this property is sometimes not very transparent in the equations for the density perturbations of this radiation. Instead, what is usually more apparent is that free-streaming particles can induce a significant anisotropic stress, which then provides a source for~$\Phi_+$ and, hence, the evolution of~$d_\gamma$ through~\eqref{eq:dgamma}. The origin of the phase shift is therefore often identified with the presence of anisotropic stress. However, in principle, one could imagine situations with significant anisotropic stress, but no supersonic propagation modes (e.g.~non-relativistic, free-streaming particles). In those cases, we would not expect a phase shift. Hence, it is the propagation speed, not the anisotropic stress itself, that makes the phase shift possible.} such as neutrinos~\cite{Bashinsky:2003tk}. In the following, we give a new derivation of this result. In Section~\ref{sec:analysis}, we will show that the CMB data is now accurate enough to detect this effect (see also~\cite{Follin:2015hya}).

Since most of the modes relevant to current and future CMB~observations entered the horizon during the era of radiation domination, our discussion in this section will ignore both the matter and baryon content of the universe. In Appendix~\ref{app:cmb-phases_appendices}, we show that this is a good approximation for high-$\ell$ modes and also discuss some of the implications of a finite matter density.\medskip

In the absence of pressure perturbations $\delta P_a = c_a^2\hskip1pt\delta\rho_a^{\phantom{0}}$, with $c_a \neq c_\gamma$, the evolution equation~\eqref{eq:Einstein0} takes the form
\begin{align}
\Phi_+'' + \frac{4}{y} \Phi_+' + \Phi_+ 	&\,=\, \tilde{\mathcal{S}}[\Phi_-] 							\nonumber \\
											&\,\equiv\, \Phi_-'' + \frac{2}{y} \Phi_-' + 3\Phi_- \, .
\end{align}
The solution for $\Phi_+$ can be written as
\beq
\Phi_+(y) = 3\Phi_{+,\mathrm{in}}\, \frac{\sin y - y \cos y}{y^3} + \int_{y_\mathrm{in}}^y \d y' \, \tilde{\mathcal{S}}[\Phi_-(y')]\,G_{\Phi_+}(y, y')\, ,	\label{eq:Phi+}
\eeq
where we introduced the Green's function 
\beq
G_{\Phi_+}(y, y') = \Theta(y-y') \frac{y'}{y^3} \Big[ (y' -y)\cos(y' -y) - (1+ y y') \sin(y' -y) \Big]\, ,
\eeq
with the Heaviside step function~$\Theta$. Following~\cite{Bashinsky:2003tk}, we will write the solution for~$\Phi_+$ as an expansion in powers of the fractional energy density contributed by the species of free-streaming particles,~$X$, as measured by the dimensionless ratio
\beq
\epsilon_X \equiv \frac{\rho_X}{\rho_{\gamma} + \rho_{X}} \, = \, \frac{\Nf}{a_\nu + \Nf} \, .
\eeq
For the Standard Model neutrinos, we have $\epsilon_\nu \approx 0.41$. We determine the superhorizon initial condition of the homogeneous solution, $\Phi_{+,\mathrm{in}}$, by matching to the constant superhorizon solution for adiabatic initial conditions~\cite{Bashinsky:2003tk},
\beq
\Phi_{+,\mathrm{in}} = \frac{20 + 4 \epsilon_X}{15 + 4 \epsilon_X}\,\zeta = \frac{4}{3} \zeta \left( 1 - \frac{1}{15}\epsilon_X + \mathcal{O}(\epsilon_X^2) \right) ,
\eeq
where~$\zeta$ is the conserved curvature perturbation.

To find the inhomogeneous part of the solution~\eqref{eq:Phi+}, we require an expression for~$\Phi_-(y)$. This is determined by the anisotropic stress~$\sigma_X$ induced by the free-streaming particles,
\beq
\Phi_-(y) = -\frac{2 k^2 \epsilon_X}{y^2} \sigma_X(y) \, ,	\label{eq:Phi-X3}
\eeq
which follows from the Einstein constraint equation~\eqref{eq:EinsteinC} with $\sigma = \epsilon_X \sigma_X$. An evolution equation for the anisotropic stress~$\sigma_X$ can be derived from the Boltzmann equation for the distribution function $f_X(\tau, \x, q, \hat \n)$ with comoving momenta $\q=q\hskip1pt\hat\n$. We separate $f_X$ into a background component $\bar f_X$ and a perturbation $\delta f_X \equiv f_X - \bar f_X$. For massless particles, it is convenient to integrate out the momentum dependence of the distribution function and define 
\beq
\int \d q\, q^3 \hskip1pt \left(\delta f_X + q \hskip1pt \partial_q \bar f_X \Psi \right) \ =\ \frac{4}{3}D_X(\tau, \x, \hat \n) \times\! \int \d q\, q^3 \bar{f}_X(q)\, .
\eeq
The linearised, collisionless Boltzmann equation is then given by
\beq
\dot D_X + i k \mu D_X = - 3 i k \mu\, \Phi_+ \, ,	\label{eq:Boltzmann}
\eeq
where we introduced $\mu = \hat\n \cdot \hat\k$. We note that~$D_X$ only depends on~$\Phi_+ $, but not on~$\Phi_- $. It is useful to expand the momentum-integrated distribution function~$D_X$ into multipole moments,
\beq
D_X = \sum_{\ell=0}^\infty (-i)^\ell (2\ell+1) D_{X,\ell} \hskip1pt P_{\ell}(\mu) \, ,
\eeq
with the Legendre polynomials $P_{\ell}(\mu)$. The monopole moment~$D_{X,0}$ determines the overdensity~$d_X$, while the quadrupole moment~$D_{X,2}$ is associated with the anisotropic stress~$\sigma_X$. To see this, one writes the perturbed stress-energy tensor in terms of the perturbed distribution function,
\beq
\delta T^\mu{}_{\nu,X} = a^{-4} \int \d\Omega_{\hat n}\, \hat n^\mu \hat n_\nu \! \int \d q\, q^3 \hskip1pt \delta f_X\, .
\eeq
Comparing this expression to~\eqref{eq:stressEnergyTensor}, we find
\beq
D_{X,0} = d_X \, , \quad \ D_{X,1} = k u_X \, , \quad \ D_{X,2} = \frac{3}{2} k^2 \sigma_X\, .
\eeq
The quadrupole moment of~\eqref{eq:Boltzmann} then provides the missing evolution equation for the anisotropic stress and we can rewrite~\eqref{eq:Phi-X3} as
\beq
\Phi_-(y) = - \frac{4}{3}\frac{\epsilon_X}{y^2} D_{X,2}(y)\, .	\label{eq:Phi-X}
\eeq
Defining $D_{X,\mathrm{in}} \equiv D_X(\tau_\mathrm{in})$ at some time~$\tau_\mathrm{in}$, the solution to~\eqref{eq:Boltzmann} is
\beq
D_X(\tau) = e^{- i k \mu (\tau-\tau_\mathrm{in})} D_{X,\mathrm{in}}- 3 i k \mu \int_{\tau_\mathrm{in}}^\tau \d \tau'\, e^{- i k\mu (\tau - \tau')} \Phi_+(\tau') \, .
\eeq
We wish to extract the quadruple moment~$D_{X,2}$ of the solution. Since $D_{X,\ell}(\tau_\mathrm{in}) \propto \tau_\mathrm{in}^\ell$, we will only keep the monopole term~$D_{X,0}(\tau_\mathrm{in})$ in the homogeneous part of the solution. This is possible because we can set the initial conditions at a sufficiently early time $\tau_\mathrm{in} \ll k^{-1}$, so that the modes with $\ell>0$ will be sub-dominant. In fact, we will take $k\tau_\mathrm{in} \to 0$ from now on. Assuming adiabatic initial conditions, i.e.~$D_{X,0}(\tau_\mathrm{in}) =d_{X,\mathrm{in}}=-3\zeta$, we get
\beq
D_{X,2}(y) = -3\zeta\, j_{2}\!\left[c_\gamma^{-1}y\right] + \frac{3}{c_\gamma} \int_0^y \d y'\, \Phi_+(y') \left\{ \frac{2}{5} j_{1}\!\left[c_\gamma^{-1}(y-y')\right] - \frac{3}{5}j_{3}\!\left[c_\gamma^{-1}(y-y')\right] \right\} ,	\label{eq:DX2}
\eeq
where the Bessel functions~$j_\ell$ arise from the Rayleigh expansion of the exponentials. Substituting this into~\eqref{eq:Phi-X} directly links the two gravitational potentials~$\Phi_+$ and~$\Phi_-$. The most important feature of the solution~\eqref{eq:DX2} is that it contains modes that travel at the speed of light. Specifically, recall that $c_\gamma^{-1} y = k \tau$ and, therefore, the Bessel functions describe oscillatory solutions with a speed of propagation of $c_s = 1$. As we have emphasized before, this is the property of the free-streaming radiation that makes a phase shift possible.\medskip

The above is a closed set of equations which we can solve perturbatively in~$\epsilon_X$:
\beq
\Phi_\pm \equiv \sum_{n} \Phi_\pm^{(n)}\, , \quad \ d_\gamma \equiv \sum_n d_\gamma^{(n)}\, ,
\eeq
where the superscripts on~$\Phi_\pm^{(n)}$ and~$d_\gamma^{(n)}$ count the order in~$\epsilon_X$. Here, we present the solution up to first order:
\begin{itemize}
\item At zeroth order in~$\epsilon_X$, we have $\Phi_-^{(0)}(y) = 0$ and, hence,~$\Phi_+^{(0)}$ is given by the homogeneous solution,
\beq
\Phi_+^{(0)}(y) = 4\zeta\, \frac{\sin y - y \cos y}{y^3}\, .	\label{eq:Phi+h}
\eeq
Inserting this into~\eqref{eq:A} and~\eqref{eq:B}, we find
\begin{align}
A^{(0)}(y) &\,=\, 2\zeta - 2\zeta\, \frac{\sin^2(y)}{y^2} \hskip12pt 	\xrightarrow{\, y\to\infty\, }\ 2\zeta \, ,	\\
B^{(0)}(y) &\,=\, 2\zeta\, \frac{y - \cos y \sin y}{y^2}  \ 			\xrightarrow{\, y\to\infty\, }\ 0 \, .
\end{align}
The result for the photon density perturbations then is
\beq
d_\gamma^{(0)}(y) \approx 3\zeta \cos y	\, .
\eeq
We conclude that in the absence of anisotropic stress, the correction due to~$\Phi_+$ is in phase with the homogeneous solution and~$B$ vanishes as expected. 

\item At first order in~$\epsilon_X$, we only need the zeroth-order solution of the anisotropic stress,~$\sigma_X^{(0)}$, since the source in~\eqref{eq:Phi-X} already comes with an overall factor of~$\epsilon_X$. We can therefore write~\eqref{eq:Phi-X} and~\eqref{eq:DX2} as
\begin{align}
\Phi_-^{(1)}(y)	&\,=\, 4 \zeta\,\frac{\epsilon_X}{y^2}\, j_{2}\!\left[c_\gamma^{-1}y\right]	\label{eq:Phi-X2}	\\[2pt]
				& \ \quad - \frac{4 }{c_\gamma} \frac{\epsilon_X}{y^2} \int_0^y \d y'\, \Phi_+^{(0)}(y') \left\{ \frac{2}{5} j_{1}\!\left[c_\gamma^{-1}(y-y')\right] - \frac{3}{5}j_{3}\!\left[c_\gamma^{-1}(y-y')\right] \right\} , \nonumber
\end{align}
where~$\Phi_+^{(0)}$ is given by~\eqref{eq:Phi+h}. Substituting~\eqref{eq:Phi-X2} into~\eqref{eq:Phi+}, we obtain
\beq
\hskip-10pt \Phi_+^{(1)}(y) = - \frac{4}{15} \zeta\, \epsilon_X\, \frac{\sin y - y \cos y}{y^3} + \int_0^y \d y' \, \tilde{\mathcal{S}}[\Phi_-^{(1)}(y')]\,G_{\Phi_+}(y, y') \, .
\eeq
Inserting this into~\eqref{eq:A} and~\eqref{eq:B}, we finally get expressions for~$A^{(1)}$ and~$B^{(1)}$. These have to be evaluated numerically and we find
\beq
A^{(1)} \approx - 0.268\, \zeta\, \epsilon_X\, ,	\qquad	B^{(1)} \approx 0.600\, \zeta\, \epsilon_X\, .
\eeq
The non-zero value of~$B^{(1)}$ corresponds to the expected phase shift. 
\end{itemize}
Using~\eqref{eq:theta} with $d_{\gamma,\mathrm{in}}=-3\zeta$, we get 
\beq
\phi \approx 0.191\pi\, \epsilon_X + \mathcal{O}(\epsilon_X^2)\, ,	\label{eq:thetaX}
\eeq
which is consistent with the result of Bashinsky\,\&\,Seljak~\cite{Bashinsky:2003tk}.\medskip

The phase shift is a clean signature of free-streaming particles and will naturally play an important role in the observational constraints discussed in Section~\ref{sec:analysis}. To put these constraints into context, let us use the analytic result of this section to relate changes in~$\Nf$ to shifts $\Delta\ell$ in the peaks of the CMB~spectra. As we show in Appendix~\ref{app:cmb-phases_appendices}, the E-mode spectrum will exhibit precisely the same phase shift as the temperature spectrum and, therefore, our analytic estimates are applicable in either case. In the small-angle approximation, a shift in angle~$\phi$ is related to a multipole shift by $\Delta\ell \simeq (\phi/\pi)\, \Delta\ell_{\text{peak}}$, where $\Delta\ell_{\text{peak}} \sim 330$~\cite{Aghanim:2015xee} is the distance between peaks in the temperature anisotropy spectrum for modes entering the horizon during radiation domination. Using~\eqref{eq:thetaX}, with $\Nf = N_\nu = 3.046$, we find that the shift of the peaks arising from ordinary neutrinos is $\Delta \ell_\nu \approx 26$ (compared to a neutrinoless universe). Similarly, small variations~$\Delta\Neff$ around the standard value $N_\nu = 3.046$ will lead to a multipole shift of order
\beq
\Delta \ell_{\Delta \Nf} \approx 5.0 \times \Delta \Nf \, ,	\label{eq:delta-ell}
\eeq 
where we have expanded to linear order in $\Delta\Neff$. While this result is likely subject to a 20 to 30~percent error, it is reliable enough to see that a sensitivity of $\sigma(\Nf) \sim 0.1$ will constrain a phase shift of order $\Delta \ell \lesssim 1$. Current constraints on $\Nf$ imply $\Delta \ell \sim \mathcal{O}(1)$. As we will see in Section~\ref{sec:analysis}, future CMB~experiments are expected to constrain, or measure, shifts of order $\Delta \ell \sim \mathcal{O}(0.1)$. This is consistent with the rough expectation from measuring $\mathcal{O}(10)$ peaks and troughs in the E-mode power spectrum.

\section{Current and Future Constraints}
\label{sec:analysis}
The CMB has the potential to distinguish between many distinct sources of BSM~physics: new free-streaming or non-free-streaming particles, isocurvature perturbations and/or non-standard thermal histories. However, ultimately the observability of the new physics depends both on the size of the effect and whether it is degenerate with other cosmological parameters. In this section, we present new constraints on the density of free-streaming and non-free-streaming radiation from the Planck satellite~\cite{Aghanim:2015xee} and then discuss the capabilities of a proposed CMB-S4 experiment~\cite{Abazajian:2016yjj}. Whenever possible, we will give some approximate analytic understanding of the qualitative origin of our results. For precise quantitative results, we will perform a full likelihood analysis.\footnote{In this section, we prefer the use of Markov chain Monte Carlo~(MCMC) techniques over Fisher matrix forecasts because Fisher matrices can underestimate the impact of degeneracies on the posterior distributions~\cite{Perotto:2006rj}. We believe this to be the origin of the (small) differences between our results and those of~\cite{Wu:2014hta}.}\medskip

As we have discussed in~\textsection\ref{sec:species_diffusionDamping}, a change in the radiation density is degenerate with a shift in the primordial helium fraction~$Y_p$ since they are anti-correlated at fixed~$\theta_s$. We therefore expect that the CMB temperature constraints on~$\Nf+\Nn$ and~$Y_p$ to weaken considerably if we allow both of these parameters to vary. Of course, we also have to break the degeneracy between~$\Nf$ and~$\Nn$, which are not distinguished by their effects on the damping tail. Fortunately, future datasets will be much less sensitive to these degeneracies for two reasons:
\begin{itemize}
\item First, as we show in Appendix~\ref{app:cmb-phases_appendices}, the amplitude of the polarization of the CMB,~$\Theta_{P,\ell}$, is proportional to~$n_e^{-1}$, but not~$H$, and, therefore, it is sensitive to~$Y_p$ alone. The key feature is that polarization is a direct measurement of the quadrupole at the surface of last-scattering, while the damping tail of the temperature spectrum is the integrated effect of the quadrupole on the monopole. This difference allows us to break the degeneracy between~$Y_p$ and~$\Nf+\Nn$.

\item Second, as we demonstrated in Section~\ref{sec:analytics}, the CMB is sensitive to the perturbations in the free-streaming particles and not just their contribution to the background evolution. This is illustrated in Fig.~\ref{fig:CMBphase} and will be explored in more detail in the next subsection. What is important here is that the phase shift associated with free-streaming particles is not expected to be degenerate with other effects and it is measurable out to very high multipoles. Furthermore, as we discuss in Appendix~\ref{app:cmb-phases_appendices}, the same phase shift appears in both temperature and polarization which means there is room for significant improvement in the sensitivity to this effect. This will largely eliminate the issues of degeneracies for~$\Nf$.
\end{itemize}

As we move towards more sensitive experiments, the perturbations in the radiation density will play an increasingly important role. In this section, we will demonstrate the potential of current and future experiments to detect the free-streaming nature of relativistic species.

\subsection{MCMC Methodology}
The following data were used in our analysis:
\begin{itemize}
\item The Planck likelihoods are described in detail in~\cite{Aghanim:2015xee}. The low-$\ell$ likelihoods ($2\leq \ell \leq 29$) include both the temperature and polarization data (even in the cases labelled ``TT-only''). For the remaining multipoles~($\ell \geq 30$), we use the \texttt{plik} joint TT+TE+EE~likelihood.\footnote{Note that the high-$\ell$ polarization data was publicly released by the Planck collaboration, but labelled preliminary due to possible unresolved systematics. It should therefore be used with caution.} This contains information about the TT~spectrum up to $\ell_\mathrm{max} = 2508$ and about the~TE and EE~spectra up to $\ell_\mathrm{max} = 1996$. The lensing potential is reconstructed in the multipole range $40\leq \ell \leq 400$ using \texttt{SMICA} temperature and polarization maps. For the TT-only constraints, we use the \texttt{plik} TT-only likelihood with range $30\leq \ell \leq 2508$ and the lensing reconstruction involves only the temperature map.

\item Our forecasts for CMB-S4 experiments assume \num{e6}~polarization-sensitive detectors with a \SI{1}{arcmin} beam and sky coverage $f_\mathrm{sky} = 0.75$. The default observing time is chosen to be five years, resulting in a sensitivity of $\sigma_T=\sigma_P/\sqrt{2}=\SI{0.558}{\muKelvin.arcmin}$ matching the most optimistic experimental setup studied in~\cite{Wu:2014hta}. Our analysis included multipoles up to $\ell_\mathrm{max}=5000$ for both temperature and polarization.\footnote{These experimental specifications are slightly more optimistic (in particular the maximum temperature multipole of $\ell_\mathrm{max}^T=5000$ which will be hard to achieve because of astrophysical foregrounds) than those currently considered by the CMB-S4 collaboration~\cite{Abazajian:2016yjj}. We present forecasts on~$\Neff$ using these specifications in Appendix~\ref{app:bao-forecast_appendices}.} We also study how our results change if we vary the beam size and the maximum multipole.
\end{itemize}

Our modification of the Boltzmann code \texttt{CLASS}~\cite{Blas:2011rf} includes an additional relativistic fluid, whose energy density is measured by the parameter~$\Nn$ defined in~\eqref{eq:Nfluid}. The equation of state and the sound speed of the fluid were fixed to $w_Y = c_Y^2 = \frac{1}{3}$, with initial conditions that were chosen to be adiabatic. With this choice, our analytic results imply that this fluid does not contribute to the phase shift in the acoustic peaks. We use \texttt{MontePython}~\cite{Audren:2012wb} to derive constraints on the parameters~$\Nf$ and~$\Nn$. Whenever the primordial helium abundance~$Y_p$ was not varied independently (which we will refer to as~``$Y_p$ fixed''), it was set to be consistent with the predictions of BBN, using the total relativistic energy density including both~$\Nf$ and~$\Nn$ in determining the expansion rate. All Monte Charlo Markov chains were run until the variation in their means was small relative to the standard deviation (using $R-1 \lesssim 0.01$ in the Gelman-Rubin criterion~\cite{Gelman:1992zz}).

\begin{figure}[t!]
	\includegraphics{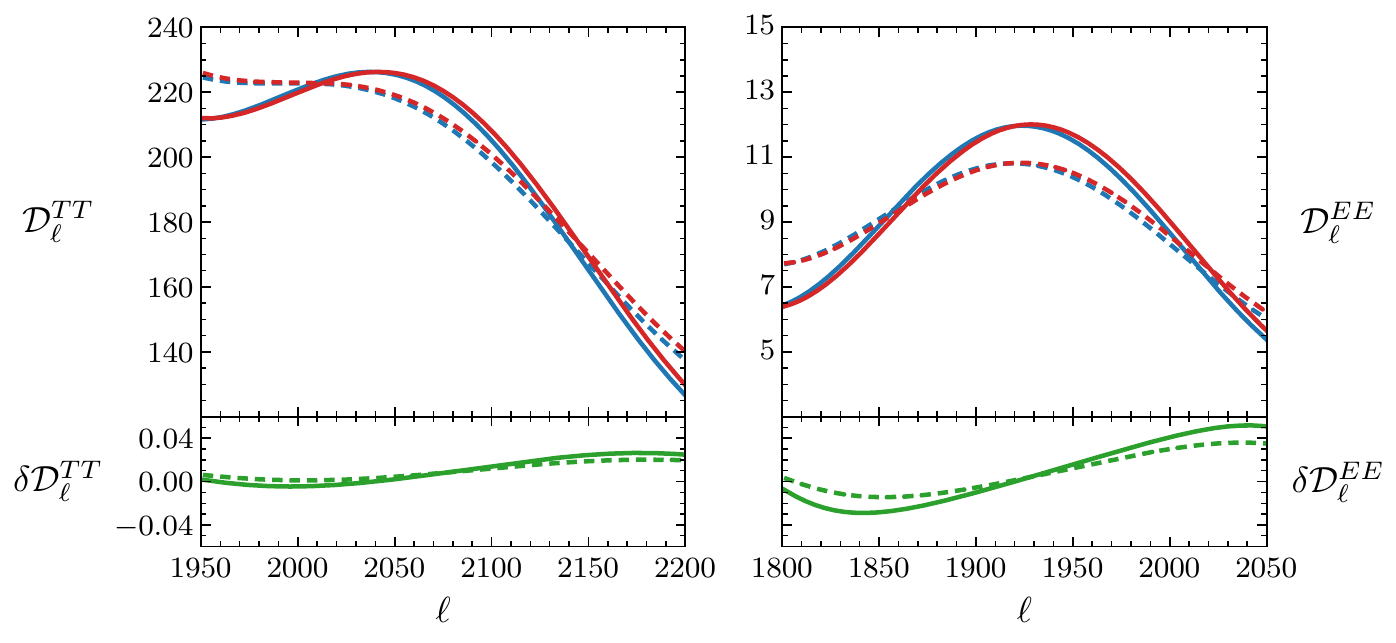}
	\caption{\textit{Top:}~Temperature spectrum~$\mathcal{D}_\ell^{TT}$~(\textit{left}) and polarization spectrum~$\mathcal{D}_\ell^{EE}$~(\textit{right}) for ($\Nf=3.046\hskip1pt, \Nn=0$)~(\textcolor{pyBlue}{blue}) and ($\Nf = 2.046\hskip1pt, \Nn=1.0$)~(\textcolor{pyRed}{red}) with $\mathcal{D}_\ell^X \equiv \ell (\ell+1) C_\ell^X / (2\pi)$ in units of~$\mu\mathrm{K}^2$. The TT~and EE~spectra represented by the red curves were rescaled by the same constant factor chosen such that the height of the seventh peak of the TT~spectrum matches for the red and blue curves. The solid and dashed lines show the unlensed and lensed data, respectively. The phase shift from~$\Nf$ and the peak smearing from lensing can be seen in both the TT and EE~spectra. \textit{Bottom:}~Illustration of the relative difference $\delta \mathcal{D}_\ell^X \equiv \Delta\mathcal{D}_\ell^X / \mathcal{D}_\ell^X$ between the ($\Nf=3.046\hskip1pt, \Nn=0$) and ($\Nf = 2.046\hskip1pt, \Nn=1.0$) spectra of the upper panels. The \textcolor{pyGreen}{green} solid and dashed lines are the differences in the unlensed and lensed data, respectively. We see that the change is largest in the unlensed EE~spectrum.} 
	\label{fig:CMBphase}
\end{figure}

Our analysis makes use of the effects of gravitational lensing of the CMB in two distinct ways. ``Lensing reconstruction'' will refer to a reconstruction of the power spectrum of the lensing potential from the measurements of the temperature and polarization four-point functions. In the case of the CMB-S4 forecasts, the power spectrum of the lensing potential was computed with~\texttt{CLASS}. CMB lensing also modifies the observed CMB power spectra (TT, TE, EE), primarily in the form of smearing the peaks~\cite{Seljak:1995ve}, as is illustrated in Fig.~\ref{fig:CMBphase}.%
``Delensing'' removes the effect of lensing on these power spectra using the reconstructed lensing potential. This is trivially implemented in forecasts in the limit of perfect delensing (we will just output spectra without computing the lensing), but is an involved procedure to implement on real data. The utility of this procedure is that lensing moves information from the power spectra to higher-point functions, but delensing moves this information back to the power spectra, so that it can easily be accounted for in our likelihood analysis (rather than through some more elaborate multi-point function likelihood).

Our Planck 2015 results use the publicly available lensing reconstruction likelihood, but do not include any delensing of the power spectra. For CMB-S4, the lensing reconstruction noise was computed using the iterated delensing method described in~\cite{Smith:2010gu} (based on~\cite{Hu:2001kj, Okamoto:2003zw, Hirata:2003ka}). Forecasts using delensed spectra assumed perfect delensing, which is a good approximation for a CMB-S4 experiment across a wide range of multipoles. Taking lensing reconstruction noise into account when computing delensed spectra would require a more careful analysis (see~\cite{Green:2016cjr, Abazajian:2016yjj} for subsequent studies).

The fiducial cosmology used for all forecasts is described by the following parameters: $\Omega_b h^2=0.022$, $\Omega_c h^2=0.120$, $h=0.67$, $\As=\num{2.42e-9}$, $\ns=0.965$, $\tau=0.078$, $\Nf=3.046$ and $\Nn = 0.0$.

\subsection{Planck 2015 Results}
\label{sec:PlanckResults}
\begin{figure}[t]
	\includegraphics{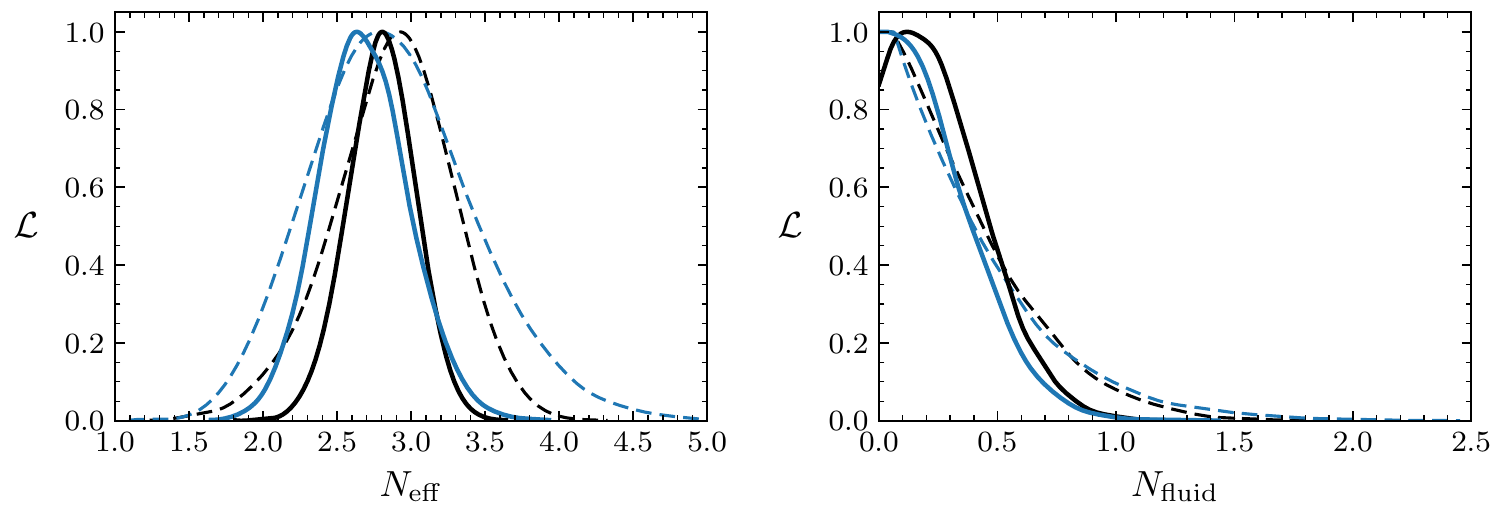}\vspace{-1pt}
	\caption{\textit{Left:} Posterior distributions for $\Nf$ from Planck~TT, TE and~EE marginalized over~$\Nn$. The \textcolor{pyBlue}{blue} curve involves the marginalization over~$Y_p$, while the black curve keeps~$Y_p$ fixed. Both likelihoods rule out $\Nf = 0$ at high significance. \textit{Right:}~Posterior distributions for~$\Nn$ from Planck~TT, TE and~EE marginalized over~$\Nf$. The \textcolor{pyBlue}{blue} curve involves the marginalization over~$Y_p$, while the black curve keeps~$Y_p$ fixed. In both panels, the likelihoods for Planck~TT-only with the same marginalizations are shown as dashed lines.\vspace{-1pt}}
	\label{fig:planckPosteriors}
\end{figure}

\begin{figure}[t]
	\definecolor{likeRed}{RGB}{224, 52, 36}
	\definecolor{likeGreen}{RGB}{34, 139, 34}
	\definecolor{likeIndigo}{RGB}{75, 0, 130}
	\definecolor{likeOlive}{RGB}{128, 128, 0}
	\definecolor{likeCyan}{RGB}{0, 139, 139}
	\definecolor{likeGray}{RGB}{112, 128, 144}
	\includegraphics{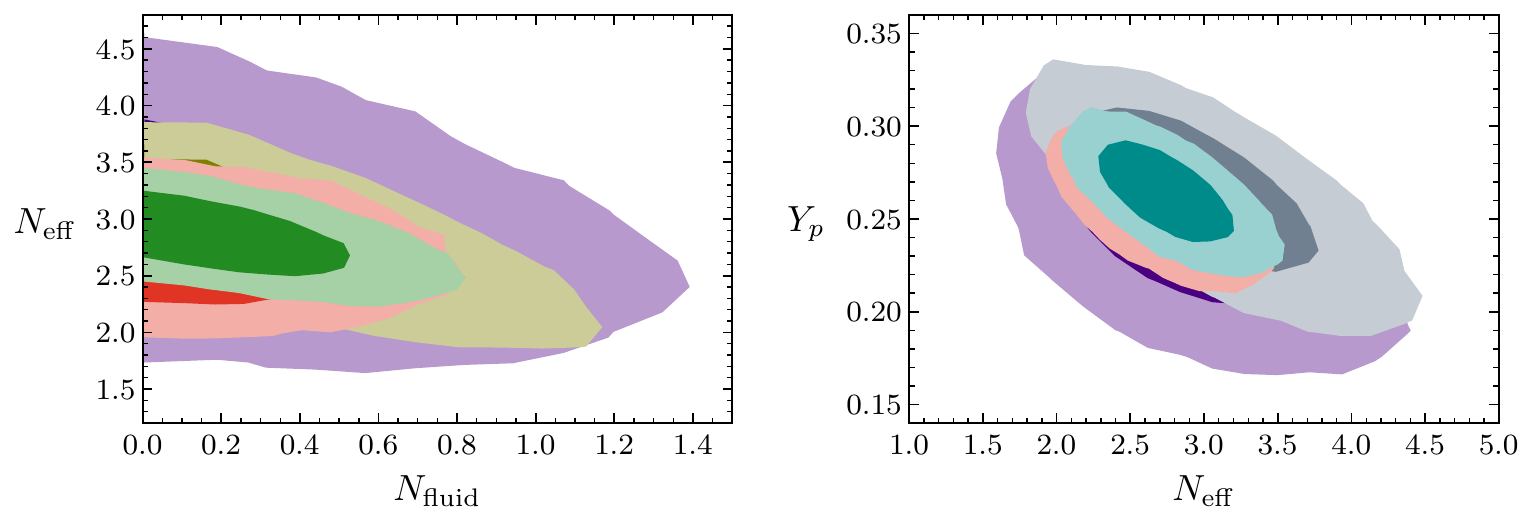}
	\caption{\textit{Left}: Constraints on $\Nf$ and $\Nn$ using the Planck~TT, TE and~EE likelihoods for varying~$Y_p$~(\textcolor{likeRed}{red}) and when~$Y_p$ is fixed~(\textcolor{likeGreen}{green}). Shown are also the Planck~TT-only results for varying~$Y_p$~(\textcolor{likeIndigo}{indigo}) and when~$Y_p$ is fixed~(\textcolor{likeOlive}{olive}). \textit{Right}:~Constraints on $\Nf$ and~$Y_p$ using the Planck~TT, TE and~EE likelihoods for varying $\Nn$~(\textcolor{likeRed}{red}) and when~$\Nn$ is fixed~(\textcolor{likeCyan}{cyan}). Shown are also the Planck~TT-only results for varying $\Nn$~(\textcolor{likeIndigo}{indigo}) and when~$\Nn$ is fixed~(\textcolor{likeGray}{gray}). In both panels, the lensing reconstruction and low-P~likelihoods were used for all of the constraints.} 
	\label{fig:NeffNfluid}
\end{figure}

The Planck 2015 results have reached an important threshold. The level of sensitivity is now sufficient to detect the free-streaming nature of the neutrinos (or any additional dark radiation). In Table~\ref{tab:Planck15}, we present marginalized constraints on~$\Nf$ and~$\Nn$, with their posterior distributions shown in Fig.~\ref{fig:planckPosteriors}. We show results both for the combined TT, TE and EE~likelihoods, and for TT~alone. The two-dimensional joint constraints are presented in Fig.~\ref{fig:NeffNfluid}. In each case, we compare the results for fixed $Y_p$ with those when~$Y_p$ is allowed to vary. These results robustly demonstrate that there is very little degeneracy between $\Nf$, $\Nn$ and $Y_p$ when using both temperature and polarization data from Planck.
\begin{table}[h!t]
	\begin{tabular}{c cc cc}
			\toprule
				& \multicolumn{2}{c}{TT, TE, EE}		 				& \multicolumn{2}{c}{TT-only}								\\
			\cmidrule(lr){2-3}\cmidrule(lr){4-5}
				& varying $Y_p$ 			& fixed $Y_p$ 				& varying $Y_p$				& fixed $Y_p$					\\
			\midrule
		$\Nf$	& $2.68^{+0.29}_{-0.33}$	& $2.80^{+0.24}_{-0.23}$ 	& $2.89^{+0.49}_{-0.62}$	& $2.87^{+0.45}_{-0.37}$		\\ 
		$\Nn$	& $< 0.64$					& $< 0.67$ 	 				& $< 1.08$					& $< 0.94$						\\
			\bottomrule
	\end{tabular}
	\caption{Best-fit values and $1\sigma$~errors for~$\Nf$ and $2\sigma$~upper limits for~$\Nn$ for the Planck~2015 data. Both~$\Nn$ and~$\Nf$ are allowed to vary in all cases. The lensing reconstruction and low-P~likelihoods were used for all of the constraints.}
	\label{tab:Planck15}
\end{table}

From the left panels in Figs.~\ref{fig:planckPosteriors} and~\ref{fig:NeffNfluid} we see that the constraints on~$\Nf$ are largely insensitive to the marginalization over~$\Nn$ and only mildly affected by the marginalization over~$Y_p$, even when the polarization data is removed. This robustness of the constraints suggests that they receive considerable constraining power from the phase shift because~$\Nf$ is degenerate with~$Y_p$ and~$\Nn$ in the damping tail. Since~$\Nf$ is the unique parameter capable of producing the phase shift, the measurement of the latter breaks the degeneracy between~$\Nf$ and both~$\Nn$ and~$Y_p$. We also see that adding polarization data leads to a large improvement in the constraint on~$\Nf$, most likely because the peaks of the E-mode spectrum are sharper, which makes the phase shift easier to measure~\cite{Bashinsky:2003tk}. This is illustrated in Fig.~\ref{fig:CMBphase}, where we show the relative differences in the TT and EE~spectra when varying~$\Nf$ and~$\Nn$. While the phase shift is visible in both cases, the size of the effect is larger in the polarization spectrum which increases the impact of the E-mode data.

Similar analyses were performed in~\cite{Bell:2005dr,Friedland:2007vv} using WMAP data (and external datasets). Their results are qualitatively similar to our TT-only analysis with~$Y_p$ fixed, although with weaker constraints on~$\Nf$ and~$\Nn$. By comparison, adding E-mode data further improves constraints in the \mbox{$\Nf$\hskip1pt-$\Nn$}~plane, also when~$Y_p$ is allowed to vary.

Figure~\ref{fig:NeffNfluid} shows that there is a significant difference in the constraints in the $\Nf$\hskip1pt-$Y_p$ plane, with and without the polarization data.%
 Without polarization, allowing $\Nn$ to vary weakens the constraints on $\Nf$ and $Y_p$. This is consistent with our discussion of degeneracies in~\textsection\ref{sec:species_diffusionDamping} and at the beginning of this section. Using only the temperature data, $Y_p$ and $\Nn$ are measured mostly from the damping tail, but their effects on the latter are degenerate. As a result, the constraints on $Y_p$ and $\Nf$ weaken when $\Nn$ is allowed to vary. The situation changes when polarization data is added. Now, there is very little difference in the constraints as we vary the marginalization over additional parameters. This is most noticeable in the right panel in Fig.~\ref{fig:NeffNfluid}, where the constraints on $Y_p$ and $\Nf$ become nearly independent of the treatment of $\Nn$. This feature was anticipated above, where we observed that polarization breaks the degeneracy between $Y_p$ and $\Nn$.\medskip

In~\cite{Follin:2015hya}, a constraint was recently placed on the effective number of free-streaming species by isolating the phase shift in the Planck 2013 temperature data: $\Nf = 2.3^{+1.1}_{-0.4}$ (\SI{68}{\percent}~c.l.)\ while keeping the damping tail fixed and $\Nf = 3.5\pm0.65$ (\SI{68}{\percent}~c.l.)\ when marginalizing over the effect on the damping tail~\cite{FollinPC}. To compare to that analysis, we remove the polarization data. We then find $\Nf =2.89^{+0.49}_{-0.62}$\hskip1pt, which is quite similar to the direct measurement of the phase shift.\footnote{Due to the marginalization over~$\Nn$ and~$Y_p$, our analysis is most comparable to the marginalized result: $\Nf = 3.5\pm0.65$~\cite{FollinPC}. One difference in our approach is that it includes information in the amplitude shift produced by the free-streaming species. However, this effect is likely sub-dominant to the phase shift due to the degeneracy with the amplitude of the primordial power spectrum. As we will discuss in the next subsection, when we allow~$Y_p$ to vary, future CMB~experiments get a considerable fraction of the sensitivity to~$\Nf$ from the phase shift and so we expect our methods to produce increasingly similar results.} When we add the TE and EE likelihoods, our constraint improves by about a factor of two to $\Nf = 2.68^{+0.29}_{-0.33}$. From the estimate\hskip1pt\footnote{We can also estimate this more schematically from the knowledge that $\Nf=3.046$ produces $\delta \ell \approx 10$ for $\ell \lesssim 3000$ relative to $\Nf = 1$~\cite{Follin:2015hya}. Current constraints allow for roughly a 10 percent variation in~$\Nf$, which would imply $\delta \ell \approx 1$. Direct measurements of the individual peak locations are given in~\cite{Aghanim:2015xee} with a similar level of precision.} in~\eqref{eq:delta-ell}, we conclude that these constraints correspond to a phase shift of about $\delta \ell \approx 1$. While this is compatible with expectations,\footnote{Forecasts using the isolated phase shift alone give $\sigma(\Nf) = 0.41$ for Planck with polarization~\cite{Follin:2015hya}.} it is nonetheless impressive that the data is sensitive to these small and subtle effects.\medskip

\subsection{CMB Stage-4 Forecasts}
While current data is already sensitive to the free-streaming nature of neutrinos, future experiments are expected to improve these constraints by at least an order of magnitude. As we have been emphasizing, an increase in the sensitivity to the $\sigma(\Nf) \sim \num{e-2}$ level probes a number of plausible BSM~scenarios that are currently unconstrained. In the following, we forecast not only the constraints on~$\Nf$, but also on~$Y_p$ and~$\Nn$ in order to identify more clearly the types of BSM~physics we might be sensitive to. We perform a full likelihood analysis (rather than a Fisher forecast) to ensure that degeneracies are treated correctly (see~\cite{Perotto:2006rj} for a discussion).\medskip

\begin{figure}[t]
	\definecolor{like2Blue}{RGB}{0, 111, 237}
	\definecolor{like2Orange}{RGB}{255, 140, 0}
	\definecolor{like2Purple}{RGB}{138, 43, 226}
	\includegraphics{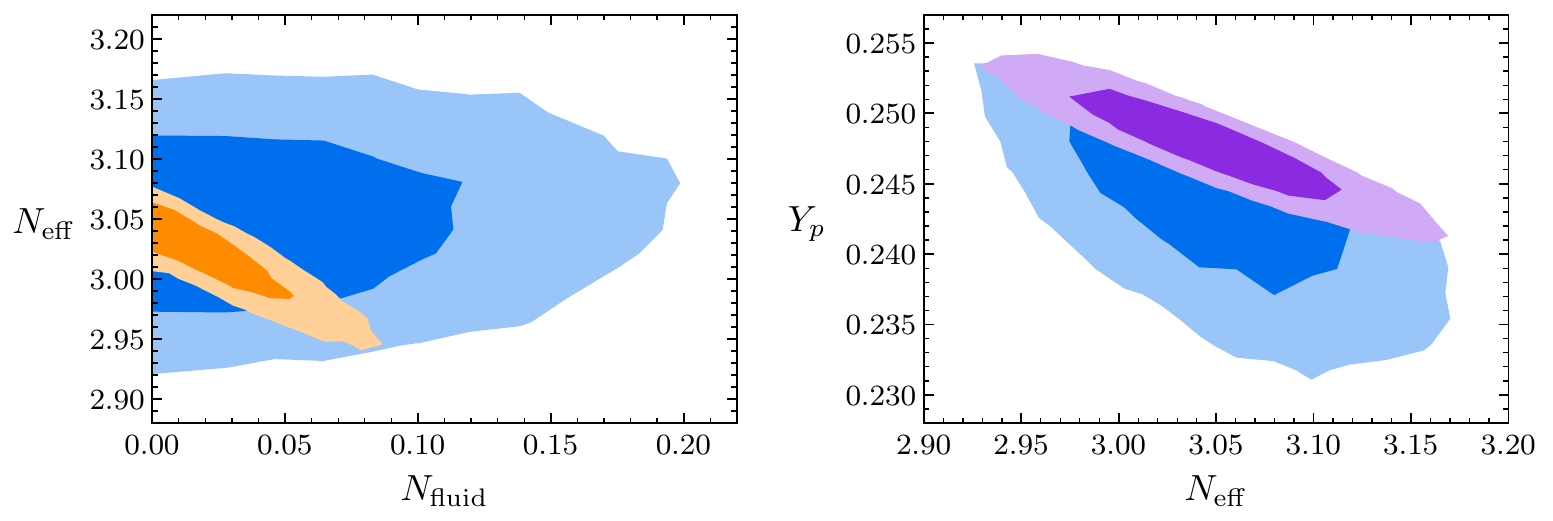}
	\caption{\textit{Left}: Forecasted CMB-S4 constraints on $\Nf$ and $\Nn$ for~$Y_p$ fixed~(\textcolor{like2Orange}{orange}) and varying~(\textcolor{like2Blue}{blue}). \textit{Right}:~Forecasted CMB-S4~constraints on $\Nf$ and~$Y_p$ for $\Nn = 0$ fixed~(\textcolor{like2Purple}{purple}) and varying~(\textcolor{like2Blue}{blue}).} 
	\label{fig:NnNf}
\end{figure}

The results of our forecasts are summarized in Table~\ref{tab:forecasts}.%
\begin{table}[t!]
	\begin{tabular}{l cc ccc} 
		\toprule
		Experiment		& Delensing	& Reconstruction	& $\sigma(\Nf)$ 	& $\sigma(Y_p)$   	& $\Nn$		\\[0.5ex] 
		\midrule
		Planck 2015 	& No 		& Yes				& 0.31\phantom{0}	& 0.019\phantom{0} 	& $< 0.64$	\\ 
		\cmidrule{1-6}
						& No 		& No				& 0.062 	   		& 0.0053			& $< 0.18$	\\ 
		CMB-S4   		& Yes		& No				& 0.054 	   		& 0.0044  			& $< 0.17$	\\
						& Yes		& Yes				& 0.050 	   		& 0.0043  			& $< 0.16$	\\
		\bottomrule
	\end{tabular}
	\caption{Marginalized likelihoods for Planck and CMB-S4. All likelihoods allow $\Nf$, $\Nn$ and~$Y_p$ to vary. The Planck likelihood uses the TT, TE and EE~power spectra as well as the lensing reconstruction likelihood. The forecasts for CMB-S4 include the possibilities of delensing and lensing reconstruction. We excluded the forecast using lensed spectra combined with lensing reconstruction since this is known to produce overly optimistic error forecasts due to a double counting of lensing information~\cite{Hu:2001fb}. The double counting can be safely ignored for Planck, but will become more important for future experiments~\cite{Schmittfull:2013uea}. Displayed are $1\sigma$~error bars for~$\Nf$ and~$Y_p$, and $2\sigma$~upper limits for~$\Nn$.}
	\label{tab:forecasts}
\end{table}
Given the constraints from Planck, it is not surprising that~$\Nf$ is easily distinguished from~$\Nn$ with CMB-S4~experiments. As before, the constraints on~$\Nf$ are significantly stronger than those on~$\Nn$, which is consistent with the interpretation that these parameters are being distinguished by differences in the perturbations for the two types of radiation. When both radiation components are included, the detailed matching of the acoustic peaks is very sensitive to the phase shifts due to~$\Nf$, but is much less affected by~$\Nn$. This is illustrated by the left panel in Fig.~\ref{fig:NnNf},%
which presents the joint constraints on~$\Nn$ and~$\Nf$, both for fixed~$Y_p$ and when it is allowed to vary. When~$Y_p$ is fixed, there is a strong degeneracy between~$\Nf$ and~$\Nn$ which is absent when~$Y_p$ is allowed to vary. Yet, in both cases, the contours close at roughly the same value of~$\Nf$, indicating that the degeneracy is broken in a way that is insensitive to the damping tail.

As we have emphasized throughout, the strong constraint on~$\Nf$ arises in part from the sharpness of the peaks of the E-mode spectrum, which leads to better measurements of the phase shift. This same intuition explains why the constraints on~$\Nf$ are strengthened by using the delensed power spectrum. One of the well-known effects of lensing is a smearing of the acoustic peaks~\cite{Seljak:1995ve} (see Fig.~\ref{fig:CMBphase}). By delensing the power spectra, we sharpen the peaks, which makes the measurements of the phase shift more precise. We illustrate the impact that delensing can have in Fig.~\ref{fig:CMBphase}, where we see that the effect of changing~$\Nf$ produces a much larger relative change on the unlensed data. This procedure is analogous to reconstructing the BAO~peak to sharpen distance measurements~\cite{Eisenstein:2006nk}. In both cases, non-linearities transfer information from the power spectrum to higher-point correlation functions. Delensing or BAO~reconstruction moves that information back to the power spectrum, so that it is more easily accounted for in these analyses. As a result, the error in the measurement of any quantity that is sensitive to the sharpness of the peak will be reduced (like the phase shift and the BAO~scale). This suggests that delensing will be a useful tool for improving constraints on cosmological parameters in these future experiments.\medskip

While much of our efforts have been devoted to understanding the degeneracies between~$\Nf$, $\Nn$ and~$Y_p$, the actual physical models we wish to constrain may not exhibit these degeneracies. First of all, many models with additional light fields still have $\Nn = 0$ (i.e.~only free-streaming radiation)~\cite{Brust:2013xpv}, which would in principle allow us to combine information from the damping tail and the phase shift to constrain~$\Nf$. Furthermore, while~$Y_p$ is often affected by BSM~physics at the time of BBN, the precise degeneracy needed to keep the damping tail fixed is unlikely to occur naturally. The constraints on such models would be significantly stronger. As we see in Table~\ref{tab:forecasts2}, a factor of 3 to 4~improvement in the constraints is possible when this degeneracy with changes in~$Y_p$ is not present.
\begin{table}[b!]
	\begin{tabular}{l S[table-format=1.3] S[table-format=1.4]} 
			\toprule
		Experiment						& {$\sigma(\Nf)$}	& {$\sigma(Y_p)$}	\\[0.5ex] 
			\midrule
		\multirow{2}{*}{Planck 2015} 	& 0.30				& 0.018			 	\\
										& 0.19				& {--}  			\\
			\cmidrule{1-3}
		\multirow{2}{*}{CMB-S4}			& 0.048 			& 0.0027			\\
										& 0.013				& {--}  	  		\\
			\bottomrule
	\end{tabular}
	\caption{Results for the marginalized $1\sigma$~errors for Planck (TT, TE, EE, lensing reconstruction) and forecasts for~CMB-S4 for~$\Nf$ and~$Y_p$, when $\Nn = 0$ is held fixed. The CMB-S4~forecasts assumed both delensing and lensing reconstruction. A dash in the $\sigma(Y_p)$~entry indicates that $Y_p$ was fixed by consistency with~BBN.}
	\label{tab:forecasts2}
\end{table}
\medskip

Finally, perhaps the most important feature of these forecasts is the dramatic improvement, relative to Planck~2015, that can be expected from the studied experimental configuration of \mbox{CMB-S4}. Our projections suggest that a factor of~5 to~10 improvements are achievable, but we should also investigate the robustness of this conclusion to changes of the experimental specifications. Here, we study variations of the beam size~($\theta_b$) and the maximal available multipole~($\ell_\mathrm{max}$). These are important for two reasons: \mbox{(\hskip-0.5pt\textit{i}\hskip1pt)}~the beam size is ultimately a choice made within the context of limited resources and \mbox{(\hskip-0.5pt\textit{ii}\hskip1pt)}~the presence of foregrounds or systematics make it difficult to predict~$\ell_\mathrm{max}$ reliably beforehand. In Table~\ref{tab:beam}%
\begin{table}[t!]
	\begin{tabular}{l ccc cc} 
		\toprule
		Parameter 				  					& $1'$ 				& $2'$ 				& $3'$ 				& $\ell_\mathrm{max} = 3000$	& $\ell_\mathrm{max} = 4000$	\\[0.5ex]
		\midrule
		$\sigma(\Nf)$ ($Y_p$ fixed, $\Nn =0$)		& 0.013 			& 0.015 			& 0.016				& 0.023 	 	 				& 0.015 						\\
		$\sigma(\Nf)$ ($Y_p$ fixed, $\Nn \ne 0$)	& 0.026 			& 0.027 			& 0.029				& 0.034 	 	 				& 0.028 						\\
		$\sigma(\Nf)$ ($Y_p$ varying, $\Nn =0$)		& 0.048 			& 0.051 			& 0.055				& 0.058 	 	 				& 0.052 						\\
		$\sigma(\Nf)$ ($Y_p$ varying, $\Nn \ne 0$)	& 0.050 			& 0.052 			& 0.055				& 0.061 	 	 				& 0.051 						\\
		\cmidrule{1-6}
		$\Nn$ ($Y_p$ varying)		  				& $<0.16\phantom{0}$& $<0.17\phantom{0}$& $<0.18\phantom{0}$& $<0.20\phantom{0}$			& $<0.17\phantom{0}$  			\\
		$\Nn$ ($Y_p$ fixed) 		  				& $<0.068$			& $<0.072$			& $<0.076$			& $<0.090$						& $<0.072$ 		 				\\
		\bottomrule
	\end{tabular}
	\caption{Forecasts for a CMB-S4~experiment with varying beam size and maximum multipole assuming \num{e6}~detectors. We take $\ell_\mathrm{max} = 5000$ as we vary the beam size and $\theta_b = \SI{1}{\arcmin}$ when we vary~$\ell_\mathrm{max}$. Displayed are $1\sigma$~error bars for~$\Nf$ and $2\sigma$~upper limits for~$\Nn$.}
	\label{tab:beam}
\end{table}
we show the forecasts for various values of~$\theta_b$ and~$\ell_\mathrm{max}$, assuming \num{e6}~detectors. 

There are two simple lessons we can draw from this table. First of all, measuring modes with $\ell > 3000$ offers only a moderate improvement on our constraints. Similarly, the benefit to reducing the beam size significantly is also limited. Most of the improvement in the measurement of the phase shift is coming from high-precision measurements of E-modes with $\ell < 3000$. In contrast, when $Y_p$ is fixed and $\Nn=0$, the sensitivity to the beam size and $\ell_\mathrm{max}$ is stronger, which suggests\hskip1pt\footnote{We have not been careful to account for foregrounds in temperature or polarization. Information in our forecasts coming from high-$\ell$ temperature data is unlikely to be available in a real experiment.} that we are gaining useful information from the damping tail at $\ell > 3000$.

We have fixed the number of detectors to be~\num{e6}, but the precise value will play a very important role in the ultimate reach of~CMB-S4, not just for~$\Nf$, but for most physics targets. For~$\Nf$ specifically, we may still improve the constraints until we reach the limit set by cosmic variance\hskip1pt\footnote{The cosmic variance-limited constraint is roughly given by $\sigma(\Nf) \sim 0.008$ (cf.\ Appendix~\ref{app:bao-forecast_appendices}).} in the E-mode power spectrum since modes with $\ell \gtrsim 3700$ ($\ell \gtrsim 2300$) are still dominated by noise for \num{e6}~(\num{e4})~detectors. As a result, within the range of~\num{e4} to \num{e6}~detectors being considered, we improve constraints significantly by further reducing the detector noise and, hence, increasing the number of detectors. Forecasts for~$\Nf$ with varying numbers of detectors were studied in~\cite{Wu:2014hta, Abazajian:2016yjj}, which confirm this intuition.

\section{Consequences for BBN and BSM}
\label{sec:consequences}
With the substantially improved constraints on $\Nf$, $\Nn$ and~$Y_p$ expected from future CMB~experiments, it is interesting to examine the possible impacts that these measurements might have on our understanding of the laws of Nature. In the following, we consider some consequences of combining these CMB~observations with those of big bang nucleosynthesis~(\textsection\ref{sec:timeEvol}) and discuss a few opportunities to probe physics beyond the Standard Model~(\textsection\ref{sec:bsmImplications}).

\subsection{Interplay Between CMB and BBN}
\label{sec:timeEvol}
One of the important benefits of measuring the fluctuations in the dark radiation is that it eliminates the degeneracy between~$Y_p$ and~$\Nf$. This has important consequences for constraints on BSM~physics, since~$Y_p$ is sensitive to the total radiation density at the time of~BBN (\SI{3}{\minute}~after the big bang), while~$\Nf$ and~$\Nn$ are related to these radiation densities around the time of recombination (\num{373000}~years after the big bang). As a result, CMB~measurements of~$Y_p$ and~$\Nf$ probe scenarios where these densities change between those two times, perhaps through the decay of a heavy particle or through some other production mechanism~\cite{Fischler:2010xz}.\medskip

It is useful to translate constraints on~$Y_p$ into bounds on the radiation density at the time of~BBN~\cite{Bernstein:1988ad},
\beq
Y_p \approx 0.247 + 0.014 \times \Delta N_\mathrm{eff+fluid}^\mathrm{BBN}\, ,	\label{eq:YpBBN}
\eeq
where $\Delta N_\mathrm{eff+fluid}^\mathrm{BBN} \equiv \Nf^\mathrm{BBN}+\Nn^\mathrm{BBN} - 3.046$. Marginalizing over~$\Nn$, we found a standard deviation in the helium fraction of $\sigma(Y_p) =0.0043$ for~CMB-S4 (see Table~\ref{tab:forecasts}), while for fixed $\Nn=0$, we get $\sigma(Y_p) = 0.0027$ (cf.\ Table~\ref{tab:forecasts2}). Using~\eqref{eq:YpBBN}, these constraints imply
\beq
\sigma(N_\mathrm{eff+fluid}^\mathrm{BBN}) = 
	\begin{dcases}
		\,0.31	& \quad \Nn \ne 0\, , \\
		\,0.19	& \quad \Nn = 0\, .
	\end{dcases}
\eeq
The last constraint is stronger than the current limit of $\sigma(N_\mathrm{eff+fluid}^\mathrm{BBN}) = 0.28$ from~BBN alone~\cite{Cyburt:2015mya}. The~CMB will therefore provide independent measurements of~$\Nf$ at two different times in a single experiment, each surpassing our current level of sensitivities from combining multiple probes.\medskip

While the constraint on $N_\mathrm{eff+fluid}^\mathrm{BBN}$ from a CMB-S4 experiment is only a modest improvement over current measurements from primordial abundances, it has the unique advantage that it is a clean measurement (i.e.~it is not affected by astrophysical processes at later times) and it can be combined with measurements of other cosmological parameters (e.g.~$\Omega_b h^2$) without combining different datasets. A common approach with current data is to combine the constraints from the CMB and primordial abundances in order to improve the overall sensitivity to $\Nf$, in the case where it is time independent. From Table~\ref{tab:forecasts2}, we see that if we do not allow a variation in $\Nf$ between BBN and the CMB, we get very strong constraints on $\Nf$ due to the lack of degeneracies in the damping tail. Given that these results are sufficiently strong, it is unlikely that including information from primordial abundances will lead to much improvement.

When discussing the time variation of $\Nf$, we have focussed only on the model-independent measurement implied by varying $\Nf$ and $Y_p$ independently. As a result, the constraints we derive are controlled primarily by the degeneracy between $\Nf$ and $Y_p$ in the damping tail. Without this degeneracy, the constraints on $\Nf$ are much stronger. In realistic models, it may be the case that both~$\Nf$ and~$Y_p$ are changed independently, but that they do not produce this degeneracy in the relevant range of parameters. For such models, a dedicated analysis of CMB data would likely offer a much larger gain over the current limits from primordial abundances.

\subsection{Implications for BSM Physics}
\label{sec:bsmImplications}
So far, we have concentrated on simple descriptions of BSM~physics in terms of the effective parameters $\Nf$, $\Nn$ and $Y_p$. These parameters capture important aspects of CMB physics and our cosmological history, which may be used to test a number of scenarios for BSM~physics (see Table~\ref{tab:bsmSummary}):%
\begin{table}[t!]
	\begin{tabular}{l ll}
	 	\toprule
	 \textit{Signature}	& \textit{Influenced by}																				& \textit{Degeneracies broken by}	\\
		\midrule\addlinespace
	 CMB damping tail		& $\Nf^\mathrm{CMB}$+$\Nn^\mathrm{CMB}$, $Y_p$, $E_\mathrm{inj}^\mathrm{CMB}$						& Phase shift, polarization			\\
	 Phase shift  			& $\Nf$, $N_\mathrm{fluid}^\mathrm{iso}$, $\cancel{\mathrm{GR}}$									& Scale dependence 					\\
	 Spectral distortions 	& $E_\mathrm{inj}^{\text{post-BBN}}$																&		 							\\
	 Primordial abundances & $\Nf^{\text{BBN}}+\Nn^\mathrm{BBN}$, $N_\nu$, $\eta^\mathrm{BBN}$, $E_\mathrm{inj}^\mathrm{BBN}$	& CMB								\\
		\addlinespace\bottomrule
	\end{tabular}
	\caption{Cosmological probes of BSM~physics and their sensitivity to free-streaming and non-free-streaming radiation ($\Nf$ and $\Nn$), the number of active neutrinos ($N_\nu$), the baryon-to-photon ratio~($\eta$) and the amount of energy injection ($E_\mathrm{inj}$). The superscripts BBN, CMB, or post-BBN denote the time at which a quantity is being probed, where post-BBN refers to redshifts of $z \lesssim \num{e6}$ when spectral distortions become possible. The parameter $\Nn^\mathrm{iso}$ abstractly stands for isocurvature fluctuations, while $\cancel{\mathrm{GR}}$ denotes modified gravity.}
	\label{tab:bsmSummary}
\end{table}

\begin{itemize}\itemsep0pt
\item As we discussed in~\textsection\ref{sec:thermalHistoryAdditional}, $\Nf$~is sensitive to the freeze-out of the particle for minimal extensions of the Standard Model with a light field. At current levels of sensitivity, $\sigma(\Nf) \gtrsim 0.1$, we can rule out some scenarios where particles freeze-out after the QCD~phase transition~\cite{Brust:2013xpv}. Freeze-out before the QCD~phase transition typically dilutes the contribution to~$\Nf$ by a factor of~10, which allows such models to easily evade current constraints. Fortunately, some of these scenarios are likely to be accessible with CMB-S4 experiments~\cite{Abazajian:2016yjj}. For these cases, we are sensitive to sufficiently early times so that BSM~physics above the \si{\tera\electronvolt}~scale may be important and can be probed along the lines of Chapter~\ref{chap:cmb-axions}.

\item Measurements of the effective number of free-streaming particles at recombination, $\Nf^\mathrm{CMB}$, are also sensitive to any energy which is injected into the Standard Model particles after the time of neutrino decoupling. Depending on the time and nature of this energy injection, it may alter the primordial abundances or introduce spectral distortions which would distinguish it from a new light field. For example, a decay to photons after~BBN would lower~$\Nf^\mathrm{CMB}$ and $\eta^\mathrm{BBN}$ (the baryon-to-photon ratio at~BBN), while keeping the radiation density at~BBN,~$\Nf^\mathrm{BBN}$, fixed~\cite{Cadamuro:2010cz}. 

\item Energy injection of many kinds is a typical byproduct of changing~$\Nf$, but may also be the dominant signature of BSM~physics. Decays during~BBN can disrupt the formation of nuclei without substantially changing the total energy in radiation. Alternatively, recombination is very sensitive to energy injection~\cite{Padmanabhan:2005es} which can alter the form of the visibility function.\footnote{The common element of both of these examples is that the tail of the Boltzmann distribution is playing a critical role (due to the large value of $\eta^{-1}$). As a result, the change to the small number of high-energy photons is more important than the total energy density.} 

\item As we discussed in Section~\ref{sec:analytics}, phase shifts of the acoustic peaks may also be produced by isocurvature perturbations (denoted by $\Nn^\mathrm{iso}$ in Table~\ref{tab:bsmSummary}). We offered a simple curvaton-like example of this effect, but we expect to be broadly sensitive to physics in the dark sector that is not purely adiabatic. Since there are many good reasons to imagine why isocurvature perturbations might arise in the dark sector, this motivates a future exploration of the observability of these effects. 

\item Finally, we have assumed the validity of the Einstein equations throughout. This enforced that $\Phi_- = 0$ in the absence of anisotropic stress. Modified theories of gravity are often parameterized in terms of their change to the Einstein constraint equation and the corresponding effect on $\Phi_-$; see e.g.~\cite{Zhang:2007nk}. Since the field $\Phi_-$ played an important role in our analysis of the phase shift, it would be interesting to explore how the result changes for specific modifications of GR. Conversely, the phase shift of the CMB spectrum may be an interesting probe of modified gravity.
\end{itemize}

\section{Summary}
\label{sec:conclusions_cmb-phases}
CMB observations have become precise enough to probe the {gravitational} imprints of BSM~physics on the perturbations of the primordial plasma. In the upcoming era of CMB polarization experiments, our sensitivity to these subtle effects will increase significantly and will offer new opportunities in the search for new physics. It is therefore timely to re-evaluate how CMB data can inform our view of the laws of physics.\medskip

In this chapter, we have explored how the phase shift of the acoustic peaks might be used as such a probe. This phase shift is particularly interesting because analytic properties of the Green's function of the gravitational potential strongly limit the possible origins of such a shift to
\begin{center}
\begin{tabular}{r l c l}
\textit{i.}		& waves propagating faster than the sound speed of the photon-baryon fluid,	\\[4pt]
\textit{ii.}	& isocurvature fluctuations.
\end{tabular}
\end{center}
\noindent
For adiabatic initial conditions, the phase shift is most easily generated by free-streaming radiation and becomes an excellent measure of the effective number of free-streaming relativistic species,~$\Nf$, at the time of recombination. Realistic models of isocurvature fluctuations typically produce a scale-dependent phase shift, which allows them to be distinguished from changes to the energy density of the radiation.\medskip

\begin{figure}[t]
	\definecolor{likeRed}{RGB}{224, 52, 36}
	\definecolor{likeGreen}{RGB}{34, 139, 34}
	\definecolor{likeIndigo}{RGB}{75, 0, 130}
	\definecolor{likeOlive}{RGB}{128, 128, 0}
	\definecolor{likeCyan}{RGB}{0, 139, 139}
	\definecolor{likeGray}{RGB}{112, 128, 144}
	\includegraphics{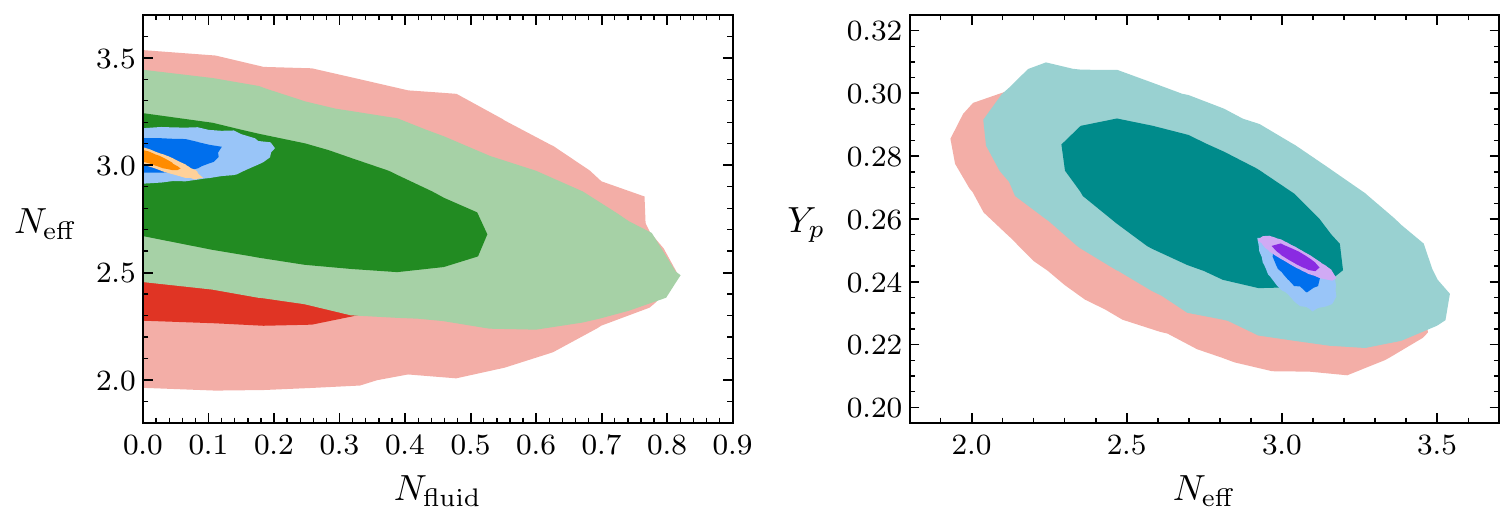}
	\caption{\textit{Left:} Planck constraints on the effective number of free-streaming and non-free-streaming relativistic species, $\Nf$ and~$\Nn$, allowing the helium fraction~$Y_p$ to vary~(\textcolor{likeRed}{red}) and keeping it fixed~(\textcolor{likeGreen}{green}). \textit{Right:}~Planck constraints on $Y_p$ and~$\Nf$ for varying $\Nn$~(\textcolor{likeRed}{red}) and when $\Nn=0$ is fixed~(\textcolor{likeCyan}{cyan}). In both plots, the contours from the CMB-S4~forecasts, presented in Fig.~\ref{fig:NnNf}, have been included to show the expected improvements in sensitivity.}
	\label{fig:cmb-phases_summary}
\end{figure}

What makes these results particularly compelling is that current and future CMB experiments are sensitive enough to detect these phase shifts at high significance~\cite{Follin:2015hya}. We have demonstrated this with an analysis of the 2015 data from the Planck satellite and forecasts for a CMB Stage-4 experiment (see Fig.~\ref{fig:cmb-phases_summary}).%
Our results highlight the important role played by the polarization data in breaking the degeneracy between the contributions from free-streaming and non-free-streaming species, $\Nf$ and $\Nn$, as well as that with the helium fraction $Y_p$. We also provide a clear detection of the free-streaming nature of neutrinos which constitutes further evidence for the cosmic neutrino background.
	\chapter{Searching for Light Relics with LSS} 
\label{chap:bao-forecast}
We have established in the previous chapter that the cosmic microwave background is a sensitive probe of light thermal relics. The forecasts for a CMB~Stage-4 experiment indicate that the possible bounds on~$\Neff$ are tantalizingly close to well-motivated theoretical targets, in particular the threshold value $\Delta\Neff=0.027$ associated with the minimal abundance of any light thermal relic. Since the initial conditions for the clustering of matter are set by the same physics that leads to the acoustic oscillations in the~CMB, we expect to find imprints of relativistic species also in the large-scale structure of the universe. In this chapter, we therefore explore to what degree the CMB~observations can be enhanced by future LSS~surveys. We carefully isolate the information encoded in the shape of the galaxy power spectrum and in the spectrum of baryon acoustic oscillations. We find that measurements of the shape of the power spectrum can significantly improve on current and near-term CMB~experiments. We also propose a modified analysis of BAO~data and show that the phase shift induced by relic neutrinos in the BAO~spectrum can be detected at high significance in future experiments.\medskip

The outline of this chapter is as follows. In Section~\ref{sec:bao-forecast_introduction}, we recapitulate the effects of free-streaming radiation on the matter power spectrum as well as the baryon acoustic oscillation signal, focussing especially on the phase shift in the BAO~spectrum. In Section~\ref{sec:forecast}, we forecast combined CMB and LSS~constraints on the number of relativistic species,~$\Neff$, for a range of future observations. In Section~\ref{sec:phaseShift}, we isolate the information encoded in the phase shift of the BAO~spectrum and study the prospects for extracting this effect in upcoming surveys. A summary is presented in Section~\ref{sec:conclusions_bao-forecast}. We refer to Appendices~\ref{app:bao-forecast_appendices} and~\ref{app:broadband+phaseShiftExtraction} for technical details of our analysis, such as the experimental specifications of the surveys, and the methods used to extract the matter broadband spectrum and the phase shift.

\section{Phases of New Physics in the BAO Spectrum}
\label{sec:bao-forecast_introduction}
Having seen that the next generation of CMB~experiments will be very sensitive to the radiation density, we will now explore the additional constraining power provided by current and future LSS~experiments, such as {(e)BOSS}~\cite{Dawson:2012va, Dawson:2015wdb}, DES~\cite{Abbott:2005bi}, DESI~\cite{Aghamousa:2016zmz}, LSST~\cite{Ivezic:2008fe} and Euclid~\cite{Laureijs:2011gra}. It was established in~\cite{Font-Ribera:2013rwa, Dodelson:2016wal, Obuljen:2017jiy} that these surveys carry information about relativistic species. We will examine how this information is stored in both the power spectrum shape and the BAO~spectrum.\medskip

\begin{figure}[h]
	\includegraphics{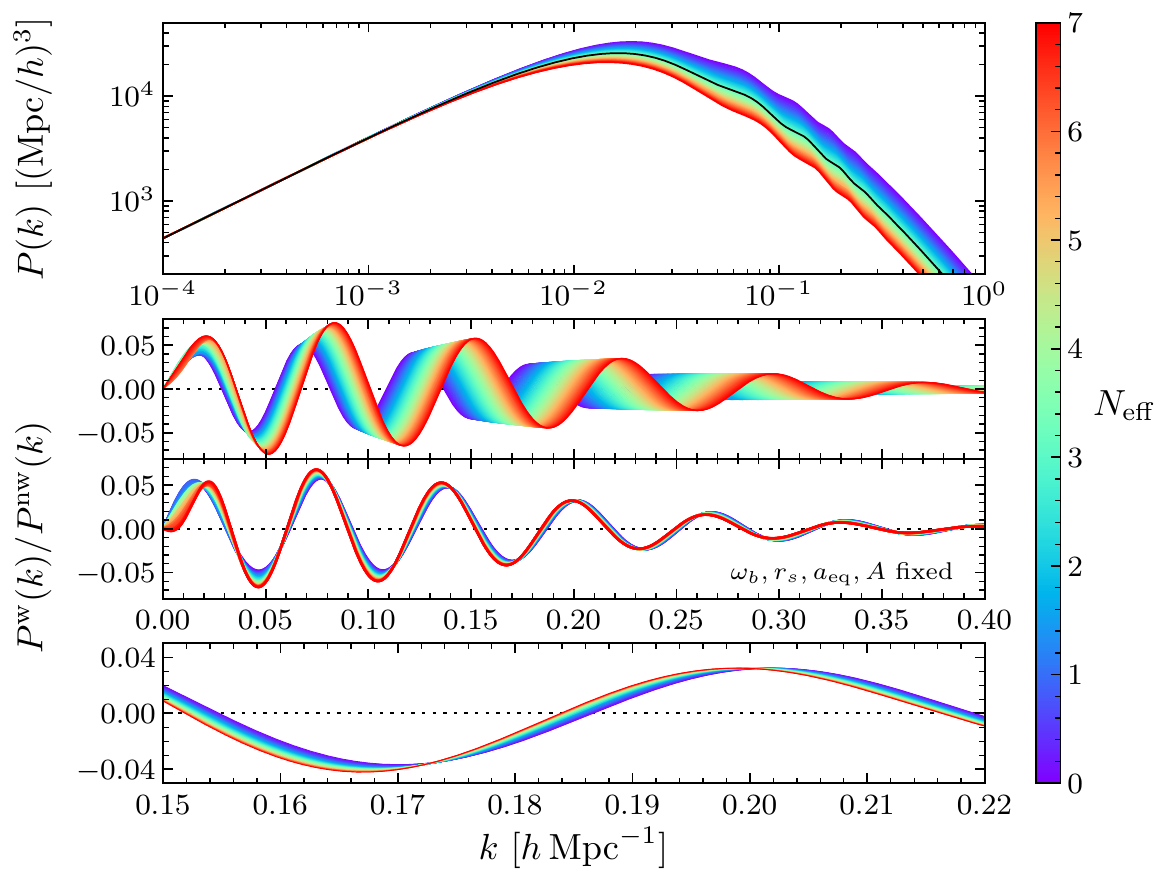}
	\caption{Variation of the matter power spectrum~$P(k)$~(\textit{top}) and the BAO~spectrum~$\Pw(k)/\Pnw(k)$~(\textit{bottom}) as a function of $\Neff$. The physical baryon density,~$\omega_b$, the physical sound horizon at the drag epoch,~$r_s$, the scale factor at matter-radiation equality, $\aeq \equiv \omega_m/\omega_r$, and the BAO~amplitude~$A$ at the fourth peak are held fixed in the second BAO~panel (as in Fig.~\ref{fig:BAOPhaseShift}). This panel and the bottom zoom-in show the remaining phase shift induced by free-streaming relativistic species.}
	\label{fig:BAOPhaseShift2}
\end{figure}
In~\textsection\ref{sec:species_equality}, we discussed the main effects of the radiation density in the early universe on the growth of structure. In brief, increasing the radiation density moves the maximum of the power spectrum to larger scales and the amplitude on small scales gets suppressed (cf.\ Fig.~\ref{fig:BAOPhaseShift2}). On the other hand, a change in the size of the sound horizon (and therefore in the BAO~frequency) has the dominant impact on the BAO~spectrum. The phase shift is clearly sub-dominant, but a distinct feature of free-streaming particles (see~\textsection\ref{sec:species_phaseShift}). Currently, of all these effects, only the change in the BAO~frequency is taken into account when analysing LSS~data. In principle, however, there is much more additional information available that can be used to improve the constraints and learn about the physics of the early universe. Although its accessibility is hampered by non-linear effects, we will demonstrate that some of these clues can still be robustly extracted from LSS~observables.\medskip

We will pay particular attention to the information about~$\Neff$ contained in the BAO~spectrum and propose a new analysis of this observable. To isolate the BAO~signal, we split the power spectrum into a smooth (`no-wiggle') part and an oscillatory (`wiggle') part,
\beq
P(k) \equiv \Pnw(k) + \Pw(k)\, ,	\label{eq:Pkdecomposition}
\eeq 
with our method for performing this separation being described in Appendix~\ref{app:broadband+phaseShiftExtraction}. We will demonstrate that the most robust information about~$\Neff$ lives in~$\Pw(k)$. The peak locations of the BAO~spectrum carry additional information about light relics that is immune to corrections to the overall shape of the power spectrum, such as those arising from non-linear gravitational evolution~\cite{Eisenstein:2006nj, Crocce:2007dt, Sugiyama:2013gza}. The reason underlying this is that the phase of the BAO~spectrum is unaffected by the effects of non-linear gravitational evolution~\cite{Baumann:2017lmt}. We claim that this information is preserved after non-linear corrections are taken into account and explore in detail how it can be isolated in the BAO~spectrum. This protected information may play a useful role in elucidating apparent discrepancies between~CMB and low-redshift measurements, and be a valuable tool in the search for exotic physics in the dark sector.

\section{Current and Future Constraints}
\label{sec:forecast}
We have been arguing throughout this thesis that measuring the radiation density at the percent level provides an interesting window into early universe cosmology and BSM~particle physics. We have already seen that upcoming CMB~experiments will considerably improve the current constraints. In the following, we will further quantify the constraining power of future cosmological observations by including LSS~surveys. We will consider two types of forecasts based on~$P(k)$ and~$\Pw(k)$, which we refer to as `$P(k)$-forecasts' and `BAO-forecasts', respectively.

\subsection{Fisher Methodology}
In this section, we will use standard Fisher information theory to forecast the constraints of future observations. While Fisher forecasts have to be used with care, they provide useful guidance for the sensitivities and design of future experiments. In the following, we recall the basic elements of the Fisher methodology and its application to galaxy surveys~\cite{Bassett:2009mm, Verde:2009tu}. The relatively standard Fisher forecasting of CMB~observations is summarized in Appendix~\ref{app:bao-forecast_appendices}, where further details on the LSS~forecasting can be found as well.\medskip

Given a likelihood function $\mathcal{L}(\vec{\theta}\hskip1pt)$ for the model parameters $\vec{\theta} \equiv \{\omega_b, \omega_c, \theta_s, \tau, \As, \ns, \Neff, Y_p\}$, we define the Fisher matrix as the average curvature of the log-likelihood around the fiducial point in parameter space,
\beq
F_{ij} = - \left\langle \frac{\partial^2 \ln\mathcal{L}}{\partial \theta_i\, \partial \theta_j} \right\rangle ,
\eeq
where the expectation value denotes an average over all possible realizations of the data. If the likelihood is Gaussian, then the inverse Fisher matrix gives the covariance matrix. This means that~$F_{ii}^{-1/2}$ is the error on the parameter~$\theta_i$, when all other parameters~$\theta_{j\ne i}$ are known, while $\sigma(\theta_i) = (F^{-1})_{ii}^{1/2}$ is the error on~$\theta_i$ after marginalizing over the other parameters. More generally, the Cram\'er-Rao bound,
\beq
\sigma(\theta_i) \ge \sqrt{(F^{-1})_{ii}} \, ,
\eeq
gives a lower limit on the marginalized constraints.\medskip

The Fisher matrix for a galaxy survey is~\cite{Tegmark:1997rp} 
\beq
F_{ij} = \int_{-1}^1 \frac{\d\mu}{2}\, \int_{\kmin}^{\kmax} \frac{\d k\, k^2}{(2\pi)^2}\, \frac{\partial \ln \Pobs(k,\mu)}{\partial \theta_i} \frac{\partial \ln \Pobs(k,\mu)}{\partial \theta_j} \,V_\mathrm{eff}(k,\mu)\, , 	\label{eq:galaxyFisherMatrix}
\eeq
where~$\Pobs(k,\mu)$ is the anisotropic galaxy power spectrum, $\mu$~is the cosine between the wavevector~$\k$ and the line-of-sight, and~$V_\mathrm{eff}$ is the effective survey volume,
\beq
V_\mathrm{eff}(k,\mu) \equiv \int\!\d^3 x \left[\frac{n_g(\x) \Pobs(k,\mu)}{n_g(\x) P_g(k,\mu)+1} \right]^2 \approx \left[\frac{\bar{n}_g \Pobs(k,\mu)}{\bar{n}_g \Pobs(k,\mu)+1} \right]^2 V \, .	\label{eq:V_eff}
\eeq
In the second equality, we have assumed that the comoving number density of galaxies is independent of position, $n_g(\x) \approx \bar{n}_g = \const$, and introduced the actual survey volume~$V$. To derive the constraints from independent redshift bins, we take~$V$ to be the volume within each bin and add the corresponding Fisher matrices. The minimum wavenumber accessible in a survey is given by the volume of the survey\hskip1pt\footnote{We assume that the survey volume has a spherical geometry. The geometry of a given redshift bin (or the full survey volume) is neither spherical nor cubic, but we have checked that all of our results are essentially unaffected by this choice.} as $\kmin = 2\pi\,\left[3V/(4\pi)\right]^{-1/3}$.

\subsubsection{Modelling the power spectrum}
\label{sec:baoModelling}
In~\textsection\ref{sec:lss}, we introduced the linear matter power spectrum~$P_\mathrm{lin}(k)$ as the main LSS~observable and discussed some of the complications that arise because we cannot observe it directly. In order to obtain semi-realistic constraints on most parameters of the cosmological model, it is often sufficient to model the observed galaxy power spectrum as $\Pobs(k) \approx b^2 P_\mathrm{lin}(k)$, where~$b$ is the linear biasing parameter. However, the constraints on extra relativistic species are particularly sensitive to the way degeneracies are broken and to the non-linear damping of the oscillatory feature, so we need to be more careful in the modelling of the signal~\cite{Blas:2016sfa, Hand:2017ilm, Ding:2017gad}. Moreover, since observations only determine the angular positions and redshifts of objects, we need to take into account the corresponding redshift space distortions~(RSD) and geometric projection effects.\medskip

Separating the spectrum into its smooth and oscillatory parts according to~\eqref{eq:Pkdecomposition}, our model for the observed galaxy power spectrum is the following remapping of the linear matter power spectrum:
\beq
\Pobs(k,\mu) = {\color{Blue}b^2 F^2(k,\mu)} {\color{darkgreen}\Pnw(k,\mu)} \Big[ 1+ {\color{Red}O(k,\mu)}\,{\color{DarkOrange}D(k,\mu)}\Big] {\color{Purple}Z(k,\mu)}\, . \label{eq:Pobs}
\eeq
All functions in this expression have an implicit redshift dependence. We now define the different elements of~\eqref{eq:Pobs}:
\begin{itemize}
\item $\color{Red}{O(k,\mu)}$: This function encodes the BAO~signal and can be written as
\beq
O(k,\mu) \equiv B(k)\,O_\mathrm{lin}(k'(k,\mu)) + A(k) \, ,	\label{eq:O}
\eeq
where $O_\mathrm{lin}(k') \equiv \Pw_\mathrm{lin}(k')/\Pnw_\mathrm{lin}(k')$ is the normalized wiggle spectrum evaluated at the rescaled wavenumbers~\cite{Ballinger:1996cd}
\beq
k' = k\, \sqrt{(1-\mu^2)/q_\perp^2 + \mu^2/q_\parallel^2}\, , \quad \text{with}\quad q_\perp \equiv \frac{D_A(z)}{D_A^\mathrm{fid}(z)}\, , \quad q_\parallel \equiv \frac{H^\mathrm{fid}(z)}{H(z)}\, .	\label{eq:rescaledWavenumbers}
\eeq
This rescaling reflects the fact that the wavenumbers~$k$ cannot be measured directly, but instead have to be derived from the measured angles and redshifts using the angular diameter distance~$D_A^\mathrm{fid}(z)$ and Hubble rate~$H^\mathrm{fid}(z)$ of a fiducial cosmology. This is often referred to as anisotropic geometric effects. In the limit of spherically-averaged clustering measurements, these become isotropic and $k' = k/q$, where $q = q_\perp^{2/3} q_\parallel^{1/3} = D_V(z)/D_V^\mathrm{fid}(z)$, with the radial BAO~dilation given by $D_V \propto (D_A^2/H)^{1/3}$.

To model uncertainties in the BAO~extraction, we have introduced two free functions~$B(k)$ and $A(k)$ in~\eqref{eq:O}, which we take to be smooth polynomials in~$k$ (see below). Ultimately, we will marginalize over these polynomials to remove any information that is not robust to the BAO~signal itself. 

\item $\color{Blue}{b(z)}$: The bias of the target galaxies (e.g.~luminous red galaxies, emission line galaxies or quasars) sets the overall amplitude of the signal in each redshift bin. We will make the common assumption that $b(z) \propto 1/D_1(z)$, where~$D_1(z)$ is the linear growth function. This means that the bias is larger at high redshifts, which implies that the galaxy power spectrum may get significant corrections from non-linear biasing even at high redshifts.

\item $\color{Blue}{F(k,\mu)}$: This function characterises the effect of redshift space distortions. Following~\cite{Kaiser:1987qv}, we write
\beq
F(k,\mu) = \frac{1}{\left(q_\perp^2 q_\parallel\right)^{1/2}} \left[ 1+\beta\hskip1pt \mu'(k,\mu)^2 R(k)\right] ,	\label{eq:linearRSD}
\eeq
where~$\beta \equiv f/b$, with the linear growth rate $f \equiv \d\hskip0.5pt\ln D_1/\!\d\hskip0.5pt\ln a$. The factors of~$q_i$ account for differences in the cosmic volume in different cosmologies. Projection effects on the angle to the line-of-sight are included as~\cite{Ballinger:1996cd}
\beq
\mu'(k,\mu) = \mu / \sqrt{ \mu^2 + (1-\mu^2) Q^2}\, ,
\eeq
where $Q \equiv q_\parallel/q_\perp$, which becomes unity in the isotropic case. BAO~reconstruction removes redshift space distortions on large scales, which we have modelled by adding the factor $R(k) = 1-\exp[-(k\Sigma_s)^2/2]$ in~\eqref{eq:linearRSD}, where the value of~$\Sigma_s$ depends on the experimental specifications, in particular the noise levels. In our baseline forecasts, we take $\Sigma_s \to \infty$, i.e.~$R \equiv 1$, but we comment on finite values of~$\Sigma_s$ in~\S\ref{sec:constraints_planned}.

\item $\color{DarkOrange}{D(k,\mu)}$: This function models the non-linear damping of the BAO~signal~\cite{Eisenstein:2006nj, Seo:2007ns}
\beq
D(k,\mu) \equiv \exp\left[- \frac{1}{2}\Big(k^2\mu^2 \Sigma^2_\parallel +k^2(1-\mu^2) \Sigma_\perp^2\Big) \right] ,	\label{eq:Dk}
\eeq
where the damping scales perpendicular and parallel to the line-of-sight are given by
\begin{align}
\Sigma_\perp(z) 	&= 9.4\,(\sigma_8(z)/0.9)\, \si{\MpcPerh}\, ,	\\
\Sigma_\parallel(z) &= (1+f(z)) \,\Sigma_\perp(z)\, ,
\end{align}
with $\sigma_8$ being the amplitude of (linear) matter fluctuations at a scale of $\SI{8}{\MpcPerh}$. We account for BAO~reconstruction by decreasing these damping scales by an appropriate factor, e.g.~$0.5$ for \SI{50}{\percent} reconstruction. Following~\cite{White:2010qd, Font-Ribera:2013rwa}, we include the degradation in the reconstruction due to shot noise using a reconstruction multiplier $r(x)$, i.e.~$\Sigma_i \to r(x) \Sigma_i$. We obtain~$r(x)$ by interpolating over the table
\begin{equation}
\begin{aligned}
r &= (1.0,\, 0.9,\, 0.8,\, 0.7,\, 0.6,\, 0.55,\, 0.52,\, 0.5)\,, \\
x &= (0.2,\, 0.3,\, 0.5,\, 1.0,\, 2.0,\, 3.0,\, 6.0,\, 10.0)\,,
\end{aligned}
\end{equation}
with $r(x<0.2)=1.0$ and $r(x>10.0)=0.5$, which depends on the number density $\bar{n}_g$ via $x \equiv \bar{n}_g \Pobs(k_0,\mu_0)/0.1734$ evaluated at $k_0 = \SI{0.14}{\hPerMpc}$ and $\mu_0 = 0.6$. This means that we assume \SI{50}{\percent}~reconstruction at high number densities and no reconstruction for low densities.

\item $\color{darkgreen}{\Pnw(k,\mu)}$: The linear no-wiggle spectrum~$\Pnw_\mathrm{lin}(k,\mu)$ is determined from the linear power spectrum using the method described in Appendix~\ref{app:broadband+phaseShiftExtraction}. Non-linear corrections to this spectrum can be parameterized as
\beq
\Pnw(k,\mu) = \tilde B(k)\Pnw_\mathrm{lin}(k'(k,\mu)) + \tilde A(k)\, ,	\label{eq:PnwX}
\eeq
where~$\tilde B(k)$ and~$\tilde A(k)$ are smooth functions (see below). For the purpose of our BAO-forecasts, $\tilde A(k)$~and~$\tilde B(k)$ are degenerate with~$A(k)$ and~$B(k)$ in~\eqref{eq:O} and it is therefore consistent to use the linear spectrum. 

\item $\color{Purple}{Z(k,\mu)}$: For photometric surveys, we take the uncertainty in the redshift determination of the targets into account through the following function:
\beq
Z(k,\mu) = \exp\left[-k^2 \mu^2 \Sigma_{z}^2 \right] ,
\eeq
where $\Sigma_z = c\,(1+z)\,\sigma_{z0} / H(z)$ is given in terms of the root-mean-square redshift error~$\sigma_{z0}$~\cite{Seo:2003pu, Zhan:2005ki}. The redshift error, which depends on the experimental specifications, reduces the effective resolution for modes along the line-of-sight. We neglect this effect for spectroscopic surveys.
\end{itemize}
When evaluating the derivatives in the Fisher matrix~\eqref{eq:galaxyFisherMatrix}, the parameters~$b(z)$, $\beta$, $R(k)$, $D(k,\mu)$ and~$Z(k,\mu)$ are always computed using the fiducial cosmology. We are assuming that, after accounting for modelling uncertainties, no relevant cosmological information can be recovered from these functions.

\subsubsection{Accounting for broadband effects}
Non-linear evolution and biasing can change the shape of the power spectrum at high wavenumbers in a way that cannot be modelled from first principles. We account for this uncertainty by marginalizing over polynomials in~$k$ in both the~$P(k)$- and BAO-forecasts. In particular, the functions introduced in~\eqref{eq:PnwX} are defined as
\beq
\tilde A(k,z_i) = \sum_{n=0}^{N_a} \tilde a_{n,i} \, k^{n} \, , \qquad
\tilde B(k,z_i) = \sum_{m=0}^{N_b} \tilde b_{m,i} \, k^{2m}\, .	\label{eq:tildeAtildeB}
\eeq
As indicated, we allow independent polynomials in each redshift bin centred around~$z_i$. The coefficients~$\tilde a_{n,i}$ and~$\tilde b_{m,i}$ are included in the list of parameters~$\theta_i$. Derivatives with respect to these parameters are determined analytically, using the fiducial values $\tilde b_{0,i} =1$ and $\tilde a_{n,i} =\tilde b_{m\neq0,i}= 0$. A more careful treatment would replace this polynomial model with a perturbative model for the dark matter and biasing, and would marginalize over the bias parameters. In practice, this has been shown to give qualitatively similar forecasts~\cite{Gleyzes:2016tdh}. Our marginalization procedure is therefore sufficient to illustrate the sensitivity of our forecasts to broadband information.\medskip

Our BAO-forecasts will marginalize over the `broadband corrections' in~\eqref{eq:O}, with~$A(k)$ and~$B(k)$ defined as in~\eqref{eq:tildeAtildeB}.\footnote{To avoid a proliferation of parameters, we will use~$a_n$ and~$b_n$ for the parameters in both~\eqref{eq:O} and~\eqref{eq:PnwX}, i.e.\ we will drop the tildes from now on. Which parameter set is meant will be clear from the context.} At the level of the Fisher matrix, marginalizing over a polynomial and an exponential are equivalent. As a result, the function $B(k)$ captures the uncertainty in the damping scales~$\Sigma_\parallel$ and~$\Sigma_\perp$ in~\eqref{eq:Dk}. This implies that our marginalization procedure will eliminate any cosmological information associated with the non-linear damping of the power spectrum, leaving the distinct information contained in the oscillating part of the spectrum~$O_\mathrm{lin}(k'(k,\mu))$. This type of procedure is used in the analysis of BAO~data to correct for errors made in the modelling of~$\Pnw(k)$, see e.g.~\cite{Beutler:2016ixs}.\medskip

We will choose various levels of marginalization in our forecasts. This will help to distinguish the information encoded in the smooth shape of the spectrum,~$\Pnw(k)$, from that contained in the frequency and phase of the BAO~spectrum,~$\Pw(k)$. In addition, these marginalizations also give a sense for the level of robustness of each type of information when accounting for the various uncertainties in modelling the data of a realistic galaxy survey.

\subsubsection{Extracting the BAO signal}
In describing the power spectrum, we introduced the idea of marginalizing over polynomials to remove the information in~$\Pobs(k)$ that is thought to be degenerate with non-linear evolution and galaxy biasing. The BAO~spectrum is known to be robust to these effects and should therefore survive any such treatment. In principle, the BAO~signal could be isolated with sufficient marginalization. However, in practice, it is more useful to extract the information associated with the BAO~signal before any marginalization. The robustness of the BAO~spectrum to non-linearities means we can be more aggressive with our choice of~$\kmax$ and less cautious with our marginalization. Consequently, it is convenient to treat the BAO~signal and the broadband information independently.\medskip

The observed BAO~spectrum is defined by 
\beq
\Oobs(k,\mu) \equiv \frac{\Pobs^\mathrm{w}(k,\mu)}{\Pobs^\mathrm{nw}(k,\mu)} = D(k,\mu) \, O(k,\mu) \, ,
\eeq
where~$D(k,\mu)$ and~$O(k,\mu)$ were introduced in~\eqref{eq:Pobs}. To derive the new Fisher matrix for the BAO spectrum directly, we first write the derivatives of~$\Pobs(k,\mu)$ as
\beq
\frac{\partial \ln \Pobs(k,\mu)}{\partial \theta_i} = \frac{1}{\Pobs^\mathrm{nw} + \Pobs^\mathrm{w}} \left( \frac{\partial \Pobs^\mathrm{nw}}{\partial \theta_i} + \frac{\partial \Pobs^\mathrm{w}}{\partial \theta_i}\right) .	\label{eq:dP}
\eeq
We then drop the term proportional to~$\partial_{\theta_i} \Pobs^\mathrm{nw}$ since it is degenerate with the marginalization over the broadband corrections. For the same reason, we write $\partial_{\theta_i} \Pobs^\mathrm{w} \approx b^2 F^2 \Pnw D\, \partial_{\theta_i} O$, i.e.\ we do not act with the derivatives on the functions~$D(k,\mu)$ and~$b F(k,\mu)$. The derivative in~\eqref{eq:dP} therefore becomes
\beq
\frac{\partial \ln \Pobs(k,\mu)}{\partial \theta_i} \approx \frac{D(k,\mu)}{1+D(k,\mu) \, O(k,\mu)} \frac{\partial O(k,\mu)}{\partial \theta_i} \, .
\eeq
While the derivatives that we have dropped are non-zero, the marginalization procedure described above is designed to remove them and the forecasts for cosmic parameters should consequently be the same. Removing this information by hand (and marginalizing) ensures that our BAO-forecasts do not include these broadband effects, as we will show in Fig.~\ref{fig:marginalizationPlanned}. The resulting Fisher matrix is then given by
\beq
F_{ij} = \int_{-1}^1 \frac{\d\mu}{2}\, \int_{\kmin}^{\kmax} \frac{\d k\, k^2}{(2\pi)^2}\, \frac{D(k,\mu)^2 }{(1+D(k,\mu) \, O(k,\mu))^2} \frac{\partial O(k,\mu)}{\partial \theta_i} \frac{\partial O(k,\mu)}{\partial \theta_j} \,V_\mathrm{eff}(k,\mu)\, .	\label{eq:baoFisherMatrix}
\eeq
We note that this Fisher matrix depends on~$\Pobs^\mathrm{nw}(k,\mu)$ only through~$V_\mathrm{eff}(k,\mu)$, which determines the signal-to-noise. For photometric surveys, we replace $V_\mathrm{eff}(k,\mu) \to Z(k,\mu)^2\, V_\mathrm{eff}(k,\mu)$ to account for the redshift error and the associated reduction of power along the line-of-sight. In principle, we should model~$\Pobs^\mathrm{nw}(k,\mu)$ using the non-linear (galaxy) power spectrum, given that we will work close to the non-linear regime. However, non-linear evolution also correlates the modes and produces a non-Gaussian covariance matrix. Since most of the surveys under consideration in this chapter are limited by shot noise, using the non-linear power spectrum without taking into account the associated mode coupling in the covariance would artificially increase the number of signal-dominated modes. To be consistent with the use of a Gaussian covariance, our forecasts will therefore use the linear broadband spectrum.

\subsection{Constraints from Planned Surveys}
\label{sec:constraints_planned}
We are now ready to forecast the constraints of current and future CMB and LSS~observations on the effective number of relativistic species~$\Neff$. Unless stated otherwise, our baseline analysis assumes a $\Lambda\mathrm{CDM}$+$\Neff$ cosmology in which the primordial helium fraction~$Y_p$ is fixed by consistency with~BBN. At the end of the section, we will also present results with~$Y_p$ as a free parameter. We will further dissect the information content of the BAO~spectrum in Section~\ref{sec:phaseShift}.\medskip
	
In Appendix~\ref{app:bao-forecast_appendices}, we present detailed forecasts for current and future CMB~experiments. The expected $1\sigma$~constraints for representative versions of the Planck satellite, a near-term CMB-S3 experiment and a future CMB-S4~mission are $\sigma(\Neff) = 0.18$, $0.054$, $0.030$, respectively.\footnote{This precise CMB-S4~value of~$\sigma(\Neff)$ differs from that in Chapter~\ref{chap:cmb-phases} because we employ slightly different experimental specifications based on~\cite{Abazajian:2016yjj}.} We also show how these constraints depend on variations of the experimental configurations in the same appendix.\medskip
	
We would like to know how much these CMB~constraints would improve with the addition of LSS~data. A number of galaxy surveys are expected to take place over the next decade. The power of these surveys to constrain~$\Neff$ is most sensitive to the survey volume, the number densities of galaxies and the redshift errors (spectroscopic versus photometric). The precise specifications of the surveys used in our analysis are given in Appendix~\ref{app:bao-forecast_appendices}, where we also present more detailed forecasts for the full set of parameters.

We will give the results of two types of forecasts based on~$P(k)$ and~$\Pw(k)$. Our $P(k)$-forecasts apply the Fisher matrix~\eqref{eq:galaxyFisherMatrix} with $\kmax = \SI{0.2}{\hPerMpc}$ and marginalize over~$b_{m\leq1}$. To be conservative about non-linear biasing, we do not increase~$\kmax$ at large redshifts, despite the (near-)linearity of the matter power spectrum. Our BAO-forecasts use the Fisher matrix~\eqref{eq:baoFisherMatrix} with $\kmax = \SI{0.5}{\hPerMpc}$ and marginalize over~$a_{n\leq4}$,~$b_{m\leq3}$. We will also show how these forecasts depend on the choice of~$\kmax$ and the level of marginalization.

\paragraph{Baseline results}
In Table~\ref{tab:CMB+Pk_Neff},%
\begin{table}[b]
	\begin{tabular}{c S[table-format=2.4] S[table-format=2.4] S[table-format=2.4] S[table-format=2.4] S[table-format=2.4] S[table-format=2.4] S[table-format=2.4]}
			\toprule
					& 		& \multicolumn{4}{c}{spectroscopic}				& \multicolumn{2}{c}{photometric}	\\
			\cmidrule(lr){3-6} \cmidrule(lr){7-8}
					& CMB 	& {BOSS}	& {eBOSS}	& {DESI}	& {Euclid}	& {DES}		& {LSST}				\\
			\midrule[0.065em] 
		Planck		& 0.18	& 0.14		& 0.13		& 0.087		& 0.079		& 0.17		& 0.14					\\
		CMB-S3		& 0.054	& 0.052		& 0.051		& 0.045		& 0.043		& 0.054		& 0.052					\\
		CMB-S4		& 0.030	& 0.030		& 0.030		& 0.028		& 0.027		& 0.030		& 0.030					\\
			\bottomrule
	\end{tabular}
	\caption{Forecasted $1\sigma$~constraints on~$\Neff$ for various combinations of current and future CMB and LSS~experiments using~$P(k)$-forecasts with $\kmax = \SI{0.2}{\hPerMpc}$.}
	\label{tab:CMB+Pk_Neff}
\end{table}
we present the $1\sigma$~constraints on~$\Neff$ for various combinations of current and future CMB and LSS~experiments using the full $P(k)$-forecast. In Table~\ref{tab:CMB+BAO_Neff},%
\begin{table}[t]
	\begin{tabular}{c S[table-format=2.4] S[table-format=2.4] S[table-format=2.4] S[table-format=2.4] S[table-format=2.4] S[table-format=2.4] S[table-format=2.4]}	
			\toprule
					& 		& \multicolumn{4}{c}{spectroscopic}				& \multicolumn{2}{c}{photometric}	\\
			\cmidrule(lr){3-6} \cmidrule(lr){7-8}
					& {CMB} & {BOSS}	& {eBOSS}	& {DESI}	& {Euclid}	& {DES}		& {LSST}				\\
			\midrule[0.065em] 
		Planck		& 0.18	& 0.15		& 0.15		& 0.14		& 0.14		& 0.16		& 0.15					\\
		CMB-S3		& 0.054	& 0.052		& 0.052		& 0.050		& 0.050		& 0.054		& 0.052					\\
		CMB-S4		& 0.030	& 0.030		& 0.030		& 0.029		& 0.029		& 0.030		& 0.030					\\
			\bottomrule
	\end{tabular}
	\caption{Forecasted $1\sigma$~constraints on~$\Neff$ for various combinations of current and future CMB and LSS~experiments using BAO-forecasts with $\kmax = \SI{0.5}{\hPerMpc}$.}
	\label{tab:CMB+BAO_Neff}
\end{table}
we compare these results to the same experiments using our BAO-forecasts. At BOSS~levels of sensitivity and number densities, the BAO~feature makes the most significant impact on constraints, particularly when combined with a CMB~experiment like Planck. In contrast, with the larger volume and redshift range of~DESI, the broadband shape carries most of the information and can lead to a significant improvement in the constraint on~$\Neff$ both for Planck and a typical CMB-S3~experiment. Finally, photometric redshift surveys like~DES and~LSST generally perform worse than spectroscopic surveys because they are effectively two-dimensional for the scales of interest. However, the employed redshift error is conservative and we do not take the full potential of these surveys into account as we are only considering observations of galaxy clustering and have not included weak gravitational lensing measurements, for instance. We expect the constraints to improve with these additional LSS~observables, but quantifying this is beyond the scope of this work.

\begin{figure}[t!]
	\includegraphics{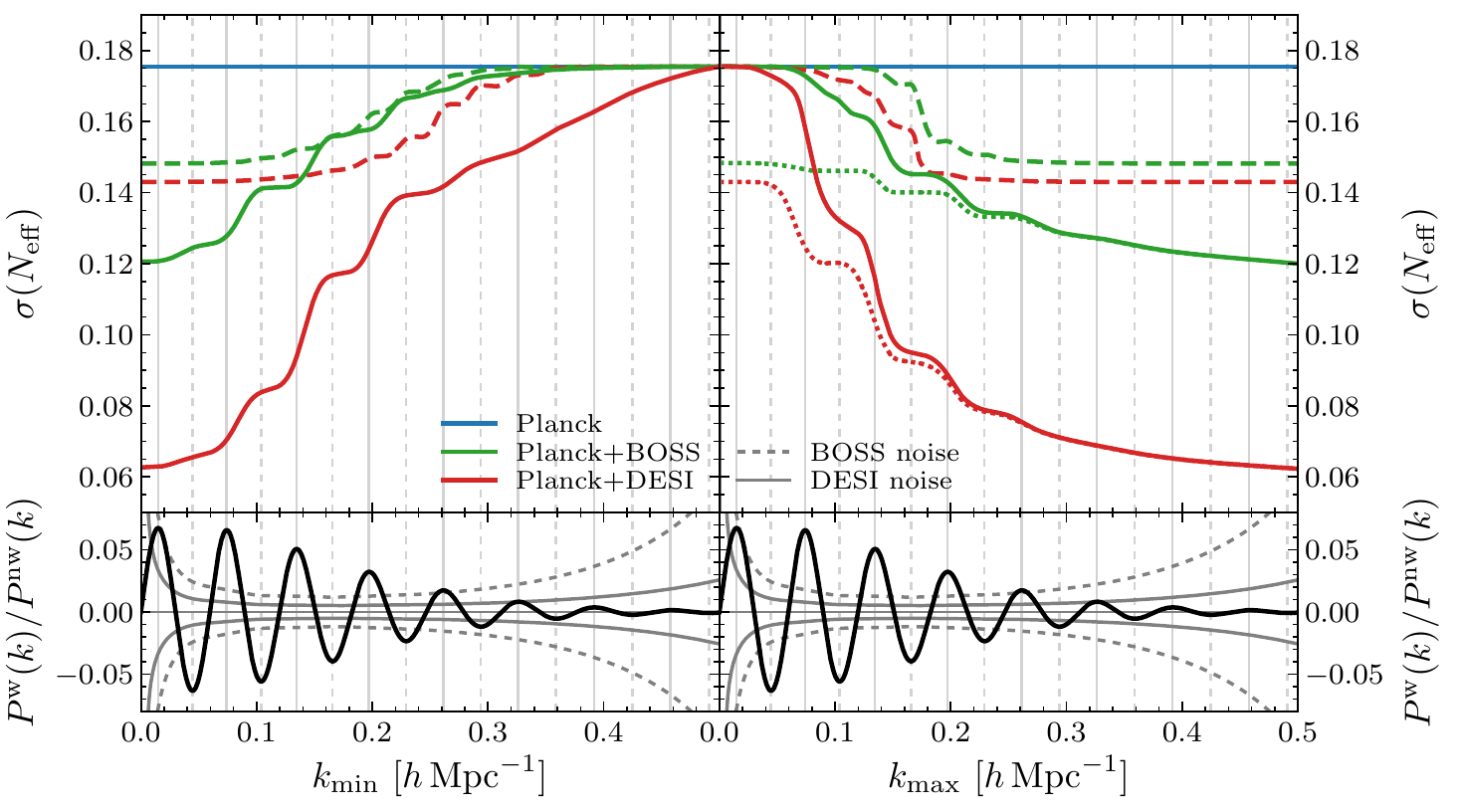}\vspace{-2pt}
	\caption{Forecasts for BOSS and DESI combined with Planck as a function of the smallest~(\textit{left}) and largest~(\textit{right}) Fourier modes used in the forecast, $\kmin$ and~$\kmax$, with $\kmax=\SI{0.5}{\hPerMpc}$ in the left panel. The solid and dashed lines indicate the constraints from the $P(k)$- and BAO-forecasts, respectively. Shown as the dotted lines are the ``optimal constraints'' as described in the main text. The lower panel displays the linear BAO~spectrum and an estimate of the noise levels.\vspace{-3pt}}
	\label{fig:kminkmax}
\end{figure}

\paragraph{Sensitivity to $\boldsymbol{k_\mathrm{max}}$}
The broadband signal is sensitive to non-linear effects and we should therefore understand how sensitive these results are to the choice of~$\kmax$. In particular, we have chosen $\kmax = \SI{0.2}{\hPerMpc}$ in Table~\ref{tab:CMB+Pk_Neff}, but the usable range of scales is uncertain. Figure~\ref{fig:kminkmax}%
shows how the constraints vary as a function of the maximal wavenumbers included in the analysis,~$\kmax$, for both the $P(k)$- and BAO-forecasts. For the BAO-forecasts, we see a clear plateau for $\kmax > \SI{0.2}{\hPerMpc}$. This behaviour is due to the damping of the oscillations at higher~$k$ relative to the smooth power spectrum. Cosmic variance is ultimately determined by the amplitude of the smooth power spectrum and one cannot recover the high-$k$~oscillations even by lowering the shot noise. In contrast, the $P(k)$-forecasts show improvements out to $\kmax > \SI{0.3}{\hPerMpc}$.\smallskip

Given that the BAO spectrum is robust to non-linear evolution, it is natural to consider an optimal combination of the~$P(k)$ and BAO~spectra that uses all the available information. This means using~$P(k)$ up to a certain~$\kmax$ and adding BAO-only information for larger~$k$. The~$\kmax$ of the $P(k)$~analysis then becomes the $\kmin$ of the BAO~analysis to avoid double counting the information. Results for this optimal combination are shown as the dotted line in Fig.~\ref{fig:kminkmax}.%
\begin{figure}[t!]
	\includegraphics{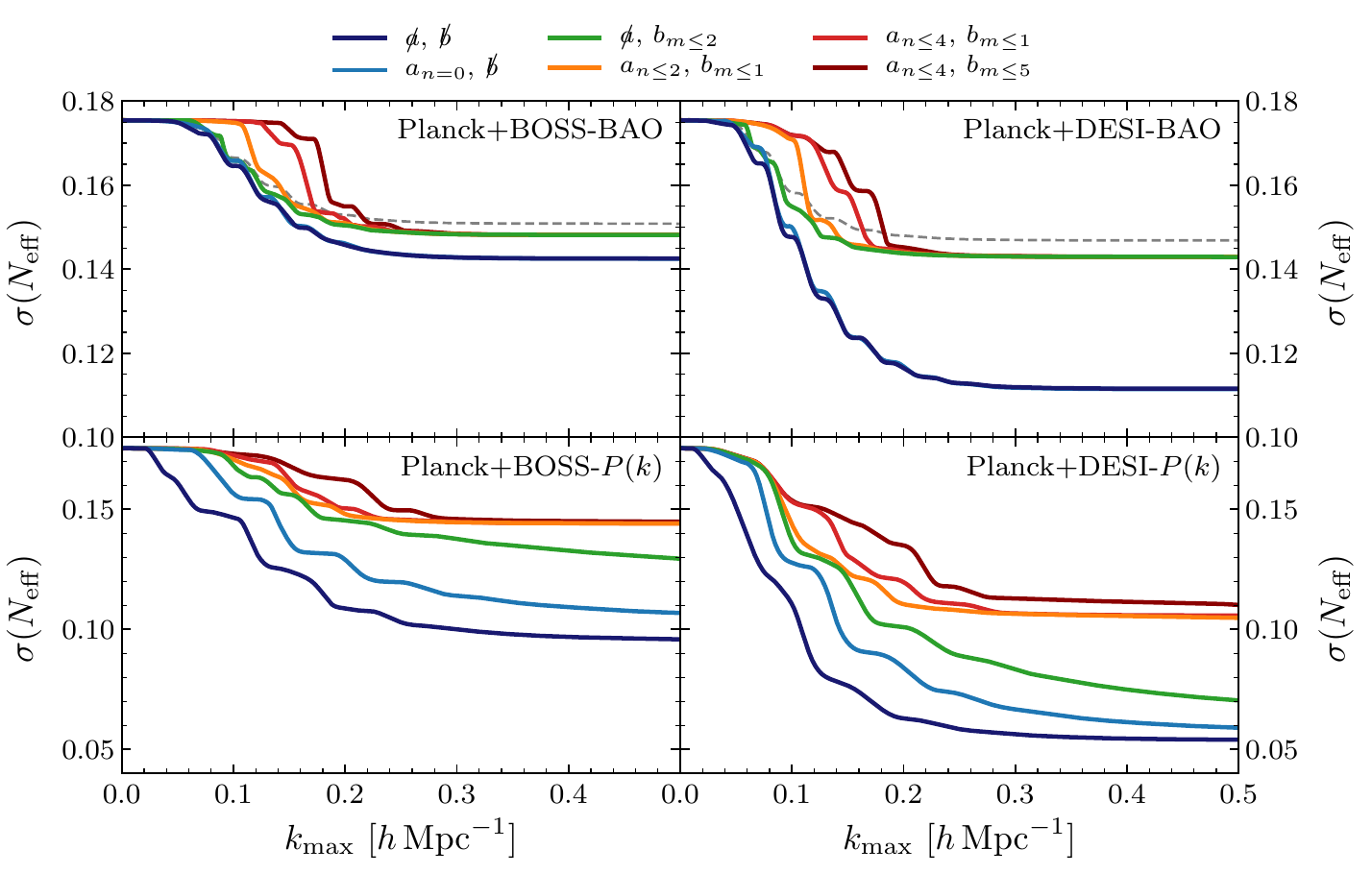}
	\caption{Forecasts for BOSS and DESI combined with Planck as a function of the largest Fourier modes used in the forecast,~$\kmax$, using various levels of both additive and multiplicative marginalization, cf.~the $a_i$ and $b_i$-terms in~\eqref{eq:tildeAtildeB}. We have varied the number of parameters in the marginalization from none~($\slashed{a}$) to five~($a_{n\leq4}$) and none~($\slashed{b}$) to six~($b_{m\leq5}$), respectively. The dashed line shows the constraints from a standard isotropic BAO~analysis for comparison.}
	\label{fig:marginalizationPlanned}
\end{figure}

\paragraph{Sensitivity to marginalization}
High-redshift galaxy surveys benefit significantly from measuring highly biased objects. These large biases can offset the growth function, $b(z) D_1(z) = \const$, and keep the amplitude of the galaxy power spectrum effectively fixed at high redshift. This boost is important for maintaining a signal above the shot noise, which we have assumed is redshift-independent. As a consequence, high-redshift and low-redshift galaxy power spectra are equally sensitive to uncertainties in the biasing coefficients. This is particularly significant when determining the largest wavenumbers that carry useful cosmological information. While taking $\kmax > \SI{0.2}{\hPerMpc}$ is appealing to maximize the constraints on~$\Neff$, we must also marginalize over successively more bias parameters. Figure~\ref{fig:marginalizationPlanned} shows how the results depend on the marginalization scheme. While both the $P(k)$- and BAO-constraints degrade significantly when going from no marginalization to a few bias parameters, the BAO-forecasts quickly become robust to the marginalization. In contrast, the $P(k)$-forecasts weaken notably with additional biasing, but always lie below the BAO-only results, as one would expect. This confirms the intuition that the information that is primarily driving the constraints derived from~$P(k)$ is present in the no-wiggle power spectrum,~$\Pnw(k)$, instead of the BAO~spectrum.\medskip

It is instructive to compare the results of our BAO-forecasts with those of a standard BAO~analysis. Specifically, it is conventional to use the BAO signal to constrain only~$q_i$, $i=\perp,\parallel$, defined in~\eqref{eq:rescaledWavenumbers} and derive parameter constraints from them.\footnote{We also compared the constraints coming from the full anisotropic treatment (cf.~\textsection\ref{sec:baoModelling}) with the isotropic approximation. The BAO-forecasts only weaken at small wavenumbers depending on the marginalization procedure, but reach the same plateau values at large wavenumbers as our baseline analysis. In contrast, the constraints on~$\Neff$ are systematically weaker in the isotropic $P(k)$-forecasts at the level of \SI{15}{\percent} for $\kmax = \SI{0.2}{\hPerMpc}$.} These derived limits on~$\Neff$ are shown as the dashed lines in Fig.~\ref{fig:marginalizationPlanned}. The fact that the standard BAO~constraints are slightly weaker than those of our full BAO-forecasts, even after marginalization, suggests there is information in the BAO~spectrum beyond the BAO~scale. Combined with our marginalization procedure, the analytic insights of Chapter~\ref{chap:cmb-phases} suggest that the improvement from the grey dashed line to the green line in the top row of Fig.~\ref{fig:marginalizationPlanned} is carried by the phase shift. The more prominent enhancement in sensitivity indicated by the blue lines should then be related to the amplitude shift induced by free-streaming particles. We will explore this further in Section~\ref{sec:phaseShift}.

\paragraph{Degeneracy with $\boldsymbol{Y_{p}}$}
To explore possible degeneracies between the effective number of relativistic species~$\Neff$ and the primordial helium fraction~$Y_p$, we now consider a $\Lambda\mathrm{CDM}$+$\Neff$+$Y_p$ cosmology. In Tables~\ref{tab:CMB+Pk_Neff+Yp}%
\begin{table}[t]
	\begin{tabular}{c c S[table-format=2.4] S[table-format=2.4] S[table-format=2.4] S[table-format=2.4] S[table-format=2.4] S[table-format=2.4] S[table-format=2.4]}	
			\toprule
					&			& 			& \multicolumn{4}{c}{spectroscopic}				& \multicolumn{2}{c}{photometric}	\\
			\cmidrule(lr){4-7} \cmidrule(lr){8-9}
					& Parameter	& {CMB}		& {BOSS}	& {eBOSS}	& {DESI}	& {Euclid}	& {DES}		& {LSST}				\\
			\midrule[0.065em] 
		\multirow{2}{*}{Planck} & $\Neff$	& 0.32		& 0.25		& 0.22		& 0.14		& 0.13		& 0.29		& 0.23		\\
								& $Y_p$		& 0.018		& 0.016		& 0.016		& 0.013		& 0.012		& 0.017		& 0.015		\\
			\midrule[0.065em]
		\multirow{2}{*}{CMB-S3}	& $\Neff$	& 0.12		& 0.12		& 0.11		& 0.094		& 0.088		& 0.12		& 0.11		\\
								& $Y_p$		& 0.0069	& 0.0068	& 0.0067	& 0.0060	& 0.0058	& 0.0069	& 0.0066	\\
			\midrule[0.065em]
		\multirow{2}{*}{CMB-S4}	& $\Neff$	& 0.081		& 0.079		& 0.078		& 0.070		& 0.067		& 0.081		& 0.078		\\
								& $Y_p$		& 0.0047	& 0.0046	& 0.0046	& 0.0043	& 0.0042	& 0.0047	& 0.0046	\\
			\bottomrule
	\end{tabular}
	\caption{Forecasted $1\sigma$~constraints on $\Neff$ and~$Y_p$ for various combinations of current and future CMB and LSS~experiments using $P(k)$-forecasts with $\kmax = \SI{0.2}{\hPerMpc}$.}
	\label{tab:CMB+Pk_Neff+Yp}
\end{table}
and~\ref{tab:CMB+BAO_Neff+Yp},%
\begin{table}
	\begin{tabular}{c c S[table-format=2.4] S[table-format=2.4] S[table-format=2.4] S[table-format=2.4] S[table-format=2.4] S[table-format=2.4] S[table-format=2.4]}	
			\toprule
					& 			& 			& \multicolumn{4}{c}{spectroscopic}				& \multicolumn{2}{c}{photometric}	\\
			\cmidrule(lr){4-7} \cmidrule(lr){8-9}
					& Parameter	& {CMB}		& {BOSS}	& {eBOSS}	& {DESI}	& {Euclid}	& {DES}		& {LSST}				\\
			\midrule[0.065em] 
		\multirow{2}{*}{Planck} & $\Neff$	& 0.32		& 0.29		& 0.29		& 0.28		& 0.28		& 0.30		& 0.29		\\
								& $Y_p$		& 0.018		& 0.018		& 0.018		& 0.018		& 0.018		& 0.018		& 0.018		\\
			\midrule[0.065em]
		\multirow{2}{*}{CMB-S3}	& $\Neff$	& 0.12		& 0.12		& 0.12		& 0.12		& 0.12		& 0.12		& 0.12		\\
								& $Y_p$		& 0.0069	& 0.0069	& 0.0069	& 0.0069	& 0.0069	& 0.0069	& 0.0069	\\
			\midrule[0.065em]
		\multirow{2}{*}{CMB-S4}	& $\Neff$	& 0.081		& 0.080		& 0.080		& 0.079		& 0.079		& 0.081		& 0.080		\\
								& $Y_p$		& 0.0047	& 0.0047	& 0.0047	& 0.0046	& 0.0046	& 0.0047	& 0.0047	\\
			\bottomrule
	\end{tabular}
	\caption{Forecasted $1\sigma$~constraints on $\Neff$ and~$Y_p$ for various combinations of current and future CMB and LSS~experiments using BAO-forecasts with $\kmax = \SI{0.5}{\hPerMpc}$.}
	\label{tab:CMB+BAO_Neff+Yp}
\end{table}
we present the $1\sigma$~constraints on~$\Neff$ and~$Y_p$ for various combinations of current and future CMB and LSS~experiments using $P(k)$-forecasts and BAO-forecasts, respectively. As expected, the CMB-only constraint on~$\Neff$ become worse due to the well-known degeneracy between $\Neff$ and~$Y_p$ in the CMB~damping tail. When broadband information is included, we find significant improvements in the constraints on both $\Neff$ and~$Y_p$. However, this improvement cannot be attributed to the phase shift as we see only modest improvements in our BAO-forecasts. The broadband shape of the matter distribution is sensitive to the expansion history and to free-streaming neutrinos, but is not significantly affected by~$Y_p$. As a result, the broadband information in~$P(k)$ can break CMB~degeneracies even without the phase shift information.

\paragraph{Comments on reconstruction}
In our baseline forecasts, we took $R \equiv 1$ in~\eqref{eq:linearRSD}, which is equivalent to taking $\Sigma_s \to \infty$. A few comments are in order regarding the effect of a finite~$\Sigma_s$. As discussed in~\cite{Seo:2015eyw}, the optimal smoothing scale~$\Sigma_s$ used in the BAO reconstruction depends on the noise levels of the experiment. Having said that, we have found only small changes in our results when going from $\Sigma_s=\infty$ to finite~$\Sigma_s$. The constraints quoted in Tables~\ref{tab:CMB+Pk_Neff} to~\ref{tab:CMB+BAO_Neff+Yp} are basically unaffected, except for DESI and Euclid in the $P(k)$-forecasts, where the impact is also mild. Changing $\Sigma_s$ from $\SI{30}{\MpcPerh}$ to $\SI{15}{\MpcPerh}$ and $\SI{10}{\MpcPerh}$, the constraint on~$\Neff$ slightly weakens from \num{0.090} to \num{0.093} and \num{0.096} for Planck+DESI (\num{0.082}, \num{0.086} and \num{0.090} for Planck+Euclid) in $\Lambda\mathrm{CDM}$+$\Neff$ compared to the quoted \num{0.087} (\num{0.079}) in Table~\ref{tab:CMB+Pk_Neff}. In practice, this roughly \SI{10}{\percent} effect has to be compared to the impact on the reconstruction efficiency.

\subsection{Designer's Guide for Future Surveys}
\label{sec:constraints_future}
One of the main benefits of a Fisher forecast is that it can inform the design of future experiments. For spectroscopic surveys, the basic parameters are the total number of objects,~$N_g$, the maximal redshift,~$\zmax$, and the sky area in square degrees,~$\Omega$. From these, we derive the survey volume,~$V$, and the comoving number density,~$\bar{n}_g$.\footnote{For simplicity, we will assume that the comoving number density can be approximated by a constant over the complete survey volume. However, very similar results are obtained for BOSS and DESI when using the specific redshift-dependent number densities.} In this section, we will explore how the constraints on~$\Neff$ depend on these parameters.\medskip

Most of the survey characteristics are encoded in the effective survey volume,\footnote{The effective survey volume also depends on the linear bias parameter~$b$ through $\bar{n}_g \Pobs \propto \bar{n}_g b^2$. This dependence is degenerate with a rescaling of $\bar{n}_g$, so we will take $b(z=0) \equiv 1$ and vary~$\bar{n}_g$. This ignores the impact that changes in~$b$ may have on redshift space distortions.} $V_\mathrm{eff}$, cf.~\eqref{eq:V_eff} and~\eqref{eq:baoFisherMatrix}. The dependence of~$V_\mathrm{eff}$ on the survey parameters is somewhat non-trivial. Increasing~$V$ (by increasing $\zmax$ and/or $\Omega$), at fixed~$N_g$, will also reduce~$\bar{n}_g$. For signal-dominated modes, $\bar{n}_g\Pobs \gg 1$, this effect is not important and the effective volume scales approximately as $V_\mathrm{eff} \propto V$. However, for $\bar{n}_g\Pobs \ll 1$, the shot noise is important and the reduction in the comoving density is more important than the increase in the volume, so that the effective volume scales as $V_\mathrm{eff} \propto V^{-1}$. This means that we will only benefit from an increase in the volume as long as the modes of interest, $k \in \SIrange[range-phrase={,}, range-units=brackets, open-bracket=[, close-bracket=]]{0.1}{0.3}{\hPerMpc}$, are signal dominated.\medskip

As mentioned before, the increased linearity of the matter distribution at high redshifts is undermined by the larger biasing. As a result, the main benefit of large~$\zmax$ is the increased survey volume and hence the total number of modes. Unfortunately, the survey volume only grows slowly with redshift for~$z>2$ and the resulting improvements in parameters is relatively modest for large increases in~$\zmax$. The situation is slightly different for the BAO spectrum as the non-linear damping factor~$D(k,\mu)$ depends on the clustering of the matter directly and is therefore less important at high redshifts. However, the BAO signal alone has a relatively modest effect on~$\Neff$\hskip1pt-forecasts in general and the change to the damping factor consequently does not make a visible difference in our forecasts.\medskip

\begin{figure}[t!]
	\includegraphics{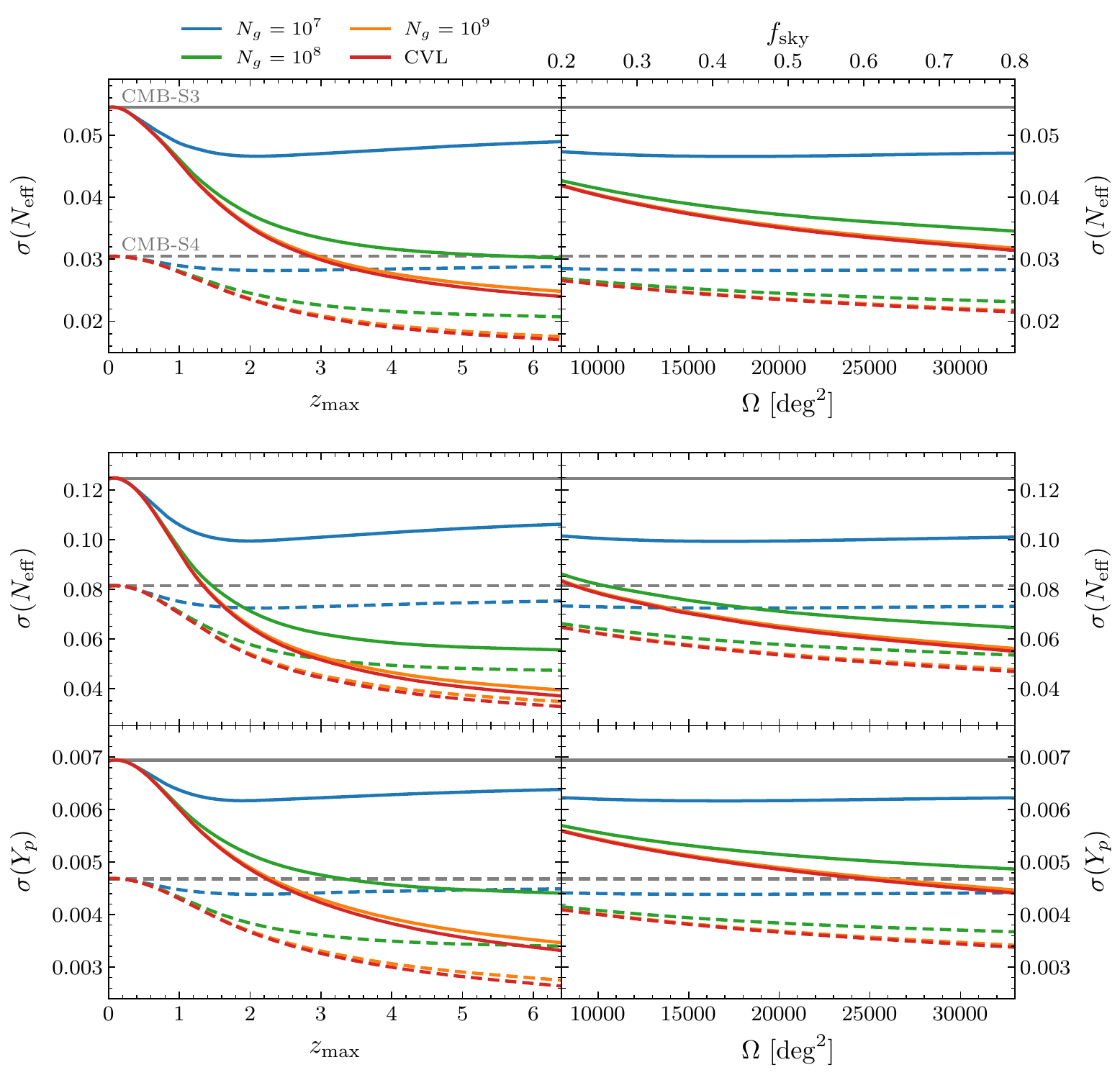}
	\caption{Future constraints for $\Lambda\mathrm{CDM}$+$\Neff$~(\textit{top}) and $\Lambda\mathrm{CDM}$+$\Neff$+$Y_p$~(\textit{bottom}) from the full galaxy power spectrum,~$\Pobs(k)$, up to $\kmax=\SI{0.2}{\hPerMpc}$ as a function of the total number of objects,~$N_g$, at fixed survey area $\Omega = \SI{20000}{deg^2}$~(\textit{left}) and as a function of the survey area~$\Omega$ (or sky fraction $\fsky$) for fixed $\zmax=2$~(\textit{right}). The comoving number density is assumed to be constant and given by the total volume of the survey. For ``CVL''~(\textcolor{pyRed}{red}), all modes in the survey are assumed to be measured up to the limit set by cosmic variance. Solid and dashed lines correspond to combining the LSS~data with CMB-S3 and CMB-S4~data, respectively. The \textcolor{pyGray}{gray} lines indicate the level of sensitivity of the respective CMB~experiments alone.}
	\label{fig:futureConstraints}
\end{figure}

In the top panel of Fig.~\ref{fig:futureConstraints},%
we present $P(k)$-forecasts for~$\Neff$ for a variety of survey configurations, assuming~$Y_p$ is fixed by BBN~consistency. We see that the largest improvement comes from increasing~$N_g$ from~\num{e7} to~\num{e8}. As we increase the number of objects further, we reach the cosmic variance limit for all modes of interest. We see that an optimistic future survey combined with a near-term CMB~experiment can provide constraints that are comparable to (or slightly stronger than) those projected for CMB-S4 alone. Having said that, it does not appear that one can push the measurement of~$\Neff$ well beyond the CMB-S4~target. Moreover, as in the case of the planned experiments, the improvements from the BAO~signal alone are rather small.\medskip

The value of LSS becomes more significant as we expand the space of parameters. The bottom panel of Fig.~\ref{fig:futureConstraints} shows $P(k)$-forecasts for $\Lambda\mathrm{CDM}$+$\Neff$+$Y_p$. We again see that the most significant jump in sensitivity arises when~$N_g$ increases from~\num{e7} to~\num{e8}. We note that a factor of two improvement in~$\sigma(\Neff)$ over CMB-S4 seems possible. We also see that the $P(k)$-forecasts for~$\Neff$ marginalized over~$Y_p$ are competitive with CMB-only forecasts with~$Y_p$ held fixed. In this sense, $P(k)$~adds robustness to the measurement of~$\Neff$ under broader extensions of~$\Lambda\mathrm{CDM}$. The improvement in~$Y_p$ is slightly weaker, but shows the same general trend.\medskip

\begin{figure}[t!]
	\includegraphics{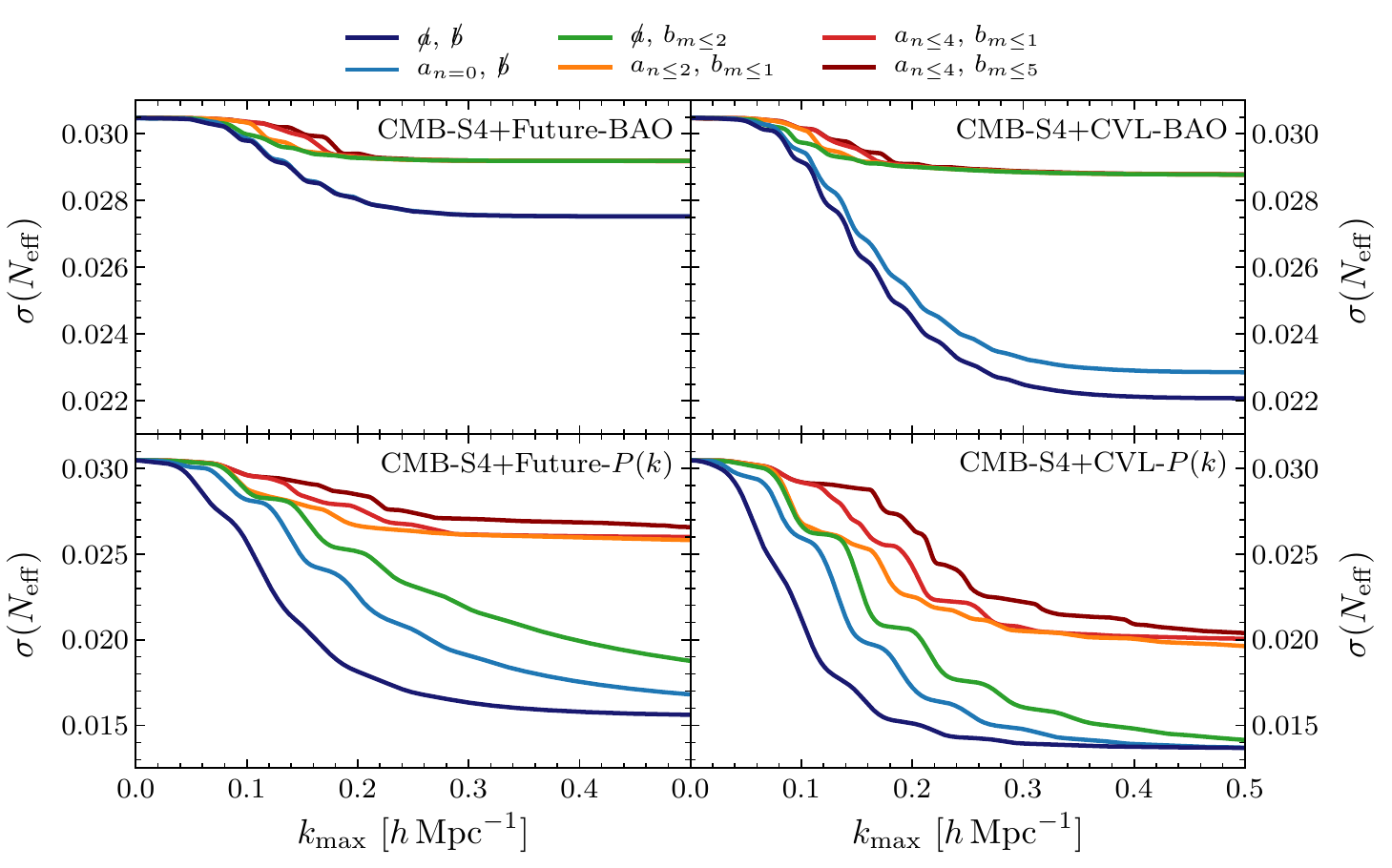}
	\caption{Forecasts for two future surveys combined with CMB-S4 as a function of the largest Fourier modes used in the forecast,~$\kmax$, using various levels of both additive and multiplicative marginalization. We have varied the number of parameters in the marginalization from none~($\slashed{a}$) to five~($a_{n \leq 4}$) and none~($\slashed{b}$) to six~($b_{n \leq 5}$), respectively. The employed experimental specifications for the ``Future''-survey are $N_g = \num{e8}$, $\zmax = 3$ and $\fsky = 0.5$, whereas the ``CVL''-survey is cosmic variance limited for all~$k$ up to~$\kmax$ over $\fsky = 0.5$ and $\zmax=6$.}
	\label{fig:marginalizationFuture}
\end{figure}

The range of accessible modes in near-term galaxy surveys is limited by their reliance on highly biased objects, but more futuristic surveys may not have the same limitations. Future surveys can also have high signal-to-noise beyond $k = \SI{0.2}{\hPerMpc}$, making it worth to consider the impact of increasing~$\kmax$. In Figure~\ref{fig:marginalizationFuture},%
we show the potential reach of two representative surveys. The first, denoted ``Future'', is characterised by $N_g = \num{e8}$, $\fsky = 0.5$ and $\zmax=3$, which is roughly the same as a spectroscopic follow-up to~LSST. The second, denoted~``CVL'', is cosmic variance limited for all $k\leq \kmax$ over $\fsky = 0.5$ and $\zmax=6$. In principle, a \SI{21}{cm}~intensity mapping survey could achieve similar performance~\cite{Obuljen:2017jiy}. We see that $\sigma(\Neff) \sim 0.015$ is achievable through the measurement of~$P(k)$ in either survey for $\kmax = \SI{0.5}{\hPerMpc}$, although the improvement with~CVL is more robust to marginalization.

\section{Measurements of the Phase Shift}
\label{sec:phaseShift}
In the previous section, we showed how much the combination of future CMB and LSS~measurements can improve the sensitivity to extra relativistic species. The dominant source of improvement came from the broadband shape of the power spectra,~$\Pnw(k)$, rather than the BAO~spectrum,~$\Pw(k)$. Nevertheless, the shift of the acoustic peaks is a particularly robust signature of free-streaming, relativistic species~\cite{Baumann:2017lmt} and is therefore an interesting observable in its own right. In this section, we will isolate the signal coming from the phase shift and forecast our ability to measure it in future surveys. Measuring the BAO phase shift provides an independent test of pre-recombination physics in a low-redshift observable. This could be used to shed light on possible discrepancies between low- and high-redshift measurements or as a discovery channel for exotic new physics.

\subsection{Isolating the Phase Shift}
\label{sec:template}
The BAO feature in Fourier space can be written as
\beq
O_\mathrm{lin}(k) = A(k) \sin\! \left[ \alpha^{-1} r_s k + \phi(k) \right] ,
\eeq 
where the parameter~$\alpha$ represents changes in the BAO scale~$r_s$, and the amplitude modulation~$A(k)$ and the phase shift~$\phi(k)$ encode a number of physical effects that alter the time evolution of the baryons. While~$\alpha$ and~$A(k)$ are implicit functions of redshift,~$\phi(k)$ is redshift independent. Relativistic species are the unique source of a constant shift in the locations of the BAO peaks in the limit of large wavenumbers, i.e.~$\phi(k \to \infty) = \phi_\infty$ (cf.~\textsection\ref{sec:examples}). In practice, however, the measurement of the BAO spectrum occurs over a relatively small range of scales with a small number of (damped) acoustic oscillations. On these scales, the $k$-dependence of the shift can be relevant. Furthermore, additional $k$-dependent shifts from other cosmological parameters may also have to be taken into account~\cite{Pan:2016zla}.\medskip

To measure the phase shift~$\phi(k)$, we will construct a template for the $k$-dependence as a function of the relevant parameters. For small variations around their fiducial values, it is a good approximation to treat the shifts arising from each cosmological parameter independently. By varying one parameter at a time and measuring the change in the peak locations, we can construct a template $\phi(k) = \sum_i \beta_i(\vec{\theta}\hskip1pt) f_i(k)$. For $\Lambda\mathrm{CDM}$+$\Neff$, the parameters $\As$, $\ns$, and $\tau$ do not affect the evolution of the baryons prior to recombination and, therefore, do not change the phase of the oscillations. The parameters~$\omega_b$ and~$\theta_s$ do alter the BAO~spectrum, but are effectively negligible for any realistic parameter range. The shifts induced by~$\omega_c$ and~$\Neff$, on the other hand, can be significant.\medskip

The parameter that is most independent of~$\Neff$ is not the dark matter density~$\omega_c$, but the scale factor at the time of matter-radiation equality,~$\aeq$. Since CMB~data essentially fixes~$\aeq$, our template model can be reduced to
\beq
\phi(k) = \beta(\Neff) f(k) \, ,	\label{eq:PhaseTemplate}
\eeq
namely the shift induced by changing~$\Neff$ at fixed~$\aeq$. This is the same choice made by Follin et~al.~\cite{Follin:2015hya} in their CMB measurement of the phase shift. Fixing~$\aeq$ also reproduces the expected constant phase shift at large wavenumbers. The template for the phase shift at fixed~$\omega_c$, in contrast, does not approach a constant at large wavenumbers, which implies that the change of~$\aeq$ to maintain constant~$\omega_c$ is introducing a phase shift of comparable size to the constant shift induced by varying~$\Neff$. For our applications, this additional shift plays no role, but it could be useful in future investigations.\medskip

\begin{figure}[t!]
	\includegraphics{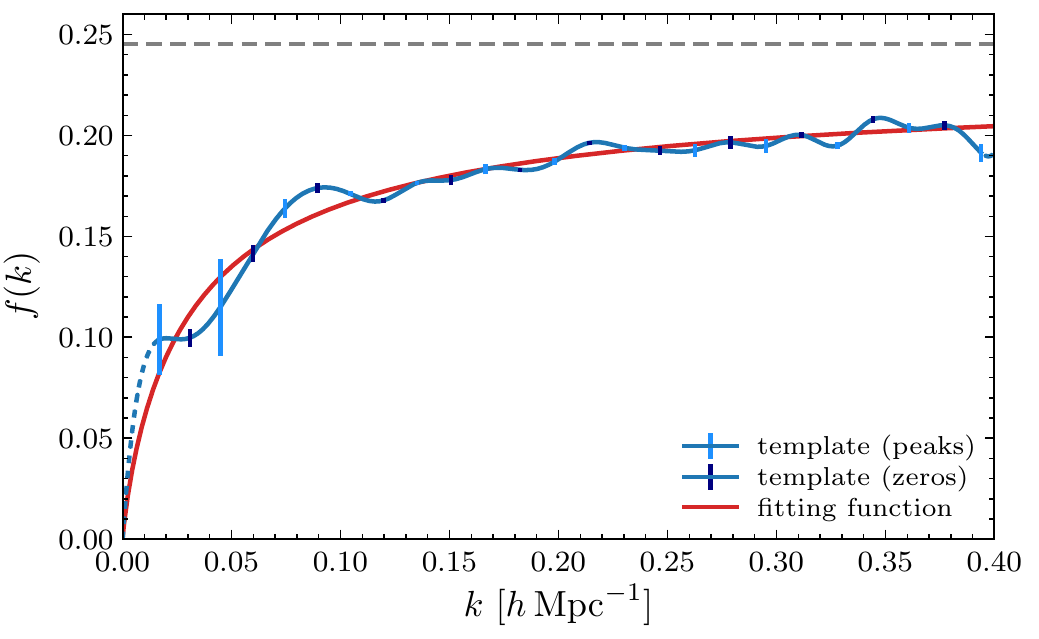}
	\caption{Template of the phase shift $f(k)$ as defined in~\eqref{eq:PhaseTemplate}. The numerical phase shifts~(\textcolor{pyBlue}{blue}) were obtained by sampling from 100 different cosmologies with varying $\Neff$ and rescaling by~$\beta(\Neff)$ as defined in~\eqref{eq:phi_norm}. The bars indicate the standard deviation in these measurements at the positions of the peaks~(\textcolor{pyLightBlue}{light blue}) and zeros~(\textcolor{pyDarkBlue}{dark blue}) compared to the fiducial BAO~spectrum. The \textcolor{pyRed}{red} line shows the fitting function defined in~\eqref{eq:phase_fit}. The dashed \textcolor{pyGray}{gray} line is the analytic approximation~\eqref{eq:thetaX} to the constant phase shift.}
	\label{fig:phaseShiftTemplate}
\end{figure}

We describe the measurement of the phase shift and the construction of the template in Appendix~\ref{app:broadband+phaseShiftExtraction}. In short, we determine the shift in the locations of the peaks/troughs and zeros of the BAO~spectrum compared to the fiducial cosmology with $\Neff=3.046$ and sample 100~different cosmologies with varying~$\Neff$ at fixed~$\aeq$. It is convenient to normalize the template~$f(k)$ such that $\beta=0\text{ and }1$ for $\Neff = 0\text{ and }3.046$, respectively. In Figure~\ref{fig:phaseShiftTemplate},%
we illustrate how the peaks/troughs and zeros of the BAO spectrum change in response to this variation in~$\Neff$. We see that the phase shift created by~$\Neff$ approaches a constant at large wavenumbers in line with physical expectations.\medskip

The measurement of the phase shift is challenging because it requires a very accurate model of the no-wiggle spectrum~$\Pnw(k)$ across a wide range of cosmological parameters. Errors in~$\Pnw(k)$ effectively change the functions~$A(k)$ and~$B(k)$ in~\eqref{eq:O} and lead to errors in the measurement of the BAO~peaks and zeros, respectively. The small size of the phase shift in Fig.~\ref{fig:phaseShiftTemplate} only exacerbates this problem. Fortunately, while the template is difficult to generate, our forecasts using the template are very stable. Furthermore, the template is well approximated by a simple fitting function,
\beq
f(k) = \frac{\phi_\infty}{1+(k_\star/k)^\xi} \, ,	\label{eq:phase_fit}
\eeq
where $\phi_\infty = 0.227$, $k_\star = \SI{0.0324}{\hPerMpc}$ and $\xi = 0.872$ were obtained by a weighted fitting procedure. From the analytic treatment in~\textsection\ref{sec:examples}, we expect $\phi_\infty = 0.191\pi\,\epsilon_\mathrm{fid} + \mathcal{O}(\epsilon_\mathrm{fid}^2) \approx 0.245$ to linear order, where $\epsilon(\Neff) = \Neff/(a_\nu +\Neff)$ is a measure of the excess radiation density, $(\rho_r-\rho_\gamma)/\rho_r$, with $a_\nu \approx 4.40$ as introduced in~\eqref{eq:anu}. This approximation overestimates the value obtained using the fitting formula by about~\SI{8}{\percent}, which is consistent with the expected corrections from higher orders in $\epsilon_\mathrm{fid} = \epsilon_\nu \approx0.41$. Around $k\sim\SI{0.1}{\hPerMpc}$, where BOSS and DESI have the largest signal-to-noise ratio, the relative difference is almost~\SI{50}{\percent}, which makes it evident that the offset from the analytic approximation has to be taken into account in an analysis such as the one proposed below, whereas the precise shape of the template plays a sub-dominant role. We also note that this template is basically independent of changes to the BAO~scale~$r_s$, for example due to changes in the dark matter density.\medskip

We use the measured phase template to write the BAO~spectrum in terms of the spectrum in the fiducial cosmology:
\beq
O(k) = O_\mathrm{fid} \big(\alpha^{-1} k + (\beta-1)\, f(k)/r_s^\mathrm{fid} \big) \, ,	\label{eq:Oscale}
\eeq
where~$\alpha\equiv\alpha(z_i)$ takes an independent value in each redshift bin centred around~$z_i$ and~$\beta$ is a single parameter for the entire survey. A measurement of~$\alpha(z_i)$ and~$\beta$ can then be translated into constraints on cosmological parameters using
\begin{align}
	\alpha(\vec{\theta}; z) &\equiv q\, r_s^\mathrm{fid}/r_s = \left[D_V(z)/r_s\right] / \left[D_V(z)/r_s\right]_\mathrm{fid} \, ,	\\
	\beta(\Neff) 			&\equiv {\epsilon/\epsilon_\mathrm{fid}} \, ,	\label{eq:phi_norm}
\end{align}
where the parameters~$q$ and~$D_V$ were introduced in~\textsection\ref{sec:baoModelling}. With this normalization, the largest possible phase shift due to~$\Neff$ is given by $\beta(\Neff\to\infty) = 2.45$.\medskip

In~\textsection\ref{sec:validation_comparison}, we will show that the forecasts produced using only these templates are in agreement with the forecasts using the full BAO spectrum. From a measurement of $\beta>0$, one gets a constraint on~$\Neff$ that is only associated to the size of the phase shift. This approach is analogous to the template-based measurement of the phase shift in the CMB by Follin et al.~\cite{Follin:2015hya}. The measurement of~$\Neff$ from the phase alone ignores the effects of~$\Neff$ on~$\alpha$, but has the advantage that any detection is unambiguously\hskip1pt\footnote{We have explicitly checked that our template gives an unbiased measurement of $\beta$, cf.~\textsection\ref{sec:validation_analysis}. In particular, we have verified that we reproduce $\beta\approx0$ for a cosmology with $\Neff=0$.} a measurement of free-streaming relativistic particles.\medskip

We will also be interested in the measurement of~$\beta$ when a prior on~$\alpha$ is included, e.g.\ from the~CMB.\footnote{We also indirectly use the CMB~data to constrain other cosmological parameters, in particular the scale factor at matter-radiation equality~$\aeq$, so that we can ignore any additional phase shifts not associated with~$\Neff$.} In a given cosmological model, the parameter~$\alpha$ is fully determined by the set of cosmological parameters, $\alpha = \alpha(\vec\theta\hskip1pt)$. As the~$\alpha(z_i)$ inferred from the~CMB are correlated between the $n$~redshift bins of a galaxy survey and~$n$~is in general larger than the number of cosmological parameters, we compute the $n$-dimensional inverse covariance matrix according to $C_\alpha^{-1} = A^T F A$, where~$F$~is the Fisher matrix and~$A$~is the pseudo-inverse of~$\nabla_{\!\vec\theta}\,\vec\alpha$. We use the CMB~Fisher matrices for the $\Lambda\mathrm{CDM}$+$\Neff$ cosmology as in Section~\ref{sec:forecast}. We can then impose the $\alpha(z_i)$-prior on the redshift-binned likelihood function $\mathcal{L}(\alpha,\beta;z_i)$ according to $\mathcal{L}(\beta) \propto \int\prod_{z_i}\!\d\alpha_i\,\prod_{z_i}\!\mathcal{L}(\alpha_i,\beta;z_i)\,\pi(\alpha_1,\ldots,\alpha_n)$, where $\alpha_i \equiv \alpha(z_i)$ and~$\pi$~is the $n$-dimensional Gaussian prior with covariance matrix $C_\alpha$. The observed posterior distribution for $\alpha(z_i)$ could also be constructed by evaluating~$\alpha(z_i)$ for each point in a given CMB~Markov chain.

\subsection{Constraints from Planned Surveys}
\label{sec:phase_planned}
We will now show how well the phase shift can be measured in planned galaxy surveys. It is useful to first understand the parameter space $\alpha$-$\beta$ without imposing a prior on~$\alpha$. Both parameters affect the locations of the acoustic peaks and are therefore quite degenerate. We will use likelihood-based forecasts to ensure accuracy. We will confirm that the posterior \mbox{distributions\hskip1pt\footnotemark}\footnotetext{Since we assume flat priors for the parameters, we can identify the posteriors with the likelihoods.} of~$\alpha$ and~$\beta$ are Gaussian, while the constraints on~$\Neff$ derived from this parametrization are significantly non-Gaussian. This suggests that a Fisher matrix forecast in terms of~$\alpha$ and~$\beta$ would be more reliable than one that starts directly from~$\Neff$.\medskip

We define the phase shift relative to the fiducial model with $\Neff = 3.046$. The broadband spectrum for the fiducial model can be isolated by using the method in Appendix~\ref{app:broadband+phaseShiftExtraction} or through the use of a fitting function along the lines of~\cite{Beutler:2016ixs}. These methods generate the BAO spectrum $O_\mathrm{fid}(k)$ and hence~$O(k)$ via~\eqref{eq:Oscale}. We compute the log-likelihood using the same noise and modelling as in the Fisher matrix~\eqref{eq:baoFisherMatrix}.

\begin{figure}[t]
	\includegraphics{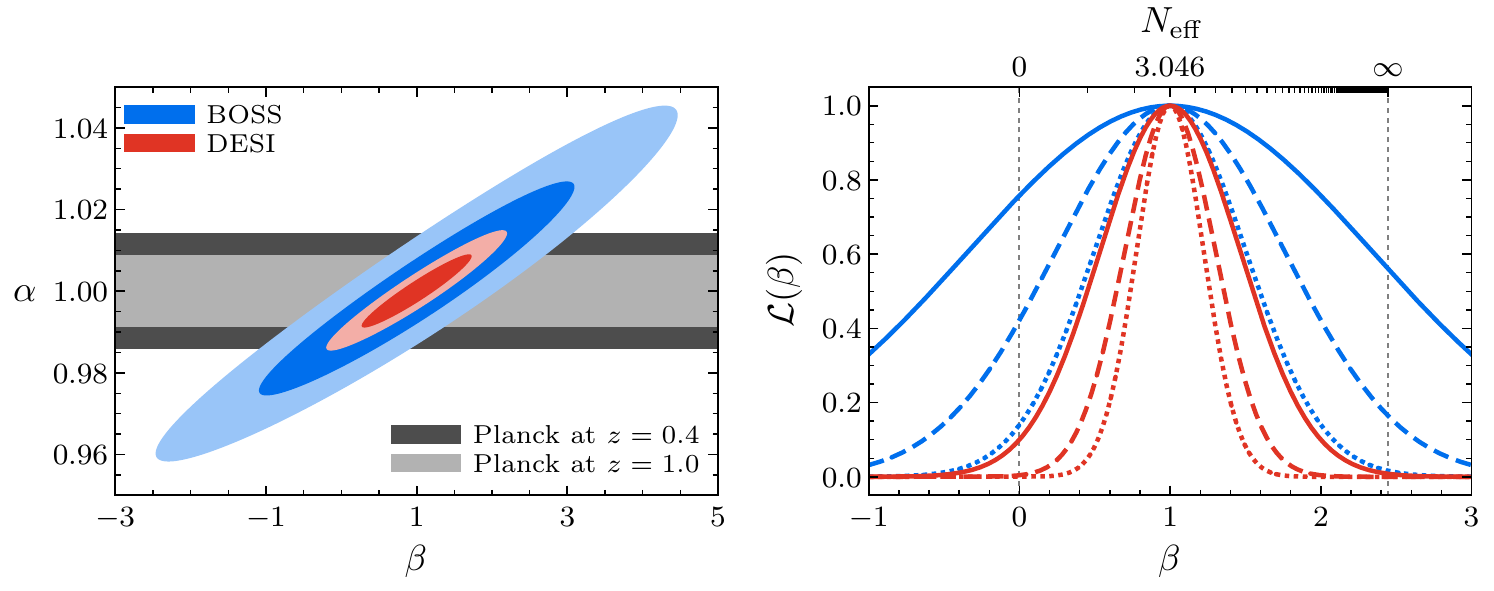}
	\caption{\textit{Left:} Contours showing~$1\sigma$ and~$2\sigma$ exclusions in the $\alpha$-$\beta$ plane for BOSS and DESI. For purpose of illustration, we have reduced these surveys to a single redshift bin (and therefore a single $\alpha$-parameter). The gray bands indicate Planck priors for~$\alpha$ assuming the median redshift is $z=0.4\text{ and }1.0$ for BOSS and DESI, respectively. \textit{Right:} One-dimensional posterior distributions of~$\beta$ for BOSS and DESI. The dashed and dotted lines indicate the use of a redshift-dependent CMB~prior on~$\alpha$ from Planck and CMB-S4, respectively.}
	\label{fig:likelihoodAlphaBeta}
\end{figure}

Forecasts for the one- and two-dimensional posteriors are shown in Fig.~\ref{fig:likelihoodAlphaBeta}%
for both BOSS and DESI. We see that for both surveys the posterior distributions are Gaussian. The best-fit Gaussian for BOSS and DESI has $\sigma(\beta) = 1.3\text{ and }0.47$, respectively, which corresponds to a rejection of $\beta=0$ at \SI{77}{\percent} and \SI{98}{\percent}~confidence. Clearly, BOSS cannot exclude $\beta=0$ (and hence $\Neff = 0$) without any prior information from the~CMB. Since the weakness of the constraint on $\beta$ is driven by the degeneracy with~$\alpha$ (see the left panel in Fig.~\ref{fig:likelihoodAlphaBeta}), we expect to get significant improvements in the constraints on $\beta$ after imposing a CMB~prior on $\alpha$. Inspection of the two-dimensional contours already shows that we will sizeably limit the range of~$\beta$. The posterior distribution with the prior from Planck (CMB-S4) is shown in the right panel of Fig.~\ref{fig:likelihoodAlphaBeta}. For~BOSS, we find $\sigma(\beta) = 0.76$~$(0.50)$ which implies that $\beta > 0$ at \SI{81}{\percent}~(\SI{95}{\percent}) confidence. For~DESI, we should find strong evidence for a phase shift with $\sigma(\beta) = 0.30$ $(0.22)$ which excludes $\beta = 0$ at $3.5\hskip1pt\sigma$ ($4.6\hskip1pt\sigma$).\medskip

\begin{figure}[b!]
	\includegraphics{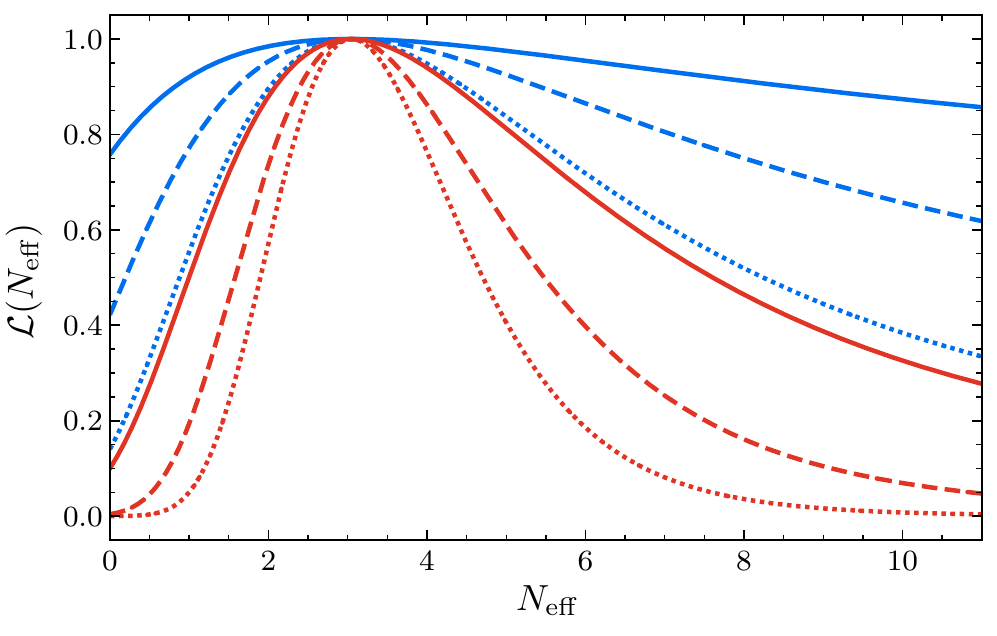}
	\caption{Posterior distributions of $\Neff$ for BOSS~(\textcolor{pyBlue2}{blue}) and DESI~(\textcolor{pyRed2}{red}) derived from the phase shift in the BAO~spectrum, i.e.\ via the measurement of~$\beta$. The dashed and dotted lines indicate that a redshift-dependent CMB prior on~$\alpha$ has been imposed using Planck and CMB-S4, respectively.}
	\label{fig:likelihoodNeff}
\end{figure}

To translate these results into constraints on~$\Neff$, we use the relationship between~$\beta$ and~$\Neff$ given in~\eqref{eq:phi_norm}. This map is non-linear over the measured range of~$\beta$ and we therefore anticipate the posteriors to be non-Gaussian. The derived $\Neff$\hskip1pt-posteriors in Fig.~\ref{fig:likelihoodNeff}%
indeed show a highly non-Gaussian distribution. As anticipated from the $\beta$-posterior for BOSS, the constraints on~$\Neff$ are relatively weak without imposing a Planck prior on~$\alpha$.

We also see that the constraining power is significantly weaker at bounding large values of~$\Neff$ than small ones. This asymmetry is simply a reflection of the fact that increasing~$\Neff$ does not produce proportionally larger phases shifts. The same asymmetry was also seen in the CMB~constraints of Follin et al.~\cite{Follin:2015hya}, likely for the same reason. Recall that we have an upper limit on the phase shift of $\beta < 2.45$, which is saturated for $\Neff \to \infty$. In practice, this means that for $\Neff \gg a_\nu \approx 4.40$, we will have an equal likelihood\hskip1pt\footnote{Realistic values of~$\Neff$ are not quite in the asymptotic regime, but still show the weakened distinguishing power for larger~$\Neff$.} for every value of~$\Neff$ because they produce identical spectra. As a result, a flat prior on~$\Neff$ (rather than~$\beta$) will lead to ill-defined results because the integral $\int^{\infty}\!\d\Neff\,\mathcal{L}(\Neff)$ will diverge. On the other hand, for highly-significant detections of $\beta>0$, a flat prior over any reasonable range of~$\Neff$ will produce stable results. We are not quite in this regime with BOSS, which is why we will only quote constraints on~$\beta$.\medskip

Table~\ref{tab:betaConstraints}%
\begin{table}[t]
	\begin{tabular}{l S[table-format=2.3] S[table-format=2.3] S[table-format=2.3] S[table-format=2.3] S[table-format=2.3] S[table-format=2.3]}
			\toprule
							& \multicolumn{4}{c}{spectroscopic}				& \multicolumn{2}{c}{photometric}	\\
			\cmidrule(lr){2-5} \cmidrule(lr){6-7}
							& {BOSS}	& {eBOSS}	& {DESI}	& {Euclid}	& {DES}		& {LSST}				\\
			\midrule[0.065em]
		BAO					& 1.3		& 1.0		& 0.47		& 0.40		& 2.6		& 1.0					\\
			\midrule[0.065em]
		+ Planck prior		& 0.76		& 0.70		& 0.30		& 0.26		& 1.1		& 0.50					\\
			\midrule[0.065em]
		+ CMB-S4 prior		& 0.50		& 0.48		& 0.22		& 0.19		& 1.0		& 0.42					\\
			\bottomrule
	\end{tabular}
	\caption{Forecasted~$1\sigma$ constraints on the amplitude of the phase shift~$\beta$ for current and future LSS~experiments. We also show the constraints on~$\beta$ after imposing a redshift-dependent prior on the BAO parameter~$\alpha$ from Planck and CMB-S4.}
	\label{tab:betaConstraints}
\end{table}
shows the projected constraints on~$\beta$ for a variety of planned surveys with and without priors from the~CMB. We see that roughly a factor of three improvement can be achieved in spectroscopic surveys going from BOSS to Euclid. Both DESI and Euclid should have sufficient sensitivity to reach a more than $5\sigma$~exclusion of $\beta=0$ when imposing a Planck prior. As before, galaxy clustering measurements in photometric surveys do not lead to competitive constraints as they are effectively two-dimensional on the relevant scales.

\subsection{Constraints from Future Surveys}
\label{sec:phase_future}
Given the robustness of the phase shift as a probe of light relics, a high-significance detection of the phase shift in~LSS would be a valuable piece of cosmological information. We have seen that current and planned surveys can detect the phase shift, but are not expected to produce constraints on~$\Neff$ that are competitive with those from the~CMB. It is natural to ask if future surveys can reach this level of sensitivity.\medskip

\begin{figure}[t]
	\includegraphics{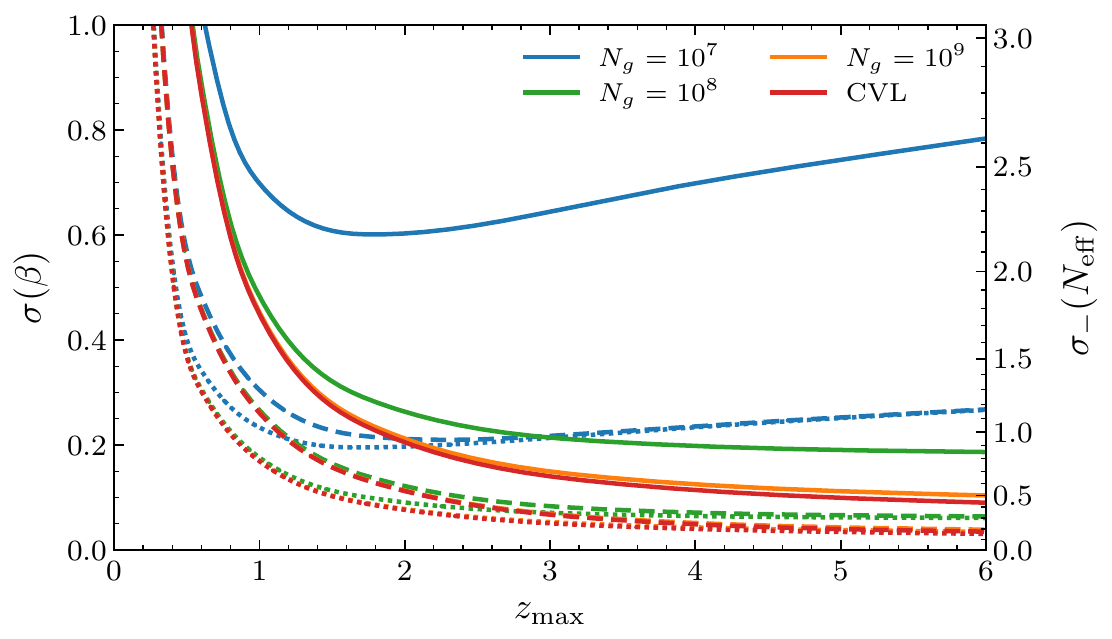}
	\caption{Future constraints on the amplitude of the phase shift~$\beta$ as a function of~$\zmax$ and~$N_g$, assuming $\fsky=0.5$. The dashed and dotted lines indicate that a CMB~prior on~$\alpha$ has been imposed using Planck and CMB-S4, respectively. The corresponding $1\sigma$~lower limit on~$\Neff$, which is $\Neff = 3.046 - \sigma_-(\Neff)$, is computed by inverting~\eqref{eq:phi_norm} and indicated by the right axis.}
	\label{fig:posteriorBetaFuture}
\end{figure}

Like the measurement of the BAO~scale, the measurement of the phase requires large signal-to-noise for $\SI{0.1}{\hPerMpc} \lesssim k \lesssim \SI{0.3}{\hPerMpc}$. As long as the number density is sufficiently large to keep the shot noise below cosmic variance, we gain primarily by increasing~$\zmax$ to achieve larger survey volumes. At larger levels of the shot noise, we only measure a few peak locations well which increases the degeneracy between~$\alpha$ and~$\beta$. Figure~\ref{fig:posteriorBetaFuture}%
shows results for a variety of possible survey configurations. As before, the constraints on~$\beta$ can be mapped into constraints on~$\Neff$ using~\eqref{eq:phi_norm}. We see that with \num{e8}~objects and $\zmax > 3$, we consistently obtain $\sigma(\Neff) < 0.5$~$(1.0)$ with (without) a prior on~$\alpha$ included.\medskip

To put these results into context, the measurement of Follin et al.~\cite{Follin:2015hya} of $N_\mathrm{eff}^\phi = 2.3_{-0.4}^{+1.1}$ from the Planck TT~spectrum is comparable to a survey with $N_g=\num{e9}$ objects out to redshift $\zmax=3$. Follin et al.~also forecasted $\sigma(N_\mathrm{eff}^\phi)=0.41$ for Planck~TT+TE+EE which is near the sensitivity of future LSS~surveys when increasing the redshift range to $\zmax=6$. Reaching this level of sensitivity will be extremely challenging with an optical survey, but could potentially be achieved with \SI{21}{cm}~intensity mapping~\cite{Obuljen:2017jiy}.

\section{Summary}
\label{sec:conclusions_bao-forecast}
Large-scale structure surveys are an untapped resource in the search for light relics of the hot big bang. The growing statistical power of these surveys will make them competitive with the~CMB in terms of the constraints they will provide on a broad range of cosmological parameters. Moreover, the combination of CMB and LSS~observations will allow powerful and robust tests of the physical laws that determined the structure and evolution of the early universe.\medskip

\begin{figure}[t!]
	\includegraphics{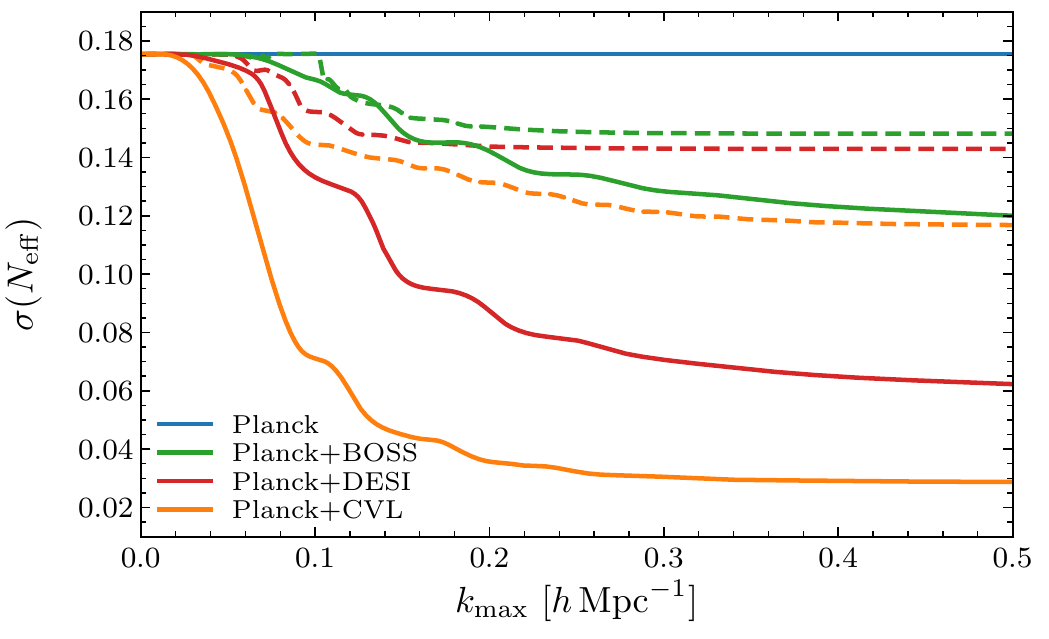}
	\caption{Sensitivity of planned and future LSS~surveys to~$\Neff$ using the galaxy power spectrum~(solid) and the BAO~spectrum~(dashed) marginalized over two bias parameters, $b_{m\leq1}$.}
	\label{fig:neffPkBAO}
\end{figure}

In this chapter, we have explored the potential impact of LSS surveys on measurements of the parameter~$\Neff$. We have found that the dominant statistical impact of future surveys lies in the shape of the galaxy power spectrum. The distribution of dark matter in the universe is altered through the gravitational influence of the free-streaming radiation, leading to changes in the shape of the power spectrum that can be detected at high significance. A summary of the reach of selected planned and future surveys is given in Fig.~\ref{fig:neffPkBAO}.%
We see that BOSS and DESI can extend results significantly beyond the current CMB~constraints. Futuristic surveys combined with a future CMB-S4~mission could achieve $\sigma(\Neff) \sim 0.015$, which is close to reaching the target of $\Delta\Neff = 0.027$ at a significance of~$2\sigma$.\medskip

Future LSS surveys will also be able to detect the coherent shift in the peak locations of the BAO spectrum. This would be an intriguing measurement as this phase shift is a highly robust and unambiguous probe of light relics and the cosmic neutrino background~\cite{Baumann:2017lmt}. The fact that the phase shift should agree between the~CMB and~BAO measurements is a highly non-trivial consequence of physics both before and after recombination, and could be an interesting test of exotic extensions of~$\Lambda\mathrm{CDM}$ without requiring the~CMB as an anchor. Being a new low-redshift observable, improved measurements of the BAO~phase shift may therefore play a useful role in elucidating apparent low-$z$/high-$z$ discrepancies in some cosmological data~\cite{Freedman:2017yms}.\medskip

In the future, the combination of different cosmological observations might significantly advance our insights into fundamental physics. In this chapter, we observed that CMB and LSS~observations can complement each other by providing tighter as well as more robust constraints on~$\Neff$. In addition, we demonstrated that the BAO~spectrum encodes cosmological information beyond the acoustic scale which can be extracted reliably. A broader exploration will likely reveal more targets that benefit from this complementarity.
	\chapter{Measurement of Neutrinos in the BAO Spectrum}
\label{chap:bao-neutrinos}
The existence of the cosmic neutrino background is a remarkable prediction of the hot big bang model. These neutrinos were a dominant component of the energy density in the early universe and, therefore, played an important role in the evolution of cosmological perturbations. In particular, fluctuations in the neutrino density produced a distinct shift in the temporal phase of sound waves in the primordial plasma, which has recently been detected in the cosmic microwave background~\cite{Follin:2015hya}. In the previous chapter, we proposed a new analysis of the baryon acoustic oscillation signal which extends the conventional BAO~analysis presented in~\cite{Beutler:2016ixs, Alam:2016hwk} by including the amplitude of the neutrino-induced phase shift as a free parameter in addition to the BAO~scale. In this chapter, we report on the first measurement of this neutrino-induced phase shift in the spectrum of baryon acoustic oscillations of the BOSS DR12~data, based on this novel approach. Constraining the acoustic scale using Planck data while marginalizing over the effects of neutrinos in the~CMB, we find evidence for a non-zero phase shift at greater than \SI{95}{\percent}~confidence. We also demonstrate the robustness of this result in simulations and forecasts. Besides providing a new measurement of the cosmic neutrino background, this work is the first application of the BAO~feature beyond its application as a standard ruler and to early universe physics.\medskip

The outline of this chapter is as follows. In Section~\ref{sec:bao-neutrinos_introduction}, we briefly review the phase shift induced by cosmic neutrinos and lay out the modified BAO~analysis pipeline which employs the proposed template-based approach. In Section~\ref{sec:validation}, we check the new method in likelihood-based forecasts and on mock catalogues. In Section~\ref{sec:bossAnalysis}, we measure the amplitude of the phase shift in Fourier space and find statistically significant evidence for the presence of the phase shift in the BOSS~DR12 dataset, in line with expectations from the mocks and forecasts. In Section~\ref{fig:cs_analysis}, we describe and perform our alternative analysis in configuration space with results that are broadly compatible with the main Fourier-space constraints. We conclude, in Section~\ref{sec:conclusions_bao-neutrinos}, with a brief summary of our results and an outlook on future improvements of our measurement.

\section{Modified BAO Analysis}
\label{sec:bao-neutrinos_introduction}
A variety of experiments have been proposed to observe the~C$\nu$B directly~\cite{Weinberg:1962zza, Ringwald:2009bg, Betts:2013uya}. However, these experiments are very challenging because neutrino interactions at low energies are extremely weak. Cosmological observations of the radiation density in the early universe, on the other hand, are making an increasingly strong case that the~C$\nu$B has already been detected indirectly (cf.~e.g.~\textsection\ref{sec:Neff}). As we showed in Chapters~\ref{chap:cmb-phases} and~\ref{chap:bao-forecast}, the effect of neutrinos on the perturbations in the primordial plasma, which have been detected in the CMB, is a particularly robust probe of the~C$\nu$B. An interesting feature of the phase shift in the BAO~spectrum is the fact that it is robust to the effects of non-linear gravitational evolution~\cite{Baumann:2017lmt}. This provides the rare opportunity of extracting a signature of primordial physics which is immune to many of the uncertainties that inflict the modelling of non-linear effects in large-scale structure observables. The BAO~phase shift therefore presents a clean imprint of early universe physics and the~C$\nu$B. In this chapter, we will provide its first measurement.\medskip

The analysis of the (isotropic) BAO~signal is usually reduced to the measurement of a single parameter, the BAO~scale. In this chapter, we consider the extension of the conventional BAO~analysis that we proposed in Chapter~\ref{chap:bao-forecast} and which takes the information contained in the phase of the spectrum into account. Since we only observe a finite number of modes, some of which evolved primarily during matter domination, we cannot simply search for a constant phase shift in the data. This means that recovering all of the accessible information requires an accurate momentum-dependent template for the phase shift that applies to the modes of interest.

\begin{figure}[t]
	\includegraphics{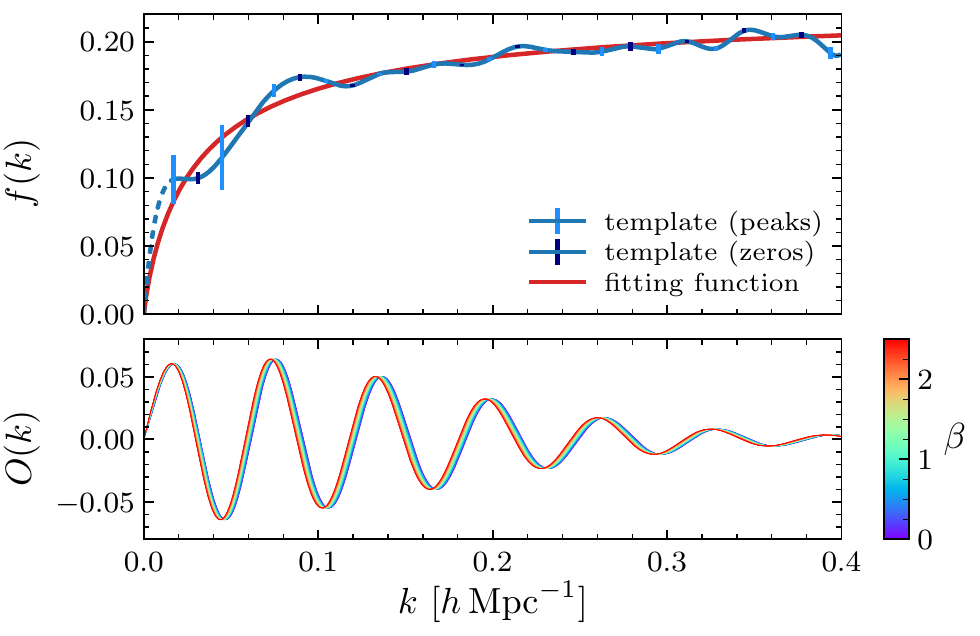}
	\caption{Template of the phase shift~$f(k)$ as defined in~\eqref{eq:PhaseTemplate}, with the fitting function~\eqref{eq:phase_fit} shown as the red curve. The bottom panel displays the linear BAO~spectrum~$O(k)$ as a function of the amplitude of the phase shift~$\beta$.}
	\label{fig:phaseShiftTemplate2}
\end{figure}

The phase shift (relative to $\Neff=0$) can be written as $\phi(k) \equiv \beta(\Neff) f(k)$, where~$\beta$ is the amplitude of the phase shift and~$f(k)$ is a function that encodes its momentum dependence (see Section~\ref{sec:phaseShift} and Appendix~\ref{app:broadband+phaseShiftExtraction} for further details). The amplitude is proportional to the fractional neutrino density, $\epsilon_\nu(\Neff) \approx \Neff/(4.4+\Neff)$, and we have chosen the normalization so that $\beta = 0\text{ and }1$ correspond to $\Neff = 0\text{ and }3.046$, respectively. We note that the parameter $\beta$ is a non-linear function of~$\Neff$ that asymptotes to $\beta \to 2.45$ for $\Neff \to \infty$ because adding more neutrinos does not change the phase shift when neutrinos become the dominant source of energy density in the universe. The template~$f(k)$ is shown in Fig.~\ref{fig:phaseShiftTemplate2}%
and is well approximated by the fitting function~\eqref{eq:phase_fit}.

The observed BAO spectrum receives various non-linear corrections. We model these contributions as in the standard BAO~analysis, e.g.~\cite{Beutler:2016ixs}, but now introduce the amplitude of the phase shift~$\beta$ as an additional free parameter. Following equation~\eqref{eq:Oscale}, we write the non-linear BAO spectrum as
\beq
O_g(k) \equiv O^\mathrm{fid}_\mathrm{lin}\big(k / \alpha + (\beta-1) f(k)/r_s^\mathrm{fid} \big) \,\mathrm{e}^{-k^2\Sigma_\mathrm{nl}^2 /2}\, ,	\label{eq:ONL}
\eeq
where $O^\mathrm{fid}_\mathrm{lin}(k)$ and~$r_s^\mathrm{fid}$ are the linear BAO~spectrum and the BAO~scale in the fiducial cosmology, which is chosen to be the same as in~\cite{Beutler:2016ixs}.\footnote{The fiducial cosmology is described by the following parameters: $\Omega_b h^2=0.022$, $\Omega_c h^2=0.119$, $h=0.676$, $\As=\num{2.18e-9}$, $\ns=0.96$, $\tau=0.08$, $\Nf=3.046$ and $\sum m_\nu = \SI{0.06}{eV}$.} The exponential factor in~\eqref{eq:ONL} describes the non-linear damping of the BAO~signal after reconstruction~\cite{Eisenstein:2006nk, Padmanabhan:2012hf}. The parameter~$\alpha$ captures the change in the apparent location of the BAO peak due to changes in the acoustic scale and the angular projection,
\beq
\alpha(\Neff) = \frac{D_V(z)\,r_s^\mathrm{fid}}{D_V^\mathrm{fid}(z)\,r_s}\, , \quad \text{with} \quad D_V(z) = \left[(1+z)^2D^2_A(z)\frac{cz}{H(z)}\right]^{1/3}\, .
\eeq
In Section~\ref{sec:validation}, we will show that this model is effectively unbiased in the sense that we recover $\beta \approx 0$ for a universe with $\Neff =0$ even when we assume a fiducial model with $\Neff=3.046$. Moreover, given the template~\eqref{eq:phase_fit}, the modelling is robust to the precise method for extracting~$O^\mathrm{fid}_\mathrm{lin}(k)$ and we will therefore use the same method as~\cite{Beutler:2016ixs}.

We model the non-linear broadband spectrum in each redshift bin as
\beq
P^\mathrm{nw}(k) = B^2 P^\mathrm{nw}_\mathrm{lin}(k) F(k,\Sigma_s) + A(k)\, .
\eeq
This includes two physical parameters: a linear bias parameter,~$B$, and a velocity damping term arising from the non-linear velocity field (``Fingers of God''),
\beq
F(k,\Sigma_s) = \frac{1}{(1 + k^2\Sigma_s^2/2)^2}\, .	\label{eq:FoG}
\eeq
In addition, we have introduced the polynomial function
\beq
A(k) = \frac{a_{1}}{k^3} + \frac{a_{2}}{k^2} + \frac{a_{3}}{k} + a_{4} + a_{5}k^2\, ,	\label{eq:fs_poly}
\eeq
whose coefficients~$a_n$ will be marginalized over. This polynomial does not represent a physical effect, but removes any residual information that is not encoded in the locations of the peaks and zeros of the BAO spectrum. With such a marginalization over broadband effects, our \mbox{$\alpha$-$\beta$}~parametrization contains essentially all of the information of the $\Lambda\mathrm{CDM}$+$\Neff$ cosmology available in the BAO~spectrum (cf.~\textsection\ref{sec:validation_comparison}). The free parameters in this model will be fitted independently in each redshift bin.

To summarize, the measured galaxy power spectrum is described by two cosmological parameters,~$\alpha$ and~$\beta$, and
a number of nuisance parameters. Except for~$\beta$, all parameters are redshift dependent, and we will fit our model to the signal in two independent redshift bins, ($0.2 < z_1 < 0.5$) and ($0.5 < z_3 < 0.75$).\footnote{The middle redshift bin ($0.4 < z_2 < 0.6$), which was used in the BOSS DR12 analysis, carries little additional information on the BAO signal since it overlaps with the other two bins.} In total, our fit to the power spectrum in the range $\SI{0.01}{\h\per\Mpc} < k < \SI{0.3}{\h\per\Mpc}$ therefore has 21~free parameters:
\beq
\beta,\, \alpha_{z_1},\, \alpha_{z_3};\, \{ B_{\mathrm{NGC},z},\, B_{\mathrm{SGC},z},\, \Sigma_{s,z},\,\Sigma_{\mathrm{nl},z},\, a_{n,z} \}_{z_1,z_3}\, ,
\eeq
where we have allowed for independent bias parameters in the North Galactic Cap~(NGC) and the South Galactic Cap~(SGC) as in~\cite{Beutler:2016ixs}. We generally employ flat priors for all parameters, in particular~$\beta$.\footnote{We note that the choice of a flat prior on~$\beta$, rather than~$\Neff$, weakens the statistical significance of the $\beta>0$ measurement compared to the analyses in the CMB, which use~$\Neff$. In other words, a flat prior on~$\Neff$ would lead to stronger constraints on the phase shift and, therefore, the~C$\nu$B.} We require the~$\alpha_z$ parameters to be between $0.8$~and~$1.2$, and the damping scales, $\Sigma_{s,z}$~and~$\Sigma_{\mathrm{nl},z}$, to be between $0$~and~$\SI{20}{\per\h\Mpc}$, while no explicit priors are imposed on the bias parameters~$B_{i,z}$, the phase parameter~$\beta$ or the polynomial terms~$a_{n,z}$. Our goal is to measure the new parameter~$\beta$, while marginalizing over all other parameters.

\section{Validation of the Method}
\label{sec:validation}
Before applying our analysis pipeline to the BOSS~data, we perform several checks and show that the new method provides reliable and consistent results. First, we explicitly demonstrate that the advocated template-based approach captures most of the information on neutrinos in the BAO~spectrum~(\textsection\ref{sec:validation_comparison}). Second, we establish in forecasts that the phase shift inferred in analyses employing the BAO~spectrum~\eqref{eq:ONL} does not depend on the employed BAO~extraction and is unbiased, in the sense that it correctly recovers the input value of the phase amplitude even if a different fiducial cosmology is assumed~(\textsection\ref{sec:validation_analysis}). Finally, we validate our modified BAO analysis using mock catalogues created for the BOSS~DR12~analysis~(\textsection\ref{sec:validation_mocks}).

\subsection{Comparison to Parameter-Based Approach}
\label{sec:validation_comparison}
We have suggested the use of a phase template to characterise the effect of neutrinos. This is a natural choice as the phase shift is the physical effect we wish to isolate. One might however worry that the template~\eqref{eq:PhaseTemplate} does not capture the entire relevant information in the BAO spectrum. For this purpose, it is instructive to compare the results of our template-based forecasts of Section~\ref{sec:phaseShift} to a more direct parameter-based approach to isolating the phase shift. In the parameter-based approach, we define two new parameters~$\thetasLSS$ and~$\NeffLSS$ that play the role of~$\theta_s$ and~$\Neff$ in the BAO~signal, but are taken to be independent of the same parameters in the~CMB. We will then fix all remaining cosmological parameters in the BAO~spectrum using the~CMB, except the physical cold dark matter density~$\omega_c$ which we traded for the scale factor at matter-radiation equality~$\aeq$. As with our template extraction, holding~$\aeq$ fixed ensures that the phase shift approaches a constant at large wavenumbers, whose value is determined by~$\NeffLSS$. Beside measuring the phase of the BAO~signal, the parameter~$\NeffLSS$ also contributes to the scale parameter~$\alpha$ and could therefore be constrained by the standard BAO-scale measurement if all the other cosmological parameters are fixed to their Planck best-fit values. Introducing the additional parameter~$\thetasLSS$ gives enough freedom to remove this effect and any constraint on~$\NeffLSS$ must be coming from the phase shift alone. This is analogous to isolating the phase shift in the~CMB by marginalizing over~$Y_p$ or any other parameters that are degenerate with the $\Neff$\hskip1pt-induced change to the damping tail. We will confirm this expectation in our forecasts.\medskip

Typically, the advantage of the parameter-based approach is that it is easy to implement. However, in this case, we found it more difficult to set up reliably. The phase shift ultimately controls the breaking of the degeneracy between~$\thetasLSS$ and~$\NeffLSS$ and, as we discussed in~\S\ref{sec:template},~$\Pnw(k)$ must therefore be determined sufficiently accurately to not produce errors in this shift. To compute the likelihood directly, we must re-compute~$\Pnw(k)$ for every value of the cosmological parameters. Producing stable results for the BAO spectrum across a wide range of parameters can be very computationally expensive and technically challenging. Simpler and faster methods can work well near the fiducial cosmology (such as the use of a fitting function), but often produce noisy results as the parameters vary significantly and typically underestimate the likelihood as we depart from the fiducial cosmology (and, hence, overestimate the constraining power).\medskip

Despite the challenge presented by a parameter-based approach, it has the advantage that it should capture all of the cosmological information available. For this reason, it is useful to compare the results of the parameter-based and template-based approaches to see if the template is missing information. Fortunately, we will see that the posterior distributions for~$\NeffLSS$ and~$\thetasLSS$ can be largely reproduced as a derived consequence of the template-based forecasts. From our results of~\textsection\ref{sec:phase_planned}, we should anticipate that the posteriors for~$\NeffLSS$ and~$\thetasLSS$ will be non-Gaussian, and will therefore require the calculation of the likelihood for~$\NeffLSS$ and~$\thetasLSS$ directly (and not only the Fisher matrix). We will follow the same approach as described in~\textsection\ref{sec:phase_planned}. Computing the full likelihood is quite involved, which is the reason why we will assume that the CMB data fixes the other cosmological parameters to their fiducial values, except for $\NeffLSS$ and $\thetasLSS$.\medskip

\begin{figure}[t!]
	\includegraphics{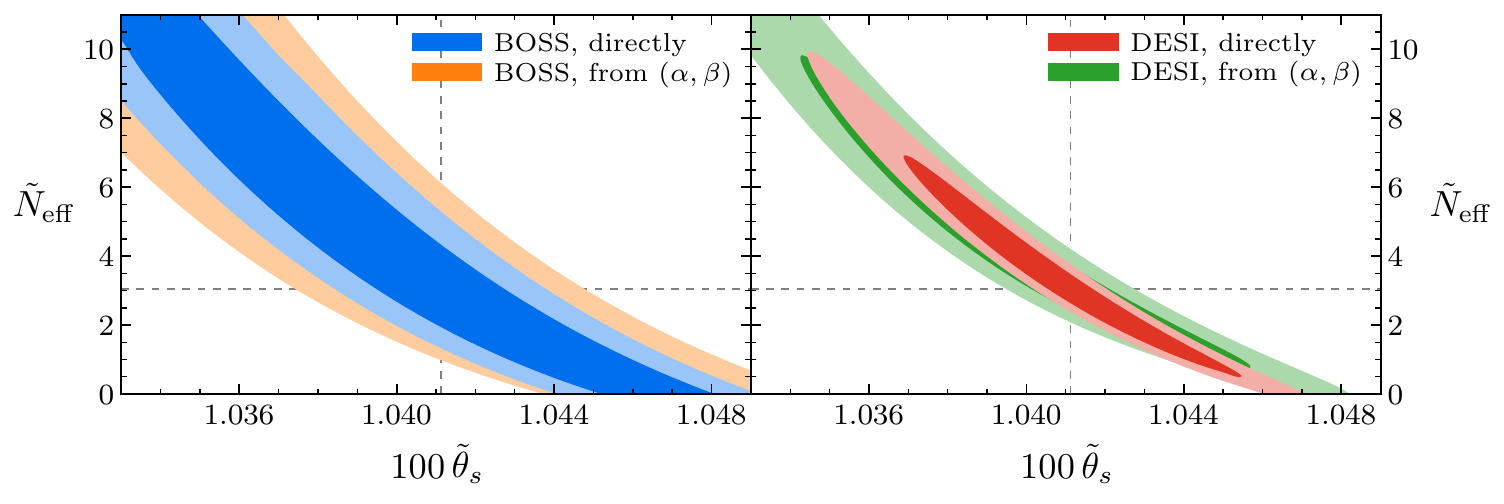}
	\caption{BOSS~(\textit{left}) and DESI~(\textit{right}) two-dimensional~$1\sigma$ and $2\sigma$~contours for~$\NeffLSS$ and~$\thetasLSS$, determined (`directly') from the likelihood for the BAO~spectrum for each value of the parameters and derived (`from $(\alpha,\beta)$') from the redshift-binned likelihood for~$\alpha$ and~$\beta$. We find good agreement between both methods, suggesting that the two-dimensional parametrization is capturing most of the relevant information. The dashed lines indicate the fiducial values.}
	\label{fig:posteriorNeffThetas}
\end{figure}

Results of the likelihood analysis in terms of these parameters for both BOSS and DESI are shown in Fig.~\ref{fig:posteriorNeffThetas}.%
We see that the results are similar, which establishes that our templates are capturing most of the information available in the BAO spectrum, in particular for~BOSS. This is an important observation because i{}t allows us to simplify the analysis to a two-parameter template without much loss of information. In fact, the conclusion that these likelihoods are the same is not easily reproduced with any method of BAO extraction, but requires the robustness and stability of a method such as the one we use (see Appendix~\ref{app:broadband+phaseShiftExtraction}). Given instead our phase shift template, one can reliably compute Fisher matrices or likelihoods for~$\alpha$ and~$\beta$, and derive the implications for cosmological parameters from them. Future surveys, such as~DESI, show somewhat larger differences between the two methods, which suggests that more information could potentially be extracted by using additional and/or alternative templates.\medskip

The doubling of cosmological parameters to treat the CMB and LSS independently, like in the case of~$\NeffLSS$ and~$\thetasLSS$, has useful conceptual advantages even if we derive constraints on these parameters from the posterior of~$\alpha$ and~$\beta$. Growing tensions between the CMB and certain \mbox{low-$z$}~measurements have garnered much attention, but lack a compelling explanation. Measuring~$\thetasLSS$ and~$\NeffLSS$ in the BAO~spectrum may provide a new perspective on this issue without the need for a CMB~anchor.

\subsection{Validation of the Modified Analysis}
\label{sec:validation_analysis}
We have seen in the left panel of Fig.~\ref{fig:posteriorNeffThetas} that our approach captures essentially all of the information in the BAO spectrum at the sensitivity levels of the BOSS~experiment. However, one may still worry that the mapping
\beq
O_\mathrm{lin}(k) \to O^\mathrm{lin}_\mathrm{fid} \big(k/\alpha + (\beta-1) f(k)/r_s^\mathrm{fid}\big)	\label{eq:model_Neff0}
\eeq
introduces additional unphysical changes to the BAO~spectrum. Since we use $\Neff = 3.046$, corresponding to $\beta=1$, as the fiducial model, a poor modelling for $\beta \neq 1$ could lead to artificially strong evidence for a phase shift and could bias the measurement of~$\beta$ if $\Neff \neq 3.046$.

\begin{figure}[t]
	\includegraphics{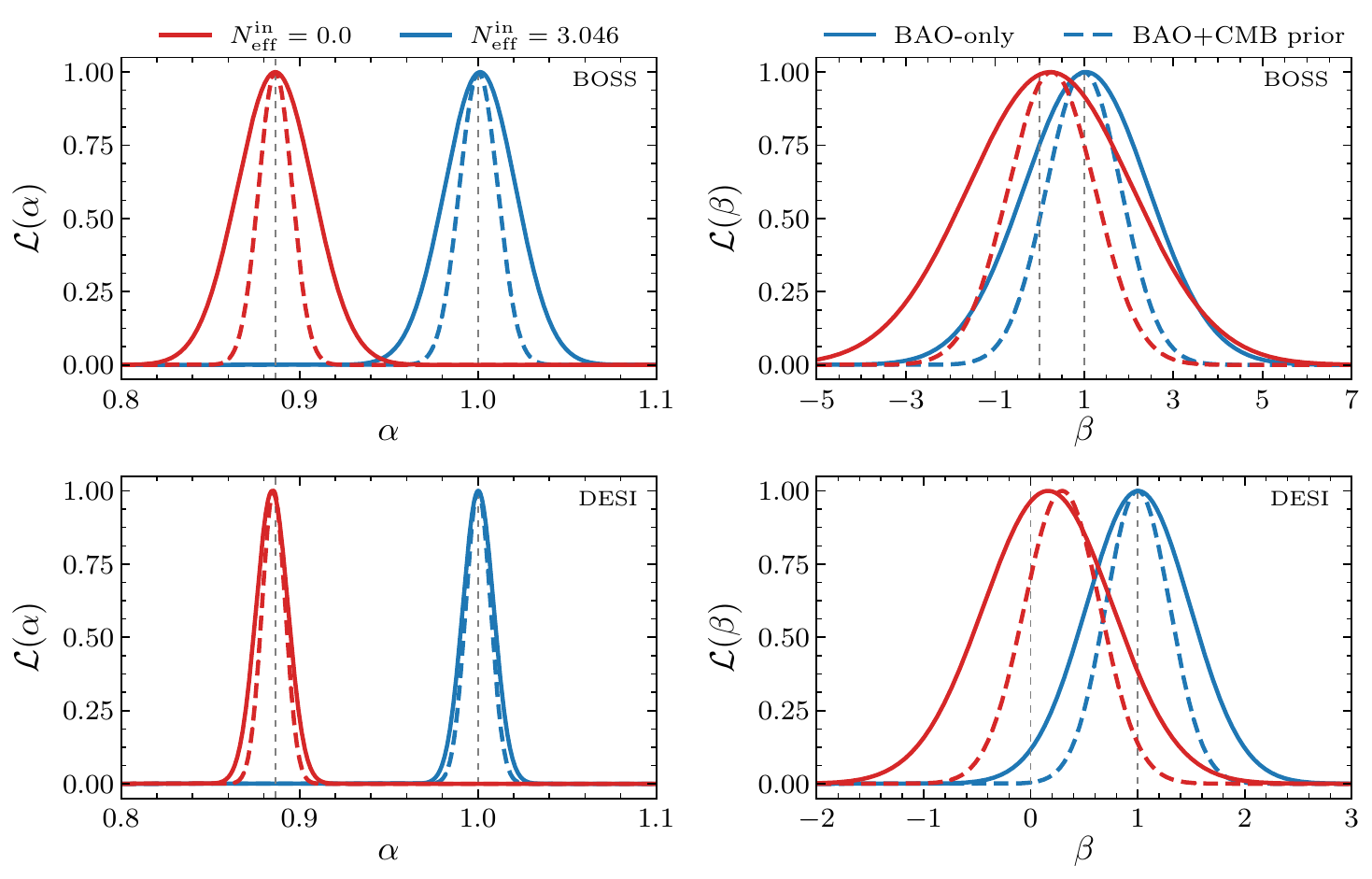}\vspace{-1pt}
	\caption{Posterior distributions for the BAO~scale parameter~$\alpha$ and the amplitude of the phase shift~$\beta$ when the mock data for BOSS~(\textit{top}) and DESI~(\textit{bottom}) were generated using $N_\mathrm{eff}^\mathrm{in} = 3.046\text{ (\textcolor{pyBlue}{blue}) and }0$~(\textcolor{pyRed}{red}), corresponding to $\beta = 1\text{ and }0$. In both cases, the model in~\eqref{eq:model_Neff0} used a fiducial cosmology with $\Neff = 3.046$ and the posteriors for~$\alpha$ are shown for one redshift bin similar to~$z_1$. The dotted lines show the posterior distributions after imposing a prior from a Planck-type CMB~experiment. We see that the posteriors reproduce the expected behaviour, which indicates that the estimation of~$\alpha_i$ and~$\beta$ is essentially unbiased.}
	\label{fig:validation}
\end{figure}

Our interest lies mostly in the exclusion of $\beta = 0$. A straightforward check that our method is reliable is to compute the posterior distribution for~$\beta$ in a cosmology with $\Neff = 0$ to see that the result is effectively unbiased. We use the same likelihood-based forecasts as in Section~\ref{sec:phaseShift} and the resulting posterior for~$\beta$ is shown in Fig.~\ref{fig:validation}.%
The expected values for~$\alpha$ and~$\beta$ are retrieved reliably in both cases. We also find good agreement when imposing the CMB~prior from~Planck with the respective input values of~$\Neff$. This test demonstrates that even though the fiducial model with $\Neff = 3.046$ is used for constructing the template, the model with $\Neff = 0$ is correctly recovered, especially for~BOSS.\footnote{In detail, the solid red curve in Fig.~\ref{fig:validation} shows a mean of $\bar\beta = 0.27$ rather than zero for a $\Neff=0$ cosmology. This level of bias is acceptably small given the much larger statistical error of $\sigma(\beta) = 0.97$. Of course, this bias should be accounted for when determining the precise statistical significance of the exclusion of $\beta=0$, but it does not affect our main conclusion in this chapter that $\beta > 0$ at \SI{95}{\percent}~confidence in the BOSS~DR12 dataset.} At higher levels of sensitivity, e.g.\ for~DESI, the expected values for~$\beta$ are recovered even more accurately for both $\Neff = 0\text{ and }3.046$. However, due to the smaller error bars and the slight difference between the parameter-based and template-based approaches around $\Neff=0$ for~DESI, the mean~$\bar{\beta}$ is found about $0.8\hskip1pt\sigma(\beta)$ too high, whereas it is excellent for the fiducial $\Neff=3.046$.\medskip

One may also be concerned that these results could depend sensitively on the method of BAO~extraction. Indeed, as we have discussed, the phase shift template~$f(k)$ is quite sensitive to the BAO~extraction and demands a method that is accurate across a wide range in~$\Neff$. In contrast, the model in~\eqref{eq:model_Neff0} only requires an accurate BAO~extraction for the fiducial cosmology. We have verified that the results in Fig.~\ref{fig:validation} do not depend on the BAO~extraction method being used.

\subsection{Tests on Mock Catalogues}
\label{sec:validation_mocks}
Finally, we test our entire analysis pipeline including the phase shift template using 999~MultiDark-Patchy mock catalogues~\cite{Kitaura:2015uqa}, which have been created for the BOSS~DR12 analysis. The Patchy mock catalogues have been calibrated to an N-body~simulation-based reference sample using analytical-statistical biasing models. The reference catalogue is extracted from one of the BigMultiDark simulations~\cite{Klypin:2014kpa}. The mock catalogues have a known issue with overdamping of the~BAO, making the signal for the traditional~BAO approximately \SI{30}{\percent}~weaker~\cite{Beutler:2016ixs}. We therefore forecast the mocks and the real data separately, taking these differences into account. For the mock forecasts, we use $\Sigma_\mathrm{nl}=\SI{7}{\per\h\Mpc}$ as the fiducial value of the non-linear damping scale.\medskip

\begin{figure}[t]
	\includegraphics{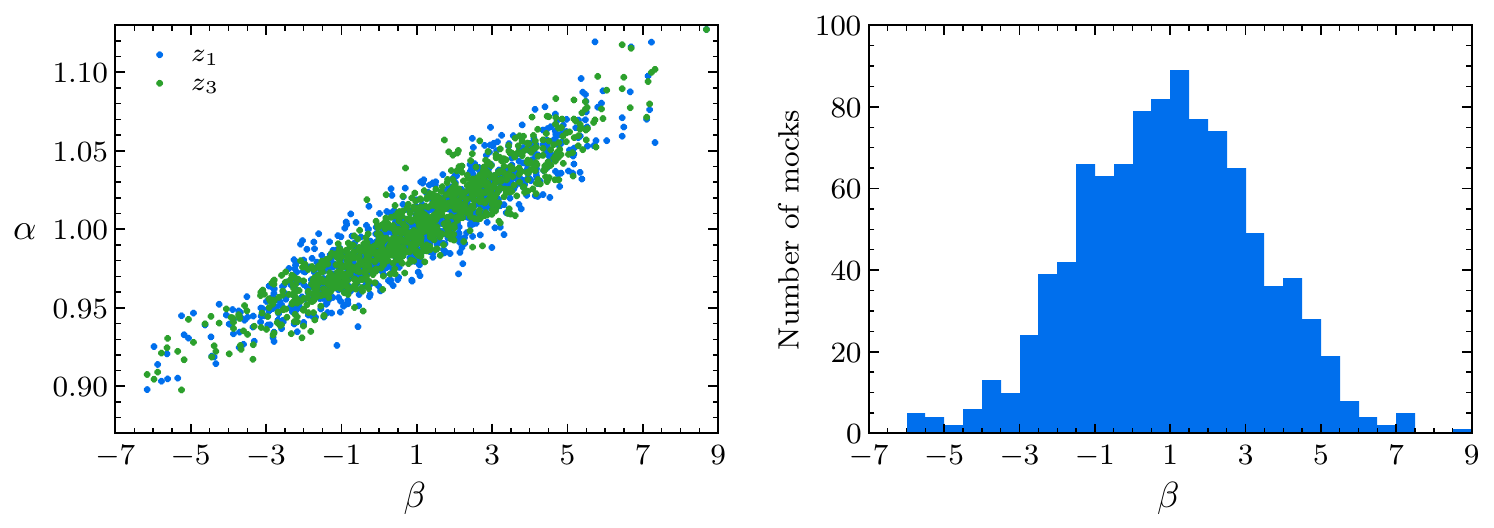}\vspace{-4pt}
	\caption{Distribution of maximum-likelihood values in the $\alpha$-$\beta$ plane for the two BOSS~redshift bins~$z_1$ and $z_3$~(\textit{left}), and for $\beta$~(\textit{right}) in 999~MultiDark-Patchy mock catalogues~\cite{Kitaura:2015uqa}, yielding $\beta = 1.0 \pm 2.4$.\vspace{-3pt}}
	\label{fig:mocks}
\end{figure}

An appealing feature of using the mock catalogues is that we can check that the performance expected from our forecasts is reproduced by the distribution of maximum-likelihood points across the catalogue. Figure~\ref{fig:mocks}%
confirms that the distributions for the parameters $\alpha$ and $\beta$ are indeed in good agreement with the fiducial value of $\beta=1$. A Gaussian fit to the distribution of maximum-likelihood values yields $\beta = 1.0 \pm 2.4$ ($\alpha_{z_1} = 1.000 \pm 0.035$, $\alpha_{z_3} = 1.000 \pm 0.035$), which is consistent with the value found from a likelihood-based forecast as in Section~\ref{sec:phaseShift}, $\sigma(\beta) = 2.1$.\medskip

As seen in the left panel of Fig.~\ref{fig:mocks}, there is a strong degeneracy between the effects of the parameters $\alpha$ and $\beta$. The origin of this degeneracy is easy to understand. If the only well-determined quantity in the data were the position of the first peak in the BAO spectrum, there would be a perfect degeneracy between phase and frequency determination. In reality, several peaks and troughs are present in the data which breaks the perfect degeneracy and allows the parameters~$\alpha$ and~$\beta$ to be constrained independently. However, one still expects them to remain significantly correlated, partly because the peaks are measured with decreasing accuracy due to damping. Since this degeneracy is a limiting factor in the measurement of $\beta$, we anticipate a significant improvement in the constraint on $\beta$ when the degeneracy with $\alpha$ is broken with additional data. Below we will see that this is indeed the case.

\section{Analysis of BOSS Data}
\label{sec:bossAnalysis}
Having demonstrated that our new method passes all the reported checks, we can apply it to the BAO signal of the final data release~(DR12) of the Baryon Oscillation Spectroscopic Survey; see~\cite{Alam:2015mbd}. As described in detail in~\cite{Reid:2015gra}, this dataset contains \num{1198006}~galaxies with spectroscopic redshifts in the range $0.2 < z < 0.75$ and covers~\SI{10252}{deg^2} of the sky.

\subsection{BAO-only Analysis}
\label{sec:fs_analysis}
\begin{figure}[t]
	\includegraphics{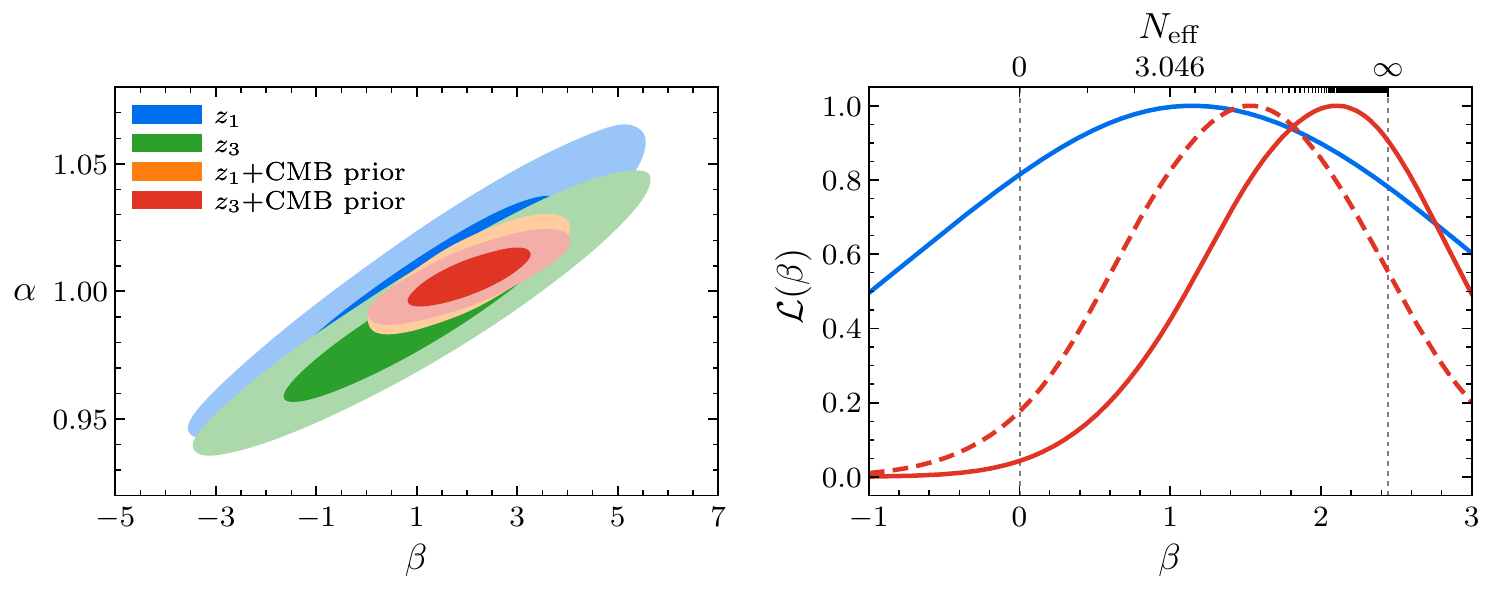}
	\caption{\textit{Left:} Contours showing $1\sigma$~and $2\sigma$~exclusions in the \mbox{$\alpha$-$\beta$}~plane for the two redshift bins~$z_1$ and~$z_3$, both from the BAO~data alone and after imposing a CMB prior on $\alpha$. \textit{Right:}~One-dimensional posterior distributions of $\beta$ without~(\textcolor{pyBlue2}{blue}) and with~(\textcolor{pyRed2}{red}) the $\alpha$-prior for the combined redshift bins. The dashed line is the result after marginalizing over the lensing amplitude $A_L$.}
	\label{fig:fs_posteriors}
\end{figure}

First, we analyse the two BOSS~redshift bins without any additional data or external prior. To explore the BAO~likelihood function, we use the Python-based, affine-invariant ensemble sampler \texttt{emcee}~\cite{ForemanMackey:2012ig} for Markov chain Monte Carlo. The convergence is determined with the Gelman-Rubin criterion~\cite{Gelman:1992zz} by comparing eight separate chains and requiring all scale-reduction parameters to be smaller than $R-1 = 0.01$. Figure~\ref{fig:fs_posteriors}
shows the posterior distribution for the parameters~$\beta$ and~$\alpha_{z_1}, \alpha_{z_3}$. The measured $\alpha$-values are in good agreement with those found in~\cite{Beutler:2016ixs}, but the errors have increased due to the degeneracy with~$\beta$. We find $\alpha_{z_1} = 1.001\pm0.025$, $\alpha_{z_3} = 0.991\pm 0.022$ and $\beta = 1.2 \pm 1.8$. These results are in good agreement with likelihood-based forecasts for the data,\footnote{These forecasted values are slightly larger than those in~\textsection\ref{sec:phase_planned} because we accounted for the roughly~\SI{30}{\percent} smaller galaxy bias measured in~\cite{Beutler:2016ixs}.}
$\sigma(\alpha_{z_1})=0.021$, $\sigma(\alpha_{z_3})=0.019$ and $\sigma(\beta) = 1.5$. A similar level of agreement between forecasts and actual performance was obtained for the measurement of~$\alpha$ in the conventional BAO analysis of BOSS DR12~\cite{Beutler:2016ixs}.

\subsection{Adding a CMB Prior}
\label{sec:fs_prior}
The BAO-only measurement of~$\beta$ is limited by the degeneracy with~$\alpha(z)$. However, in a given cosmology, the allowed range of~$\alpha(z)$ has to be consistent with constraints on the cosmological parameters. Our interest is to measure the neutrino-induced phase shift in the BAO~signal assuming a background cosmology that is consistent with the Planck CMB~constraints. We infer the prior on~$\alpha(z)$ from the Planck~2015 temperature and polarization data\hskip1pt\footnote{We use the low-$\ell$ ($2 \leq \ell \leq 29$) temperature and LFI~polarization data, and the high-$\ell$ ($30 \leq \ell \leq 2508$) \texttt{plik}~cross half-mission temperature and polarization spectra. In~``TT-only'', we omit the high-$\ell$~polarization spectra.}~\cite{Ade:2015xua}, while marginalizing over any additional cosmological information (including all effects of~$\Neff$). If available, we directly employ the Markov chains supplied by the Planck collaboration, which were computed using \texttt{CAMB}~\cite{Lewis:1999bs} and \texttt{CosmoMC}~\cite{Lewis:2002ah} with the publicly released priors and settings. In particular for the $\Lambda\mathrm{CDM}$+$\Neff$+$A_L$ prior cosmology, we sample the data using the same codes and priors. At each point in the Monte Carlo Markov chains obtained from the Planck likelihood for a certain background cosmology, we compute the values of~$\alpha_{z_1}$ and~$\alpha_{z_3}$ associated with the given set of cosmological parameters. In this way, we obtain the two-dimensional (Gaussian) posterior for~\mbox{$\alpha_{z_1}$-$\alpha_{z_3}$}. We confirmed on the mock catalogues that a Gaussian prior with the expected mean values and the Planck $\Lambda\mathrm{CDM}$+$\Neff$ covariance matrix results in an unbiased measurement of $\beta = 1.00 \pm 0.85$ (see also~\textsection\ref{sec:validation_analysis} for the equivalent forecasts with unbiased posterior distributions). On the data, we impose the Planck posterior on~$\alpha$ by importance-sampling our BAO-only Monte Carlo Markov chains.

The right panel of Fig.~\ref{fig:fs_posteriors} shows the marginalized posterior distributions for the parameter~$\beta$. We see that including the $\alpha$-posterior from the Planck $\Lambda\mathrm{CDM}$+$\Neff$ chains as a prior sharpens the distribution significantly. Having obtained a constraint of $\beta = 2.05\pm 0.81$ on the phase shift amplitude, we want to evaluate the statistical significance of an exclusion of $\beta = 0$, corresponding to no phase shift and no free-streaming neutrinos. For this purpose, we extract the fraction of Monte Carlo samples which have $\beta>\beta_0$. To be cautious about the small bias found in~\textsection\ref{sec:validation_analysis}, we employ $\beta_0 = 0.27$ instead of $\beta_0 = 0$.\footnote{We explicitly checked that the computation based on likelihood ratios leads to essentially the same confidence levels, which is expected since the posterior distributions are very close to Gaussian.} In this and other aspects of the analysis, we have therefore made intentionally conservative choices in stating our statistical significance. The measurement of $\beta = 2.05\pm 0.81$ consequently corresponds to an exclusion of $\beta = 0$ at greater than \SI{99}{\percent}~confidence. The statistical error of the measurement is in good agreement with the forecasted value of $\sigma(\beta) = 0.77$. On the other hand, the central value is more than a $1\sigma$~fluctuation away from the expected Standard Model value $\beta = 1$. Any upward fluctuation adds to the confidence of our exclusion, provided that it is simply a statistical fluctuation. We tested the stability of this upward fluctuation to changes in the cosmological model and the CMB~likelihood (see Table~\ref{tab:bao+priorResults}).%
\begin{table}[t]
	\centering
	\begin{tabular}{l S[table-format=1.2, separate-uncertainty, table-figures-uncertainty=2]}
			\toprule
		Prior Cosmology							& {$\beta$}		\\
			\midrule[0.065em]
		None (BAO-only)							& 1.2  \pm 1.8	\\
			\midrule[0.02em]
			\rowcolor[gray]{0.9}
		$\Lambda\mathrm{CDM}$+$\Neff$			& 2.05 \pm 0.81	\\
		$\Lambda\mathrm{CDM}$					& 1.97 \pm 0.73	\\
			\midrule[0.02em]
		$\Lambda\mathrm{CDM}$+$\Neff$ (TT-only)	& 1.6  \pm 1.1	\\
		$\Lambda\mathrm{CDM}$ (TT-only)			& 1.87 \pm 0.89	\\
			\midrule[0.02em]
		$\Lambda\mathrm{CDM}$+$\Neff$+$A_L$		& 1.53 \pm 0.83	\\
		$\Lambda\mathrm{CDM}$+$A_L$				& 1.49 \pm 0.76	\\
			\bottomrule 
	\end{tabular}
	\caption{Constraints on the amplitude of the phase shift~$\beta$ with and without a Planck prior on the BAO~scale, assuming various underlying cosmologies. Our baseline result uses the $\Lambda\mathrm{CDM}$+$\Neff$ prior, marginalizing over all of the effects of~$\Neff$ in the~CMB. We see that this result is robust to including or excluding~$\Neff$ and~$A_L$ in the prior cosmology. Finally, we show the large central value of~$\beta$ also appears when using TT-only spectra and is therefore not solely a consequence of the polarization data.}
	\label{tab:bao+priorResults}
\end{table}
The statistical significance of the result is largely insensitive to the choice of cosmology and likelihood. The largest deviation from~$\Lambda\mathrm{CDM}$ within the Planck data alone is the preference for a larger lensing amplitude~$A_L$~\cite{Aghanim:2016sns}. To estimate the impact of this upward fluctuations on our analysis, we marginalized over~$A_L$ in the implementation of the $\alpha$-prior. The dashed posterior curve in Fig.~\ref{fig:fs_posteriors} shows the result obtained from the $\Lambda\mathrm{CDM}$+$\Neff$+$A_L$ prior cosmology, which corresponds to $\beta = 1.53 \pm 0.83$. We see that marginalizing over~$A_L$ indeed brings the central value of~$\beta$ into closer agreement with $\beta=1$, suggesting that part of our large central value is due to a known upward fluctuation of the Planck data. Having said that, even with this marginalization, we find evidence for a positive phase shift, $\beta>0$, at greater than \SI{94}{\percent}~confidence.\footnote{Note that we marginalized over~$A_L$ because it experiences a large fluctuation in the Planck data. The statistical significance of the corresponding result should therefore not be compared to the results of our blind analysis.} In summary, while the precise significance of the phase shift measurement depends on the implementation of the CMB~prior, the exclusion of $\beta=0$ at greater than \SI{95}{\percent}~confidence is stable to all choices of the prior that we have considered.

\section{Analysis in Configuration Space} 
\label{fig:cs_analysis}
\begin{figure}[t]
	\includegraphics{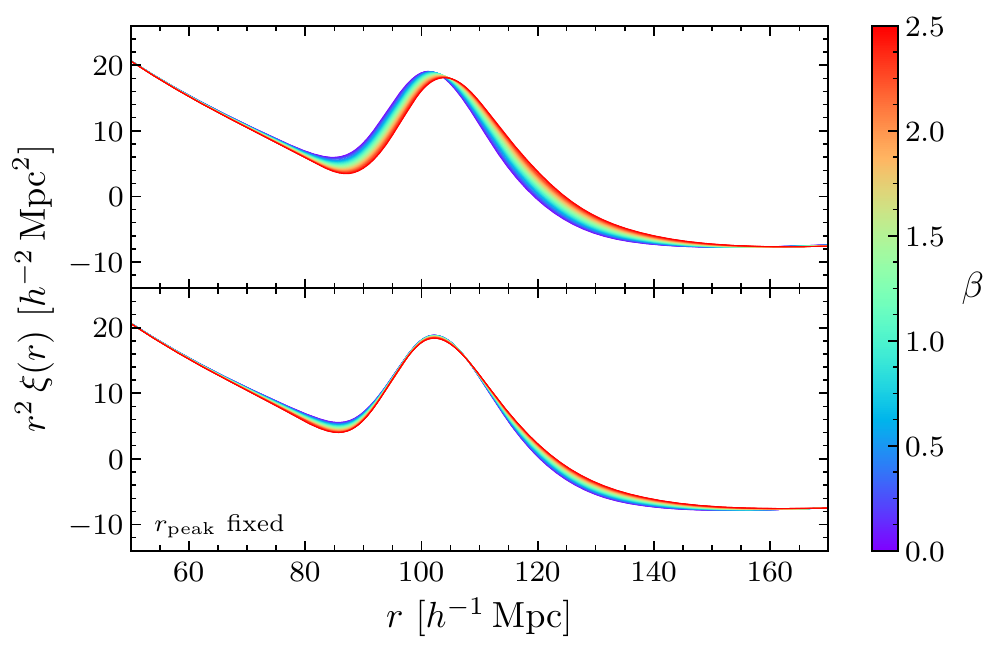}
	\caption{Rescaled linear correlation function~$r^2 \xi(r)$ as a function of the amplitude of the~phase shift~$\beta$. The upper panel keeps $\alpha = 1$ fixed, while~$\alpha$ is varied in the lower panel to fix the position of the peak,~$r_\mathrm{peak}$. This illustrates the degeneracy between~$\alpha$ and~$\beta$ in configuration space.}
	\label{fig:bao+xi_betaDependence}
\end{figure}

The neutrino-induced phase shift is characteristically a Fourier-space~(FS) quantity. By contrast, the BAO~frequency is more commonly described in configuration space~(CS) as the scale of the BAO~feature in the two-point correlation function. Consequently, the measurement of the BAO~scale is often depicted as the determination of the BAO~peak location in~CS~\cite{Ross:2016gvb, Vargas-Magana:2016imr}. The phase shift modifies the shape of the BAO peak and manifests itself in~CS as a transfer of correlations around the peak position from small to large scales (see Fig.~\ref{fig:bao+xi_betaDependence}). Given that the BAO~scale measurement is known to give compatible results in~CS and~FS (see e.g.~\cite{Alam:2016hwk}), we anticipate the same to be true of the phase shift. We will therefore implement a modified version of the CS~method used in~\cite{Vargas-Magana:2016imr} as a cross-check of our main FS~analysis.%

In the following, we first describe how we have incorporated the phase shift into the configuration-space analysis of the BAO~signal~(\textsection\ref{sec:cs_modified}). Similar to the validation in~FS, we also check the modified method on mock catalogues. We then apply the analysis pipeline to the BOSS~DR12~data and impose the same Planck priors on~$\alpha$ as in~FS, and find results that are consistent with those of the Fourier-space analysis~(\textsection\ref{sec:cs_results}).

\subsection{Modified Pipeline and Cross-Checks}
\label{sec:cs_modified}
Our non-linear model for the correlation function starts from the processed matter power spectrum
\beq
P(k) = F(k,\Sigma_s) P_\mathrm{lin}^\mathrm{nw}(k) \left[1+O_g(k)\right] ,
\eeq
where~$O_g(k)$ is the template-based non-linear BAO~spectrum defined in~\eqref{eq:ONL} and $F(k,\Sigma_s)$ is given by~\eqref{eq:FoG}. The two-point galaxy correlation function is then modelled as
\beq
\xi_g(r) = B^2 \int \!\d\log k\, \frac{k^3}{2\pi^2}P(k)\, j_0(kr) + A(r)\, ,
\eeq
where~$j_0(kr)$ is a spherical Bessel function. We introduced the constant bias parameter~$B$ and the polynomial function~$A(r)$, taken to have the same form as in~\cite{Vargas-Magana:2016imr},\vspace{-3pt}
\beq
A(r) = \frac{a_1}{r^2} + \frac{a_2}{r} + a_3 \, ,	\vspace{-3pt}
\eeq
where the coefficients~$a_n$ are marginalized over. While the constant bias matches the same parameter in the FS~analysis, the polynomial~$A(r)$ is \textit{not} equivalent to the polynomial~$A(k)$ in~\eqref{eq:fs_poly}. This is one of the notable differences between the~FS and CS~analyses. Except for the amplitude of the phase shift~$\beta$, all parameters are redshift dependent. Since the scale~$\Sigma_s$ is held fixed to the best-fit value obtained on the mock catalogues, we fit the following 13~parameters to the correlation function in the range $r\in\SIrange[open-bracket=[, close-bracket=],]{55}{160}{\per\h\Mpc}$:\hskip1pt\footnote{We employ flat priors on the cosmological parameters, requiring~$\beta$ to be between $-10$ and $10$, and $\alpha_z$ to be between $0.5$ and $1.5$, but do not impose explicit priors for the other ten parameters. On the data, we speed up the analysis by analytically marginalizing over the broadband parameters~$a_{n,z}$ in each step.}\vspace{-4pt}
\beq
\beta,\, \alpha_{z_1},\, \alpha_{z_3};\, \{B_z,\, \Sigma_{\mathrm{nl},z},\, a_{n,z} \}_{z_1,z_3}\, ,	\vspace{-4pt}
\eeq
for the same two redshift bins as in Fourier space.\medskip

\begin{figure}
	\includegraphics{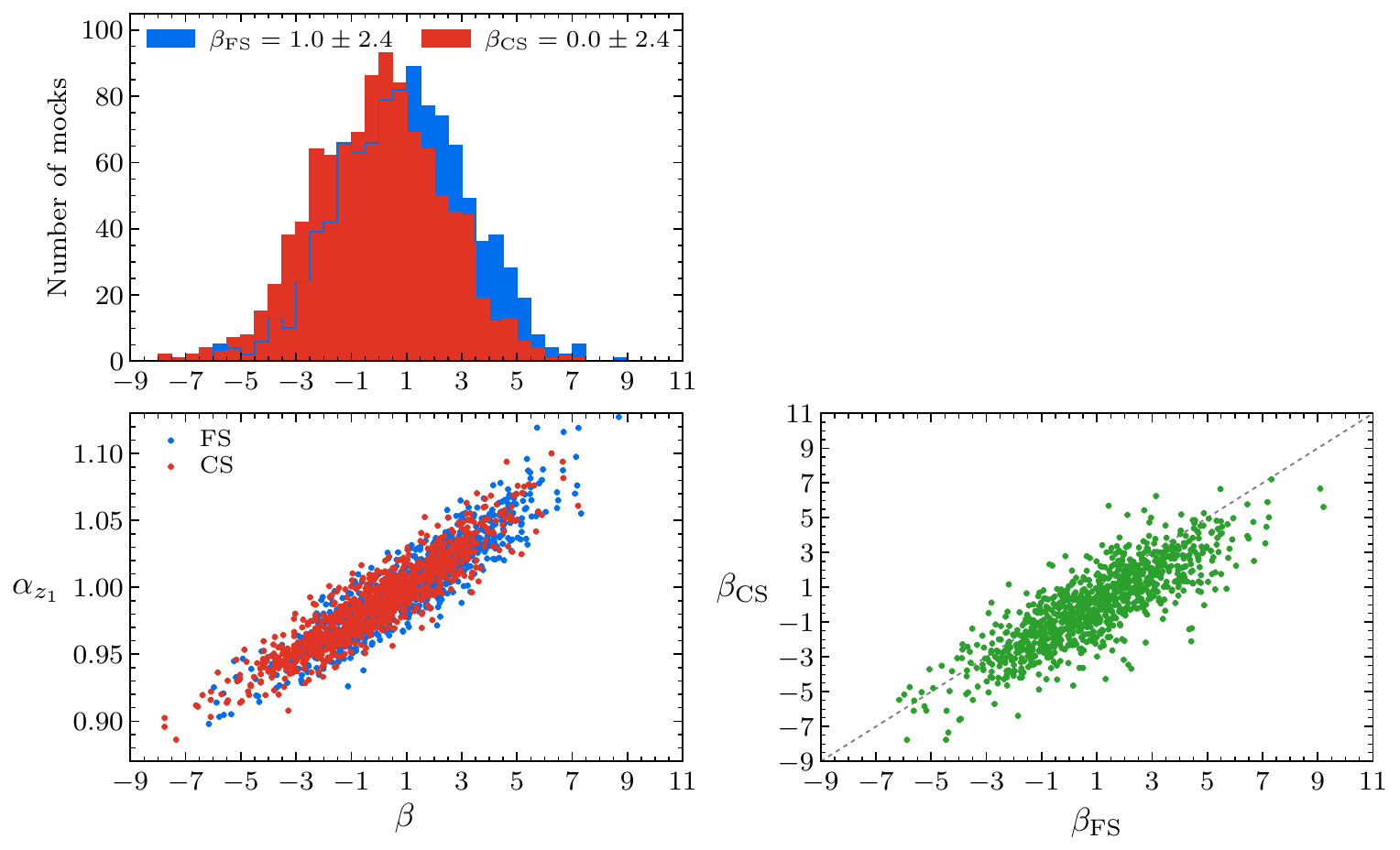}\vspace{-5pt}
	\caption{Comparison of the distribution of maximum-likelihood values in 999~mock catalogues~\cite{Kitaura:2015uqa} for the Fourier-space~(FS, \textcolor{pyBlue}{blue}) and configuration-space~(CS, \textcolor{pyRed}{red}) analyses. We also show the correlation between the inferred phase shift amplitudes in the two analyses~(\textcolor{pyGreen}{green}).\vspace{-4pt}}
	\label{fig:cs_mocks}
\end{figure}

We apply the same pipeline as in~\cite{Vargas-Magana:2016imr} to the MultiDark-Patchy mock catalogues~\cite{Kitaura:2015uqa} and determine the distributions of maximum-likelihood values for the parameters~$\alpha$ and~$\beta$. The results are shown in Fig.~\ref{fig:cs_mocks}%
and correspond to $\beta_\mathrm{CS} = 0.0 \pm 2.4$ ($\alpha_{z_1} = 0.989 \pm 0.033$, $\alpha_{z_3} = 0.990 \pm 0.034$). Comparing these distributions with the FS~mock catalogue~analysis of~\textsection\ref{sec:validation_mocks}, we observe a strong correlation with correlation coefficient $r = 0.84$, but a statistically significant bias of about~$1/3$ of a standard deviation for both~$\alpha_i$ and~$\beta$, albeit with approximately the same standard deviations. When including the CMB~prior, the mean shifts upwards and gives $\beta_\mathrm{CS} = 0.75 \pm 0.89$, corresponding to a bias of about~$1/4$ of a standard deviation which is also slightly larger than in~FS. These values demonstrate good statistical agreement between the CS~and FS~analyses, and demonstrate that CS~provides a useful cross-check of the FS~analysis. While CS~does show larger biases, they are sufficiently small that they should not meaningfully affect the statistical significance of our results. On the other hand, we noticed that the precise choice of the broadband polynomial~$A(r)$ altered both the mean and standard deviation, while being consistent with the fiducial cosmology. These features of the CS~analysis will be explored in future work. The shifts seen in~CS further highlight the remarkable robustness of the phase shift in~FS.

\subsection{Analysis of BOSS Data}
\label{sec:cs_results}
\begin{figure}
	\includegraphics{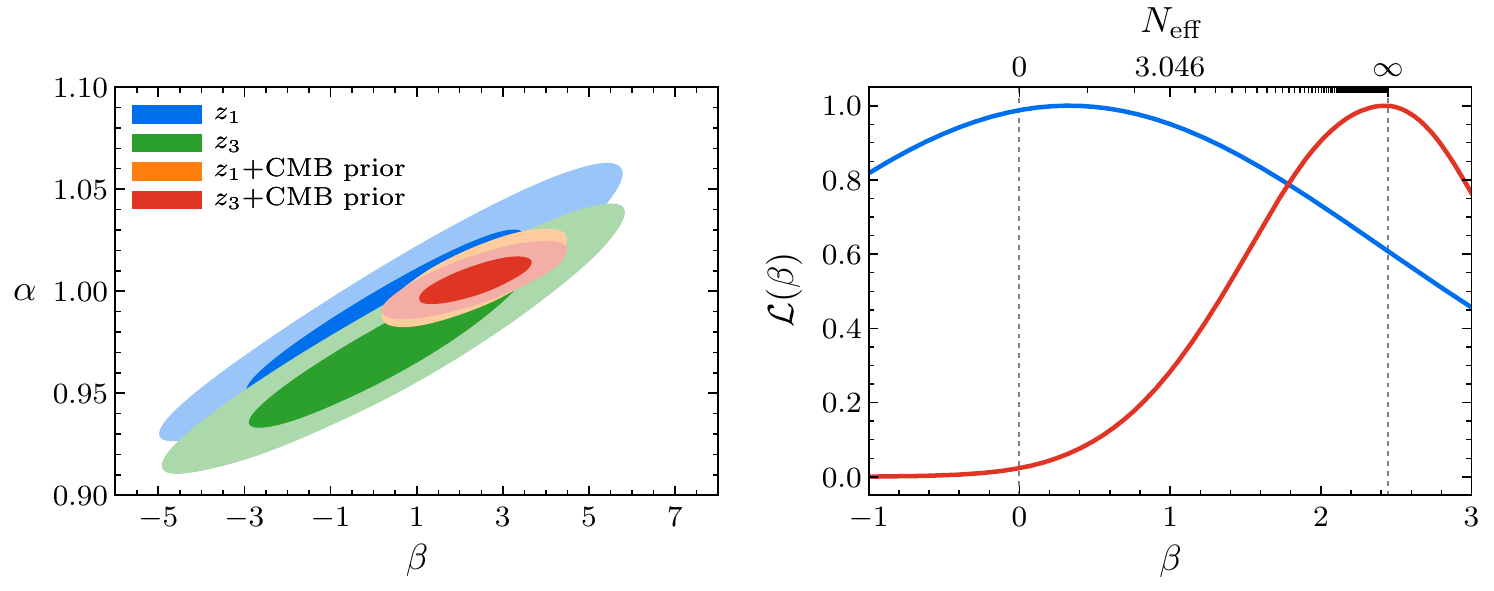}\vspace{-5pt}
	\caption{\textit{Left:} Contours showing $1\sigma$ and $2\sigma$ exclusions in the $\alpha$-$\beta$ plane for the two redshift bins $z_1$ and $z_3$ in configuration space, both from the BAO data alone and after imposing a CMB prior on $\alpha$. \textit{Right:}~One-dimensional posterior distributions of $\beta$ without~(\textcolor{pyBlue2}{blue}) and with~(\textcolor{pyRed2}{red}) the $\alpha$-prior for the combined redshift bins.\vspace{-5pt}}
	\label{fig:cs_posteriors}
\end{figure}

With these caveats in mind, we apply the CS~pipeline to the BOSS~DR12 dataset. The posterior distributions for the parameters~$\alpha_{z_1}$, $\alpha_{z_3}$ and~$\beta$ are presented in Fig.~\ref{fig:cs_posteriors},%
and correspond to measurements of $\alpha_{z_1} = 0.991 \pm 0.027$, $\alpha_{z_3} = 0.973 \pm 0.026$ and $\beta_\mathrm{CS} = 0.4\pm 2.1$. These mean values of~$\alpha_i$ are about~$1/4$ of a standard deviation lower than the ones found in the standard BAO~analysis~\cite{Vargas-Magana:2016imr}. In addition, the error bars increased, mainly related to the degeneracy between~$\alpha$ and~$\beta$ discussed in~\textsection\ref{sec:validation_mocks}. The value of $\bar\beta$ is $0.3\hskip1pt\sigma$~lower than in~FS with a \SI{16}{\percent}~larger error. When adding a Planck prior to break the degeneracy, we measure $\beta_\mathrm{CS} = 2.36 \pm 0.86$ which is larger than in~FS because of the mentioned bias in~$\alpha_i$ towards lower values. Nevertheless, these CS~measurements are statistically consistent with the main FS~results, with similar shifts in the mean values as observed in the mock analysis. Given that the broadband modelling and peak isolation in configuration and Fourier space are distinct, an agreement between the two analyses was not guaranteed, although the change to the BAO~peak is simply the inverse Fourier transform of the phase shift. Having said that, this analysis confirms that a measurement can also be made in configuration space and, despite the discussed differences, is comparable to the main analysis in Fourier space.

\section{Summary}
\label{sec:conclusions_bao-neutrinos}
In this chapter, we have reported on the first measurement of the neutrino-induced phase shift in the BAO~spectrum. This is the first evidence for the cosmic neutrino background in the clustering of galaxies and the first application of the BAO~signal beyond its use as a standard ruler.\medskip

To extract the phase information, we modified the conventional BAO~data analysis by allowing the amplitude of the phase shift to be an additional free parameter. We determined this new parameter to be non-zero at greater than \SI{95}{\percent}~confidence, even allowing for very conservative marginalization over corrections to the broadband spectrum. Our measurement is a nontrivial confirmation of the standard cosmological model at low redshifts and a proof of principle that there is additional untapped information in the phase of the BAO~spectrum. Since this phase information is protected from the effects of non-linear gravitational evolution~\cite{Baumann:2017lmt}, it is a particularly robust probe of early universe physics.\medskip

\begin{figure}[h]
	\includegraphics{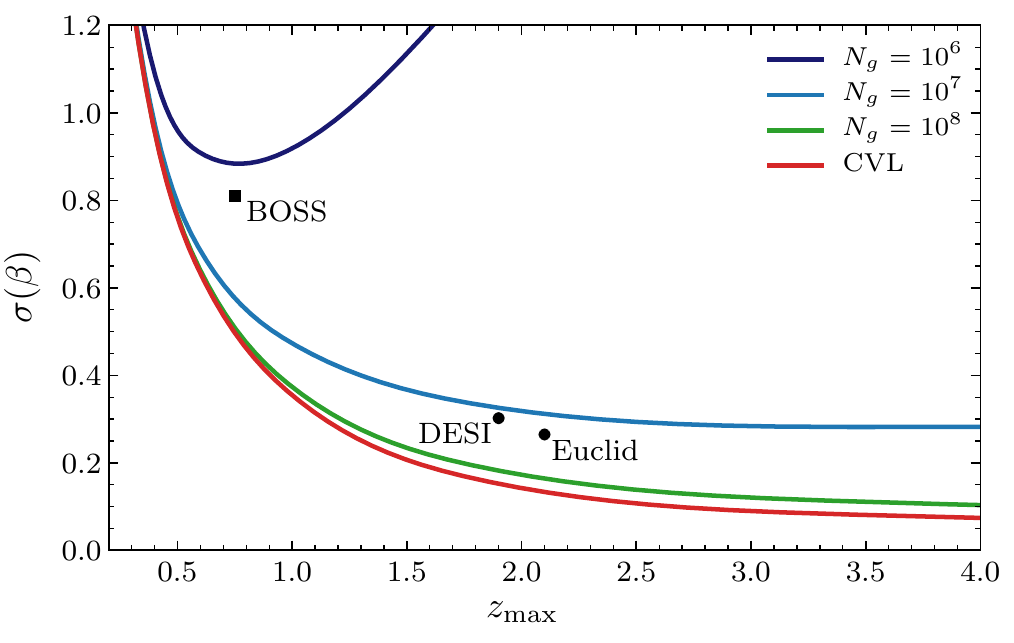}\vspace{-1pt}
	\caption{Constraints on the amplitude of the phase shift~$\beta$ including a CMB~prior on the BAO~scale parameter~$\alpha$ from Planck. The lines are forecasted constraints as a function of the maximum redshift~$z_\mathrm{max}$ and the number of objects~$N_g$ of a cosmological survey observing a sky fraction of $f_\mathrm{sky}=0.5$ (see~\textsection\ref{sec:phase_future} for details). Shown is also the cosmic variance limit. The square indicates the result obtained in this work. The dots mark projected constraints for~DESI and Euclid assuming~$z_\mathrm{max}$ to be given by the largest redshift bin used to define the survey in~\cite{Font-Ribera:2013rwa}.} 
	\label{fig:bao-neutrinos_summary}
\end{figure}
A number of galaxy surveys are planned over the next decade which have the potential to significantly improve on our measurement of the neutrino background (see Fig.~\ref{fig:bao-neutrinos_summary}). The Dark Energy Spectroscopic Instrument, for example, should be sensitive to the~C$\nu$B at more than~$3\sigma$, making the BAO~phase shift measurement more comparable to current limits from the~CMB~\cite{Follin:2015hya}. Combining Euclid with a prior from a next-generation CMB experiment would allow a $5\sigma$~detection of the~C$\nu$B. Moreover, having shown that there is valuable information in the phase of the BAO spectrum, we should ask what else can be learned from it beyond the specific application to light relics. As the observed BAO feature is the result of the combined dynamics of the dark matter and baryons, it is broadly sensitive to new physics in these sectors. The BAO phase shift is one particularly clean probe of this physics and we hope that our work will inspire new ideas for exploring the early universe at low redshifts.
	\chapter{Conclusions and Outlook}
\label{chap:conclusions}
The wealth of cosmic microwave background and large-scale structure data has transformed the field of cosmology. Remarkably, these observations have not only become precise enough to answer questions about the universe at large, but also to start addressing puzzles in the microscopic description of Nature. Cosmological measurements are particularly well suited to study Standard Model neutrinos and to shed light on the possible existence of other light relics beyond the Standard Model. In this thesis, we have contributed to this endeavour by uncovering new constraints and identifying robust signatures of these particles in cosmological observables. We established the free-streaming nature of cosmic neutrinos in both the cosmic microwave background anisotropies and the clustering of galaxies, and paved the way to more efficiently using the wealth and precision of cosmological datasets in the future. In this final chapter, we summarize the main results of this work and make a few remarks on future directions.

\subsection*{Summary}
We employed cosmological observations to probe fundamental physics in two domains: within and beyond the Standard Model of particle physics. We focussed on providing new insights into the neutrino sector of the Standard Model as well as some of its extensions containing extra light and weakly-coupled species. In particular, we presented the first measurement of the cosmic neutrino background in the clustering of galaxies, obtained new CMB~constraints on additional forms of radiation and derived novel bounds on light scalar particles, such as axions. Along the way, we highlighted the power of a subtle phase shift in the acoustic oscillations of the primordial plasma as a robust probe of neutrinos and other free-streaming relativistic species.\medskip

Based on a detailed analytical understanding of the phase shift, we presented new evidence for the cosmic neutrino background. We achieved this by establishing the free-streaming nature of these particles separately in cosmic microwave background observations of the Planck~satellite and in the large-scale structure mapped by the Baryon Oscillation Spectroscopic Survey. In the BAO~analysis, we established a new method to extract the neutrino-induced phase shift. This allowed us to perform the first measurement of the imprint of neutrinos in a low-redshift observable. At the same time, this investigation marked the first application of the BAO~signal to early universe physics and illustrated that the spectrum of baryon acoustic oscillations carries more accessible cosmological information than only the acoustic scale. The gravitational effect of the cosmic neutrino background has now been observed in the damping and the phase shift of both the CMB~anisotropies and the BAO~spectrum of galaxy clustering. Cosmological observations have therefore been able to provide new tests of the least understood part of the Standard Model.

Our forecasts indicate that future observations are guaranteed to explore regimes of BSM~physics which have so far been inaccessible. We showed that future CMB~measurements have the potential to probe the energy density in neutrinos and other relativistic species at the one-percent level, corresponding to an order of magnitude improvement over current bounds. In addition, we established that there is further, currently untapped information in the LSS~data which will help us to push constraints on the effective number of relativistic species,~$\Neff$, below well-motivated theoretical targets. This might have far-reaching consequences for new light particles, such as axions, which are predicted in many interesting SM~extensions. These BSM~species are hard to detect in terrestrial experiments due to their weak couplings, but the large number densities in the early universe make it possible to measure their gravitational effects. We demonstrated that reaching the sensitivity of the minimal thermal contribution of one scalar particle, $\Delta\Neff = 0.027$, would have important implications: We could either detect any light particles that have ever been in thermal equilibrium with the Standard Model, or put strong bounds on their SM~interactions. We exemplified this for axions and other scalar BSM~particles, and found that, in many cases, existing constraints from astrophysical and terrestrial searches could be surpassed by orders of magnitude. This result and the target $\Delta\Neff=0.027$ have now been adopted by the CMB~Stage-4 collaboration as one of their main science targets~\cite{Abazajian:2016yjj,}.

\subsection*{Outlook}
We have shown that future CMB and LSS~observations could have a significant impact on our understanding of fundamental physics. In order to harvest all of the hidden clues stored in the observables, our ability to extract information from cosmological measurements should however further increase in this age of data-driven cosmology. In addition, we have argued that, in the case of~$\Neff$, these observations can play complementary roles by both enhancing the raw sensitivity and adding to the robustness of the measurement. Our forecasts also suggest that the constraints on certain cosmological parameters, such as the effective number of relativistic species or the matter density, may be tightened when moving beyond the standard BAO~analysis. A broader exploration, including additional observables such as weak gravitational lensing and the cross-correlation of different probes, will likely reveal more aspects of the early and/or late universe that could be discovered in this way. It would also be intriguing in this respect if we could uncover further quantities like the phase shift that are robust against non-linear gravitational evolution.\medskip

With the increasing precision of cosmological surveys, we are not only measuring the homogeneous background evolution of the universe, but have become sensitive to differences in the evolution of perturbations. This motivates revisiting the predictions of specific models of BSM~physics. When we studied the impact of future $\Neff$~measurements on the SM couplings of axions, we provided one such avenue within an effective field theory of light scalar species. Since the theoretical threshold for~$\Delta\Neff$ is larger for particles with spin and might soon be reached by CMB~polarization experiments together with LSS~surveys, the EFT~of light species should be revisited. Moreover, these effective models could also include massive particles which are abundant in well-motivated SM~extensions and might decay. The type of energy injection from such decays may then be observable over a wide range of times, even in the absence of new light fields. Including other cosmological observables, such as CMB~spectral distortions or further late-time LSS~probes, could therefore jointly constrain these models. Furthermore, the detailed predictions for~$\Neff$ may be significantly altered from the minimal case, which can change the impact of future constraints. Finally, the investigation of light dark matter models, whose SM~interactions must be mediated by new light fields to satisfy current thermal abundance limits, could link the search for light relics with the hunt for the (particle) nature of dark matter.

\bigskip\noindent
The prospects of probing the early universe with the future influx of cosmological data are very bright. At the same time, the possible implications on particle physics cannot be overstated. These observations will provide an opportunity to probe physics beyond the Standard Model at a much more precise level than was previously possible and in a regime that is inaccessible to terrestrial experiments. We are optimistic that this will teach us something interesting. We will either discover a whole new world of dark physics, or learn to what remarkable degree it is decoupled from the rest of physics. In the meantime, we remain curious what future cosmological measurements will tell us about the universe and the underlying laws of Nature.
	\begin{myappendix}
		\addtocontents{toc}{\protect\setcounter{tocdepth}{1}}

\chapter{Goldstone Production and Decay}
\label{app:cmb-axions_appendices}
In this appendix, we provide supplemental material to Chapter~\ref{chap:cmb-axions}. We compute the production rates of axions, familons and majorons via their Standard Model interactions~(\textsection\ref{app:rates}), and discuss the effects of the possible decays of these Goldstone bosons to photons and neutrinos~(\textsection\ref{app:decays}).

\section{Production Rates}
\label{app:rates}
In this section, we derive the rates of Goldstone boson production used in the main text. We consider separately the couplings to gauge fields and to matter fields.

\subsection{Couplings to Gauge Fields}
Above the scale of electroweak symmetry breaking, the coupling of the Goldstone boson to the Standard Model gauge sector is
\beq
\L_{\phi\mathrm{EW}} = - \frac{1}{4}\frac{\phi}{\Lambda} \left( c_1 \hskip1pt B_{\mu\nu} \tilde{B}^{\mu\nu} + c_2 \hskip1pt W_{\mu\nu}^a \tilde{W}^{\mu\nu,a} +c_3 \hskip1pt G_{\mu\nu}^a \tilde{G}^{\mu\nu,a} \right) .	\label{eq:LEW}
\eeq
The dominant processes leading to the production of the axion~$\phi$ are illustrated in Fig.~\ref{fig:diagramsGluon}.%
\begin{figure}[b]
	\centering
	\subcaptionbox{Primakoff process.}[\widthof{(a) Primakoff process.}]{
		\includegraphics*{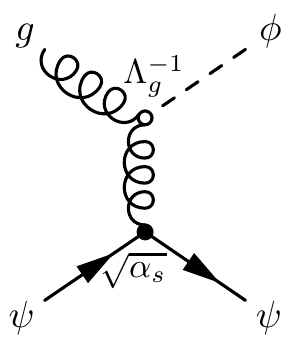}
	}\hfill
	\subcaptionbox{Fermion annihilation.}{
		\includegraphics*{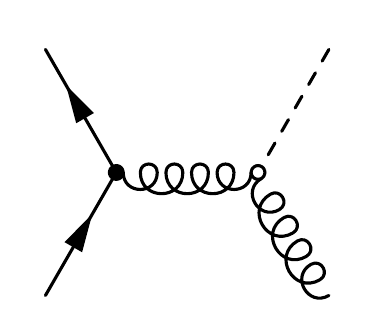}
	}\hfill
	\subcaptionbox{Gluon fusion (representative diagrams).}{
		\includegraphics*{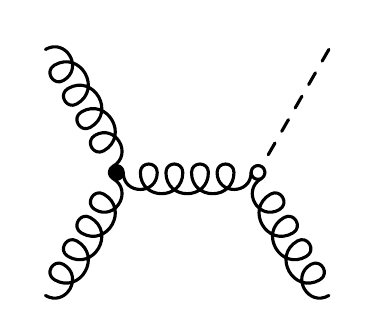}
		\hspace{-0.8cm}
		\includegraphics*{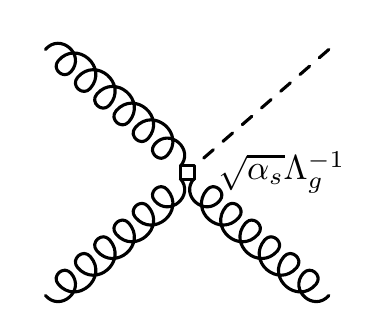}
	}
	\caption{Feynman diagrams for the dominant Goldstone production via the gluon coupling. For gluon fusion, there are $t$-~and $u$-channel diagrams in addition to the presented~$s$-channel diagram. Similar diagrams apply for the couplings to the electroweak gauge bosons.}
	\label{fig:diagramsGluon}
\end{figure}
In the limit of massless gauge bosons, the cross sections for some of these processes have infrared~(IR) divergences. The results therefore depend slightly on how these divergences are regulated; see e.g.~\cite{Braaten:1991dd, Bolz:2000fu, Masso:2002np, Graf:2010tv, Salvio:2013iaa}. The most detailed analysis has been performed in~\cite{Salvio:2013iaa}, where the total production rate was found to be	
\beq
\Gamma = \frac{T^3}{8\pi \Lambda^2} \Big[ c_1^2\hskip1pt F_1(T) + 3 c_2^2 \hskip1pt F_2(T) + 8 c_3^2 \hskip1 pt F_3(T) \Big]\, ,	\label{eq:TotalRate}
\eeq
where the functions $F_n(T)$ were derived numerically. We extracted~$F_n(T)$ from Fig.~1 of~\cite{Salvio:2013iaa}, together with the one-loop running of the gauge couplings~$\alpha_i(T)$.

\subsubsection{Coupling to gluons}
To isolate the effect of the coupling to gluons, we write $c_1=c_2 \equiv 0$ and define $\Lambda_g \equiv \Lambda/c_3$. In this case, the production rate~\eqref{eq:TotalRate} becomes
\beq
\Gamma_g(T) = \frac{F_3(T)}{\pi} \frac{T^3}{\Lambda_g^2} \equiv \gamma_g(T) \frac{T^3}{\Lambda_g^2}\, ,	\label{eq:Gg}
\eeq
with $\gamma_g(\SI{e10}{GeV}) = 0.41$. The function~$\gamma_g(T)$ is presented in the left panel of Fig.~\ref{fig:rateGluon}.%
\begin{figure}[t]
	\centering
	\includegraphics{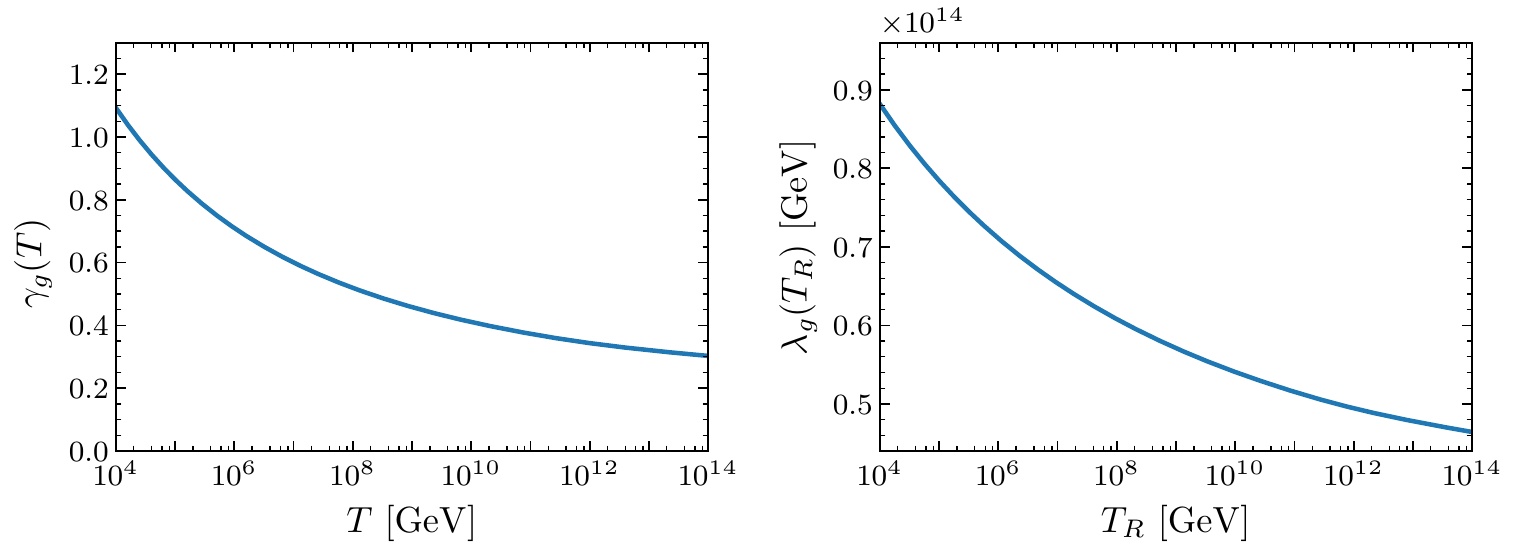}
	\caption{\textit{Left}:~Axion production rate associated with the coupling to gluons as parametrized by~$\gamma_g(T)$ in~\eqref{eq:Gg}. \textit{Right}:~Constraint on the axion-gluon coupling~$\Lambda_g$ as parametrized by~$\lambda_g(T_R)$ in~\eqref{eq:Lambg}.}
	\label{fig:rateGluon}
\end{figure}
The freeze-out bound on the gluon coupling then is
\beq
\Lambda_g \, >\, \left(\frac{\pi^2}{90} g_{*,R}\right)^{\!-1/4} \sqrt{\gamma_{g,R} \hskip1pt \Tr\hskip1pt \Mp} \,\, \equiv \, \lambda_g(\Tr) \left(\frac{\Tr}{\SI{e10}{GeV}}\right)^{\!1/2}\, ,	\label{eq:Lambg}
\eeq
where $g_{*,R} \equiv g_*(\Tr)$ and $\gamma_{g,R}\equiv \gamma_g(\Tr)$. The bound in~\eqref{eq:Lambg} is illustrated in the right panel of Fig.~\ref{fig:rateGluon}. In the main text, we used $\lambda_g(\SI{e10}{GeV}) = \SI{5.4e13}{GeV}$.

\subsubsection{Coupling to photons}
To isolate the coupling to the electroweak sector, we set $c_3=0$. In this case, the Lagrangian~\eqref{eq:LEW} can be written as
\beq
\L_{\phi\mathrm{EW}} = - \frac{1}{4}\frac{\phi}{\Lambda} \left( c_a \hskip1pt B_{\mu\nu} \tilde{B}^{\mu\nu} + s_a \hskip1pt W_{\mu\nu}^a \tilde{W}^{\mu\nu,a} \right) ,	\label{eq:A5}
\eeq
where we have defined
\beq
\Lambda \to \frac{\Lambda}{\sqrt{c_1^2+c_2^2}} \quad \mathrm{and} \quad c_a \equiv \frac{c_1}{\sqrt{c_1^2 +c_2^2}} \ , \ \ s_a \equiv \frac{c_2}{\sqrt{c_1^2 +c_2^2}}\, .
\eeq
Note that $c_a^2 + s_a^2 =1$, so we can use~$\Lambda$ and~$c_a$ as the two free parameters. The production rate~\eqref{eq:TotalRate} is then given by
\beq
\Gamma = \frac{[ c_a^2 F_1(T) + 3 s_a^2 \hskip1pt F_2(T) ]}{8\pi } \frac{T^3}{\Lambda^2}\equiv \gamma(T,c_a)\hskip2pt \frac{T^3}{\Lambda^2}\, .	\label{eq:GGgamma}
\eeq
The function~$\gamma(T,c_a)$ is shown in the left panel of Fig.~\ref{fig:ratePhoton}.%
\begin{figure}
	\centering
	\includegraphics{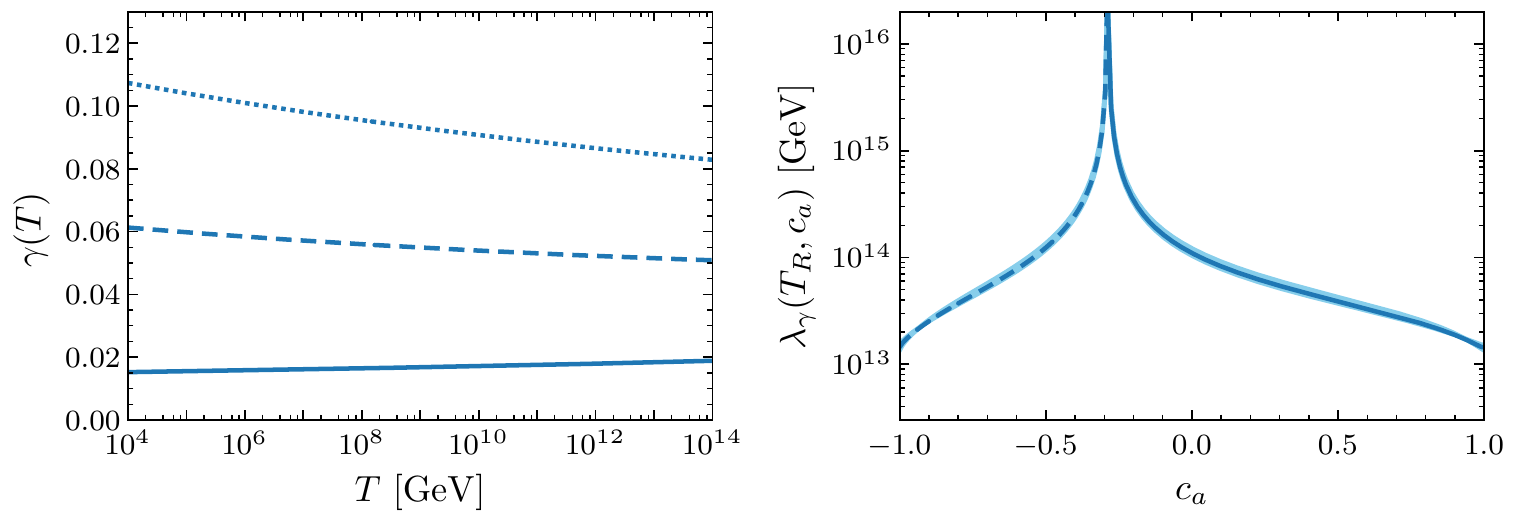}
	\caption{\textit{Left}:~Axion production rate associated with the coupling to the electroweak gauge bosons as parametrized by~$\gamma(T,c_a)$ in~\eqref{eq:GGgamma} for $c_a=0$~(dotted line), $1/\sqrt{2}$~(dashed line) and~$1$~(solid line). \textit{Right}:~Constraint on the axion-photon coupling~$\Lambda_\gamma$ as parametrized by~$\lambda_\gamma(\Tr, c_a)$ in~\eqref{eq:Lambgamma}. The solid and dashed lines correspond to bounds on positive and negative~$\Lambda_\gamma$ for $\Tr =\SI{e10}{GeV}$. The band displays the change for reheating temperatures between \SI{e4}{GeV}~(upper edge) and \SI{e15}{GeV}~(lower edge).}
	\label{fig:ratePhoton}
\end{figure}
In the main text, we employed $\gamma(\SI{e10}{GeV},1) = 0.017$. The freeze-out bound on the coupling then is
\beq
\Lambda(c_a) > \left(\frac{\pi^2}{90} g_{*,R}\right)^{\!-1/4} \sqrt{\gamma_{R}(c_a) \hskip1pt \Tr\hskip1pt \Mp}\, ,	\label{eq:Lca}
\eeq
with $\gamma_R(c_a)\equiv \gamma(\Tr,c_a)$. We wish to relate this bound to the couplings below the EWSB~scale.\medskip

At low energies, the axion couplings to the electroweak sector become
\beq
\L_{\phi\mathrm{EW}} =- \frac{1}{4} \left( \frac{\phi}{\Lambda_\gamma} \hskip1pt F_{\mu \nu} \tilde{F}^{\mu \nu} + \frac{\phi}{\Lambda_{Z}} Z_{\mu \nu} \tilde{Z}^{\mu \nu} + \frac{\phi}{\Lambda_{Z\gamma}} \hskip1pt Z_{\mu \nu} \tilde{F}^{\mu \nu} + \frac{\phi}{\Lambda_{W}} W^+_{\mu \nu} \tilde{W}^{-}{}^{\mu \nu} \right) ,	\label{eq:LphiEW}
\eeq
where $F_{\mu \nu}$, $Z_{\mu \nu}$ and $W^{\pm}_{\mu\nu}$ are the field strengths for the photon, $Z$ and $W^\pm$, respectively. Here, we have dropped additional (non-Abelian) terms proportional to $c_2$ which are cubic in the gauge fields. In order to match the high-energy couplings in~\eqref{eq:A5} to the low-energy couplings in~\eqref{eq:LphiEW}, we define
\begin{align}
\Lambda_\gamma^{-1}		&= \left( c^2_w \hskip1pt c_a + s_w^2 \hskip1pt s_a \right) \Lambda^{-1} \, ,	\label{eq:Lg}	\\ 
\Lambda_Z^{-1}			&= \left(c^2_w \hskip1pt s_a + s^2_w\hskip1pt c_a\right) \Lambda^{-1} \, , 						\\
\Lambda_{Z \gamma}^{-1}	&= 2 s_w c_w \left(s_a - c_a \right) \Lambda^{-1} \, ,											\\
\Lambda_{W}^{-1}		&= s_a \hskip 1pt \Lambda^{-1} \, ,
\end{align}
where $\{c_w,s_w\}\equiv \{\cos \theta_w, \sin \theta_w\}$, with Weinberg's mixing angle $\theta_w \approx \SI{30}{\degree}$. Using~\eqref{eq:Lg}, we can write~\eqref{eq:Lca} as a bound on the photon coupling, 
\begin{align}
\Lambda_\gamma(c_a) \,>\,	&\ \left( c^2_w \hskip1pt c_a + s_w^2 \hskip1pt s_a \right)^{\!-1} \times \left(\frac{\pi^2}{90} g_{*,R}\right)^{\!-1/4} \sqrt{\gamma_{R}(c_a) \hskip1pt \Tr\hskip1pt \Mp} \nonumber\\[2pt]
							&\ \equiv\, \lambda_\gamma(\Tr,c_a) \left(\frac{\Tr}{\SI{e10}{GeV}}\right)^{\!1/2} \,.	\label{eq:Lambgamma}
\end{align}
This bound is illustrated in the right panel of Fig.~\ref{fig:ratePhoton}. We see that we get the most conservative constraint by setting $s_a \equiv 0$, for which we have $\lambda_\gamma(\SI{e10}{GeV},1) = \SI{1.4e13}{GeV}$.

\subsection{Couplings to Matter Fields}
The calculation of the Goldstone production rates associated with the couplings to the SM~fermions is somewhat less developed. In this section, we calculate the relevant rates following the procedure outlined in~\cite{Masso:2002np}.

\subsubsection{Preliminaries}
The integrated Boltzmann equation for the evolution of the number density of the Goldstone boson takes the form
\beq
\frac{\d n_\phi}{\d t} + 3 H n_\phi = \Gamma (n^\mathrm{eq}_\phi - n_\phi) \, ,
\eeq
where $n^\mathrm{eq}_\phi = \zeta(3) T^3/\pi^2$ is the equilibrium density of a relativistic scalar. In order to simplify the analysis, we will replace the integration over the phase space of the final states with the centre-of-mass cross section,~$\sigma_\mathrm{cm}$, or the centre-of-mass decay rate,~$\Gamma_\mathrm{cm}$. While this approach is not perfectly accurate, it has the advantage of relating the vacuum amplitudes to the thermal production rates in terms of relatively simple integrals.

\begin{itemize}
\item For a two-to-two process, $1+2 \to 3+4$, we have
\beq
\Gamma_{2\to 2} \simeq \frac{1}{n_{\phi}^\mathrm{eq}} \int \frac{\d^3 p_1}{(2\pi)^3} \frac{\d^3 p_2}{(2\pi)^3} \frac{f_1(p_1)}{2 E_1} \frac{f_2(p_2)}{2 E_2} \big[1\pm f_3\big] \big[1\pm f_4\big] \,2 s \sigma_\mathrm{cm}(s)\, ,
\eeq
where~$f_{1,2}$ are the distribution functions of the initial states and $s\equiv (p_1+p_2)^2$ is the Mandelstam variable. We have included simplified Bose enhancement and Pauli blocking terms, $\big[1\pm f_3\big] \big[1\pm f_4\big] \to \frac{1}{2} \big( [1\pm f_3(p_1)] [1\pm f_4(p_2)] +\{ p_1 \leftrightarrow p_2 \} \big)$, which is applicable in the centre-of-mass frame where the initial and final momenta are all equal.\footnote{These Pauli blocking and Bose enhancement terms were not included in~\cite{Masso:2002np}, as they complicate the rate calculations. We have included them to ensure that the rates computed for both the forward and backward processes give the same results.} For $s \gg m^2_i$, the centre-of-mass cross section is given by 
\beq
\sigma_\mathrm{cm}(s) \simeq \frac{1}{32 \pi} \int\! \d\cos \theta\, \frac{\sum |\M|^2(s, \theta)}{s} \, ,	\label{eq:SigmaCM}
\eeq
where~$\sum |\M|^2$ is the squared scattering amplitude including the sum over spins and charges, and~$\theta$ is the azimuthal angle in the centre-of-mass frame. For all models of freeze-out considered in the main text, the centre-of-mass cross section is independent of~$s$. Moreover, we will only encounter fermion-boson scattering or fermion annihilation in this section. With the enhancement/blocking terms, one finds that the numerical pre-factors in both cases agree to within 10~percent. To simplify the calculations, we will therefore use the fermion annihilation rate throughout,
\beq
\Gamma_{2\to2} \simeq \sigma_\mathrm{cm}\,T^3 \left(\frac{7}{8}\right)^2 \frac{\zeta(3) }{ \pi^2} \approx 0.093 \, \sigma_\mathrm{cm} \,T^3\, .	\label{eq:Gamma2to2}
\eeq
The advantage of this approach is that we can relate the centre-of-mass cross section directly to the production rate with minimal effort and reasonable accuracy.

\item For a one-to-two process, $1 \to 2+3$, the decay rate in the centre-of-mass frame is
\beq
\Gamma_\mathrm{cm} \simeq \frac{1}{32 \pi \, m_1} \int\! \d \cos\theta\, \sum |\M|^2 \, ,
\eeq
where we have taken the two final particles to be massless. Since~$\Gamma_\mathrm{cm}$ is independent of energy, the rate only depends on whether the initial state is a fermion or boson. Transforming this rate to a general frame gives 
\beq
\Gamma_{1\to 2} \simeq \frac{1}{n_{\phi}^\mathrm{eq}} \int\! \frac{\d^3 p_1}{(2\pi)^3} \,f_1(p_1) \big[1\pm f_2(p_1/2)\big]\big[1\pm f_3(p_1/2)\big]\frac{m_1}{E_{1}} \Gamma_\mathrm{cm} \, ,	\label{eq:Gamma12}
\eeq
where~$f_1$ is the distribution function of the decaying particle (not necessarily~$\phi$). We are mostly interested in the limit $T \gg m_1$, in which case the rate~\eqref{eq:Gamma12} reduces to
\beq
\Gamma_{1\to2} \simeq \frac{m_1}{T} \frac{\pi^2}{16 \hskip1pt \zeta(3)} \,\Gamma_\mathrm{cm} \times 
	\begin{dcases}
		\,1-\frac{4}{\pi^2}	& \qquad \mathrm{fermion}, \\
		\,1 				& \qquad \mathrm{boson},
	\end{dcases}
\label{eq:decayRate}
\eeq
where the dependence on the number of degrees of freedom of the decaying particle has been absorbed into~$\Gamma_\mathrm{cm}$ through the sum over spins and charges. Note that, in equilibrium, the rates for decay and inverse decay are equal.
\end{itemize}

\subsubsection{Coupling to charged fermions}
We consider the following coupling between a Goldstone boson and charged fermions:
\beq
\L_{\phi \psi} = \frac{\phi}{\Lambda_\psi} \bigg(\ii H \,\bar\psi_{L,i} \!\left[ (\lambda_i - \lambda_j) g_V^{ij} + (\lambda_i + \lambda_j) g_A^{ij} \right]\! \psi_{R,j} + \mathrm{h.c.} \bigg) \, ,	\label{eq:Lferm}
\eeq
where~$H$ is the Higgs doublet, $\psi_{L,R} \equiv \tfrac{1}{2}(1\mp \gamma^5) \psi$, and the~$SU(2)_L$ and $SU(3)_C$~structures have been left implicit. Distinct processes dominate in the various limits of interest:

\begin{itemize}
\item \textbf{Freeze-out} \hskip6pt At high energies, the Goldstone boson is produced through the following two processes (see Fig.~\ref{fig:diagramsFermionsHiggs}):%
\begin{figure}[t]
	\centering
	\subcaptionbox{Fermion annihilation.}{ 		
		\hspace{0.6cm}\includegraphics*{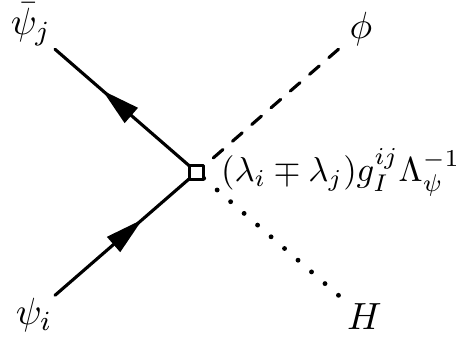}
	}\hspace{1.5cm}
	\subcaptionbox{Fermion-Higgs scattering.}{
		\hspace{0.6cm}\includegraphics*{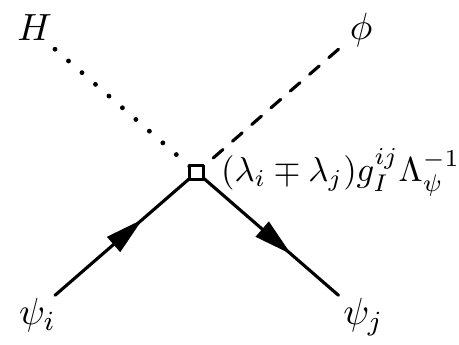}
	}
	\caption{Feynman diagrams for the dominant Goldstone production via the coupling to charged fermions above the electroweak scale. For the vector and axial vector couplings, $I \in \{V,A\}$, the~`$-$' and `$+$'~signs apply, respectively.}
	\label{fig:diagramsFermionsHiggs}
\end{figure}
(\hskip-0.9pt\textit{a})~$\psi_i +\bar \psi_j \to H + \phi$ and (\hskip-0.9pt\textit{b})~$\psi_i +H \to \psi_j +\phi$. Summing over the spins and charges, we get
\begin{align}
\sum |\M|^2_{(a)} &= 4 N_\psi \, s \, \frac{(\lambda_i - \lambda_j)^2 (g_V^{ij})^2 + (\lambda_i + \lambda_j)^2 (g_A^{ij})^2}{\Lambda_\psi^2} \, ,					\\
\sum |\M|^2_{(b)} &= 4 N_\psi \, s (1- \cos \theta) \, \frac{(\lambda_i - \lambda_j)^2 (g_V^{ij})^2 + (\lambda_i + \lambda_j)^2 (g_A^{ij})^2}{\Lambda_\psi^2} \, ,
\end{align}
where we have combined fermion and anti-fermion scattering in the sum over charges and introduced
\beq
N_\psi \equiv
	\begin{dcases}
		\,1	& \qquad \psi = \mathrm{lepton}, \\
		\,3 & \qquad \psi = \mathrm{quark}.
	\end{dcases}	
\eeq
We also find it convenient to define $\Lambda_{ij}^I \equiv \Lambda_\psi/g^{ij}_I$, with $I \in \{V, A\}$. Using~\eqref{eq:SigmaCM} and~\eqref{eq:Gamma2to2}, and treating the vector and axial-vector couplings separately, we obtain
\beq
\Gamma^{I}_{ij} = N_\psi \left(\frac{7}{8}\right)^2 \frac{ 4\hskip1pt \zeta(3)}{\pi^2} \frac{(\lambda_i \mp \lambda_j)^2}{8 \pi } \, \frac{ T^3}{(\Lambda_{ij}^{I})^2} \, \simeq\, 0.19 \hskip1pt N_\psi \frac{(\lambda_i \mp \lambda_j)^2}{8 \pi } \, \frac{ T^3}{(\Lambda_{ij}^{I})^2}\, ,
\eeq
where the~`$-$' and `$+$'~signs apply to $I=V$ and $I=A$, respectively.

\item\textbf{Freeze-in} \hskip6pt Below the scale of~EWSB, the Lagrangian~\eqref{eq:Lferm} becomes
\beq
\L_{\phi \psi} = \ii \frac{\phi}{\Lambda_\psi} \bar \psi_i \left[(m_i-m_j) g_V^{ij} +(m_i+m_j) g_A^{ij} \gamma^5 \right] \psi_j\, ,
\eeq
with $m_i \equiv \sqrt{2}\lambda_i/v$. The Goldstone production processes associated with these couplings are shown in Fig.~\ref{fig:diagramsFermions}.\medskip

\textit{Diagonal couplings.}---We first consider the diagonal part of the interaction, which takes the form~$\ii \tilde \epsilon_{ii} \, \phi \hskip1pt \bar\psi_i \gamma^5 \psi_i$, with $\tilde \epsilon_{ii} \equiv 2 m_i g_{A}^{ii} / \Lambda_\psi$. Kinematical constraints require us to include at least one additional particle in order to get a non-zero amplitude. The two leading processes are (\hskip-0.9pt\textit{a})~$\psi_i + \{\gamma,g\}\to \psi_i + \phi$ (cf.~Fig.~\ref{fig:fermionComptonProcessS}) and (\hskip-0.9pt\textit{b})~$\psi_i + \bar{\psi}_i \to \phi + \{\gamma,g\}$ (cf.~Fig.~\ref{fig:fermionFermionAnnihilation}), where~$\{\gamma,g\}$ is either a photon or gluon depending on whether the fermion is a lepton or quark, respectively. 
\begin{figure}[t]
	\centering
	\subcaptionbox{\label{fig:fermionComptonProcessS}Compton-like process.}{
		\includegraphics*{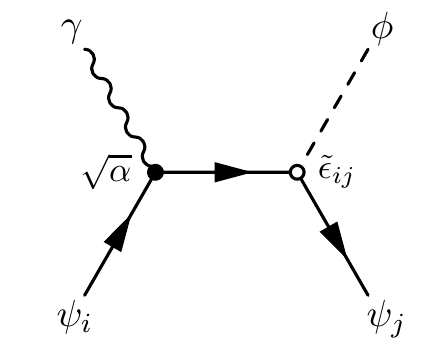}
	}\hspace{1.0cm}
	\subcaptionbox{\label{fig:fermionFermionAnnihilation}Fermion annihilation.}[\widthof{(b) Fermion annihilation.}]{
		\includegraphics*{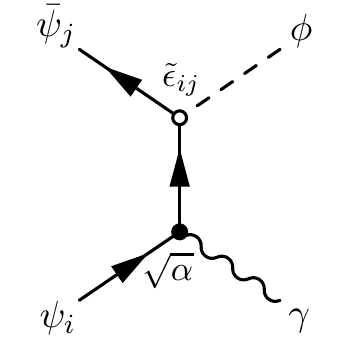}
	}\hspace{0.6cm}
	\subcaptionbox{\label{fig:fermionFermionDecay}Fermion decay.}{
		\includegraphics*{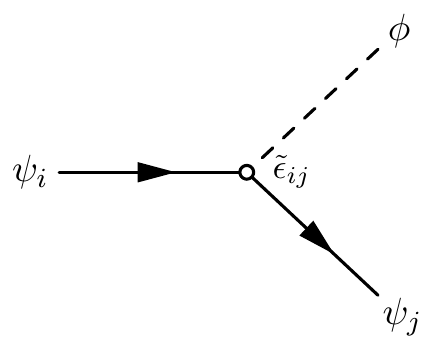}
	}
	\caption{Feynman diagrams for the dominant Goldstone production via the coupling to charged fermions below the electroweak scale. For quarks, the coupling to photons is replaced by that to gluons. In addition to the displayed $s$-~and~$t$-channel diagrams for the Compton-like process and fermion annihilation, there are $u$-channel diagrams which are not shown.}
	\label{fig:diagramsFermions}
\end{figure} 
Summing over spins and charges, we obtain
\begin{align}
\sum |\M|^2_{(a)} &= 16 \pi \hskip1pt A_\psi \, |\tilde \epsilon_{ii}|^2 \frac{s^2}{(m_i^2-t)(m_i^2-u)} \, ,	\\
\sum |\M|^2_{(b)} &= 16 \pi \hskip1pt A_\psi \, |\tilde \epsilon_{ii}|^2 \frac{t^2}{(s-m_i^2)(m_i^2-u)} \, ,
\end{align}
where~$s$, $t$ and~$u$ are the Mandelstam variables and
\beq
A_\psi \equiv 
		\begin{dcases}
			\,\alpha		& \qquad \psi = \mathrm{lepton}, \\
			\,4 \alpha_s	& \qquad \psi = \mathrm{quark}.
		\end{dcases}
\eeq
In the massless limit, the cross section has IR~divergences in the $t$-~and~$u$-channels from the exchange of a massless fermion. The precise production rate therefore depends on the treatment of the soft modes. Regulating the IR~divergence with the fermion mass and taking the limit $s \gg m_i^2$, we find
\beq
\sigma_\mathrm{cm}(s) \simeq \frac{1}{s} \, A_\psi\, |\tilde \epsilon_{ii}|^2 \left[3 \log \frac{s}{m_i^2} - \frac{3}{2 } \right] .
\eeq
At high temperatures, the fermion mass is controlled by the thermal mass $m_i^2 \to m_T^2 = \frac{1}{2} \pi A_\psi T^2$ and the production rate becomes
\beq
\tilde \Gamma_{ii} = \frac{3\pi^3}{64\zeta(3)} A_\psi\, \frac{|\tilde \epsilon_{ii}|^2}{8 \pi} T \left[ \log\frac{2}{ \pi A_\psi} +2 \log 2 -\frac{3}{2} \right] .	\label{eqn:chargeddiagonal}
\eeq
This formula is expected to break down at $T \lesssim m_i$, but will be sufficient at the level of approximation being used in this work. A proper treatment of freeze-in at $T\sim m_i$ should go beyond $\Gamma = H$ and fully solve the Boltzmann equations. However, this level of accuracy is not needed for estimating the constraint on the coupling~$\tilde\epsilon_{ii}$.

The result~\eqref{eqn:chargeddiagonal} will be of limited utility for the coupling to quarks. This is due to the fact that, for $T \lesssim \SI{30}{GeV}$, the QCD~coupling becomes large and our perturbative calculation becomes unreliable.\footnote{These effects are computable using the techniques of~\cite{Salvio:2013iaa}, but this is beyond the scope of the present work.} In fact, we see that the production rate~\eqref{eqn:chargeddiagonal} becomes negative in this regime. While the top quark is sufficiently heavy to be still at weak coupling, its mass is close to the electroweak phase transition and, therefore, the assumption $s \gg m_t^2$ is not applicable. For these reasons, we will not derive bounds on the quark couplings from these production rates.

\textit{Off-diagonal couplings.}---When the coupling of~$\phi$ is off-diagonal in the mass basis, the dominant process at low energies is the decay $\psi_i \to \psi_j +\phi$, cf.~Fig.~\ref{fig:fermionFermionDecay}. Since the mass splittings of the SM~fermions are large and $m_\phi \ll m_\psi$, the centre-of-mass decay rate is well approximated by
\beq
\Gamma_\mathrm{cm} = \frac{N_\psi}{8\pi} \frac{m_i^3}{\Lambda_{ij}^2} \, ,
\eeq
where $\Lambda_{ij} \equiv \big[(g_V^{ij})^2 + (g_A^{ij})^2\big]^{-1/2}\, \Lambda_\psi$. Using~\eqref{eq:decayRate}, we get 
\beq
\tilde \Gamma_{ij} = \frac{(\pi^2-4) }{16\hskip1pt \zeta(3)} \frac{N_\psi}{8\pi} \frac{1}{T} \frac{m_i^4}{\Lambda_{ij}^2} \,\simeq\, 0.31 N_\psi \frac{|\tilde \epsilon_{ij}|^2}{8\pi} \frac{m_i^2}{T} \, ,
\eeq
with $\tilde \epsilon_{ij} \approx m_i/\Lambda_{ij}$. In addition to this decay, we also have production with a photon or gluon, given by~\eqref{eqn:chargeddiagonal} with $\tilde \epsilon_{ii} \to \tilde \epsilon_{ij}$. We will neglect this contribution as it is suppressed by a factor of~$\alpha$ or~$\alpha_s$ for $T \sim m_i$.
\end{itemize}

\subsubsection{Coupling to neutrinos}
The coupling between the Goldstone boson and neutrinos is
\beq
\L_{\phi \nu}= - \frac{1}{2}\left( \ii \hskip1pt \tilde \epsilon_{ij} \phi \nu_i \nu_j - \frac{1}{2\Lambda_\nu} \epsilon_{ij} \phi^2 \nu_i \nu_j +\cdots \right) + \mathrm{h.c.} \, ,	\label{eq:LN}
\eeq
where we have written the Majorana neutrinos in two-component notation. The first term in~\eqref{eq:LN} will control freeze-in and the second will determine freeze-out:

\begin{itemize}
\item\textbf{Freeze-out} \hskip6pt At high energies, the dominant production mechanism is $\nu_i + \nu_j\to \phi + \phi$ (cf.~Fig.~\ref{fig:neutrinoNeutrinoAnnihilation}) through the second term in~$\L_{\nu \phi}$. 
\begin{figure}[t]
	\centering
	\subcaptionbox{\label{fig:neutrinoNeutrinoAnnihilation}Neutrino annihilation.}{
		\includegraphics*{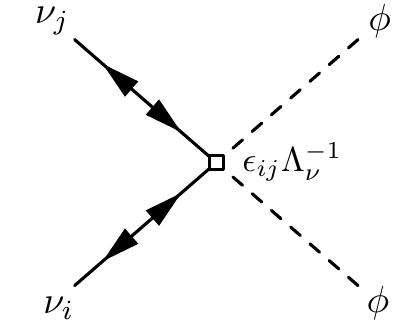}
	}\hspace{1.0cm}
	\subcaptionbox{\label{fig:neutrinoNeutrinoDecay}Neutrino decay.}{
		\includegraphics*{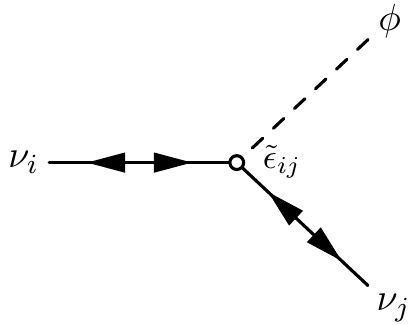}
	}\hspace{1.0cm}
	\subcaptionbox{\label{fig:neutrinoInverseDecay}Inverse decay.}{
		\includegraphics*{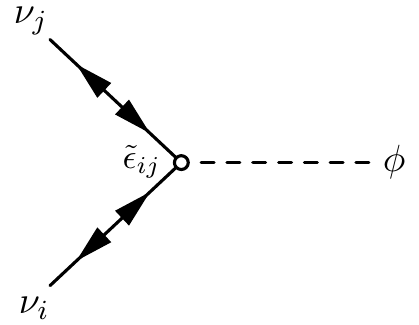}
	}
	\caption{Feynman diagrams for the dominant Goldstone production via the coupling to neutrinos. The double arrows denote the sum over the spinor index structure for two-component fermions~\cite{Dreiner:2008tw}.}
	\label{fig:diagramsNeutrinos}
\end{figure}
The spin-summed amplitude squared is
\beq
\sum |\M|^2 = |\epsilon_{ij}|^2 \frac{2 s}{\Lambda_\nu^2}\, ,
\eeq 
which results in the production rate
\beq
\Gamma_{ij} = \frac{1}{2}s_{ij}\left(\frac{7}{8}\right)^2 \frac{\zeta(3)}{\pi^2}\, \frac{|\epsilon_{ij}|^2}{8 \pi} \frac{T^3}{\Lambda_\nu^2} \,\simeq\, 0.047 \hskip1pt s_{ij}\, \frac{|\epsilon_{ij}|^2}{8 \pi} \frac{T^3}{\Lambda_\nu^2}\, ,
\eeq
where the factors of~$\frac{1}{2}$ and $s_{ij} \equiv 1-\frac{1}{2} \delta_{ij}$ are the symmetry factors for identical particles in the initial and final states, respectively. The contribution to the rate from higher-order terms in~\eqref{eq:LN} is suppressed by further powers of~$T^2/\Lambda_\nu^2$.

\item\textbf{Freeze-in} \hskip6pt Unlike for charged fermions, the freeze-in abundance from the coupling to neutrinos arises only through decays. Below the scale of~EWSB, the couplings of neutrinos to the rest of the~SM are suppressed by the weak scale and are irrelevant. The only freeze-in processes that are allowed by kinematics are therefore three-body decays.\medskip

\textit{Low-mass regime.}---For $m_\phi \ll m_i-m_j$, with $m_i > m_j$, the off-diagonal linear coupling allows the decay $\nu_i \to \nu_j +\phi$, cf.~Fig.~\ref{fig:neutrinoNeutrinoDecay}. The decay rate in the centre-of-mass frame is
\beq
\Gamma_\mathrm{cm} = \frac{1}{8\pi} \frac{m_i^2-m_j^2}{m_i^3} \left(|\tilde \epsilon_{ij}|^2 (m_i^2+m_j^2) + 2\,\mathrm{Re}\!\left[\!(\tilde \epsilon_{ij})^2\right] m_i m_j \right) .
\eeq
In order to simplify the calculations in the main text, we take $m_i \gg m_j$ which is guaranteed for the minimal mass normal hierarchy. Since the decaying particle is a fermion, the thermal production rate in~\eqref{eq:decayRate} becomes
\beq
\tilde \Gamma_{ij} = \frac{\pi^2-4}{16\hskip1pt\zeta(3)} \frac{|\tilde \epsilon_{ij}|^2 }{8\pi} \frac{m_{i}^2}{T} \,\simeq\, 0.31 \frac{|\tilde \epsilon_{ij}|^2 }{8\pi} \frac{m_{i}^2}{T} \, .
\eeq
Notice that the off-diagonal decay rate is the same for charged leptons and neutrinos even though the neutrinos have a Majorana mass.\medskip

\textit{High-mass regime.}---For $m_\phi \gg m_i \ge m_j$, the Goldstone boson decays to fermions, $\phi \to \nu_i + \nu_j$, both through the diagonal and off-diagonal couplings. The inverse decay $\nu_i + \nu_j \to \phi$ (see Fig.~\ref{fig:neutrinoInverseDecay}) is therefore a production channel. The decay rate is given by
\beq
\Gamma_\mathrm{cm} = \frac{|\tilde \epsilon_{ij}|^2 }{8\pi} m_{\phi} \, ,
\eeq
which, in equilibrium, is equal to the rate for the inverse decay. Since the decaying particle is a boson, the thermal production rate in~\eqref{eq:decayRate} becomes
\beq
\tilde \Gamma_{ij} = s_{ij} \, \frac{\pi^2}{16\hskip1pt \zeta(3)} \frac{|\tilde \epsilon_{ij}|^2 }{8\pi} \frac{m_{\phi}^2}{T} \,\simeq\, 0.51 \hskip1pt s_{ij}\, \frac{|\tilde \epsilon_{ij}|^2 }{8\pi} \frac{m_{\phi}^2}{T} \, .
\eeq
The rate is somewhat enhanced compared to the decay in the low-mass regime because the decaying particle is a boson.
\end{itemize}
The dominant Goldstone production mechanism through the couplings to neutrinos is quite sensitive to kinematics. For $m_\phi \lesssim m_{\nu}$, the diagonal decay is forbidden and the dominant Goldstone production is through freeze-out. In addition, when $m_\phi \sim m_\nu$ there are additional kinematic constraints for both diagonal and off-diagonal couplings. As a result, the limits on the interaction scale~$\Lambda_\nu$ (or the dimensionless couplings~$\epsilon_{ij}$ and~$\tilde \epsilon_{ij}$) are sensitive to~$m_\phi$.

\section{Comments on Decays}
\label{app:decays}
Throughout Chapter~\ref{chap:cmb-axions}, we treated each of the operators which couple the pseudo-Nambu Goldstone bosons to the Standard Model independently. For computing the production rates, this is justified since the amplitudes for the different processes that we consider do not interfere and the couplings therefore add in quadrature. One may still ask, however, if the interplay between several operators can affect the cosmological evolution after the production. In particular, one might worry that some operators would allow for the decay of the~pNGBs and that this might evade the limits on~$\Neff$. In the following, we will address this concern. We are assuming that $m_\phi < \SI{1}{MeV}$, so that the only kinematically allowed decays are to photons and neutrinos.

\subsection{Decay to photons}
If the decay occurs after recombination, then the pNGBs are effectively stable as far as the CMB is concerned and our treatment in the main text applies directly. To see when this is the case, we computed the decay temperature~$\Td$ associated with the decay mediated by the coupling to photons~\eqref{eq:phiF}:
\beq
\frac{\Td}{\Trec} \,\approx\, \num{9.5e-10} \left(\frac{\Lambda_\gamma}{\SI{e10}{GeV}} \right)^{\!-4/3} \left(\frac{m_\phi}{\Trec}\right)^{\!2} \, .
\eeq
Recalling the stellar cooling bound, $\Lambda_\gamma >\SI{1.3e10}{GeV}$~\cite{Friedland:2012hj}, we see that the~pNGBs are effectively stable as long as $m_\phi\lesssim \SI{10}{keV}$. For comparison, a stable particle with $m_\phi \gtrsim \SI{100}{eV} $ produces $\Omega_m >1$ and is therefore excluded by constraints on the dark matter abundance. For $m_\phi>\SI{10}{keV}$, the decay to photons does affect the phenomenology and must be considered explicitly. Nevertheless, in the regime of interest, the~pNGBs are non-relativistic and, therefore, carry a large energy density, $\rho_\phi \simeq m_\phi n_\phi$. As a result, this region is highly constrained by current cosmological observations~\cite{Cadamuro:2011fd, Millea:2015qra}.

\subsection{Decay to neutrinos}
Depending on the mass of the~pNGB, the decay to neutrinos leads to the following three scenarios:
\begin{itemize}
\item For $m_\phi < \Trec$, the implications of the decays are relatively easy to characterise. As discussed in \S\ref{sec:nufreezein}, the phenomenology is only modified if $\Tfl>\Trec$. In this case, strong interactions between the~pNGBs and the neutrinos imply that the neutrinos are no longer free-streaming particles, which is ruled out by current CMB~observations, cf.\ Section~\ref{sec:analysis}.

\item For $\Td> m_\phi > \Trec$, the~pNGBs are brought into equilibrium with the neutrinos at $T \sim \Td$ and then become Boltzmann suppressed for $T\lesssim m_\phi$. This process leads to a contribution to $\Neff$, even if the~pNGBs have negligible energy density to begin with. To estimate the size of the effect, we first note that the freeze-in at~$\Td$ conserves the total energy density in neutrinos and~pNGBs, 
\beq
(g_{*,\nu} + g_{*,\phi}) (a_1 T_1)^4 = g_{*,\nu} (a_0 T_0)^4\, ,
\eeq
where~$T_0$ and~$T_1$ are the initial and final temperatures during the equilibration, and~$g_{*,\nu}$ and $g_{*,\phi} = 1$ are the effective numbers of degrees of freedom in~$\nu$ and~$\phi$, respectively. When the temperature drops below the mass of the~pNGBs, their energy density is converted to neutrinos. This process conserves the comoving entropy density,
\beq
(g_{*,\nu} + g_{*,\phi}) (a_1 T_1)^3 =g_{*,\nu} (a_2 T_2 )^3 \, ,
\eeq
where $T_2 \ll m_\phi$ is some temperature after the pNGB~population has decayed. The final energy density of the neutrinos becomes
\beq
a_2^4 \hskip1pt \rho_{\nu,2} = \left(\frac{g_{*,\nu} + g_{*,\phi}}{g_{*,\nu}} \right)^{\!1/3} a_0^4\hskip1pt \rho_{\nu, 0} \, ,
\eeq
where $\rho_{\nu, i}\equiv \rho_\nu(a_i)$. Using the definition of~$\Neff$ in~\eqref{eq:Neff} together with~\eqref{eq:neutrinoTemperature} as well as~\eqref{eq:anu}, and $\rho_\gamma a^4 = \const$, we find 
\beq
\Neff =\left(\frac{g_{*,\nu} + g_{*,\phi}}{g_{*,\nu}} \right)^{\!1/3} N_{\mathrm{eff},0}\, .
\eeq
Considering the coupling to a single neutrino flavour (rather than all three), i.e.~$N_{\mathrm{eff},0} \simeq 1$ and $g_{*,\nu} = 7/4$, we then get
\beq
\Delta\Neff = \left(1+ \frac{4}{7}\right)^{\!1/3} -1 \,\simeq\, 0.16\, ,
\eeq
where $\Delta\Neff \equiv \Neff-N_{\mathrm{eff},0}$. Coupling to more than one neutrino flavour and including a non-zero initial temperature for the~pNGBs would increase this number slightly, so that we will use $\Delta\Neff \geq 0.16$.

\item The production of~pNGBs through the freeze-in process is avoided if $m_\phi > \Td > \Trec$, in which case the~pNGBs decay to neutrinos out of equilibrium. To a good approximation, this decay conserves the energy density, which is therefore simply transferred from~$\phi$ to~$\nu$ at the time of the decay. The contribution to~$\Delta\Neff$ is enhanced by the amount of time that~$\phi$ is non-relativistic before its decay, which may be a large effect for $m_\phi \gg 1$ eV (see e.g.~\cite{Fischler:2010xz} for a related discussion).
\end{itemize}
In summary, operators that allow the Goldstone bosons to decay do not substantially alter the predictions presented in the main text. On the one hand, decays to photons cannot occur early enough to impact the~CMB. On the other hand, decays to neutrinos typically increase the contributions to~$\Delta\Neff$ and would therefore strengthen our bounds.
		\addtocontents{toc}{\protect\setcounter{tocdepth}{1}}

\chapter{Further Aspects of the Phase Shift}
\label{app:cmb-phases_appendices}
In this appendix, we provide supplemental material to Chapter~\ref{chap:cmb-phases}. We estimate the effects of matter domination on the neutrino-induced phase shift~(\textsection\ref{app:matter}) and derive that the phase shift in the CMB~polarization spectrum is the same as in the temperature spectrum~(\textsection\ref{app:polarization}).

\section{Comments on Matter}
\label{app:matter}
In the main text, we computed the phase shift of the photon density fluctuations assuming a radiation-dominated background. While this simplification made an analytic treatment possible, we may wonder if it misses important effects. In this appendix, we will bridge this gap to the degree which is possible without using numerics, focussing on the contributions from free-streaming radiation.\medskip

There are several reasons why we want to understand the contributions to the phase shift from modes in the matter era. First, recombination occurs during matter domination and, in principle, it could therefore be important for every mode in the~CMB.\footnote{We derive the effects of matter on the phase shift in the acoustic peaks of the~CMB. However, it should be clear from our discussion in Chapter~\ref{chap:bao-forecast} that this treatment equally applies to the phase shift in the BAO~spectrum.} Second, modes corresponding to large angular scales (small~$\ell$) enter the horizon during (or near) matter domination and their complete evolution is consequently governed by the physics in the matter era. Finally, ref.~\cite{Follin:2015hya} found a logarithmic dependence of the phase shift on the multipoles~$\ell$ for observable modes. One may be tempted to interpret this as an effect of the finite matter density. Our goal in this appendix is to further clarify these effects, by studying the limits $\ell \to \infty$ and $\ell \to 0$, accounting for the contributions from matter. We will study these limits in turn:

\begin{itemize}
\item We first consider modes which entered the horizon during the radiation era. These correspond to small angular scales in the~CMB anisotropy spectra. We begin by writing~\eqref{eq:B}~as
\beq
B(y) = \underbrace{\int^{y_\mathrm{eq}}_0 \d y'\, \Phi_+(y')\hskip1pt\cos y' }_{\displaystyle \equiv B_\mathrm{rad}} \, \ +\, \ \underbrace{\int_{y_\mathrm{eq}}^y \d y'\, \Phi_+(y')\hskip1pt\cos y'}_{\displaystyle \equiv B_\mathrm{mat}} \ ,	\label{eq:BBmatter}
\eeq
where $y_\mathrm{eq} = c_\gamma k \tau_\mathrm{eq}$ corresponds to the moment of matter-radiation equality. Modes that entered the horizon long before $\tau_\mathrm{eq}$ correspond to $y_\mathrm{eq} \gg 1$. For these modes, the first term in~\eqref{eq:BBmatter} can be approximated as
\beq
B_\mathrm{rad} \simeq \int^{\infty}_0 \d y'\, \Phi^\mathrm{(rad)}_+(y')\hskip1pt\cos y' \, , 
\eeq
which is precisely the result computed in Section~\ref{sec:analytics}. The main correction from the matter era then is the second term in~\eqref{eq:BBmatter}:
\beq
B_\mathrm{mat} \simeq \int_{y_\mathrm{eq}}^y \d y'\, \Phi_+^\mathrm{(mat)}(y') \hskip1pt\cos y' \, .
\eeq
To estimate this effect, we simply have to repeat the discussion of \textsection\ref{sec:fs} for the matter era. The important difference is that $\epsilon_X \equiv \bar \rho_X/\bar \rho$ now is not a constant, but scales as $a^{-1} \propto \tau^{-2}$. Setting~$\tau_\mathrm{in}$ in the matter era, we have $\epsilon_X = \epsilon_{X,\mathrm{in}}\hskip1pt \tau_\mathrm{in}^2/\tau^2$ and eq.~\eqref{eq:Phi-X} becomes
\beq
\Phi_-(y) = - \frac{8k^2}{y^4} \epsilon_{X,\mathrm{in}} \hskip1pt y_\mathrm{in}^2\, \sigma_X(y) = - \frac{16}{3}\frac{1}{y^4} \epsilon_{X,\mathrm{in}} \hskip1pt y_\mathrm{in}^2\, D_{X,2}(y) \, .	\label{eq:phimmatter}
\eeq
To determine~$\Phi_-(y)$ to first order in~$\epsilon_X$, we only need the quadrupole moment~$D_{X,2}(y)$ to zeroth order. From~\eqref{eq:DX2}, we get
\begin{align}
D_{X,2}^{(0)}(y) \,=\, \	& d_{X,\mathrm{in}} \, j_{2}\!\left[c_\gamma^{-1}(y-y_\mathrm{in})\right] \nonumber \\
							&+ \frac{3}{c_\gamma} \Phi_{+,\mathrm{in}} \int_{y_\mathrm{in}}^y \d y'\, \left\{ \frac{2}{5} j_{1}\!\left[c_\gamma^{-1}(y-y')\right] - \frac{3}{5}j_{3}\!\left[c_\gamma^{-1}(y-y')\right] \right\} \, ,
\end{align}
where we have used that $\Phi_{+}^{(0)} = \const$ during the matter era. In the limit, $y \gg 1$, this leads~to
\beq
\Phi_-^{(1)}(y) \,\simeq\, 16\epsilon_{X,\mathrm{in}} \, y_\mathrm{in}^2\, \frac{\sin (c_\gamma^{-1} y)}{c_\gamma^{-1}y^5} \Big( \Phi_{+,\mathrm{in}} + \frac{1}{3} d_{X,\mathrm{in}} \Big) \, .
\eeq
Hence, we get $\Phi_- \propto y^{-5} \to 0$ in the limit $y \to \infty$. At late times, $\Phi_+$ is therefore no longer sourced by~$\Phi_-$ and will be given by the homogeneous solution (with coefficients that may depend on~$\epsilon_X$). Importantly, the value of~$B$ will be the same as that predicted in Section~\ref{sec:analytics}. At high~$k$ (and thus~$\ell$), the phase shift of the acoustic oscillations will consequently be equal to the value in a radiation-dominated universe.

\item Next, let us study modes which entered the horizon during the matter era, corresponding to large angular scales in the~CMB. In this case, it is more challenging to cleanly separate the result into a correction to the amplitude of oscillations and a phase shift. Our primary goal will be to understand how the result scales with wavenumber~$k$ in the limit $k \to 0$. Fortunately, this scaling is the same for the amplitude correction and the phase shift, and is easy to understand analytically. 

Intuitively, we expect the contributions from dark radiation (including neutrinos) to vanish as $k \to 0$. As we lower~$k$, the time of horizon entry increases compared to the time of matter-radiation equality and, therefore, the radiation energy density should be diluted relative to the matter. Since this radiation only affects observations through its gravitational influence, its role in the evolution of the modes should become negligible.

We can confirm this intuition by returning to~\eqref{eq:phimmatter} and noticing that $y_\mathrm{in} = c_\gamma k \tau_\mathrm{in}$, where $\tau_\mathrm{in}$ is a fixed time which is independent of~$k$, e.g.\ we may choose~$\tau_\mathrm{in}$ to be the time of matter-radiation equality. We therefore have $\Phi_-^{(1)} = \epsilon_{X,\mathrm{in}} c_\gamma^2 k^2 \tau_\mathrm{in}^2\, g(y)$ and the correction to~$d_\gamma$ at linear order in~$\epsilon_X$ will take the form
\beq
d^{(1)}_\gamma (\tau) = \epsilon_{X,\mathrm{in}}\, c_\gamma^2 k^2 \tau_\mathrm{in}^2 \int^{y}_{y_\mathrm{in}} \d y' f(y,y') \, .
\eeq
If the integral converges as $y \to \infty$, it is clear that $d^{(1)}_\gamma \propto k^2 \to 0$. In fact, if the integral diverges as $\tau \to \infty$, the result will be suppressed by additional powers of~$k$, due to the scaling of the upper limit of integration ($y=c_\gamma k \tau$ at fixed~$\tau$). Hence, we conclude that the amplitude and phase corrections from neutrinos (or any dark radiation) will vanish \textit{at least} as fast as~$k^2$.
\end{itemize}

\noindent
From these asymptotic scaling arguments, we draw the following conclusions:
\begin{itemize}
\item For $k \to \infty$, the phase shift due to free-streaming particles approaches a constant.
\item For $k \to 0$, the phase and amplitude corrections scale at least as~$k^2$.
\end{itemize}
In the flat-sky limit, these results translate approximately to $\ell \simeq k (\tau_0-\tau_{\rec})$, where~$\tau_0$ is the conformal time today. We therefore expect a constant phase shift at high~$\ell$. Given that matter-radiation equality corresponds to relatively low~$\ell$, we do not expect our asymptotic formula for $k \to 0$ to be more than a rough guide. The primary purpose of this discussion was to highlight that a power law is the likely behaviour, simply due to the power law decay of the energy density of the extra radiation. As a result, the phase shift per~$\ell$ should be some function that interpolates between a power law and a constant, and is unlikely to follow the ansatz of~\cite{Follin:2015hya} in detail (although the logarithmic dependence appears to work well enough on intermediate scales).

\section{Comments on Polarization}
\label{app:polarization}
The analytic discussions in the main text were phrased in terms of the temperature anisotropy, but as we saw in Section~\ref{sec:analysis}, CMB~polarization plays a crucial role in present and future data analyses. In this appendix, we show that the phase shift of the polarization spectrum is the same as that of the temperature spectrum.\medskip

Following~\cite{Zaldarriaga:1995gi}, we write the Boltzmann equation for the amplitude of polarized anisotropies,~$\Theta_P$, as
\beq
\dot \Theta_P + i k \mu \Theta_P \,=\, - \dot \kappa \left[-\Theta_P + \frac{1}{2} \left(1-P_{2}(\mu) \right) \Pi \right] \, ,	\label{eq:A8}
\eeq
where $\Pi \equiv \Theta_{2} + \Theta_{P,0} + \Theta_{P,2}$ and $\dot \kappa =-n_e \sigma_T a$ is the time derivative of the optical depth~$\kappa$ (to avoid confusion with the conformal time~$\tau$). The temperature quadrupole is determined by the photon anisotropic stress, $\Theta_{2} \equiv \frac{1}{2} k^2 \sigma_\gamma $. Equation~\eqref{eq:A8} admits a solution as a line-of-sight integral,
\beq
\Theta_P(\tau_0) = \int^{\tau_0}_{\tau_\mathrm{in}} \d\tau \, e^{i k \mu (\tau-\tau_0) - \kappa(\tau)} \left(\frac{3}{4} \dot \kappa(\tau) \hskip2pt (\mu^2-1)\hskip2pt \Pi(\tau) \right) \, .	\label{eq:ThetaP}
\eeq
The integral in~\eqref{eq:ThetaP} is proportional to the visibility function $-\dot\kappa \hskip1pt e^{-\kappa}$ and is therefore peaked at the surface of last-scattering. In the limit of instantaneous recombination, $-\dot \kappa e^{-\kappa} \simeq \delta_D(\tau-\tau_\rec)$, we get
\beq
\Theta_P(\tau_0) \simeq e^{i k \mu (\tau_\rec -\tau_0)} \frac{3}{4} \big(1-\mu^2\big) \Pi (\tau_\rec) \, .	\label{TP}
\eeq 
Solving for~$\Pi$ to leading order in $\dot\kappa \gg 1$, one finds $\Pi \simeq \tfrac{5}{2} \Theta_2 \simeq -\frac{10}{9} k \, \dot \kappa^{-1} \,\Theta_1$ (using the collision term in the Boltzmann equation for temperature). Applying the continuity equation, $\dot d_\gamma = - 3 k \Theta_1$, and performing a multipole expansion, one then finds 
\beq
\Theta_{P,\ell}(\tau_0)\simeq \frac{5}{18} \,\dot d_\gamma(k,\tau_\rec)\, \dot\kappa^{-1}(\tau_\rec)\left(1+ \frac{\partial^2}{\partial (k\tau_0)^2}\right) j_{\ell}(k \tau_0)\, .
\eeq
Two facts should be noticed about this result: 
\begin{itemize}
\item $\Theta_{P,\ell} \propto \dot d_\gamma$.---Since the time derivative will not affect the phase shift from dark radiation, we see that the locations of the acoustic peaks in the polarization spectrum are affected by the fluctuations in the dark radiation in the same way as in the temperature spectrum.
\item $\Theta_{P,\ell} \propto \dot \kappa^{-1} \propto n_e^{-1}$.---This is important because it allows the degeneracy between~$H$ and~$n_e$ (or~$Y_p$) in the damping tail (which scales as~$(n_e H)^{-1}$; cf.~\textsection\ref{sec:species_diffusionDamping}) to be broken.
\end{itemize}
		\addtocontents{toc}{\protect\setcounter{tocdepth}{1}}

\chapter{Details of the Fisher Forecasts}
\label{app:bao-forecast_appendices}
In this appendix, we provide supplemental material to Chapter~\ref{chap:bao-forecast}. We first describe our CMB~Fisher forecasts and present results for a range of experimental configurations~(\textsection\ref{app:CMB}). We then provide details of our LSS~forecasts, define the specifications for the employed galaxy surveys, and report results for a range of data combinations and cosmologies~(\textsection\ref{app:LSS}). Finally, we show a few of the convergence tests that we performed to establish the stability of our numerical computations~(\textsection\ref{app:convergence}).

\section{Forecasting CMB Constraints} 
\label{app:CMB}
Forecasting the sensitivities of future CMB~observations is by now a standard exercise; see e.g.~\cite{Wu:2014hta, Galli:2014kla, Allison:2015qca, Abazajian:2016yjj}. For completeness, this appendix collects the basic elements of our CMB~Fisher analysis as well as the specifications of the CMB~experiments that were used in our forecasts of Section~\ref{sec:forecast}.

\subsection{Fisher Matrix}
The Fisher matrix for CMB experiments can be written as
\beq
F_{ij} = \sum_{X,Y}\, \sum_{\ell=\lmin}^{\lmax} \frac{\partial C_\ell^{X}}{\partial \theta_i} \left[\mathbf{C}_\ell^{XY}\right]^{\!-1}\frac{\partial C_\ell^{Y}}{\partial \theta_j} \, .
\eeq
The covariance matrix $\mathbf{C}_\ell^{XY}$ for each multipole $\ell$ and $X=ab$, $Y=cd$, with $a,b,c,d=T,E,B$, is defined by
\beq
\mathbf{C}_\ell^{ab\hskip1ptcd} = \frac{1}{(2\ell+1) \fsky} \left[ (C_\ell^{ac} + N_\ell^{ac})(C_\ell^{bd} + N_\ell^{bd}) + (C_\ell^{ad} + N_\ell^{ad})(C_\ell^{bc} + N_\ell^{bc})\right] ,
\eeq
where~$C_\ell^X$ are the theoretical CMB~power spectra,~$N_\ell^X$ are the (Gaussian) noise spectra of a given experiment and~$\fsky$ is the effective sky fraction that is used in the cosmological analysis. We employ perfectly delensed power spectra and omit the lensing convergence for simplicity as it is sufficient for our purposes. We however comment on the effects of these assumptions below. The noise power spectra are
\beq
N_\ell^X = (\Delta X)^2 \exp\left\{ \frac{\ell (\ell+1)\, \theta_b^2}{8\ln2} \right\} ,	\label{eq:noiseSpectrum}
\eeq
with the map sensitivities for temperature and polarization spectra $\Delta X = \Delta T,\Delta P$, respectively, and the beam width~$\theta_b$ (taken to be the full width at half maximum). Note that we set $N^{TE}_\ell\equiv0$ as we assume the noise in temperature and polarization to be uncorrelated. For a multi-frequency experiment, the noise spectrum~\eqref{eq:noiseSpectrum} applies for each frequency channel separately. The effective noise after combining all channels is\vspace{-6pt}
\beq
N_\ell^X = \left[\sum_\nu \left(N_\ell^{X,\nu}\right)^{\!-1}\right]^{\!-1}\, ,	\vspace{-6pt}
\eeq
where~$N_\ell^{X,\nu}$ are the noise power spectra for the separate frequency channels~$\nu$.

\subsection{Experimental Specifications}
Our specifications for the Planck satellite are collected in Table~\ref{tab:planckSpecs}. The adopted configuration is the same as that used in the CMB-S4~Science Book~\cite{Abazajian:2016yjj}. For the low-$\ell$ data, we use the unlensed TT~spectrum with $\lmin = 2$, $\lmax=29$ and $\fsky = 0.8$. We do not include low-$\ell$~polarization data, but instead impose a Gaussian prior on the optical depth, with $\sigma(\tau) = 0.01$. For the high-$\ell$ data, we use the unlensed~TT, TE, EE~spectra with $\lmin=30$, $\lmax = 2500$ and $\fsky = 0.44$. Since the low-$\ell$ and high-$\ell$ modes are independent, we simply add the corresponding Fisher matrices.
\begin{table}
	\begin{tabular}{l S[table-format=4.1] S[table-format=4.1] S[table-format=4.1] S[table-format=4.1] S[table-format=4.1] S[table-format=4.1] S[table-format=4.1]}
			\toprule
		Frequency [\si{\giga\hertz}]		& 30	& 44	& 70	& 100	& 143	& 217	& 353	\\
			\midrule[0.065em]
		$\theta_b$ [\si{arcmin}]			& 33	& 23	& 14	& 10	& 7		& 5		& 5		\\
		$\Delta T$ [$\si{\muKelvin.arcmin}$]	& 145	& 149	& 137	& 65	& 43	& 66	& 200	\\
		$\Delta P$ [$\si{\muKelvin.arcmin}$]	& {--}	& {--}	& 450	& 103	& 81	& 134	& 406	\\
			\bottomrule 
	\end{tabular}
	\caption{Specifications for the Planck-like experiment used in~\cite{Allison:2015qca} and in the CMB-S4~Science Book~\cite{Abazajian:2016yjj}. The dashes in the first two columns for~$\Delta P$ indicate that those frequency channels are not sensitive to polarization.\vspace{-3pt}}
	\label{tab:planckSpecs}
\end{table}

We parametrize future CMB~experiments in terms of a single effective frequency with noise level~$\Delta T$, beam width~$\theta_b$ and sky fraction~$\fsky$. We will present constraints as a function of these three parameters. We take $\theta_b=\SI{3}{\arcmin}$, $\Delta T=\SI{5}{\muKelvin.arcmin}$ and $\fsky=0.3$ as the fiducial configuration of a CMB-S3-like experiment. For a representative CMB-S4~mission, we adopt the same configuration as in the CMB-S4~Science Book~\cite{Abazajian:2016yjj}: $\theta_b=\SI{2}{\arcmin}$, $\Delta T=\SI{1}{\muKelvin.arcmin}$ and $\fsky=0.4$. For both experiments, we use unlensed temperature and polarization spectra with $\lmin=30$, $\lmax^T = 3000$, $\lmax^P = 5000$. We add the low-$\ell$ Planck data as described above, include high-$\ell$~Planck data with $\fsky=0.3$ and $\fsky=0.2$ for CMB-S3 and~CMB-S4, respectively, and impose the same Gaussian prior on the optical depth~$\tau$ as for Planck. In addition, we forecast a cosmic variance-limited experiment with $\lmin=2$, $\lmax^T = 3000$, $\lmax^P = 5000$ and $\fsky=0.75$ to estimate the potential reach.

Unlike the CMB-S4 Science Book, we do not include delensing of the T-~and E-modes. For $\Neff$\hskip1pt-forecasts, this was shown to have a negligible impact~\cite{Green:2016cjr}, while using unlensed spectra overestimates the constraining power of the~CMB by roughly~\SI{30}{\percent} for $\Neff$+$Y_p$. We are primarily interested in the improvement in parameters from adding LSS~data, which should be robust to these relatively small differences. We also ignore the lensing convergence as it basically does not impact the constraints on these parameters.

\subsection{Future Constraints}
As a point of reference and for comparison with the results of Section~\ref{sec:analysis}, we present constraints derived from CMB~observations alone. In Table~\ref{tab:cmbForecast_LCDM_LCDM+Neff},%
\begin{table}[t]
	\begin{tabular}{c S[table-format=2.4] S[table-format=2.4] S[table-format=2.4] S[table-format=2.4] S[table-format=2.4] S[table-format=2.4] S[table-format=2.4] S[table-format=2.4]}
			\toprule
		Parameter				& {Planck}	& {\!CMB-S3}& {CMB-S4\!}& {CVL}		& {Planck}	& {\!CMB-S3}& {CMB-S4\!}& {CVL}		\\
			\midrule[0.065em]
		$\num{e5}\,\omega_b$	& 16		& 5.1		& 2.7		& 0.97		& 26		& 8.3		& 3.8		& 1.3		\\
		$\num{e4}\,\omega_c$	& 16		& 8.3		& 7.1		& 4.5		& 26		& 10		& 7.9		& 4.5		\\
		$\num{e7}\,\theta_s$	& 29		& 9.4		& 5.9		& 3.5		& 44		& 13		& 6.7		& 3.5		\\
		$\!\ln(\num{e10}\As)\!$	& 0.020		& 0.020		& 0.020		& 0.0041	& 0.021		& 0.020		& 0.020		& 0.0041	\\
		$\ns$					& 0.0040	& 0.0023	& 0.0020	& 0.0012	& 0.0093	& 0.0040	& 0.0030	& 0.0020	\\
		$\tau$					& 0.010		& 0.010		& 0.010		& 0.0020	& 0.010		& 0.010		& 0.010		& 0.0020	\\
		$\Neff$					& {--}		& {--}		& {--}		& {--}		& 0.18		& 0.054		& 0.030		& 0.011		\\
			\bottomrule
	\end{tabular}
	\caption{Forecasted sensitivities of Planck, CMB-S3, CMB-S4 and a CVL experiment for the parameters of $\Lambda\mathrm{CDM}$ and $\Lambda\mathrm{CDM}$+$\Neff$.}
	\label{tab:cmbForecast_LCDM_LCDM+Neff}
\end{table}
we show the $1\sigma$~constraints for Planck, the described representative configurations of CMB-S3 and CMB-S4, and the mentioned CVL~experiment. In Table~\ref{tab:cmbForecast_LCDM+Yp_LCDM+Neff+Yp}, we display how these constraints vary when we allow the helium fraction~$Y_p$ to be an additional free parameter. The differences in the forecasted sensitivities for Planck%
\begin{table}[h!]
	\begin{tabular}{c S[table-format=2.4] S[table-format=2.4] S[table-format=2.4] S[table-format=2.4] S[table-format=2.4] S[table-format=2.4] S[table-format=2.4] S[table-format=2.4]}
			\toprule
		Parameter				& {Planck}	& {\!CMB-S3}& {CMB-S4\!}& {CVL}		& {Planck}	& {\!CMB-S3}& {CMB-S4\!}& {CVL}		\\
			\midrule[0.065em]
		$\num{e5}\,\omega_b$	& 24		& 8.2		& 3.8		& 1.4		& 26		& 8.4		& 3.8		& 1.4		\\
		$\num{e4}\,\omega_c$	& 17		& 8.6		& 7.2		& 5.0		& 49		& 21		& 14		& 8.4		\\
		$\num{e7}\,\theta_s$	& 33		& 9.9		& 6.3		& 3.8		& 89		& 27		& 15		& 6.9		\\
		$\!\ln(\num{e10}\As)\!$	& 0.020		& 0.020		& 0.020		& 0.0041	& 0.022		& 0.020		& 0.020		& 0.0042	\\
		$\ns$					& 0.0082	& 0.0038	& 0.0029	& 0.0019	& 0.0093	& 0.0040	& 0.0030	& 0.0020	\\
		$\tau$					& 0.010		& 0.010		& 0.010		& 0.0020	& 0.010		& 0.010		& 0.010		& 0.0020	\\
		$\Neff$					& {--}		& {--}		& {--}		& {--}		& 0.32		& 0.12		& 0.081		& 0.045		\\
		$Y_p$					& 0.012		& 0.0037	& 0.0021	& 0.0008	& 0.018		& 0.0069	& 0.0047	& 0.0026	\\
			\bottomrule
	\end{tabular}
	\caption{Forecasted sensitivities of Planck, CMB-S3, CMB-S4 and a CVL experiment for the parameters of $\Lambda\mathrm{CDM}$+$Y_p$ and $\Lambda\mathrm{CDM}$+$\Neff$+$Y_p$.}
	\label{tab:cmbForecast_LCDM+Yp_LCDM+Neff+Yp}
\end{table}
compared to the constraints published in~\cite{Ade:2015xua} can be attributed entirely to the improvement in~$\sigma(\tau)$ which arises from the imposed prior on the optical depth~$\tau$. The forecast of~$\Neff$ for~CMB-S3 is a rough estimate and will be subject to the precise specifications of the respective experiment. While the precise design of~CMB-S4 is also undetermined at this point, $\sigma(\Neff) = 0.03$ is a primary science target and is therefore more likely to be a reliable estimate of the expected performance. For a CVL experiment, some improvement is expected when including the lensing convergence, with constraints possibly reaching $\sigma(\Neff) \lesssim 0.008$.

In Figure~\ref{fig:cmbNeffFsky},%
\begin{figure}[t]
	\includegraphics{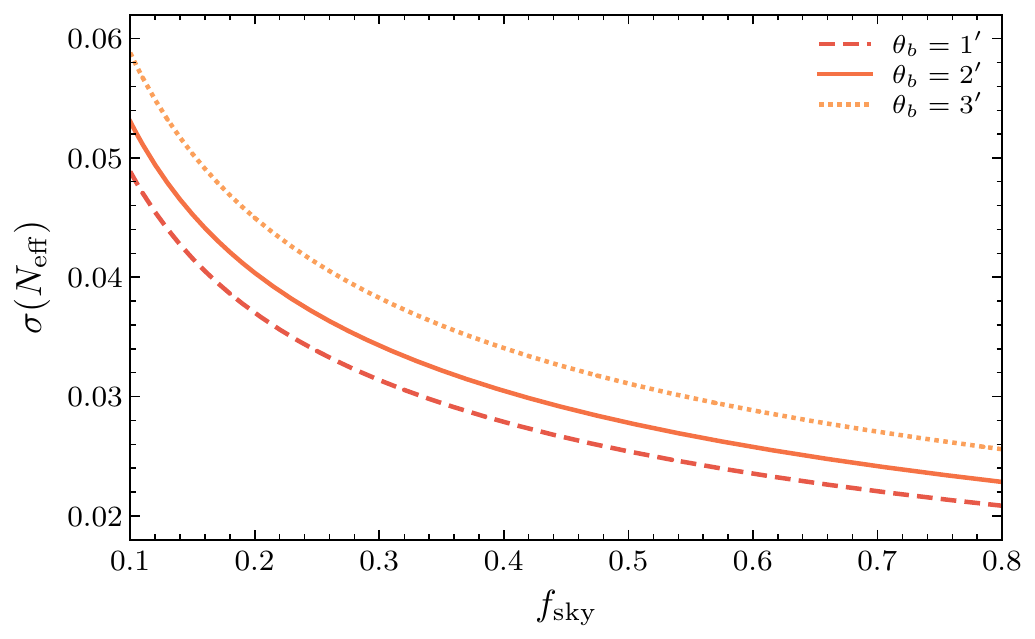}\vspace{-3pt}
	\caption{Marginalized constraints on~$\Neff$ as a function of the sky fraction~$\fsky$ for three values of the beam width~$\theta_b$ and fixed noise level $\Delta T=\SI{1}{\muKelvin.arcmin}$.\vspace{-2pt}}
	\label{fig:cmbNeffFsky}
\end{figure}
we demonstrate how the constraints on~$\Neff$ depend on the sky fraction~$\fsky$, for~three different values of~$\theta_b$ and fixed noise level $\Delta T = \SI{1}{\muKelvin.arcmin}$. When varying the total sky fraction, we also appropriately change the contribution of the included high-$\ell$~Planck data. In Figure~\ref{fig:cmbNeffThetabDeltaT},%
\begin{figure}[h!t]
	\includegraphics{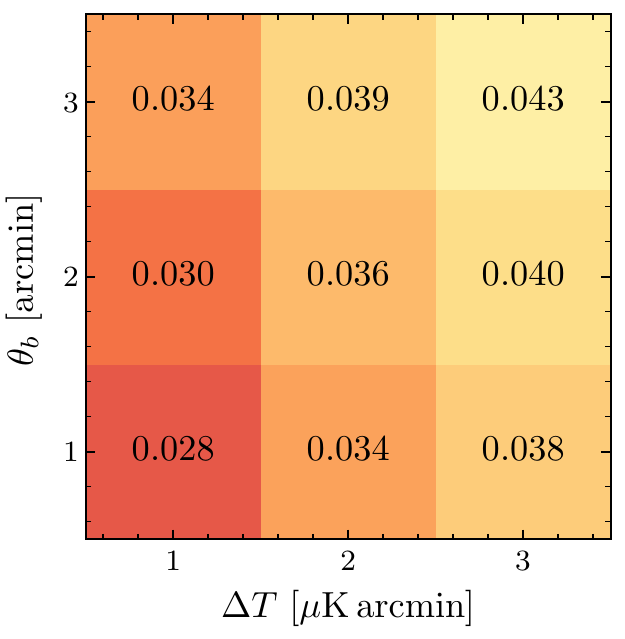}\hspace{0.7cm}
	\includegraphics{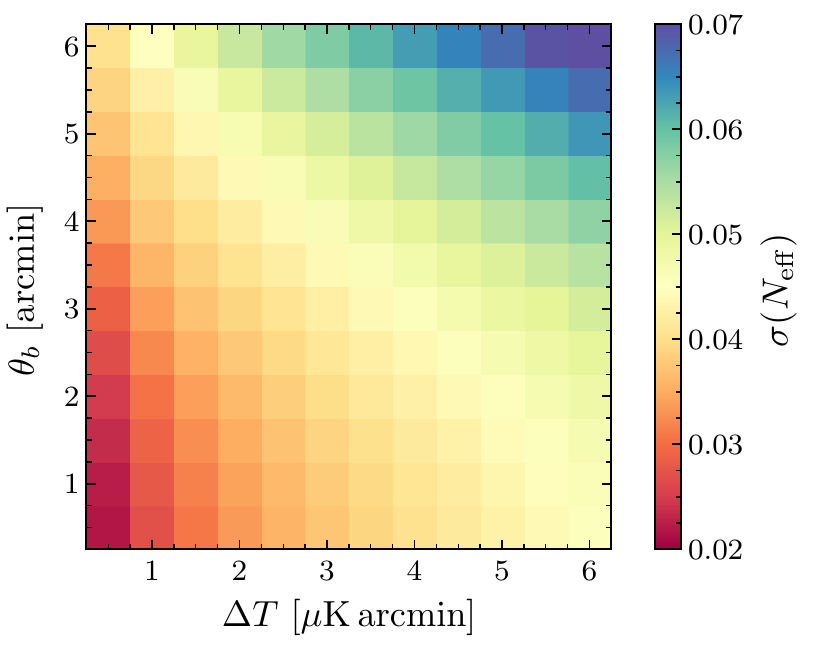}\vspace{-3pt}
	\caption{Marginalized constraints on~$\Neff$ as a function of the beam size~$\theta_b$ and the temperature noise level~$\Delta T$, for fixed sky fraction $\fsky = 0.4$.}
	\label{fig:cmbNeffThetabDeltaT}
\end{figure}
we illustrate the constraint on~$\Neff$ as a function of the beam size~$\theta_b$ and the noise level~$\Delta T$, for fixed sky fraction $\fsky = 0.4$. Comparing Figure~\ref{fig:cmbNeffThetabDeltaT} to the equivalent figure in the CMB-S4 Science Book~\cite{Abazajian:2016yjj} (Fig.~22), we see that the difference between the two forecasts is $\Delta \sigma(\Neff) \approx 0.002$. This can be attributed to the effects of imperfect delensing and is completely negligible for our purposes.

\section{Forecasting LSS Constraints}
\label{app:LSS}
In this appendix, we collect the specific information regarding the planned LSS surveys which we used in our Fisher and likelihood forecasts. We also provide the full set of constraints on all of the cosmological parameters and cosmologies that are studied in Chapter~\ref{chap:bao-forecast}.

\subsection{Survey Specifications}
Below, we provide the experimental specifications for the galaxy surveys used in our forecasts. We have slightly simplified the details compared to~\cite{Font-Ribera:2013rwa}, for example. In particular, we group different types of tracers (e.g.\ luminous red galaxies, emission line galaxies or quasars) into a single effective number density and bias. We find our results to be fairly insensitive to many of these details and well approximated by a fixed number of objects distributed with a constant comoving number density over the same redshift range.\medskip

The employed parametrization of the spectroscopic redshift surveys BOSS, eBOSS, DESI and Euclid are provided in Tables~\ref{tab:specsBOSS} to~\ref{tab:specsEuclid}.%
\begin{table}[b]
	\renewcommand{\arraystretch}{1.15}
	\begin{tabular}{l S[table-format=1.4] S[table-format=1.4] S[table-format=1.4] S[table-format=1.4] S[table-format=1.4] S[table-format=1.4] S[table-format=1.4] S[table-format=1.4]}
		\toprule
		$\bar{z}$													& 0.05		& 0.15	& 0.25	& 0.35	& 0.45	& 0.55	& 0.65	& 0.75		\\
		\midrule[0.065em]
		$b$ 														& 1.79		& 1.90	& 1.98	& 2.09	& 2.32	& 2.26	& 2.38	& 3.09		\\
		$\num{e3}\,\bar{n}_g\ [\si{\h\tothe{3}\per\Mpc\tothe{3}}]$ 	& 0.289		& 0.290	& 0.300	& 0.304	& 0.276	& 0.323	& 0.120	& 0.0100	\\
		$V\ [\si{\per\h\tothe{3}\Gpc\tothe{3}}]$ 					& 0.0255	& 0.164	& 0.402	& 0.704	& 1.04	& 1.38	& 1.72	& 2.04		\\
		\bottomrule
	\end{tabular}
	\caption{Basic specifications for BOSS derived from~\cite{Font-Ribera:2013rwa} with a sky area of $\Omega=\SI{10000}{deg^2}$ resulting in roughly \num{1.4e6}~objects in a volume of about~$\SI{7.5}{\per\h\tothe{3}\Gpc\tothe{3}}$.}
	\label{tab:specsBOSS}
\end{table}
\begin{table}[t]
	\renewcommand{\arraystretch}{1.15}
	\begin{tabular}{l S[table-format=2.3] S[table-format=2.3] S[table-format=2.3] S[table-format=2.3] S[table-format=2.4] S[table-format=2.3] S[table-format=2.3] S[table-format=2.3]}
		\toprule
		$\bar{z}$													& 0.55	& 0.65	& 0.75	& 0.85	& 0.95		& 1.05	& 1.15	& 1.25	\\
		\midrule[0.065em]
		$b$ 														& 3.07	& 2.07	& 1.57	& 1.57	& 1.61		& 3.51	& 1.98	& 2.35	\\
		$\num{e5}\,\bar{n}_g\ [\si{\h\tothe{3}\per\Mpc\tothe{3}}]$	& 0.463	& 21.3	& 35.5	& 23.6	& 5.40		& 0.563	& 1.53	& 1.48	\\
		$V\ [\si{\per\h\tothe{3}\Gpc\tothe{3}}]$ 					& 0.208	& 0.258	& 0.307	& 0.352	& 0.392		& 0.429	& 0.461	& 0.489	\\
		\midrule[0.065em]
		$b$ 														& 3.07	& 2.42	& 2.45	& 2.56	& 7.84		& 3.51	& 1.98	& 2.35	\\
		$\num{e5}\,\bar{n}_g\ [\si{\h\tothe{3}\per\Mpc\tothe{3}}]$	& 0.463	& 13.5	& 7.02	& 3.35	& 0.0412	& 0.563	& 1.53	& 1.48	\\
		$V\ [\si{\per\h\tothe{3}\Gpc\tothe{3}}]$ 					& 0.830	& 1.03	& 1.23	& 1.41	& 1.57		& 1.71	& 1.84	& 1.96	\\
		\bottomrule
	\end{tabular}\\[4pt]
	\begin{tabular}{l S[table-format=2.3] S[table-format=2.3] S[table-format=1.3] S[table-format=1.3] S[table-format=1.3] S[table-format=1.3] S[table-format=1.3] S[table-format=1.3] S[table-format=1.3]}
		\toprule
		$\bar{z}$																& 1.35	& 1.45	& 1.55	& 1.65	& 1.75	& 1.85	& 1.95	& 2.05	& 2.15	\\
		\midrule[0.065em]
		$b$ 																	& 3.65	& 2.40	& 2.42	& 2.08	& 2.10	& 3.33	& 3.35	& 1.72	& 1.73	\\
		$\num{e5}\,\bar{n}_g\ [\si{\h\tothe{3}\per\Mpc\tothe{3}}]\hskip1.5pt$	& 0.664	& 1.66	& 1.76	& 2.03	& 2.15	& 0.912	& 0.965	& 2.91	& 3.07	\\
		$V\ [\si{\per\h\tothe{3}\Gpc\tothe{3}}]$								& 0.513	& 0.533	& 0.551	& 0.565	& 0.577	& 0.587	& 0.594	& 0.600	& 0.604	\\
		\midrule[0.065em]
		$b$ 																	& 3.65	& 2.40	& 2.42	& 2.08	& 2.10	& 3.33	& 3.35	& 1.72	& 1.73	\\
		$\num{e5}\,\bar{n}_g\ [\si{\h\tothe{3}\per\Mpc\tothe{3}}]$				& 0.664	& 1.66	& 1.76	& 2.03	& 2.15	& 0.912	& 0.965	& 2.91	& 3.07	\\
		$V\ [\si{\per\h\tothe{3}\Gpc\tothe{3}}]$ 								& 2.05	& 2.13	& 2.20	& 2.26	& 2.31	& 2.35	& 2.38	& 2.40	& 2.42	\\
		\bottomrule
	\end{tabular}
	\caption{Basic specifications for eBOSS derived from~\cite{Font-Ribera:2013rwa}. The redshift range is covered twice, first showing the survey covering $\Omega=\SI{1500}{deg^2}$ that will include emission line galaxies (resulting in roughly \num{3.8e5}~objects in a volume of about~$\SI{8.0}{\si{\per\h\tothe{3}\Gpc\tothe{3}}}$), and then the survey with $\Omega=\SI{6000}{deg^2}$ that will not (resulting in roughly \num{7.2e5}~objects in a volume of about~$\SI{32}{\per\h\tothe{3}\Gpc\tothe{3}})$.}
	\label{tab:specseBOSS} 
\end{table}
\begin{table}[h!b]
	\renewcommand{\arraystretch}{1.15}
	\begin{tabular}{l S[table-format=1.3] S[table-format=1.3] S[table-format=1.3] S[table-format=1.3] S[table-format=1.3] S[table-format=1.3] S[table-format=1.4] S[table-format=1.4] S[table-format=1.5, group-digits=false]}
		\toprule
		$\bar{z}$													& 0.15	& 0.25	& 0.35	& 0.45	& 0.55	& 0.65	& 0.75	& 0.85	& 0.95		\\
		\midrule[0.065em]
		$b$ 														& 1.13	& 1.39	& 1.64	& 1.81	& 1.87	& 1.89	& 1.90	& 1.82	& 1.53		\\
		$\num{e3}\,\bar{n}_g\ [\si{\h\tothe{3}\per\Mpc\tothe{3}}]$	& 2.38	& 1.07	& 0.684	& 0.568	& 0.600	& 0.696	& 0.810	& 0.719	& 0.558		\\
		$V\ [\si{\per\h\tothe{3}\Gpc\tothe{3}}]$					& 0.229	& 0.563	& 0.985	& 1.45	& 1.94	& 2.41	& 2.86	& 3.28	& 3.66		\\
		\bottomrule
	\end{tabular}\\[4pt]
	\begin{tabular}{l S[table-format=1.3] S[table-format=1.3] S[table-format=1.3] S[table-format=1.3] S[table-format=1.3] S[table-format=1.3] S[table-format=1.4] S[table-format=1.4] S[table-format=1.5, group-digits=false]}
		\toprule
		$\bar{z}$													& 1.05	& 1.15	& 1.25	& 1.35	& 1.45	& 1.55	& 1.65		& 1.75		& 1.85		\\
		\midrule[0.065em]
		$b$ 														& 1.47	& 1.49	& 1.58	& 1.62	& 1.73	& 2.01	& 1.98		& 2.56		& 4.17		\\
		$\num{e3}\,\bar{n}_g\ [\si{\h\tothe{3}\per\Mpc\tothe{3}}]$	& 0.522	& 0.506	& 0.454	& 0.356	& 0.242	& 0.127	& 0.0736	& 0.0289	& 0.00875	\\
		$V\ [\si{\per\h\tothe{3}\Gpc\tothe{3}}]$					& 4.00	& 4.30	& 4.56	& 4.79	& 4.98	& 5.14	& 5.28		& 5.39		& 5.48		\\
		\bottomrule
	\end{tabular}
	\caption{Basic specifications for DESI derived from~\cite{Font-Ribera:2013rwa}, covering a sky area $\Omega=\SI{14000}{deg^2}$ and resulting in roughly \num{2.3e7}~objects in a volume of about~$\SI{61}{\per\h\tothe{3}\Gpc\tothe{3}}$.}
	\label{tab:specsDESI} 
\end{table}
\begin{table}
	\renewcommand{\arraystretch}{1.15}
	\begin{tabular}{l S[table-format=1.3] S[table-format=1.3] S[table-format=1.3] S[table-format=1.3] S[table-format=1.3] S[table-format=1.3] S[table-format=1.3] S[table-format=1.3]}
		\toprule
		$\bar{z}$														& 0.65	& 0.75	& 0.85	& 0.95	& 1.05	& 1.15	& 1.25	& 1.35	\\
		\midrule[0.065em]
		$b$ 															& 1.06	& 1.11	& 1.16	& 1.21	& 1.27	& 1.33	& 1.38	& 1.44	\\
		$\num{e3}\,\bar{n}_g\ [\si{\h\tothe{3}\per\Mpc\tothe{3}}]\,$	& 0.637	& 1.46	& 1.63	& 1.50	& 1.33	& 1.14	& 1.00	& 0.837	\\
		$V\ [\si{\per\h\tothe{3}\Gpc\tothe{3}}]$						& 2.58	& 3.07	& 3.52	& 3.92	& 4.29	& 4.61	& 4.89	& 5.13	\\
		\bottomrule
	\end{tabular}\\[4pt]
	\begin{tabular}{l S[table-format=1.4] S[table-format=1.4] S[table-format=1.4] S[table-format=1.4] S[table-format=1.4] S[table-format=1.4] S[table-format=1.4]}
		\toprule
		$\bar{z}$ 														& 1.45	& 1.55	& 1.65	& 1.75	& 1.85	& 1.95		& 2.05		\\
		\midrule[0.065em]
		$b$ 															& 1.51	& 1.54	& 1.63	& 1.70	& 1.85	& 1.90		& 1.26		\\
		$\num{e3}\,\bar{n}_g\ [\si{\h\tothe{3}\per\Mpc\tothe{3}}]$		& 0.652	& 0.512	& 0.357	& 0.246	& 0.149	& 0.0904	& 0.0721	\\
		$V\ [\si{\per\h\tothe{3}\Gpc\tothe{3}}]$						& 5.33	& 5.51	& 5.65	& 5.77	& 5.87	& 5.94		& 6.00		\\
		\bottomrule
	\end{tabular}
	\caption{Basic specifications for Euclid derived from~\cite{Font-Ribera:2013rwa}, covering a sky area $\Omega=\SI{15000}{deg^2}$ and resulting in roughly \num{5.0e7}~objects in a volume of about~$\SI{72}{\per\h\tothe{3}\Gpc\tothe{3}}$.}
	\label{tab:specsEuclid}
\end{table}
For eBOSS, we combine BOSS and the two eBOSS configurations of Table~\ref{tab:specseBOSS} into one survey neglecting the small overlap. We effectively treat each redshift bin with mean redshift $\bar{z}$ as an independent three-dimensional survey. Our Fisher matrix is the sum of the Fisher matrices associated with each bin, $F = \sum_{\bar z} F_{\bar{z}}$. We translated the survey specifications used in~\cite{Font-Ribera:2013rwa} into three numbers per redshift bin: the linear galaxy bias~$b$, the comoving number density of galaxies $\bar{n}_g$ and the bin volume $V$. This is sufficient to fully specify the Fisher matrix in each bin. The spherical bin volume is given by 
\beq
V = \frac{4\pi}{3}\, \fsky \left[d_c(\zmax)^3 - d_c(z_\mathrm{min})^3\right] , \qquad d_c(z) = \int_0^z \!\d z\, \frac{c}{H(z)}\, ,
\eeq
where $\fsky$ is the sky fraction, $d_c(z)$ is the comoving distance to redshift $z$, and $z_\mathrm{min} = \bar{z} - \Delta z/2$ and $\zmax = \bar{z} + \Delta z/2$ are the minimum and maximum redshift of the respective bin. We use redshift bins of width $\Delta z = 0.1$ throughout.\medskip

For the photometric surveys DES and LSST, we follow~\cite{Font-Ribera:2013rwa} and define the surveys by using $(\alpha, \beta, z_*, N_\mathrm{tot}, b_0) = (1.25, 2.29, 0.88, \SI{12}{arcmin^{-2}}, 0.95)$ and $(2.0, 1.0, 0.3, \SI{50}{arcmin^{-2}}, 0.95)$, respectively. These parameters are related to those employed in our forecasts as follows:
\begin{align}
\bar{n}_g(\bar{z})	&= \frac{N_\mathrm{tot}}{V} \frac{\beta/z_*}{\Gamma\left[(\alpha+1)/\beta\right]} \int_{z_\mathrm{min}}^{\zmax} \!\d z\, \left(z/z_*\right)^\alpha \exp\left\{-\left(z/z_*\right)^\beta\right\} ,	\\[2pt]
b(\bar{z}) 			&= \frac{D_1(0)}{D_1(\bar{z}_i)}\, b_0\, ,
\end{align}
with gamma function~$\Gamma$ and linear growth function~$D_1(z)$.\medskip

For DES, we employ a survey area of $\Omega=\SI{5000}{deg^2}$ and a redshift coverage of $0.1 \leq z \leq 2.0$, while we take~\SI{20000}{deg^2} and $0.1 \leq z \leq 3.5$ for~LSST. This results in approximately \num{1.4e8} and \num{5.9e8}~objects in a total survey volume of about $\SI{24}{\per\h\tothe{3}\Gpc\tothe{3}}$ and $\SI{215}{\per\h\tothe{3}\Gpc\tothe{3}}$ for the two surveys, respectively. We neglect the spectroscopic redshift error as it is expected to be comparable to (or smaller than) the longitudinal damping scale~$\Sigma_\parallel$, but use a conservative root-mean-square photometric redshift error of $\sigma_{z0}=0.05$ for both DES and~LSST. Finally, we reiterate that, by considering galaxy clustering alone, we only take a subset of the cosmological observables into account, in particular for photometric surveys, and we therefore expect to underestimate the full power of these experiments.

\subsection{Future Constraints}
Using these specifications, we generated forecasts for all of the cosmological parameters discussed in the main text in combination with the Fisher matrices for Planck, CMB-S3 and~CMB-S4. We include both $P(k)$- and BAO-forecasts for $\Lambda\mathrm{CDM}$~(Table~\ref{tab:CMB+LSS_LCDM_full}), $\Lambda\mathrm{CDM}$+$\Neff$~(Table~\ref{tab:CMB+LSS_Neff_full}), $\Lambda\mathrm{CDM}$+$Y_p$~(Table~\ref{tab:CMB+LSS_Yp_full}) and $\Lambda\mathrm{CDM}$+$\Neff$+$Y_p$~(Table~\ref{tab:CMB+LSS_Neff+Yp_full}).%
\begin{table}
	\scriptsize\renewcommand{\arraystretch}{1.08}
	\subcaptionbox{Planck + $P(k)$}{
		\begin{tabular}{c S[table-format=2.4] S[table-format=2.4] S[table-format=2.4] S[table-format=2.4] S[table-format=2.4] S[table-format=2.4] S[table-format=2.4]}	
			\toprule
			Parameter				& {Planck}	& {BOSS}	& {eBOSS}	& {DESI}	& {Euclid}	& {DES}		& {LSST}	\\
			\midrule[0.065em]
			$\num{e5}\,\omega_b$	& 16		& 13		& 13		& 12		& 11		& 14		& 12		\\
			$\num{e4}\,\omega_c$	& 16		& 8.9		& 7.7		& 4.6		& 4.3		& 13		& 8.2		\\
			$\num{e7}\,\theta_s$	& 29		& 28		& 27		& 27		& 27		& 29		& 28		\\
			$\ns$					& 0.0040	& 0.0033	& 0.0032	& 0.0028	& 0.0027	& 0.0037	& 0.0033	\\
			\bottomrule
		\end{tabular}\vspace{-1pt}
	}\\[7pt]
	\subcaptionbox{CMB-S3 + $P(k)$}{
		\begin{tabular}{c S[table-format=2.4] S[table-format=2.4] S[table-format=2.4] S[table-format=2.4] S[table-format=2.4] S[table-format=2.4] S[table-format=2.4]}	
			\toprule
			Parameter				& {CMB-S3}	& {BOSS}	& {eBOSS}	& {DESI}	& {Euclid}	& {DES}		& {LSST}	\\
			\midrule[0.065em]
			$\num{e5}\,\omega_b$	& 5.1		& 4.9		& 4.9		& 4.7		& 4.6		& 5.0		& 4.8		\\
			$\num{e4}\,\omega_c$	& 8.3		& 6.7		& 6.1		& 4.0		& 3.7		& 7.8		& 6.3		\\
			$\num{e7}\,\theta_s$	& 9.4		& 9.1		& 9.0		& 8.7		& 8.6		& 9.3		& 9.1		\\
			$\ns$					& 0.0023	& 0.0021	& 0.0021	& 0.0019	& 0.0019	& 0.0022	& 0.0021	\\
			\bottomrule
		\end{tabular}\vspace{-1pt}
	}\\[7pt]
	\subcaptionbox{S4 + $P(k)$}{
		\begin{tabular}{c S[table-format=2.4] S[table-format=2.4] S[table-format=2.4] S[table-format=2.4] S[table-format=2.4] S[table-format=2.4] S[table-format=2.4]}	
			\toprule
			Parameter				& {CMB-S4}	& {BOSS}	& {eBOSS}	& {DESI}	& {Euclid}	& {DES}		& {LSST}	\\
			\midrule[0.065em]
			$\num{e5}\,\omega_b$	& 2.7		& 2.7		& 2.7		& 2.6		& 2.6		& 2.7		& 2.6		\\
			$\num{e4}\,\omega_c$	& 7.1		& 6.0		& 5.6		& 3.9		& 3.6		& 6.8		& 5.8		\\
			$\num{e7}\,\theta_s$	& 5.9		& 5.7		& 5.6		& 5.3		& 5.2		& 5.9		& 5.7		\\
			$\ns$					& 0.0020	& 0.0018	& 0.0018	& 0.0016	& 0.0016	& 0.0019	& 0.0018	\\
			\bottomrule
		\end{tabular}\vspace{-1pt}
	}\\[7pt]
	\subcaptionbox{Planck + BAO}{
		\begin{tabular}{c S[table-format=2.4] S[table-format=2.4] S[table-format=2.4] S[table-format=2.4] S[table-format=2.4] S[table-format=2.4] S[table-format=2.4]}	
			\toprule
			Parameter				& {Planck}	& {BOSS}	& {eBOSS}	& {DESI}	& {Euclid}	& {DES}		& {LSST}	\\
			\midrule[0.065em]
			$\num{e5}\,\omega_b$	& 16		& 13		& 13		& 13		& 13		& 15		& 14		\\
			$\num{e4}\,\omega_c$	& 16		& 8.7		& 8.0		& 5.1		& 5.5		& 13		& 9.4		\\
			$\num{e7}\,\theta_s$	& 29		& 27		& 27		& 27		& 26		& 29		& 27		\\
			$\ns$					& 0.0040	& 0.0031	& 0.0031	& 0.0028	& 0.0028	& 0.0037	& 0.0032	\\
			\bottomrule
		\end{tabular}\vspace{-1pt}
	}\\[7pt]
	\subcaptionbox{CMB-S3 + BAO}{
		\begin{tabular}{c S[table-format=2.4] S[table-format=2.4] S[table-format=2.4] S[table-format=2.4] S[table-format=2.4] S[table-format=2.4] S[table-format=2.4]}	
			\toprule
			Parameter				& {CMB-S3}	& {BOSS}	& {eBOSS}	& {DESI}	& {Euclid}	& {DES}		& {LSST}	\\
			\midrule[0.065em]
			$\num{e5}\,\omega_b$	& 5.1		& 5.0		& 5.0		& 4.9		& 4.9		& 5.1		& 5.0		\\
			$\num{e4}\,\omega_c$	& 8.3		& 6.5		& 6.2		& 4.4		& 4.6		& 7.9		& 6.8		\\
			$\num{e7}\,\theta_s$	& 9.4		& 9.0		& 8.9		& 8.6		& 8.6		& 9.3		& 9.0		\\
			$\ns$					& 0.0023	& 0.0021	& 0.0020	& 0.0019	& 0.0019	& 0.0022	& 0.0021	\\
			\bottomrule
		\end{tabular}\vspace{-1pt}
	}\\[7pt]
	\subcaptionbox{S4 + BAO}{
		\begin{tabular}{c S[table-format=2.4] S[table-format=2.4] S[table-format=2.4] S[table-format=2.4] S[table-format=2.4] S[table-format=2.4] S[table-format=2.4]}	
			\toprule
			Parameter				& {CMB-S4}	& {BOSS}	& {eBOSS}	& {DESI}	& {Euclid}	& {DES}		& {LSST}	\\
			\midrule[0.065em]
			$\num{e5}\,\omega_b$	& 2.7		& 2.7		& 2.7		& 2.7		& 2.7		& 2.7		& 2.7		\\
			$\num{e4}\,\omega_c$	& 7.1		& 5.9		& 5.7		& 4.2		& 4.3		& 6.8		& 6.1		\\
			$\num{e7}\,\theta_s$	& 5.9		& 5.6		& 5.6		& 5.2		& 5.2		& 5.9		& 5.7		\\
			$\ns$					& 0.0020	& 0.0018	& 0.0018	& 0.0016	& 0.0016	& 0.0019	& 0.0018	\\
			\bottomrule
		\end{tabular}\vspace{-1pt}
	}
	\caption{Full set of forecasted $1\sigma$~constraints in a $\Lambda\mathrm{CDM}$~cosmology for current and future LSS~surveys in combination with the CMB~experiments Planck, CMB-S3 and~CMB-S4. We do not quote the sensitivities to~$\ln(\num{e10}\As)$ and~$\tau$ as they are the same as in Table~\ref{tab:cmbForecast_LCDM_LCDM+Neff} for all combinations.}
	\label{tab:CMB+LSS_LCDM_full}
\end{table}
\begin{table}
	\scriptsize\renewcommand{\arraystretch}{1.08}
	\subcaptionbox{Planck + $P(k)$}{
		\begin{tabular}{c S[table-format=2.4] S[table-format=2.4] S[table-format=2.4] S[table-format=2.4] S[table-format=2.4] S[table-format=2.4] S[table-format=2.4]}	
			\toprule
			Parameter				& {Planck}	& {BOSS}	& {eBOSS}	& {DESI}	& {Euclid}	& {DES}		& {LSST}	\\
			\midrule[0.065em]
			$\num{e5}\,\omega_b$	& 26		& 19		& 18		& 15		& 15		& 24		& 20		\\
			$\num{e4}\,\omega_c$	& 26		& 23		& 21		& 15		& 13		& 25		& 19		\\
			$\num{e7}\,\theta_s$	& 44		& 41		& 40		& 35		& 34		& 43		& 39		\\
			$\ns$					& 0.0093	& 0.0068	& 0.0061	& 0.0039	& 0.0035	& 0.0085	& 0.0069	\\
			$\Neff$					& 0.18		& 0.14		& 0.13		& 0.087		& 0.079		& 0.17		& 0.14		\\
			\bottomrule
		\end{tabular}\vspace{-1.5pt}
	}\\[6pt]
	\subcaptionbox{CMB-S3 + $P(k)$}{
		\begin{tabular}{c S[table-format=2.4] S[table-format=2.4] S[table-format=2.4] S[table-format=2.4] S[table-format=2.4] S[table-format=2.4] S[table-format=2.4]}	
			\toprule
			Parameter				& {CMB-S3}	& {BOSS}	& {eBOSS}	& {DESI}	& {Euclid}	& {DES}		& {LSST}	\\
			\midrule[0.065em]
			$\num{e5}\,\omega_b$	& 8.3		& 7.9		& 7.8		& 7.3		& 7.1		& 8.2		& 8.0		\\
			$\num{e4}\,\omega_c$	& 10		& 9.6		& 9.2		& 7.8		& 7.5		& 10		& 8.8		\\
			$\num{e7}\,\theta_s$	& 13		& 12		& 12		& 12		& 12		& 12		& 12		\\
			$\ns$					& 0.0040	& 0.0037	& 0.0036	& 0.0029	& 0.0028	& 0.0039	& 0.0037	\\
			$\Neff$					& 0.054		& 0.052		& 0.051		& 0.045		& 0.043		& 0.054		& 0.052		\\
			\bottomrule
		\end{tabular}\vspace{-1.5pt}
	}\\[6pt]
	\subcaptionbox{S4 + $P(k)$}{
		\begin{tabular}{c S[table-format=2.4] S[table-format=2.4] S[table-format=2.4] S[table-format=2.4] S[table-format=2.4] S[table-format=2.4] S[table-format=2.4]}	
			\toprule
			Parameter				& {CMB-S4}	& {BOSS}	& {eBOSS}	& {DESI}	& {Euclid}	& {DES}		& {LSST}	\\
			\midrule[0.065em]
			$\num{e5}\,\omega_b$	& 3.8		& 3.7		& 3.7		& 3.6		& 3.6		& 3.8		& 3.7		\\
			$\num{e4}\,\omega_c$	& 7.9		& 7.1		& 6.8		& 5.5		& 5.3		& 7.6		& 6.7		\\
			$\num{e7}\,\theta_s$	& 6.7		& 6.6		& 6.5		& 6.2		& 6.2		& 6.7		& 6.5		\\
			$\ns$					& 0.0030	& 0.0029	& 0.0028	& 0.0025	& 0.0024	& 0.0030	& 0.0029	\\
			$\Neff$					& 0.030		& 0.030		& 0.030		& 0.028		& 0.027		& 0.030		& 0.030		\\
			\bottomrule
		\end{tabular}\vspace{-1.5pt}
	}\\[6pt]
	\subcaptionbox{Planck + BAO}{
		\begin{tabular}{c S[table-format=2.4] S[table-format=2.4] S[table-format=2.4] S[table-format=2.4] S[table-format=2.4] S[table-format=2.4] S[table-format=2.4]}	
			\toprule
			Parameter				& {Planck}	& {BOSS}	& {eBOSS}	& {DESI}	& {Euclid}	& {DES}		& {LSST}	\\
			\midrule[0.065em]
			$\num{e5}\,\omega_b$	& 26		& 18		& 18		& 17		& 17		& 22		& 19		\\
			$\num{e4}\,\omega_c$	& 26		& 26		& 26		& 26		& 26		& 26		& 26		\\
			$\num{e7}\,\theta_s$	& 44		& 43		& 43		& 43		& 43		& 44		& 44		\\
			$\ns$					& 0.0093	& 0.0065	& 0.0063	& 0.0059	& 0.0059	& 0.0081	& 0.0067	\\
			$\Neff$					& 0.18		& 0.15		& 0.15		& 0.14		& 0.14		& 0.16		& 0.15		\\
			\bottomrule
		\end{tabular}\vspace{-1.5pt}
	}\\[6pt]
	\subcaptionbox{CMB-S3 + BAO}{
		\begin{tabular}{c S[table-format=2.4] S[table-format=2.4] S[table-format=2.4] S[table-format=2.4] S[table-format=2.4] S[table-format=2.4] S[table-format=2.4]}	
			\toprule
			Parameter				& {CMB-S3}	& {BOSS}	& {eBOSS}	& {DESI}	& {Euclid}	& {DES}		& {LSST}	\\
			\midrule[0.065em]
			$\num{e5}\,\omega_b$	& 8.3		& 7.8		& 7.7		& 7.4		& 7.4		& 8.2		& 7.8		\\
			$\num{e4}\,\omega_c$	& 10		& 10		& 9.9		& 9.6		& 9.6		& 10		& 10		\\
			$\num{e7}\,\theta_s$	& 13		& 13		& 13		& 13		& 13		& 13		& 13		\\
			$\ns$					& 0.0040	& 0.0035	& 0.0035	& 0.0031	& 0.0032	& 0.0039	& 0.0036	\\
			$\Neff$					& 0.054		& 0.052		& 0.052		& 0.050		& 0.050		& 0.054		& 0.052		\\
			\bottomrule
		\end{tabular}\vspace{-1.5pt}
	}\\[6pt]
	\subcaptionbox{S4 + BAO}{
		\begin{tabular}{c S[table-format=2.4] S[table-format=2.4] S[table-format=2.4] S[table-format=2.4] S[table-format=2.4] S[table-format=2.4] S[table-format=2.4]}	
			\toprule
			Parameter				& {CMB-S4}	& {BOSS}	& {eBOSS}	& {DESI}	& {Euclid}	& {DES}		& {LSST}	\\
			\midrule[0.065em]
			$\num{e5}\,\omega_b$	& 3.8		& 3.7		& 3.7		& 3.7		& 3.7		& 3.8		& 3.7		\\
			$\num{e4}\,\omega_c$	& 7.9		& 7.2		& 7.1		& 6.5		& 6.5		& 7.8		& 7.3		\\
			$\num{e7}\,\theta_s$	& 6.7		& 6.6		& 6.6		& 6.5		& 6.5		& 6.7		& 6.6		\\
			$\ns$					& 0.0030	& 0.0028	& 0.0028	& 0.0025	& 0.0025	& 0.0030	& 0.0028	\\
			$\Neff$					& 0.030		& 0.030		& 0.030		& 0.029		& 0.029		& 0.030		& 0.030		\\
			\bottomrule
		\end{tabular}\vspace{-1.5pt}
	}
	\caption{As in Table~\ref{tab:CMB+LSS_LCDM_full}, but for an extended $\Lambda\mathrm{CDM}$+$\Neff$ cosmology.}
	\label{tab:CMB+LSS_Neff_full}
\end{table}
\begin{table}
	\scriptsize\renewcommand{\arraystretch}{1.08}
	\subcaptionbox{Planck + $P(k)$}{
		\begin{tabular}{c S[table-format=2.4] S[table-format=2.4] S[table-format=2.4] S[table-format=2.4] S[table-format=2.4] S[table-format=2.4] S[table-format=2.4]}	
			\toprule
			Parameter				& {Planck}	& {BOSS}	& {eBOSS}	& {DESI}	& {Euclid}	& {DES}		& {LSST}	\\
			\midrule[0.065em]
			$\num{e5}\,\omega_b$	& 24		& 19		& 19		& 17		& 16		& 22		& 20		\\
			$\num{e4}\,\omega_c$	& 17		& 8.9		& 7.8		& 4.7		& 4.3		& 13		& 8.6		\\
			$\num{e7}\,\theta_s$	& 33		& 30		& 29		& 27		& 27		& 32		& 30		\\
			$\ns$					& 0.0082	& 0.0066	& 0.0063	& 0.0048	& 0.0044	& 0.0077	& 0.0068	\\
			$Y_p$					& 0.012		& 0.011		& 0.0100	& 0.0087	& 0.0082	& 0.011		& 0.011		\\
			\bottomrule
		\end{tabular}\vspace{-1.5pt}
	}\\[6pt]
	\subcaptionbox{CMB-S3 + $P(k)$}{
		\begin{tabular}{c S[table-format=2.4] S[table-format=2.4] S[table-format=2.4] S[table-format=2.4] S[table-format=2.4] S[table-format=2.4] S[table-format=2.4]}	
			\toprule
			Parameter				& {CMB-S3}	& {BOSS}	& {eBOSS}	& {DESI}	& {Euclid}	& {DES}		& {LSST}	\\
			\midrule[0.065em]
			$\num{e5}\,\omega_b$	& 8.2		& 7.9		& 7.8		& 7.5		& 7.4		& 8.1		& 7.9		\\
			$\num{e4}\,\omega_c$	& 8.6		& 6.8		& 6.3		& 4.0		& 3.7		& 8.1		& 6.6		\\
			$\num{e7}\,\theta_s$	& 9.9		& 9.5		& 9.4		& 8.9		& 8.8		& 9.8		& 9.5		\\
			$\ns$					& 0.0038	& 0.0035	& 0.0034	& 0.0030	& 0.0029	& 0.0037	& 0.0036	\\
			$Y_p$					& 0.0037	& 0.0036	& 0.0036	& 0.0034	& 0.0033	& 0.0037	& 0.0036	\\
			\bottomrule
		\end{tabular}\vspace{-1.5pt}
	}\\[6pt]
	\subcaptionbox{S4 + $P(k)$}{
		\begin{tabular}{c S[table-format=2.4] S[table-format=2.4] S[table-format=2.4] S[table-format=2.4] S[table-format=2.4] S[table-format=2.4] S[table-format=2.4]}	
			\toprule
			Parameter				& {CMB-S4}	& {BOSS}	& {eBOSS}	& {DESI}	& {Euclid}	& {DES}		& {LSST}	\\
			\midrule[0.065em]
			$\num{e5}\,\omega_b$	& 3.8		& 3.8		& 3.8		& 3.7		& 3.7		& 3.8		& 3.8		\\
			$\num{e4}\,\omega_c$	& 7.2		& 6.1		& 5.7		& 3.9		& 3.6		& 6.9		& 5.9		\\
			$\num{e7}\,\theta_s$	& 6.3		& 6.0		& 5.9		& 5.5		& 5.4		& 6.2		& 6.0		\\
			$\ns$					& 0.0029	& 0.0028	& 0.0028	& 0.0025	& 0.0024	& 0.0029	& 0.0028	\\
			$Y_p$					& 0.0021	& 0.0021	& 0.0021	& 0.0020	& 0.0020	& 0.0021	& 0.0021	\\
			\bottomrule
		\end{tabular}\vspace{-1.5pt}
	}\\[6pt]
	\subcaptionbox{Planck + BAO}{
		\begin{tabular}{c S[table-format=2.4] S[table-format=2.4] S[table-format=2.4] S[table-format=2.4] S[table-format=2.4] S[table-format=2.4] S[table-format=2.4]}	
			\toprule
			Parameter				& {Planck}	& {BOSS}	& {eBOSS}	& {DESI}	& {Euclid}	& {DES}		& {LSST}	\\
			\midrule[0.065em]
			$\num{e5}\,\omega_b$	& 24		& 19		& 19		& 18		& 18		& 22		& 19		\\
			$\num{e4}\,\omega_c$	& 17		& 8.7		& 8.0		& 5.5		& 5.7		& 14		& 9.4		\\
			$\num{e7}\,\theta_s$	& 33		& 29		& 29		& 28		& 28		& 31		& 29		\\
			$\ns$					& 0.0082	& 0.0063	& 0.0062	& 0.0059	& 0.0059	& 0.0075	& 0.0065	\\
			$Y_p$					& 0.012		& 0.011		& 0.011		& 0.0100	& 0.011		& 0.011		& 0.011		\\
			\bottomrule
		\end{tabular}\vspace{-1.5pt}
	}\\[6pt]
	\subcaptionbox{CMB-S3 + BAO}{
		\begin{tabular}{c S[table-format=2.4] S[table-format=2.4] S[table-format=2.4] S[table-format=2.4] S[table-format=2.4] S[table-format=2.4] S[table-format=2.4]}	
			\toprule
			Parameter				& {CMB-S3}	& {BOSS}	& {eBOSS}	& {DESI}	& {Euclid}	& {DES}		& {LSST}	\\
			\midrule[0.065em]		
			$\num{e5}\,\omega_b$	& 8.2		& 7.8		& 7.8		& 7.6		& 7.6		& 8.1		& 7.9		\\
			$\num{e4}\,\omega_c$	& 8.6		& 6.6		& 6.3		& 4.4		& 4.6		& 8.2		& 6.9		\\
			$\num{e7}\,\theta_s$	& 9.9		& 9.3		& 9.3		& 8.8		& 8.9		& 9.8		& 9.4		\\
			$\ns$					& 0.0038	& 0.0034	& 0.0034	& 0.0031	& 0.0031	& 0.0037	& 0.0035	\\
			$Y_p$					& 0.0037	& 0.0036	& 0.0036	& 0.0035	& 0.0035	& 0.0037	& 0.0036	\\
			\bottomrule
		\end{tabular}\vspace{-1.5pt}
	}\\[6pt]
	\subcaptionbox{S4 + BAO}{
		\begin{tabular}{c S[table-format=2.4] S[table-format=2.4] S[table-format=2.4] S[table-format=2.4] S[table-format=2.4] S[table-format=2.4] S[table-format=2.4]}	
			\toprule
			Parameter				& {CMB-S4}	& {BOSS}	& {eBOSS}	& {DESI}	& {Euclid}	& {DES}		& {LSST}	\\
			\midrule[0.065em]
			$\num{e5}\,\omega_b$	& 3.8		& 3.8		& 3.8		& 3.8		& 3.8		& 3.8		& 3.8		\\
			$\num{e4}\,\omega_c$	& 7.2		& 5.9		& 5.7		& 4.2		& 4.3		& 6.9		& 6.1		\\
			$\num{e7}\,\theta_s$	& 6.3		& 5.9		& 5.8		& 5.4		& 5.5		& 6.2		& 5.9		\\
			$\ns$					& 0.0029	& 0.0027	& 0.0027	& 0.0025	& 0.0025	& 0.0029	& 0.0028	\\
			$Y_p$					& 0.0021	& 0.0021	& 0.0021	& 0.0021	& 0.0021	& 0.0021	& 0.0021	\\
			\bottomrule
		\end{tabular}\vspace{-1.5pt}
	}
	\caption{As in Table~\ref{tab:CMB+LSS_LCDM_full}, but for an extended $\Lambda\mathrm{CDM}$+$Y_p$ cosmology. The constraints on~$\ln(\num{e10}\As)$ and~$\tau$ are the same as in Table~\ref{tab:cmbForecast_LCDM+Yp_LCDM+Neff+Yp} for all combinations.}
	\label{tab:CMB+LSS_Yp_full}
\end{table}
\begin{table}
	\scriptsize\renewcommand{\arraystretch}{1.08}
	\subcaptionbox{Planck + $P(k)$}{
		\begin{tabular}{c S[table-format=2.4] S[table-format=2.4] S[table-format=2.4] S[table-format=2.4] S[table-format=2.4] S[table-format=2.4] S[table-format=2.4]}	
			\toprule
			Parameter				& {Planck}	& {BOSS}	& {eBOSS}	& {DESI}	& {Euclid}	& {DES}		& {LSST}	\\
			\midrule[0.065em]
			$\num{e5}\,\omega_b$	& 26		& 20		& 19		& 17		& 16		& 24		& 21		\\
			$\num{e4}\,\omega_c$	& 49		& 40		& 35		& 23		& 21		& 45		& 34		\\
			$\num{e7}\,\theta_s$	& 89		& 76		& 70		& 53		& 50		& 84		& 69		\\
			$\ns$					& 0.0093	& 0.0069	& 0.0065	& 0.0048	& 0.0045	& 0.0086	& 0.0071	\\
			$\Neff$					& 0.32		& 0.25		& 0.22		& 0.14		& 0.13		& 0.29		& 0.23		\\
			$Y_p$					& 0.018		& 0.016		& 0.016		& 0.013		& 0.012		& 0.017		& 0.015		\\
			\bottomrule
		\end{tabular}\vspace{-2.5pt}
	}\\[2.76pt]
	\subcaptionbox{CMB-S3 + $P(k)$}{
		\begin{tabular}{c S[table-format=2.4] S[table-format=2.4] S[table-format=2.4] S[table-format=2.4] S[table-format=2.4] S[table-format=2.4] S[table-format=2.4]}	
			\toprule
			Parameter				& {CMB-S3}	& {BOSS}	& {eBOSS}	& {DESI}	& {Euclid}	& {DES}		& {LSST}	\\
			\midrule[0.065em]
			$\num{e5}\,\omega_b$	& 8.4		& 8.0		& 7.9		& 7.5		& 7.5		& 8.3		& 8.1		\\
			$\num{e4}\,\omega_c$	& 21		& 20		& 19		& 15		& 14		& 20		& 18		\\
			$\num{e7}\,\theta_s$	& 27		& 26		& 26		& 22		& 21		& 27		& 25		\\
			$\ns$					& 0.0040	& 0.0037	& 0.0036	& 0.0030	& 0.0029	& 0.0039	& 0.0037	\\
			$\Neff$					& 0.12		& 0.12		& 0.11		& 0.094		& 0.088		& 0.12		& 0.11		\\
			$Y_p$					& 0.0069	& 0.0068	& 0.0067	& 0.0060	& 0.0058	& 0.0069	& 0.0066	\\
			\bottomrule
		\end{tabular}\vspace{-2.5pt}
	}\\[2.76pt]
	\subcaptionbox{S4 + $P(k)$}{
		\begin{tabular}{c S[table-format=2.4] S[table-format=2.4] S[table-format=2.4] S[table-format=2.4] S[table-format=2.4] S[table-format=2.4] S[table-format=2.4]}	
			\toprule
			Parameter				& {CMB-S4}	& {BOSS}	& {eBOSS}	& {DESI}	& {Euclid}	& {DES}		& {LSST}	\\
			\midrule[0.065em]
			$\num{e5}\,\omega_b$	& 3.8		& 3.8		& 3.8		& 3.7		& 3.7		& 3.8		& 3.8		\\
			$\num{e4}\,\omega_c$	& 14		& 14		& 13		& 12		& 11		& 14		& 13		\\
			$\num{e7}\,\theta_s$	& 15		& 15		& 14		& 13		& 13		& 15		& 14		\\
			$\ns$					& 0.0030	& 0.0029	& 0.0028	& 0.0025	& 0.0024	& 0.0030	& 0.0029	\\
			$\Neff$					& 0.081		& 0.079		& 0.078		& 0.070		& 0.067		& 0.081		& 0.078		\\
			$Y_p$					& 0.0047	& 0.0046	& 0.0046	& 0.0043	& 0.0042	& 0.0047	& 0.0046	\\
			\bottomrule
		\end{tabular}\vspace{-2.5pt}
	}\\[2.76pt]
	\subcaptionbox{Planck + BAO}{
		\begin{tabular}{c S[table-format=2.4] S[table-format=2.4] S[table-format=2.4] S[table-format=2.4] S[table-format=2.4] S[table-format=2.4] S[table-format=2.4]}	
			\toprule
			Parameter				& {Planck}	& {BOSS}	& {eBOSS}	& {DESI}	& {Euclid}	& {DES}		& {LSST}	\\
			\midrule[0.065em]
			$\num{e5}\,\omega_b$	& 26		& 19		& 19		& 18		& 18		& 23		& 20		\\
			$\num{e4}\,\omega_c$	& 49		& 49		& 49		& 48		& 48		& 49		& 49		\\
			$\num{e7}\,\theta_s$	& 89		& 87		& 87		& 87		& 87		& 88		& 88		\\
			$\ns$					& 0.0093	& 0.0066	& 0.0065	& 0.0060	& 0.0061	& 0.0081	& 0.0068	\\
			$\Neff$					& 0.32		& 0.29		& 0.29		& 0.28		& 0.28		& 0.30		& 0.29		\\
			$Y_p$					& 0.018		& 0.018		& 0.018		& 0.018		& 0.018		& 0.018		& 0.018		\\
			\bottomrule
		\end{tabular}\vspace{-2.5pt}
	}\\[2.76pt]
	\subcaptionbox{CMB-S3 + BAO}{
		\begin{tabular}{c S[table-format=2.4] S[table-format=2.4] S[table-format=2.4] S[table-format=2.4] S[table-format=2.4] S[table-format=2.4] S[table-format=2.4]}	
			\toprule
			Parameter				& {CMB-S3}	& {BOSS}	& {eBOSS}	& {DESI}	& {Euclid}	& {DES}		& {LSST}	\\
			\midrule[0.065em]
			$\num{e5}\,\omega_b$	& 8.4		& 7.9		& 7.9		& 7.6		& 7.7		& 8.3		& 8.0		\\
			$\num{e4}\,\omega_c$	& 21		& 21		& 21		& 21		& 21		& 21		& 21		\\
			$\num{e7}\,\theta_s$	& 27		& 27		& 27		& 27		& 27		& 27		& 27		\\
			$\ns$					& 0.0040	& 0.0035	& 0.0035	& 0.0032	& 0.0032	& 0.0039	& 0.0036	\\
			$\Neff$					& 0.12		& 0.12		& 0.12		& 0.12		& 0.12		& 0.12		& 0.12		\\
			$Y_p$					& 0.0069	& 0.0069	& 0.0069	& 0.0069	& 0.0069	& 0.0069	& 0.0069	\\
			\bottomrule
		\end{tabular}\vspace{-2.5pt}
	}\\[2.76pt]
	\subcaptionbox{S4 + BAO}{
		\begin{tabular}{c S[table-format=2.4] S[table-format=2.4] S[table-format=2.4] S[table-format=2.4] S[table-format=2.4] S[table-format=2.4] S[table-format=2.4]}	
			\toprule
			Parameter				& {CMB-S4}	& {BOSS}	& {eBOSS}	& {DESI}	& {Euclid}	& {DES}		& {LSST}	\\
			\midrule[0.065em]
			$\num{e5}\,\omega_b$	& 3.8		& 3.8		& 3.8		& 3.8		& 3.8		& 3.8		& 3.8		\\
			$\num{e4}\,\omega_c$	& 14		& 14		& 14		& 14		& 14		& 14		& 14		\\
			$\num{e7}\,\theta_s$	& 15		& 15		& 15		& 15		& 15		& 15		& 15		\\
			$\ns$					& 0.0030	& 0.0028	& 0.0028	& 0.0025	& 0.0026	& 0.0030	& 0.0028	\\
			$\Neff$					& 0.081		& 0.080		& 0.080		& 0.079		& 0.079		& 0.081		& 0.080		\\
			$Y_p$					& 0.0047	& 0.0047	& 0.0047	& 0.0046	& 0.0046	& 0.0047	& 0.0047	\\
			\bottomrule
		\end{tabular}\vspace{-2.5pt}
	}
	\caption{As in Table~\ref{tab:CMB+LSS_Yp_full}, but for an extended $\Lambda\mathrm{CDM}$+$\Neff$+$Y_p$ cosmology.}
	\label{tab:CMB+LSS_Neff+Yp_full}
\end{table}
As in~\textsection\ref{sec:constraints_planned}, the $P(k)$-forecasts use wavenumbers up to $\kmax=\SI{0.2}{\hPerMpc}$ and marginalize over the $b_{m\leq1}$-terms of~\eqref{eq:tildeAtildeB}. For the BAO-forecasts, we set $\kmax=\SI{0.5}{\hPerMpc}$ and marginalize over~$a_{n\leq4}$ and~$b_{m\leq3}$ in each redshift bin. As we marginalize over galaxy bias, our forecasts show no improvements beyond the~CMB for~$\ln(\num{e10}\As)$ and~$\tau$. We therefore do not include these two parameters in the following tables.\medskip

Apart from the improvements in the constraints on~$\Neff$ and~$Y_p$, which we already discussed in~\textsection\ref{sec:constraints_planned}, we see that mainly~$\omega_b$ and~$\omega_c$ benefit from combining the considered LSS~surveys with CMB~experiments. The sensitivities may be enhanced by factors of three~(two) and more compared to Planck~(CMB-S3). We note that the DESI~specifications of Table~\ref{tab:specsDESI} are slightly more optimistic overall than what was considered in~\cite{Aghamousa:2016zmz} resulting in roughly the same BAO-forecasts and up to about \SI{5}{\percent}~better $P(k)$-forecasts.\medskip

Comparing our forecasts with the ones obtained from the BAO~scale alone (combined with Planck), we see that the BOSS~analysis for $\Lambda\mathrm{CDM}$ is nearly optimal, but improvements on the constraints of more than~\SI{10}{\percent} can be achieved in extended cosmologies. For instance, the constraints on~$\omega_b$, $\ns$ and~$\Neff$ improve by~\SI{3}{\percent} or more, and~$\omega_c$ in $\Lambda\mathrm{CDM}$+$Y_p$ even by \SI{12}{\percent}. For~DESI, the obtained sensitivities can generally be increased by a larger amount, e.g.\ up to~\SI{15}{\percent} for~$\omega_b$ and~$\ns$ in $\Lambda\mathrm{CDM}$+$\Neff$, and for~$\omega_c$ in $\Lambda\mathrm{CDM}$+$Y_p$.

\section{Convergence and Stability Tests}
\label{app:convergence}
One of the motivations for including our full list of forecasts in Appendix~\ref{app:LSS} is to make the results reproducible. It is therefore also important that we explain how the numerical derivatives were computed in the Fisher matrix, including the employed step sizes. In this appendix, we provide this information and demonstrate that the step sizes are appropriate for the convergence and stability of our calculations.\medskip

The numerical derivatives in~\eqref{eq:galaxyFisherMatrix} and~\eqref{eq:baoFisherMatrix} are computed using a symmetric difference quotient or two-point stencil, $f'(\theta) = \left[f(\theta+h)-f(\theta-h)\right]\!/\!\left(2h\right)$, with fiducial parameter value~$\theta$ and absolute step size~$h$. For each parameter, we choose the step sizes given in Table~\ref{tab:parameterSpacings} resulting in relative step sizes, $h_\mathrm{rel} = h/\theta$, that are generally of order~$\mathcal{O}\!\left(\num{e-2}\right)$.\medskip

In Figures~\ref{fig:convergencePk}%
\begin{table}[h]
	\begin{tabular}{c S[table-format=1.4] S[table-format=1.1e-1]}
		\toprule
		Parameter						& {$h$}		& {$h_\mathrm{rel}$}	\\
		\midrule[0.065em]
		$\omega_b$						& 0.0008	& 3.6e-2				\\
		$\omega_c$						& 0.002		& 1.7e-2				\\
		$100\,\theta_s$					& 0.002		& 1.9e-3				\\
		$\ln(\num{e10}\As)$				& 0.05		& 1.6e-2				\\
		$\ns$							& 0.01		& 1.0e-2				\\
		$\tau$							& 0.02		& 3.0e-1				\\
		\midrule[0.065em]
		$\Neff$							& 0.08		& 2.6e-2				\\
		$Y_p$							& 0.005		& 2.0e-2				\\
		\bottomrule 
	\end{tabular}
	\caption{Absolute and relative step sizes,~$h$ and~$h_\mathrm{rel}$, used when computing the derivatives in the Fisher matrices.}
	\label{tab:parameterSpacings}
\end{table}
\begin{figure}[t!]
	\includegraphics{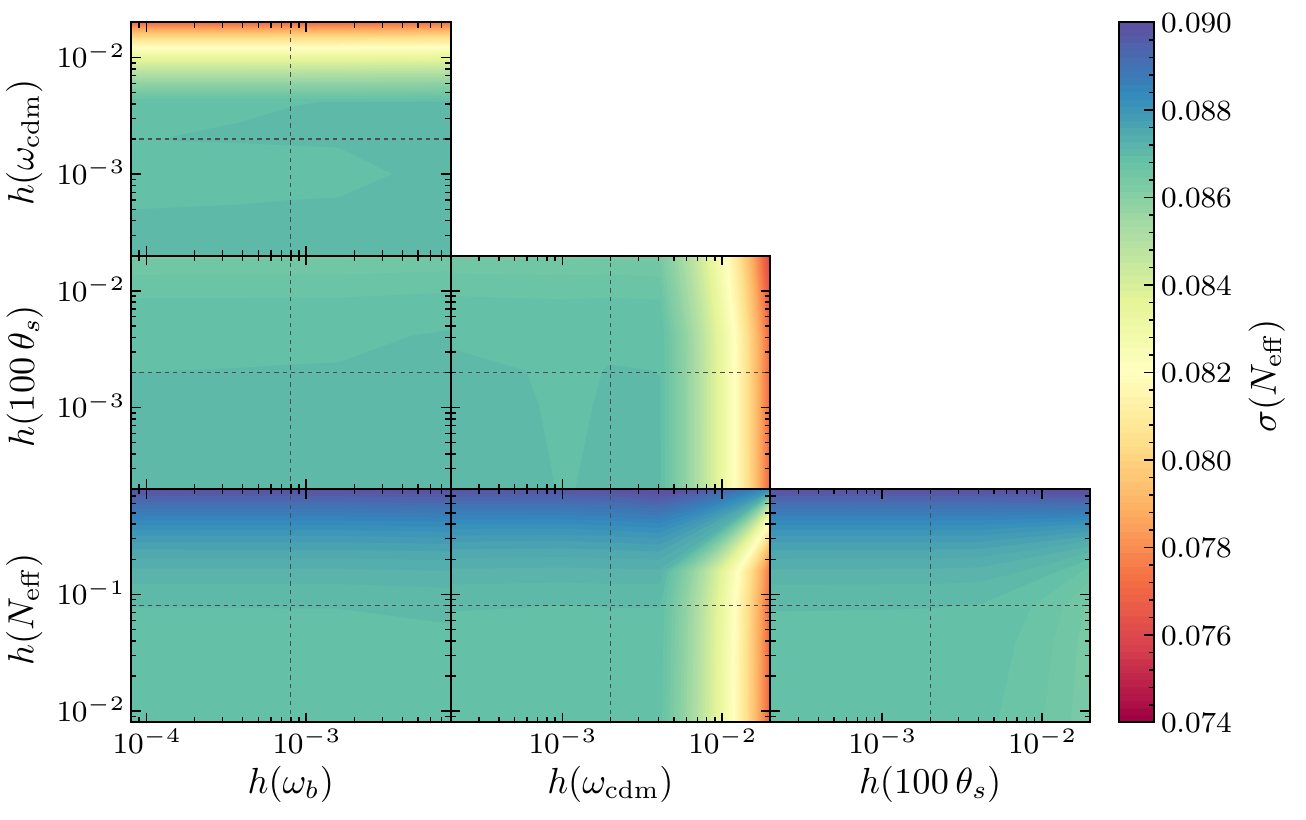}
	\caption{Results of the convergence test for the $P(k)$-forecasts of~DESI in the fiducial $\Lambda\mathrm{CDM}$+$\Neff$ cosmology. The spectra for the numerical derivatives were computed using a high-accuracy setting of~\texttt{CLASS}. The dashed lines indicate the step sizes employed in our forecasts.}
	\label{fig:convergencePk}
\end{figure}
and~\ref{fig:convergenceBAO},%
\begin{figure}[b!]
	\includegraphics{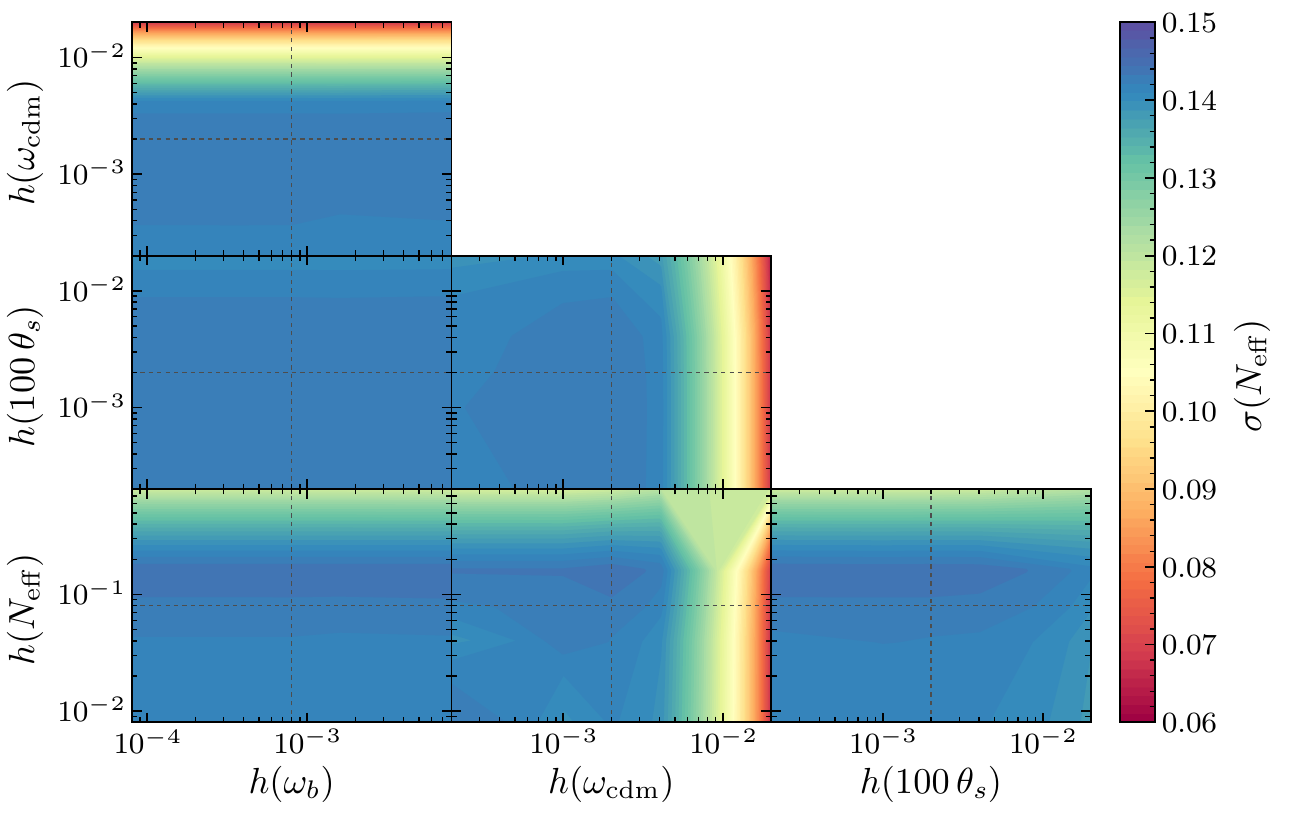}
	\caption{As in Figure~\ref{fig:convergencePk}, but for the BAO-forecasts of~DESI.}
	\label{fig:convergenceBAO}
\end{figure}
we show that our results are converged for both the $P(k)$- and BAO-forecasts. The results in these figures (as in the rest of our forecasts) use~\texttt{CLASS} with a high-accuracy setting. We have also checked that the forecasted constraints are converged when employing the standard accuracy setting, but note that the results are slightly less stable to changes away from these values. For the $P(k)$-forecasts, we see that a sufficiently small step size is needed, but a further decrease in the step size still leads to converged results. The BAO-forecasts, by contrast, show islands of convergence where performance decreases both when the step size is increased and when it is decreased. This behaviour is more noticeable using the standard accuracy setting of~\texttt{CLASS}, but likely reflects the fact that the BAO~feature is itself a small effect and small step sizes are therefore more likely to produce effects comparable to numeric or modelling errors.
		\addtocontents{toc}{\protect\setcounter{tocdepth}{1}}

\chapter{Broadband and Phase Extraction} 
\label{app:broadband+phaseShiftExtraction}
In this appendix, we provide supplemental material to Chapters~\ref{chap:bao-forecast} and~\ref{chap:bao-neutrinos}. We describe our implementation of a robust method to extract the matter broadband power spectrum~(\textsection\ref{app:broadbandExtraction}) and the computation of the phase shift template~(\textsection\ref{app:phaseMeasurement}).

\section{Broadband Extraction}
\label{app:broadbandExtraction}
The split of the matter power spectrum into a broadband (`no-wiggle') part and an oscillatory (`wiggle') part, $P(k) = \Pnw(k) + \Pw(k)$, is not uniquely defined, but depends on the method that is being used. In the following, we describe our procedure for extracting the broadband spectrum which is robust and stable over a very large parameter space.\medskip

Computationally, it is easier to identify a bump over a smooth background than to find the zeros of oscillations on top of a smooth background. This suggests that it is convenient to sine transform the matter power spectrum to discrete real space where the oscillations map to a localized bump. We then remove this bump and inverse transform back to Fourier space.\medskip

An algorithm for the discrete spectral method was outlined in~\textsection A.1 of~\cite{Hamann:2010pw}. Concretely, the relevant steps of our implementation are:
\begin{enumerate}
\item Provide~$P(k)$: Compute the theoretical matter power spectrum~$P(k)$ using \texttt{CLASS} for discrete wavenumbers~$k$ up to a chosen~$\kmax$ and log-log interpolate using cubic splines.

\item Sample~$\log[k P(k)]$: Sample~$\log[k P(k)]$ in $2^n$~points for an integer number~$n$. These points are chosen equidistant in~$k$.

\item Fast sine transform: Perform a fast sine transform of the $\log[k P(k)]$-array using the orthonormalized type-II sine transform. Denoting the index of the resulting array by~$i$, split the even and odd entries, i.e.\ those entries with even~$i$ and odd~$i$, into separate arrays.

\item Interpolate arrays: Linearly interpolate the two arrays separately using cubic splines.

\item Identify baryonic bumps: Compute the second derivative separately for the interpolated even and odd arrays, and average over next-neighbouring array entries to minimize noise. Choose $i_\mathrm{min} = i_* - 3$, where~$i_*$ is the array index of the first minimum of the second derivative. Set $i_\mathrm{max} = i^* + \Delta i$, where~$i^*$ is the array index of the second maximum of the second derivative, and $\Delta i = 10\text{ and }20$ for the even and odd array, respectively. These shifts were obtained empirically, but are found to give reliable and stable results for a large range of~$n$ and~$\kmax$.

\item Cut baryonic bumps: Having found the location of the bumps within $[i_\mathrm{min}, i_\mathrm{max}]$ for the even and odd arrays, respectively, remove the elements within this range from the arrays. Then, fill the gap by interpolating the arrays rescaled by a factor of~$(i+1)^2$ using cubic splines. This is analogous to interpolating~$r^2\,\xi(r)$ instead of the correlation function~$\xi(r)$ at separation~$r$.

\item Inverse fast sine transform: Merge the two arrays containing the respective elements without the bumps, and without the rescaling factor of $(i+1)^2$, and inversely fast sine transform. This leads to a discretized version of~$\log[k \Pnw(k)]$.

\item Provide~$\Pnw(k)$ and~$\Pw(k)$: In order to cut off numerical noise at low and high wavenumbers, perform two cuts at~$k_1$ and~$k_2$, where $k_1 = 3 \cdot 2^{-n}$ and the value of~$k_2$ is found as the trough of~$|P(k)-\Pnw(k)|/\Pnw(k)$ following the smallest maximum (before the oscillation amplitude increases again due to the numerical artefacts intrinsic to the procedure). The reliably extracted no-wiggle spectrum~$\Pnw(k)$ is then valid for $k \in [k_1,k_2]$. In practice, choose~$n$ and~$\kmax$ large enough initially so that~$k_{1,2}$ are outside the range of wavenumbers of interest, e.g.\ those covered by a survey. The wiggle spectrum in this range is then given by $\Pw(k) = P(k) - \Pnw(k)$.
\end{enumerate}

\noindent
Examples of the broadband extraction using this procedure are shown in Fig.~\ref{fig:broadbandExtraction}.%
\begin{figure}[t]
	\includegraphics{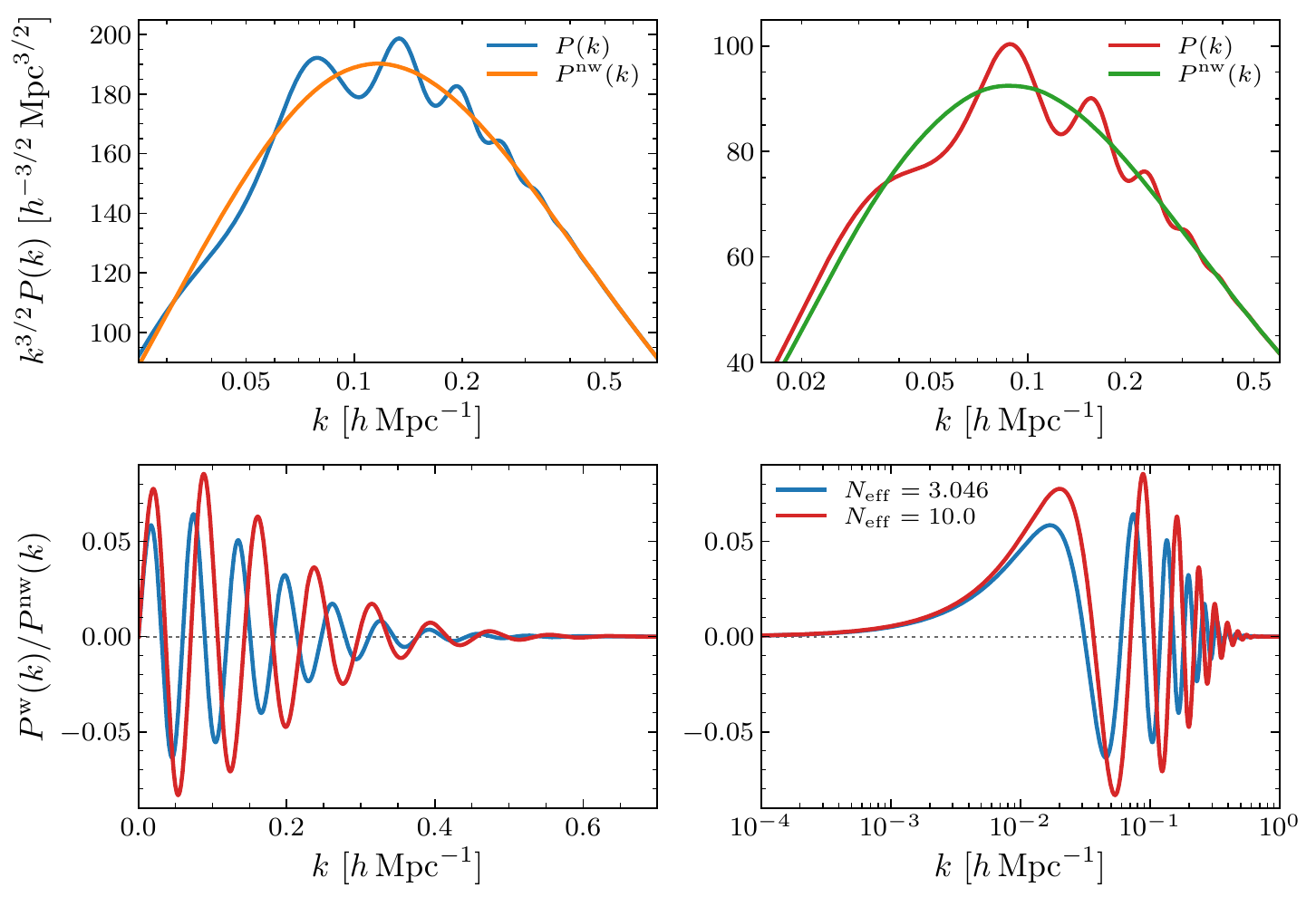}
	\caption{\textit{Top}:~Extracted broadband spectrum~$\Pnw(k)$ compared to the full power spectrum~$P(k)$ for $\Neff=3.046\text{ (\textit{left}) and }10$~(\textit{right}). The spectra are rescaled by~$k^{3/2}$ to exaggerate any oscillations. \textit{Bottom}:~Extracted BAO~spectrum~$\Pw(k)/\Pnw(k)$ for $\Neff=3.046\text{ and }10$ with linear~(\textit{left}) and logarithmic~(\textit{right}) $k$-axis.}
	\label{fig:broadbandExtraction}
\end{figure}
We see that the extraction method is unbiased, i.e.\ the resulting wiggle spectrum both oscillates around zero and asymptotes to zero for large wavenumbers. In addition, it is robust and stable over a large parameter space at small computation time (depending on~$n$). As the position of the first BAO~peak is close to the peak of the matter power spectrum, it is sensitive to how exactly the baryonic bump is removed. However, we have checked that the computed constraints on cosmological parameters are insensitive to this uncertainty. The same holds for varying the parameters~$n$ and~$\kmax$ with fixed shifts in step~5 as long as~$k_{1,2}$ are outside the range of wavenumbers of interest.

\section{Phase Shift Measurement}
\label{app:phaseMeasurement}
In the following, we describe our method for computing the phase shift template used in the likelihood forecasts of Section~\ref{sec:phaseShift} and in the BOSS DR12~data analysis of Chapter~\ref{chap:bao-neutrinos}.\medskip

First, we compute the BAO~spectrum using~\texttt{CLASS} and the broadband extraction method detailed above for a given value of~$\Neff$. In practice, we set the primordial helium fraction~$Y_p$ to the fiducial value, but the final template is independent of this choice. As discussed in~\textsection\ref{sec:template}, we keep the time of matter-radiation equality fixed at its fiducial value by changing the dark matter density~$\omega_c$ according to
\beq
\omega_c = \frac{a_\nu + \Neff}{a_\nu + N_\mathrm{eff}^\mathrm{fid}} \left(\omega_c^\mathrm{fid}+\omega_b^\mathrm{fid}\right) - \omega_b^\mathrm{fid}\, ,
\eeq
where~$a_\nu$ is defined in~\eqref{eq:anu}. We then fit the following envelope function to the maxima of the absolute value of the BAO~spectrum:
\beq
a(k) \equiv e(k) d(k)\, , \quad \text{where} \quad\hskip5pt	
			\begin{aligned}	
				e(k) &\equiv 1 - A_e \exp\left\{-a_e\, (k/k_e)^{\kappa_e}\right\} , \\
				d(k) &\equiv A_d \exp\left\{-a_d \, (k/k_d)^{\kappa_d}\right\} .
			\end{aligned}
\eeq
The parameters~$A_i$, $a_i$, $\kappa_i$, with $i=d,e$, are fitting parameters, while~$k_e$ is the location of the peak of~$\Pnw(k)$ and~$k_d$ is the wavenumber associated with the mean squared diffusion distance. These fitting functions are motivated by the modelling in~\cite{Follin:2015hya, Baldauf:2015xfa}. We define the `undamped spectrum' as 
\beq
\mathcal{O}(k) \equiv a(k)^{-1} \Pw(k)/\Pnw(k)\, .
\eeq
For the fiducial cosmology, for instance, we find the following parameters: $A_e\approx0.141$, $a_e\approx0.0035$, $\kappa_e\approx5.5$, $k_e\approx\SI{0.016}{\hPerMpc}$, and $A_d\approx0.072$, $a_d\approx0.32$, $\kappa_d\approx1.9$, $k_d\approx\SI{0.12}{\hPerMpc}$.\medskip

Before we can measure the phase shift, we have to match the sound horizon at the drag epoch,~$r_s$, to that in the fiducial cosmology to remove the change to the BAO frequency induced by~$\Neff$. By rescaling the wavenumbers as $k \to r_s^\mathrm{fid}/r_s\,k$, we fix~$r_s k$ to the fiducial model for all wavenumbers~$k$. For convenience, we also normalize the spectrum such that the amplitude of the fourth peak is the same as in the fiducial cosmology.\medskip

Finally, we can extract the phase shift as the shift of the peaks/troughs and zeros of~$\mathcal{O}(k)$ relative to the fiducial cosmology, $\delta k_* = k_* - k_*^\mathrm{fid}$. To obtain the template~$f(k)$, we sample 100~cosmologies with varying $\Neff \in [0,3.33]$,\footnote{We restrict to this range of values of~$\Neff$ as we observed a small, but unexpected jump in the peak locations around $\Neff\sim3.33$. Below and above, the peak locations change coherently with~$\Neff$. This range was then chosen as we are mostly interested in smaller~$\Neff$. However, we expect the template to also be valid for larger~$\Neff$ outside the sampling range.} and define
\beq
f(k) \equiv \left\langle \frac{1}{1-\beta(\Neff)}\, \frac{\delta k_*(k;\Neff)}{r_s^\mathrm{fid}} \right\rangle_{\!\Neff} \, ,
\eeq
where~$\beta(\Neff)$ is the normalization introduced in~\eqref{eq:phi_norm}. The bars in Fig.~\ref{fig:phaseShiftTemplate} indicate the locations of the peaks/troughs/zeros of the fiducial spectrum~$\mathcal{O}(k)$ and their length shows the standard deviation in these measurements which is generally small.
	\end{myappendix}
	\addtocontents{toc}{\protect\setcounter{tocdepth}{0}}

\cleardoublepageusingstyle{plain}
\printbibliography[heading=bibintoc, title={References}]
\end{document}